BeginDviLaserDoc
CLRP 300 3600 RES
1000 BP 39600 30600 PM 0 0 XY
XP /F48 /cmbx12 300 598 597.758 128 [-3 -12 55 36] PXLNF RP
XP /F48 83 374 3 0 26 33 34 24 0
<03FE0C 0FFF9C 1FFFFC 3F03FC 7C00FC 7C007C F8003C F8003C F8001C
 FC001C FE0000 FF0000 FFF000 7FFF00 7FFFE0 3FFFF0 1FFFF8 0FFFFC
 03FFFE 007FFE 0003FF 0000FF 00003F 00003F E0001F E0001F E0001F
 F0001F F8003E FC007E FF80FC FFFFF8 E7FFF0 C0FFC0>
PXLC RP
4223 5813 XY F48(S)S
XP /F48 79 505 3 0 37 33 34 40 0
<0007FC0000 003FFF8000 00FE0FE000 03F803F800 07E000FC00 0FC0007E00
 1FC0007F00 1F80003F00 3F80003F80 3F80003F80 7F00001FC0 7F00001FC0
 7F00001FC0 FF00001FE0 FF00001FE0 FF00001FE0 FF00001FE0 FF00001FE0
 FF00001FE0 FF00001FE0 FF00001FE0 FF00001FE0 7F00001FC0 7F80003FC0
 7F80003FC0 3F80003F80 3FC0007F80 1FC0007F00 0FE000FE00 07F001FC00
 03F803F800 00FE0FE000 003FFF8000 0007FC0000>
PXLC RP
4597 5813 XY F48(O)S
XP /F48 76 404 2 0 30 33 34 32 0
<FFFFE000 FFFFE000 FFFFE000 07F00000 07F00000 07F00000 07F00000
 07F00000 07F00000 07F00000 07F00000 07F00000 07F00000 07F00000
 07F00000 07F00000 07F00000 07F00000 07F00000 07F00000 07F00038
 07F00038 07F00038 07F00038 07F00078 07F00078 07F00078 07F000F0
 07F001F0 07F003F0 07F00FF0 FFFFFFF0 FFFFFFF0 FFFFFFF0>
PXLC RP
5102 5813 XY F48(L)S
XP /F48 65 508 2 0 38 33 34 40 0
<0000700000 0000F80000 0000F80000 0000F80000 0001FC0000 0001FC0000
 0003FE0000 0003FE0000 0003FE0000 00077F0000 00077F0000 000F7F8000
 000E3F8000 000E3F8000 001E3FC000 001C1FC000 001C1FC000 00380FE000
 00380FE000 00780FF000 007007F000 007007F000 00FFFFF800 00FFFFF800
 01FFFFFC00 01C001FC00 01C001FC00 03C001FE00 038000FE00 038000FE00
 0700007F00 FFF00FFFF8 FFF00FFFF8 FFF00FFFF8>
PXLC RP
5506 5813 XY F48(A)S
XP /F48 82 502 2 0 40 33 34 40 0
<FFFFFE0000 FFFFFFC000 FFFFFFF000 07F007F800 07F001FC00 07F001FE00
 07F000FE00 07F000FF00 07F000FF00 07F000FF00 07F000FF00 07F000FF00
 07F000FE00 07F001FE00 07F001FC00 07F007F000 07FFFFC000 07FFFF0000
 07F01FC000 07F00FE000 07F007F000 07F003F000 07F003F800 07F003F800
 07F003F800 07F003FC00 07F003FC00 07F003FC00 07F003FC00 07F003FE0E
 07F003FE0E FFFF81FF1E FFFF80FFFC FFFF801FF8>
PXLC RP
6014 5813 XY F48(R)S
XP /F48 78 526 2 0 40 33 34 40 0
<FFF8003FFE FFFC003FFE FFFE003FFE 07FF0001C0 07FF0001C0 077F8001C0
 077FC001C0 073FE001C0 071FE001C0 070FF001C0 070FF801C0 0707FC01C0
 0703FE01C0 0701FE01C0 0700FF01C0 0700FF81C0 07007FC1C0 07003FE1C0
 07001FE1C0 07000FF1C0 07000FF9C0 070007FDC0 070003FDC0 070001FFC0
 070000FFC0 070000FFC0 0700007FC0 0700003FC0 0700001FC0 0700001FC0
 0700000FC0 FFF80007C0 FFF80003C0 FFF80001C0>
PXLC RP
6740 5813 XY F48(N)S
XP /F48 69 441 2 0 33 33 34 32 0
<FFFFFFFC FFFFFFFC FFFFFFFC 07F001FC 07F0007C 07F0003C 07F0001C
 07F0001E 07F0001E 07F01C0E 07F01C0E 07F01C0E 07F01C0E 07F03C00
 07F07C00 07FFFC00 07FFFC00 07FFFC00 07F07C00 07F03C00 07F01C07
 07F01C07 07F01C07 07F01C0E 07F0000E 07F0000E 07F0001E 07F0001E
 07F0003E 07F0007E 07F001FC FFFFFFFC FFFFFFFC FFFFFFFC>
PXLC RP
7265 5813 XY F48(E)S
XP /F48 85 517 2 0 39 33 34 40 0
<FFFF807FFC FFFF807FFC FFFF807FFC 07F0000380 07F0000380 07F0000380
 07F0000380 07F0000380 07F0000380 07F0000380 07F0000380 07F0000380
 07F0000380 07F0000380 07F0000380 07F0000380 07F0000380 07F0000380
 07F0000380 07F0000380 07F0000380 07F0000380 07F0000380 07F0000380
 07F0000380 07F0000780 03F0000780 03F8000F00 01FC000F00 01FE003E00
 00FF80FC00 003FFFF800 001FFFE000 0001FF8000>
PXLC RP
7707 5813 XY F48(U)S
XP /F48 84 468 2 0 35 33 34 40 0
<7FFFFFFF80 7FFFFFFF80 7FFFFFFF80 7E03F81F80 7C03F80780 7803F80780
 7003F80380 F003F803C0 F003F803C0 E003F801C0 E003F801C0 E003F801C0
 E003F801C0 0003F80000 0003F80000 0003F80000 0003F80000 0003F80000
 0003F80000 0003F80000 0003F80000 0003F80000 0003F80000 0003F80000
 0003F80000 0003F80000 0003F80000 0003F80000 0003F80000 0003F80000
 0003F80000 03FFFFF800 03FFFFF800 03FFFFF800>
PXLC RP
8224 5813 XY F48(TR)S
XP /F48 73 250 1 0 19 33 34 24 0
<FFFFE0 FFFFE0 FFFFE0 03F800 03F800 03F800 03F800 03F800 03F800
 03F800 03F800 03F800 03F800 03F800 03F800 03F800 03F800 03F800
 03F800 03F800 03F800 03F800 03F800 03F800 03F800 03F800 03F800
 03F800 03F800 03F800 03F800 FFFFE0 FFFFE0 FFFFE0>
PXLC RP
9193 5813 XY F48(INOS)S
XP /F48 58 187 4 0 11 21 22 8 0
<3C 7E FF FF FF FF 7E 3C 00 00 00 00 00 00 3C 7E FF FF FF FF 7E 3C>
PXLC RP
10848 5813 XY F48(:)S
XP /F48 87 695 1 0 55 33 34 56 0
<FFFF07FFF83FFE FFFF07FFF83FFE FFFF07FFF83FFE 07F0003F8001C0
 07F0003F8001C0 07F8003F8003C0 03F8007FC00380 03F8007FC00380
 03FC00FFE00780 01FC00EFE00700 01FC00EFE00700 00FE01EFF00E00
 00FE01C7F00E00 00FE01C7F00E00 007F03C7F81C00 007F0383F81C00
 007F8383F83C00 003F8783FC3800 003F8701FC3800 003FC701FC7800
 001FCE00FE7000 001FCE00FE7000 001FEE00FEF000 000FFC007FE000
 000FFC007FE000 000FFC007FE000 0007F8003FC000 0007F8003FC000
 0003F8003F8000 0003F0001F8000 0003F0001F8000 0001F0001F0000
 0001E0000F0000 0001E0000F0000>
PXLC RP
11259 5813 XY F48(W)S
XP /F48 72 526 2 0 40 33 34 40 0
<FFFF83FFFE FFFF83FFFE FFFF83FFFE 07F0001FC0 07F0001FC0 07F0001FC0
 07F0001FC0 07F0001FC0 07F0001FC0 07F0001FC0 07F0001FC0 07F0001FC0
 07F0001FC0 07F0001FC0 07F0001FC0 07FFFFFFC0 07FFFFFFC0 07FFFFFFC0
 07F0001FC0 07F0001FC0 07F0001FC0 07F0001FC0 07F0001FC0 07F0001FC0
 07F0001FC0 07F0001FC0 07F0001FC0 07F0001FC0 07F0001FC0 07F0001FC0
 07F0001FC0 FFFF83FFFE FFFF83FFFE FFFF83FFFE>
PXLC RP
11953 5813 XY F48(HERE)S 225 x(WE)S 224 x(ARE)S
XP /F48 44 187 4 -10 12 7 18 16 0
<3C00 7F00 FF00 FF80 FF80 FF80 7F80 3D80 0180 0380 0380 0300 0700
 0E00 0E00 1C00 3800 3000>
PXLC RP
16899 5813 XY F48(,)S 224 x(WHERE)S 224 x(WE)S 224 x(ARE)S
XP /F48 71 529 3 0 40 33 34 40 0
<0003FF00C0 003FFFE1C0 00FFFFF7C0 01FF80FFC0 07FC003FC0 0FF8001FC0
 1FE0000FC0 1FC00007C0 3FC00003C0 7F800003C0 7F800003C0 7F800001C0
 FF000001C0 FF00000000 FF00000000 FF00000000 FF00000000 FF00000000
 FF00000000 FF00000000 FF000FFFFC FF000FFFFC 7F800FFFFC 7F80001FC0
 7F80001FC0 3FC0001FC0 1FC0001FC0 1FE0001FC0 0FF8001FC0 07FC003FC0
 01FF80FFC0 00FFFFFFC0 003FFFF3C0 0003FF80C0>
PXLC RP
23174 5813 XY F48(GOING)S
XP /F43 /cmr12 300 598 597.758 128 [-3 -13 47 36] PXLNF RP
XP /F43 74 301 1 -1 21 33 35 24 0
<03FFF8 03FFF8 000F80 000F80 000F80 000F80 000F80 000F80 000F80
 000F80 000F80 000F80 000F80 000F80 000F80 000F80 000F80 000F80
 000F80 000F80 000F80 000F80 000F80 000F80 000F80 300F80 FC0F80
 FC0F80 FC0F80 FC1F80 F81F00 703F00 7C7E00 3FF800 0FE000>
PXLC RP
12113 6961 XY F43(J)S
XP /F43 79 455 3 -1 33 34 36 32 0
<000FE000 007FFC00 00F83E00 03E00F80 078003C0 0F0001E0 0F0001E0
 1E0000F0 3E0000F8 3C000078 7C00007C 7C00007C 7800003C 7800003C
 F800003E F800003E F800003E F800003E F800003E F800003E F800003E
 F800003E F800003E 7C00007C 7C00007C 7C00007C 3C000078 3E0000F8
 1E0000F0 0F0001E0 0F8003E0 07C007C0 03E00F80 00F83E00 007FFC00
 000FE000>
PXLC RP
12413 6961 XY F43(O)S
XP /F43 72 439 2 0 33 33 34 32 0
<FFFE7FFF FFFE7FFF 07C003E0 07C003E0 07C003E0 07C003E0 07C003E0
 07C003E0 07C003E0 07C003E0 07C003E0 07C003E0 07C003E0 07C003E0
 07C003E0 07FFFFE0 07FFFFE0 07C003E0 07C003E0 07C003E0 07C003E0
 07C003E0 07C003E0 07C003E0 07C003E0 07C003E0 07C003E0 07C003E0
 07C003E0 07C003E0 07C003E0 07C003E0 FFFE7FFF FFFE7FFF>
PXLC RP
12869 6961 XY F43(H)S
XP /F43 78 439 2 0 33 33 34 32 0
<FFE00FFF FFE00FFF 07F001F8 07F00060 07F80060 06FC0060 06FC0060
 067E0060 063E0060 063F0060 061F8060 061F8060 060FC060 0607C060
 0607E060 0603F060 0603F060 0601F860 0600F860 0600FC60 06007E60
 06007E60 06003F60 06001FE0 06001FE0 06000FE0 06000FE0 060007E0
 060003E0 060003E0 060001E0 1F8001E0 FFF000E0 FFF00060>
PXLC RP
13307 6961 XY F43(N)S 195 x(N)S
XP /F43 46 163 4 0 8 4 5 8 0
<70 F8 F8 F8 70>
PXLC RP
14380 6961 XY F43(.)S
XP /F43 66 414 2 0 30 33 34 32 0
<FFFFF800 FFFFFF00 07C00F80 07C007C0 07C003E0 07C001E0 07C001F0
 07C001F0 07C001F0 07C001F0 07C001F0 07C001E0 07C003E0 07C007C0
 07C00F80 07FFFF00 07FFFE00 07C00F80 07C007C0 07C003E0 07C001F0
 07C000F0 07C000F8 07C000F8 07C000F8 07C000F8 07C000F8 07C000F8
 07C001F0 07C001F0 07C003E0 07C00FC0 FFFFFF00 FFFFFC00>
PXLC RP
14738 6961 XY F43(B)S
XP /F43 65 439 2 0 33 34 35 32 0
<00018000 0003C000 0003C000 0003C000 0007E000 0007E000 0007E000
 000FF000 000DF000 000DF000 001DF800 0018F800 0018F800 0038FC00
 00307C00 00307C00 00607E00 00603E00 00603E00 00C03F00 00C01F00
 00C01F00 01801F80 01FFFF80 01FFFF80 030007C0 030007C0 030007C0
 060003E0 060003E0 060003E0 0E0001F0 1F0003F0 FFC01FFF FFC01FFF>
PXLC RP
15152 6961 XY F43(AH)S
XP /F43 67 423 3 -1 30 34 36 32 0
<000FF030 007FFC30 00FC1E70 03F00770 07C003F0 0F8001F0 1F0001F0
 1F0000F0 3E0000F0 3C000070 7C000070 7C000070 7C000030 F8000030
 F8000030 F8000000 F8000000 F8000000 F8000000 F8000000 F8000000
 F8000000 F8000030 7C000030 7C000030 7C000030 3E000070 3E000060
 1F0000E0 1F0000C0 0F8001C0 07C00380 03F00700 00FC1E00 007FFC00
 000FF000>
PXLC RP
16030 6961 XY F43(CA)S
XP /F43 76 366 2 0 26 33 34 32 0
<FFFF0000 FFFF0000 07C00000 07C00000 07C00000 07C00000 07C00000
 07C00000 07C00000 07C00000 07C00000 07C00000 07C00000 07C00000
 07C00000 07C00000 07C00000 07C00000 07C00000 07C00000 07C00000
 07C00180 07C00180 07C00180 07C00180 07C00380 07C00380 07C00300
 07C00700 07C00700 07C00F00 07C03F00 FFFFFF00 FFFFFF00>
PXLC RP
16891 6961 XY F43(LL)S
XP /F46 /cmti12 300 598 597.758 128 [-2 -13 54 36] PXLNF RP
XP /F46 73 225 3 0 23 33 34 24 0
<00FFF8 00FFF8 000F00 000F00 001F00 001E00 001E00 001E00 003E00
 003C00 003C00 003C00 007C00 007800 007800 007800 00F800 00F000
 00F000 00F000 01F000 01E000 01E000 01E000 03E000 03C000 03C000
 03C000 07C000 078000 078000 0F8000 FFF800 FFF800>
PXLC RP
9472 7683 XY F46(I)S
XP /F46 110 329 3 0 27 20 21 32 0
<1F07E000 3F9FF800 33FC7800 73F03C00 63E03C00 63E03C00 E3C07C00
 07C07800 07807800 07807800 0780F800 0F80F000 0F00F000 0F01F180
 0F01E380 1F01E300 1E03E300 1E03C700 1E01CE00 3E01FC00 1C00F800>
PXLC RP
9697 7683 XY F46(n)S
XP /F46 115 239 3 0 19 20 21 24 0
<00FE00 03FF80 078380 070780 0E0780 0E0780 0F0300 0FC000 0FF800
 0FFE00 07FE00 03FF00 003F00 000F00 700F00 F00E00 F00E00 E01E00
 F07C00 7FF800 1FC000>
PXLC RP
10026 7683 XY F46(s)S
XP /F46 116 194 4 0 16 30 31 16 0
<00E0 01E0 01E0 01E0 03E0 03E0 03C0 03C0 07C0 07C0 FFF8 FFF8 0F80
 0F80 0F00 0F00 1F00 1F00 1E00 1E00 3E00 3E00 3C00 3C18 7C38 7C30
 7870 7860 7DE0 3FC0 1F00>
PXLC RP
10265 7683 XY F46(t)S
XP /F46 105 179 3 0 15 32 33 16 0
<0078 0078 0078 0070 0000 0000 0000 0000 0000 0000 0000 0000 0F80
 1FC0 39E0 31E0 71E0 63E0 E3C0 03C0 07C0 0780 0F80 0F00 0F00 1F18
 1E38 1E30 3E30 3C70 1CE0 1FC0 0F80>
PXLC RP
10459 7683 XY F46(it)S
XP /F46 117 314 3 0 26 20 21 24 0
<0F803C 1FC03C 39E07C 31E078 71E078 63E078 E3C0F8 03C0F0 07C0F0
 0780F0 0781F0 0F81E0 0F01E0 0F01E3 0F03E3 0F03C7 0F03C6 0F07CE
 0F9FEC 07FDFC 03F0F8>
PXLC RP
10833 7683 XY F46(ut)S
XP /F46 101 269 6 0 20 20 21 16 0
<01F8 07FE 0F8E 1E06 3C06 7806 780E F83C FFF8 FFC0 F000 F000 E000
 E000 E000 E006 E00E F01C 7878 3FF0 1FC0>
PXLC RP
11341 7683 XY F46(e)S
XP /F46 102 179 -2 -10 21 34 45 24 0
<00007E 0000FF 0001EF 0003CF 0003CE 0003C0 000780 000780 000780
 000780 000F80 000F00 000F00 000F00 01FFF8 01FFF8 001E00 001E00
 001E00 001E00 003E00 003C00 003C00 003C00 003C00 007C00 007800
 007800 007800 007800 00F800 00F000 00F000 00F000 00F000 01F000
 01E000 01E000 01E000 01C000 73C000 F38000 F78000 FF0000 7E0000>
PXLC RP
11819 7683 XY F46(f)S
XP /F46 111 299 5 0 23 20 21 24 0
<007E00 01FF80 07C7C0 0F83C0 1F01E0 1E01E0 3C01E0 3C01E0 7801E0
 7801E0 7803E0 F803C0 F003C0 F003C0 F00780 F00F80 700F00 781E00
 3C7C00 1FF000 0FC000>
PXLC RP
11998 7683 XY F46(o)S
XP /F46 114 247 3 0 22 20 21 24 0
<1F0FC0 3F9FE0 33FC70 73F070 63E0F0 63C0F0 E3C0E0 07C000 078000
 078000 078000 0F8000 0F0000 0F0000 0F0000 1F0000 1E0000 1E0000
 1E0000 3E0000 1C0000>
PXLC RP
12297 7683 XY F46(r)S
XP /F46 65 434 3 0 32 34 35 32 0
<00000300 00000700 00000700 00000F00 00000F80 00001F80 00001F80
 00003780 00007780 00006780 0000C780 0000C780 00018780 00018780
 00030780 00030780 00060780 000E0780 000C0780 00180780 00180780
 00300780 003FFFC0 007FFFC0 006003C0 00C003C0 01C003C0 018003C0
 030003C0 030003C0 060003C0 0E0003C0 1F0007C0 FFC03FFC FFC03FFC>
PXLC RP
12753 7683 XY F46(A)S
XP /F46 100 299 5 0 27 34 35 24 0
<00003E 0003FE 0003FC 00007C 00007C 00007C 000078 000078 0000F8
 0000F8 0000F0 0000F0 0001F0 0001F0 00FDE0 03FFE0 07CFE0 0F07E0
 1E03C0 1E03C0 3C07C0 7C07C0 780780 780780 780F80 F80F80 F00F00
 F00F18 F01F18 F01F38 F03E30 707E70 78FF60 3FCFE0 1F87C0>
PXLC RP
13157 7683 XY F46(d)S
XP /F46 118 269 3 0 22 20 21 24 0
<0F80E0 1FC1F0 39E1F0 31E1F0 71E0F0 63E070 E3C070 03C060 07C060
 078060 0780E0 0F80C0 0F00C0 0F01C0 0F0180 0F0180 0F0300 0F0700
 0F8E00 07FC00 01F800>
PXLC RP
13456 7683 XY F46(v)S
XP /F46 97 299 5 0 25 20 21 24 0
<00FDE0 03FFE0 07CFE0 0F07C0 1E03C0 1E03C0 3C07C0 7C0780 780780
 780780 780F80 F80F00 F00F00 F00F18 F01F18 F01E38 F03E30 707E70
 78FF60 3FCFE0 1F87C0>
PXLC RP
13725 7683 XY F46(an)S
XP /F46 99 269 5 0 21 20 21 24 0
<007E00 01FF80 07C380 0F0780 1E0780 1E0780 3C0000 3C0000 780000
 780000 780000 F80000 F00000 F00000 F00000 F00300 700700 780E00
 3C3C00 3FF800 0FE000>
PXLC RP
14353 7683 XY F46(c)S -30 x(e)S -30 x(d)S
XP /F46 83 329 3 -1 29 34 36 32 0
<0003F860 000FFCE0 001E1FC0 007807C0 007003C0 00E003C0 01E00380
 01C00380 01C00380 03C00380 03C00300 03C00000 03C00000 03E00000
 03F80000 01FF8000 01FFE000 00FFF800 003FFC00 0007FC00 00007C00
 00003E00 00001E00 00001E00 00001E00 30001C00 30001C00 30001C00
 30003C00 70003800 78007000 7800F000 7C01E000 EF07C000 E7FF0000
 C1FC0000>
PXLC RP
15339 7683 XY F46(Stud)S
XP /F46 121 284 3 -10 24 20 31 24 0
<0F803C 1FC03C 39E07C 31E078 71E078 63E078 E3C0F8 03C0F0 07C0F0
 0780F0 0781F0 0F81E0 0F01E0 0F01E0 0F03E0 0F03C0 0F03C0 0F07C0
 0F9FC0 07FF80 03F780 000780 000F80 000F00 3C0F00 3C1E00 3C3E00
 387C00 38F800 3FE000 1F8000>
PXLC RP
16475 7683 XY F46(y)S
XP /F46 44 179 3 -10 9 4 15 8 0
<0E 1E 1E 1E 1E 06 0E 0C 0C 1C 38 30 70 E0 C0>
PXLC RP
16759 7683 XY F46(,)S
XP /F46 79 448 7 -1 37 34 36 32 0
<0000FE00 0007FF80 000F03E0 003C01F0 007800F8 00F00078 01E0003C
 03C0003C 0780003C 0F00003E 1F00003E 1E00003E 3E00003E 3E00003E
 7C00003E 7C00003E 7C00003E F800007C F800007C F800007C F8000078
 F80000F8 F00000F8 F00001F0 F00001F0 F00003E0 F00003C0 F80007C0
 F8000F80 78000F00 7C001E00 3C007C00 1E00F800 0F83E000 07FF8000
 01FC0000>
PXLC RP
17148 7683 XY F46(O)S
XP /F46 108 149 4 0 15 34 35 16 0
<01F0 1FF0 1FE0 03E0 03E0 03E0 03C0 03C0 07C0 07C0 0780 0780 0F80
 0F80 0F00 0F00 1F00 1F00 1E00 1E00 3E00 3E00 3C00 3C00 7C00 7C00
 7800 7980 F980 FB80 F300 F300 FF00 7E00 3E00>
PXLC RP
17596 7683 XY F46(lden)S
XP /F46 76 367 3 0 29 33 34 32 0
<00FFFC00 00FFFC00 000F8000 000F0000 001F0000 001E0000 001E0000
 001E0000 003E0000 003C0000 003C0000 003C0000 007C0000 00780000
 00780000 00780000 00F80000 00F00000 00F00000 00F00000 01F00000
 01E00060 01E00060 01E000E0 03E000C0 03C001C0 03C001C0 03C00380
 07C00380 07800780 07801F00 0F807F00 FFFFFF00 FFFFFE00>
PXLC RP
18851 7683 XY F46(L)S -30 x(ane,)S
XP /F46 80 397 3 0 34 33 34 32 0
<00FFFFE0 00FFFFF8 000F007C 001F001E 001F001F 001E000F 001E000F
 003E000F 003E001F 003C001F 003C001F 007C001E 007C003E 0078003C
 00780078 00F800F0 00F803E0 00FFFF80 00FFFE00 01F00000 01F00000
 01E00000 01E00000 03E00000 03E00000 03C00000 03C00000 07C00000
 07C00000 07800000 07800000 0F800000 FFF80000 FFF80000>
PXLC RP
12210 8405 XY F46(Princ)S -30 x(eton,)S
XP /F46 78 434 3 0 40 33 34 40 0
<00FF801FFC 00FF801FFC 000F8007E0 000FC00380 001FC00380 001BC00300
 001BE00300 001BE00700 0039E00700 0031F00600 0031F00600 0030F00E00
 0070F80E00 0060F80C00 0060780C00 0060781C00 00E07C1C00 00C03C1800
 00C03C1800 00C03E3800 01C01E3800 01801E3000 01801F3000 01800F7000
 03800F7000 03000FE000 030007E000 030007E000 070007E000 060003C000
 0E0003C000 1F0003C000 FFE001C000 FFE0018000>
PXLC RP
15080 8405 XY F46(N)S
XP /F46 74 307 5 -1 29 33 35 32 0
<000FFF80 001FFF80 0000F800 0000F800 0000F000 0000F000 0000F000
 0001F000 0001E000 0001E000 0001E000 0003E000 0003C000 0003C000
 0003C000 0007C000 00078000 00078000 00078000 000F8000 000F8000
 000F0000 000F0000 001F0000 001F0000 001E0000 301E0000 703E0000
 F03C0000 F03C0000 C0780000 E0F80000 71F00000 7FC00000 1F000000>
PXLC RP
15515 8405 XY F46(J)S
XP /F46 48 299 6 -1 25 32 34 24 0
<000F80 003FC0 00F0E0 01C070 038070 038070 070070 0F0070 0E0070
 1E0070 1E0070 1C00F0 3C00F0 3C00F0 3C00F0 7801E0 7801E0 7801E0
 7801E0 F003C0 F003C0 F003C0 F00380 E00780 E00780 E00700 E00F00
 E00E00 E01C00 E01C00 F03800 78F000 3FE000 1F0000>
PXLC RP
16031 8405 XY F46(0)S
XP /F46 56 299 5 -1 26 32 34 24 0
<000FC0 003FF0 00F8F8 01E038 01C038 03803C 03803C 078038 070038
 078078 078070 07C0F0 07E1E0 03F7C0 03FF00 01FE00 01FF00 07FF80
 0F1FC0 1E0FC0 3C07C0 7803C0 7001C0 F001C0 E001C0 E001C0 E003C0
 E00380 E00780 E00F00 F01E00 787C00 3FF000 0FC000>
PXLC RP
16330 8405 XY F46(8)S
XP /F46 53 299 5 -1 26 32 34 24 0
<00C00C 00F03C 00FFF8 01FFF0 01FFE0 01FF00 018000 038000 030000
 030000 030000 070000 067F00 06FF80 07C3C0 0F81C0 0E01E0 0C01E0
 0001E0 0001E0 0001E0 0001E0 0003E0 7003E0 F003C0 F003C0 F00780
 C00780 C00F00 C01E00 E03C00 78F800 3FF000 1FC000>
PXLC RP
16628 8405 XY F46(5)S
XP /F46 52 299 2 -10 22 32 43 24 0
<000078 000078 0000F8 0000F0 0000F0 0000F0 0001F0 0001E0 0001E0
 0003E0 0003C0 0003C0 000780 000780 000700 000F00 000E00 001E00
 001C00 003C00 003800 0071C0 00F3E0 00E3C0 01C3C0 0383C0 0787C0
 0F0780 1E0780 3FC780 7FFF80 F07F30 C01FF0 000FE0 001F00 001E00
 001E00 001E00 003E00 003C00 003C00 003C00 003800>
PXLC RP
16927 8405 XY F46(40)S
XP /F43 97 293 2 0 22 20 21 24 0
<1FF000 7FFC00 7C3E00 7C1F00 7C0F00 380F00 000F00 000F00 01FF00
 0FCF00 3E0F00 7C0F00 780F00 F00F00 F00F18 F00F18 F01F18 F81F18
 7C7FB8 3FE7F0 1F83C0>
PXLC RP
14396 9127 XY F43(a)S
XP /F43 110 325 1 0 24 20 21 24 0
<0F1F80 FF3FE0 FF61E0 1FC1F0 0F80F0 0F80F0 0F00F0 0F00F0 0F00F0
 0F00F0 0F00F0 0F00F0 0F00F0 0F00F0 0F00F0 0F00F0 0F00F0 0F00F0
 0F00F0 FFF3FF FFF3FF>
PXLC RP
14689 9127 XY F43(n)S
XP /F43 100 325 2 0 24 34 35 24 0
<0001E0 001FE0 001FE0 0003E0 0001E0 0001E0 0001E0 0001E0 0001E0
 0001E0 0001E0 0001E0 0001E0 0001E0 03F9E0 0FFFE0 1F0FE0 3E03E0
 7C01E0 7801E0 7801E0 F001E0 F001E0 F001E0 F001E0 F001E0 F001E0
 F001E0 7001E0 7801E0 7803E0 3C07E0 1F0FF0 0FFDFE 03F1FE>
PXLC RP
15014 9127 XY F43(d)S 723 y 5225 X F46(Institut)S 2 x(e)S 209 x(for)S
209 x(Nucle)S -29 x(ar)S
XP /F46 84 418 9 0 37 33 34 32 0
<1FFFFFF8 1FFFFFF8 1F03C0F8 3C03C038 3807C038 30078038 70078030
 60078030 E00F8030 C00F0030 C00F0030 C00F0030 001F0000 001E0000
 001E0000 001E0000 003E0000 003C0000 003C0000 003C0000 007C0000
 00780000 00780000 00780000 00F80000 00F00000 00F00000 00F00000
 01F00000 01E00000 01E00000 03E00000 FFFF8000 FFFF0000>
PXLC RP
10667 9850 XY F46(T)S
XP /F46 104 299 3 0 25 34 35 24 0
<00F800 0FF800 0FF000 01F000 01F000 01F000 01E000 01E000 01E000
 03E000 03C000 03C000 03C000 07C000 079F80 07FFE0 07F1E0 0FC0F0
 0F80F0 0F80F0 0F01F0 1F01E0 1E01E0 1E01E0 1E03E0 3E03C0 3C03C0
 3C07C6 3C078E 7C078C 780F8C 780F1C 780738 F807F0 7003E0>
PXLC RP
11086 9850 XY F46(he)S -30 x(ory,)S
XP /F46 85 434 9 -1 40 33 35 32 0
<3FFE07FF 3FFE07FF 03C001F8 07C000E0 07C000E0 078000C0 078000C0
 0F8001C0 0F800180 0F000180 0F000180 1F000380 1F000300 1E000300
 1E000300 3E000700 3E000600 3C000600 3C000600 7C000E00 7C000C00
 78000C00 78000C00 78001C00 F8001800 F0001800 F0003800 F0003000
 70006000 7800E000 7801C000 3C038000 1E1F0000 0FFC0000 07F00000>
PXLC RP
12842 9850 XY F46(University)S 210 x(of)S
XP /F46 87 584 10 -1 54 33 35 48 0
<FFF03FF81FF8 FFE07FF83FF0 1F000FC007C0 1E000F800780 1E000F800700
 1E001F800600 1E001F800C00 1E0037800C00 1E0037801800 1E0067801800
 1E0067803000 1E00C7803000 1E00C7806000 1F0187806000 0F018780C000
 0F030780C000 0F0307818000 0F0607818000 0F0607830000 0F0C07830000
 0F0C07860000 0F1807860000 0F38078C0000 0F30079C0000 0F6007980000
 0F6007B80000 0FC007B00000 0FC007E00000 0F8007E00000 0F8007C00000
 0F0007C00000 0F0007800000 0E0007800000 0E0007000000 0C0007000000>
PXLC RP
16362 9850 XY F46(Washin)S
XP /F46 103 269 2 -10 22 20 31 24 0
<001F78 007FF8 00F3F8 01E1F8 03C0F0 0780F0 0781F0 0F81F0 0F01E0
 0F01E0 0F03E0 1F03E0 1E03C0 1E03C0 1E07C0 1E07C0 1E0F80 0E0F80
 0F3F80 07FF80 03EF00 000F00 001F00 001F00 001E00 001E00 703C00
 F07C00 F0F800 FFF000 7FC000>
PXLC RP
18291 9850 XY F46(gton,)S 209 x(Se)S -30 x(attle,)S 211 x(W)S -45 x
(A)S
XP /F46 57 299 5 -1 25 32 34 24 0
<000FC0 003FE0 0078F0 00F078 01E038 03C038 078038 078038 0F0038
 0F0038 0F0078 1F0078 1E0078 1E0078 1E00F8 1E00F0 1C01F0 1E03F0
 1E03F0 0F0FE0 0FFDE0 07F9E0 00C3C0 0003C0 000780 000780 000F00
 700E00 F01E00 F03C00 E07800 E1F000 FFC000 3F0000>
PXLC RP
23016 9850 XY F46(98)S
XP /F46 49 299 6 0 22 32 33 24 0
<000080 000180 000380 000780 001F00 003F00 03FF00 07DF00 001E00
 001E00 003E00 003E00 003C00 003C00 007C00 007C00 007800 007800
 00F800 00F800 00F000 00F000 01F000 01F000 01E000 01E000 03E000
 03E000 03C000 03C000 07C000 FFFC00 FFFC00>
PXLC RP
23613 9850 XY F46(195)S
XP /F34 /cmr10 329 546 545.454 128 [-3 -12 44 33] PXLNF RP
XP /F34 65 410 1 0 31 31 32 32 0
<00038000 00038000 00038000 0007C000 0007C000 0007C000 000FE000
 000FE000 000FE000 0019F000 0019F000 0019F000 0030F800 0030F800
 0030F800 0060FC00 00607C00 00607C00 00E07E00 00C03E00 00C03E00
 01FFFF00 01FFFF00 01801F00 03801F80 03000F80 03000F80 07000FC0
 070007C0 0F8007C0 FFE07FFE FFE07FFE>
PXLC RP
13330 11172 XY F34(A)S
XP /F34 66 387 1 0 27 30 31 32 0
<FFFFF000 FFFFFC00 07C01F00 07C00F00 07C00F80 07C007C0 07C007C0
 07C007C0 07C007C0 07C007C0 07C00780 07C00F80 07C01F00 07C03E00
 07FFFC00 07FFFE00 07C01F00 07C00780 07C007C0 07C003C0 07C003E0
 07C003E0 07C003E0 07C003E0 07C003E0 07C003C0 07C007C0 07C00F80
 07C01F00 FFFFFE00 FFFFF800>
PXLC RP
13739 11172 XY F34(B)S
XP /F34 83 303 3 -1 20 31 33 24 0
<07E180 1FF980 3E3F80 780F80 780780 F00380 F00380 F00180 F00180
 F00180 F80000 FC0000 FE0000 7FE000 7FFC00 3FFE00 1FFF00 0FFF80
 00FF80 001FC0 0007C0 0007C0 0003C0 C003C0 C003C0 C003C0 E003C0
 E00380 F00780 F80F00 FE1F00 CFFE00 C3F800>
PXLC RP
14125 11172 XY F34(S)S
XP /F34 84 395 2 0 29 30 31 32 0
<7FFFFFE0 7FFFFFE0 7C0F81E0 700F80E0 600F8060 600F8060 E00F8070
 C00F8030 C00F8030 C00F8030 C00F8030 000F8000 000F8000 000F8000
 000F8000 000F8000 000F8000 000F8000 000F8000 000F8000 000F8000
 000F8000 000F8000 000F8000 000F8000 000F8000 000F8000 000F8000
 000F8000 07FFFF00 07FFFF00>
PXLC RP
14428 11172 XY F34(T)S
XP /F34 82 403 1 -1 31 30 32 32 0
<FFFFC000 FFFFF800 07C07C00 07C01E00 07C01F00 07C00F00 07C00F80
 07C00F80 07C00F80 07C00F80 07C00F00 07C01F00 07C01E00 07C07C00
 07FFF800 07FFF000 07C0F800 07C07C00 07C07E00 07C03E00 07C03E00
 07C03E00 07C03E00 07C03E00 07C03E00 07C03E00 07C03E06 07C03E06
 07C01F0E FFFE1F8E FFFE0FFC 000003F8>
PXLC RP
14822 11172 XY F34(RA)S
XP /F34 67 395 3 -1 28 31 33 32 0
<001FC0C0 00FFF0C0 01F83DC0 07E00FC0 0F8007C0 0F0007C0 1F0003C0
 3E0001C0 3C0001C0 7C0001C0 7C0000C0 7C0000C0 F80000C0 F8000000
 F8000000 F8000000 F8000000 F8000000 F8000000 F8000000 F8000000
 7C0000C0 7C0000C0 7C0000C0 3C0001C0 3E000180 1F000180 0F000380
 0F800700 07E00E00 01F83C00 00FFF800 001FE000>
PXLC RP
15618 11172 XY F34(CT)S 12222 Y 6165 X(T)S
XP /F34 104 303 0 0 23 31 32 24 0
<0F0000 FF0000 FF0000 0F0000 0F0000 0F0000 0F0000 0F0000 0F0000
 0F0000 0F0000 0F0000 0F1F80 0F3FE0 0F61E0 0FC1F0 0F80F0 0F80F0
 0F00F0 0F00F0 0F00F0 0F00F0 0F00F0 0F00F0 0F00F0 0F00F0 0F00F0
 0F00F0 0F00F0 0F00F0 FFF3FF FFF3FF>
PXLC RP
6559 12222 XY F34(h)S
XP /F34 101 242 1 0 17 19 20 24 0
<03F000 0FFC00 1E1E00 3C0F00 780700 780780 F00780 F00380 FFFF80
 FFFF80 F00000 F00000 F00000 F80000 780000 7C0180 3E0380 1F8700
 0FFE00 03F800>
PXLC RP
6862 12222 XY F34(e)S
XP /F34 114 214 0 0 15 19 20 16 0
<0F7C FFFF FFDF 0F9F 0F9F 0F0E 0F00 0F00 0F00 0F00 0F00 0F00 0F00
 0F00 0F00 0F00 0F00 0F00 FFF0 FFF0>
PXLC RP
7104 12222 XY F34(re)S
XP /F34 97 273 2 0 21 19 20 24 0
<1FF000 7FFC00 7C7E00 7C1E00 7C1F00 380F00 000F00 000F00 03FF00
 0FCF00 3E0F00 7C0F00 F80F00 F00F30 F00F30 F01F30 F81F30 7C7FF0
 3FE7E0 1F83C0>
PXLC RP
7713 12222 XY F34(are)S
XP /F34 102 167 0 0 16 31 32 24 0
<007E00 01FF00 03EF80 07CF80 0F8F80 0F0F80 0F0000 0F0000 0F0000
 0F0000 0F0000 0F0000 FFF000 FFF000 0F0000 0F0000 0F0000 0F0000
 0F0000 0F0000 0F0000 0F0000 0F0000 0F0000 0F0000 0F0000 0F0000
 0F0000 0F0000 0F0000 FFF000 FFF000>
PXLC RP
8595 12222 XY F34(f)S
XP /F34 111 273 1 0 20 19 20 24 0
<01F800 07FE00 1E0780 3C03C0 3801C0 7000E0 7000E0 F000F0 F000F0
 F000F0 F000F0 F000F0 F000F0 7000E0 7801E0 3801C0 3C03C0 1E0780
 07FE00 01F800>
PXLC RP
8762 12222 XY F34(o)S
XP /F34 117 303 0 0 23 19 20 24 0
<0F00F0 FF0FF0 FF0FF0 0F00F0 0F00F0 0F00F0 0F00F0 0F00F0 0F00F0
 0F00F0 0F00F0 0F00F0 0F00F0 0F00F0 0F00F0 0F01F0 0F03F0 0F87F0
 07FEFF 01F8FF>
PXLC RP
9034 12222 XY F34(ur)S
XP /F34 105 152 0 0 10 30 31 16 0
<0E00 1F00 1F00 1F00 0E00 0000 0000 0000 0000 0000 0000 0F00 7F00
 7F00 0F00 0F00 0F00 0F00 0F00 0F00 0F00 0F00 0F00 0F00 0F00 0F00
 0F00 0F00 0F00 FFE0 FFE0>
PXLC RP
9704 12222 XY F34(i)S
XP /F34 109 456 0 0 37 19 20 40 0
<0F1FC0FE00 FF3FF1FF80 FF71F38F80 0FC0FE07C0 0F807C03C0 0F807C03C0
 0F007803C0 0F007803C0 0F007803C0 0F007803C0 0F007803C0 0F007803C0
 0F007803C0 0F007803C0 0F007803C0 0F007803C0 0F007803C0 0F007803C0
 FFF3FF9FFC FFF3FF9FFC>
PXLC RP
9855 12222 XY F34(m)S
XP /F34 112 303 0 -9 21 19 29 24 0
<0F1F80 FFFFE0 FFE1F0 0F80F0 0F00F8 0F0078 0F007C 0F003C 0F003C
 0F003C 0F003C 0F003C 0F003C 0F007C 0F0078 0F80F8 0F80F0 0FE3E0
 0F7FC0 0F3F80 0F0000 0F0000 0F0000 0F0000 0F0000 0F0000 0F0000
 FFF000 FFF000>
PXLC RP
10310 12222 XY F34(p)S 15 x(or)S
XP /F34 116 212 1 0 13 27 28 16 0
<0600 0600 0600 0600 0E00 0E00 1E00 3E00 FFF8 FFF8 1E00 1E00 1E00
 1E00 1E00 1E00 1E00 1E00 1E00 1E00 1E18 1E18 1E18 1E18 1E18 1F30
 0FF0 07E0>
PXLC RP
11114 12222 XY F34(ta)S
XP /F34 110 303 0 0 23 19 20 24 0
<0F1F80 FF3FE0 FF61E0 0FC1F0 0F80F0 0F80F0 0F00F0 0F00F0 0F00F0
 0F00F0 0F00F0 0F00F0 0F00F0 0F00F0 0F00F0 0F00F0 0F00F0 0F00F0
 FFF3FF FFF3FF>
PXLC RP
11599 12222 XY F34(n)S -15 x(t)S 153 x(fa)S
XP /F34 99 242 2 0 17 19 20 16 0
<03F8 0FFE 1E3E 3C3E 783E 781C F000 F000 F000 F000 F000 F000 F000
 F800 7800 7C03 3E07 1F0E 0FFC 03F0>
PXLC RP
12692 12222 XY F34(ct)S
XP /F34 115 215 2 0 14 19 20 16 0
<1FB0 7FF0 70F0 E070 E030 E030 F000 FF00 7FC0 7FF0 3FF0 07F8 00F8
 C078 C038 E038 F078 F8F0 FFE0 CFC0>
PXLC RP
13146 12222 XY F34(s)S 153 x(a)S
XP /F34 98 303 0 0 21 31 32 24 0
<0F0000 FF0000 FF0000 0F0000 0F0000 0F0000 0F0000 0F0000 0F0000
 0F0000 0F0000 0F0000 0F1F80 0FFFE0 0FE1F0 0F80F0 0F0078 0F0078
 0F003C 0F003C 0F003C 0F003C 0F003C 0F003C 0F003C 0F003C 0F0078
 0F8078 0F80F0 0FE1E0 0E7FC0 0C3F80>
PXLC RP
13787 12222 XY F34(b)S 15 x(out)S 153 x(so)S
XP /F34 108 152 0 0 11 31 32 16 0
<0F00 FF00 FF00 0F00 0F00 0F00 0F00 0F00 0F00 0F00 0F00 0F00 0F00
 0F00 0F00 0F00 0F00 0F00 0F00 0F00 0F00 0F00 0F00 0F00 0F00 0F00
 0F00 0F00 0F00 0F00 FFF0 FFF0>
PXLC RP
15534 12222 XY F34(lar)S 151 x(neutrinos)S
XP /F34 46 152 4 0 8 4 5 8 0
<70 F8 F8 F8 70>
PXLC RP
18541 12222 XY F34(.)S 152 x(The)S
XP /F34 121 288 1 -9 21 19 29 24 0
<FFC7F8 FFC7F8 1F03E0 0F01C0 0F0180 0F0180 078300 078300 078300
 03C600 03C600 03EE00 01EC00 01EC00 01FC00 00F800 00F800 007000
 007000 007000 006000 006000 006000 70C000 F8C000 C18000 E38000
 7F0000 3C0000>
PXLC RP
19785 12222 XY F34(y)S 153 x(are)S 152 x(liste)S
XP /F34 100 303 2 0 23 31 32 24 0
<0003C0 003FC0 003FC0 0003C0 0003C0 0003C0 0003C0 0003C0 0003C0
 0003C0 0003C0 0003C0 07F3C0 0FFFC0 1E1FC0 3C07C0 7803C0 7803C0
 F003C0 F003C0 F003C0 F003C0 F003C0 F003C0 F003C0 F003C0 7803C0
 7807C0 3C0FC0 3E1FC0 1FFBFC 07E3FC>
PXLC RP
22080 12222 XY F34(d)S 153 x(in)S 152 x(order)S 677 y 5347 X(of)S 
188 x(imp)S 14 x(ortance)S 190 x(in)S 188 x(this)S 189 x(abstract)S
189 x(and)S 189 x(discussed)S 190 x(more)S 188 x(in)S 189 x(the)S 
189 x(te)S
XP /F34 120 288 0 0 22 19 20 24 0
<7FE7FC 7FE7FC 0F83E0 078380 03C300 01E600 01EE00 00FC00 007800
 007800 003C00 007E00 00FE00 01CF00 018780 038780 0703C0 1F83E0
 FFCFFE FFCFFE>
PXLC RP
19560 12899 XY F34(xt)S 189 x(of)S 188 x(the)S 190 x(tal)S
XP /F34 107 288 0 0 21 31 32 24 0
<0F0000 FF0000 FF0000 0F0000 0F0000 0F0000 0F0000 0F0000 0F0000
 0F0000 0F0000 0F0000 0F0FF8 0F0FF8 0F07C0 0F0780 0F0F00 0F1E00
 0F3C00 0F7800 0FFC00 0FFC00 0FFE00 0FBE00 0F1F00 0F1F80 0F0F80
 0F07C0 0F07C0 0F03E0 FFE7FC FFE7FC>
PXLC RP
22460 12899 XY F34(k.)S
XP /F34 70 357 1 0 26 30 31 32 0
<FFFFFF80 FFFFFF80 07C00F80 07C00380 07C00180 07C00180 07C001C0
 07C000C0 07C000C0 07C060C0 07C060C0 07C06000 07C06000 07C0E000
 07FFE000 07FFE000 07C0E000 07C06000 07C06000 07C06000 07C06000
 07C00000 07C00000 07C00000 07C00000 07C00000 07C00000 07C00000
 07C00000 FFFF0000 FFFF0000>
PXLC RP
23089 12899 XY F34(First)S
XP /F34 44 152 4 -9 9 4 14 8 0
<70 F8 FC FC 7C 0C 0C 0C 1C 18 38 70 F0 60>
PXLC RP
24237 12899 XY F34(,)S 678 y 5347 X(solar)S 199 x(neutrinos)S 201 x
(ha)S
XP /F34 118 288 1 0 21 19 20 24 0
<FFC7F8 FFC7F8 1F03E0 0F01C0 0F0180 0F0180 078300 078300 078300
 03C600 03C600 03EE00 01EC00 01EC00 01FC00 00F800 00F800 007000
 007000 007000>
PXLC RP
9652 13577 XY F34(v)S -15 x(e)S 201 x(b)S 15 x(een)S 202 x(detect)S
2 x(ed)S 201 x(in)S 201 x(four)S 200 x(exp)S 15 x(erimen)S -15 x(ts)S
XP /F34 119 395 1 0 30 19 20 32 0
<FF9FF3FC FF9FF3FC 1E07C0F0 1E03C0E0 0F03C0C0 0F07C0C0 0F07E0C0
 0787E180 078CE180 078CF180 03CCF300 03D87300 03D87B00 03F87F00
 01F87E00 01F03E00 01F03E00 00F03C00 00E01C00 00E01C00>
PXLC RP
18758 13577 XY F34(with)S 201 x(appro)S -16 x(xim)S -2 x(ately)S 
201 x(the)S 677 y 5347 X(ener)S
XP /F34 103 273 1 -10 20 20 31 24 0
<0001E0 07F3F0 1FFF70 3E3EF0 3C1E70 780F00 780F00 780F00 780F00
 780F00 3C1E00 3E3E00 3FFC00 37F000 300000 300000 380000 3FFE00
 3FFF80 1FFFE0 3FFFE0 7803F0 F000F0 E00070 E00070 E00070 F000F0
 7801E0 3E07C0 1FFF80 03FC00>
PXLC RP
6348 14254 XY F34(gies)S 231 x(and)S
XP /F34 13 303 0 0 23 31 32 24 0
<001FF0 00FFF0 03F1F0 07C1F0 0781F0 0F01F0 0F00F0 0F00F0 0F00F0
 0F00F0 0F00F0 0F00F0 FFFFF0 FFFFF0 0F00F0 0F00F0 0F00F0 0F00F0
 0F00F0 0F00F0 0F00F0 0F00F0 0F00F0 0F00F0 0F00F0 0F00F0 0F00F0
 0F00F0 0F00F0 0F00F0 FFE7FF FFE7FF>
PXLC RP
8570 14254 XY F34(\015uxes)S 231 x(predicted)S 232 x(b)S -15 x(y)S
230 x(the)S 231 x(standard)S 231 x(solar)S 229 x(mo)S 14 x(del,)S 
231 x(con)S
XP /F34 12 303 0 0 23 31 32 24 0
<001FC0 00FFE0 03F0F0 07C1F0 0781F0 0F01F0 0F01F0 0F0000 0F0000
 0F0000 0F0000 0F0000 FFFFF0 FFFFF0 0F00F0 0F00F0 0F00F0 0F00F0
 0F00F0 0F00F0 0F00F0 0F00F0 0F00F0 0F00F0 0F00F0 0F00F0 0F00F0
 0F00F0 0F00F0 0F00F0 FFC3FF FFC3FF>
PXLC RP
20714 14254 XY F34(\014rmi)S -2 x(ng)S 231 x(the)S 231 x(h)S -15 x
(y)S
XP /F34 45 182 1 9 10 11 3 16 0
<FFC0 FFC0 FFC0>
PXLC RP
24207 14254 XY F34(-)S 678 y 5347 X(p)S 15 x(othesis)S 248 x(that)S
248 x(the)S 249 x(energy)S 249 x(source)S 248 x(for)S 247 x(the)S 
249 x(solar)S 247 x(luminosi)S -2 x(t)S -15 x(y)S 248 x(is)S 248 x
(the)S 249 x(fusion)S 247 x(of)S 248 x(ligh)S -16 x(t)S 677 y 5347 X
(elemen)S -15 x(ts.)S 190 x(Second,)S 191 x(the)S 191 x(measured)S
190 x(ev)S -15 x(en)S -14 x(t)S 190 x(rates)S 191 x(are)S 190 x
(signi\014can)S -16 x(tly)S 190 x(less)S 191 x(than)S 190 x(the)S 
191 x(ev)S -15 x(en)S -14 x(t)S 678 y 5347 X(rates)S 133 x
(predicted)S 133 x(b)S -15 x(y)S 133 x(the)S 134 x(com)S -16 x
(bined)S 134 x(standard)S 132 x(solar)S 132 x(and)S 133 x(electro)S
-15 x(w)S -15 x(eak)S 133 x(mo)S 14 x(dels)S 133 x(in)S 133 x(all)S
132 x(four)S 677 y 5347 X(exp)S 15 x(erimen)S -15 x(ts.)S 156 x
(Third,)S 156 x(a)S 157 x(compariso)S -2 x(n)S 157 x(of)S 157 x(the)S
158 x(ev)S -15 x(en)S -15 x(t)S 157 x(rates)S 157 x(measured)S 157 x
(in)S 157 x(the)S 157 x(c)S -14 x(hlori)S -2 x(ne)S 158 x(ex-)S 678 y 
5347 X(p)S 15 x(erimen)S -16 x(t)S
XP /F34 40 212 3 -12 13 33 46 16 0
<00E0 01E0 03C0 0380 0700 0F00 0E00 1E00 1C00 3C00 3800 3800 7800
 7800 7000 7000 F000 F000 F000 F000 F000 F000 F000 F000 F000 F000
 F000 F000 F000 F000 7000 7000 7800 7800 3800 3800 3C00 1C00 1E00
 0E00 0F00 0700 0380 03C0 01E0 00E0>
PXLC RP
7702 17642 XY F34(\(threshold)S
XP /F34 48 273 2 -1 19 29 31 24 0
<03F000 0FFC00 1E1E00 1C0E00 380700 780780 700380 700380 700380
 F003C0 F003C0 F003C0 F003C0 F003C0 F003C0 F003C0 F003C0 F003C0
 F003C0 F003C0 F003C0 F003C0 700380 700380 700380 780780 380700
 1C0E00 1E1E00 0FFC00 03F000>
PXLC RP
10363 17642 XY F34(0.)S
XP /F34 56 273 2 -1 19 29 31 24 0
<03F000 0FFC00 1E1E00 380F00 380780 700380 700380 700380 780380
 7C0780 7F0700 3FCE00 3FFC00 1FF800 0FFC00 0FFE00 1FFF00 3C7F80
 781F80 700FC0 F007C0 E003C0 E001C0 E001C0 E001C0 F00180 700380
 780700 3E1F00 1FFC00 07F000>
PXLC RP
10788 17642 XY F34(8)S
XP /F34 77 501 1 0 39 30 31 40 0
<FFC0000FFE FFE0001FFE 07E0001FC0 07E0001FC0 06F00037C0 06F00037C0
 06F00037C0 06780067C0 06780067C0 06780067C0 063C00C7C0 063C00C7C0
 063C00C7C0 061E0187C0 061E0187C0 060F0307C0 060F0307C0 060F0307C0
 06078607C0 06078607C0 06078607C0 0603CC07C0 0603CC07C0 0603CC07C0
 0601F807C0 0601F807C0 0601F807C0 0600F007C0 0F00F007C0 FFF0F0FFFE
 FFF060FFFE>
PXLC RP
11293 17642 XY F34(Mev)S
XP /F34 41 212 3 -12 13 33 46 16 0
<E000 F000 7800 3800 1C00 1E00 0E00 0F00 0700 0780 0380 0380 03C0
 03C0 01C0 01C0 01E0 01E0 01E0 01E0 01E0 01E0 01E0 01E0 01E0 01E0
 01E0 01E0 01E0 01E0 01C0 01C0 03C0 03C0 0380 0380 0780 0700 0F00
 0E00 1E00 1C00 3800 7800 F000 E000>
PXLC RP
12323 17642 XY F34(\))S 233 x(and)S 232 x(the)S 233 x
(neutrino-electron)S 232 x(scattering)S 233 x(exp)S 15 x(erimen)S 
-15 x(t)S 677 y 5347 X(\()S
XP /F34 75 425 1 0 31 30 31 32 0
<FFFE0FFE FFFE0FFE 07C003F0 07C003C0 07C00780 07C00F00 07C01E00
 07C03C00 07C07800 07C07000 07C0E000 07C1C000 07C3C000 07C7C000
 07CFE000 07DFE000 07FBF000 07F1F800 07E1F800 07C0FC00 07C07C00
 07C07E00 07C03F00 07C03F00 07C01F80 07C00F80 07C00FC0 07C007E0
 07C00FF0 FFFE3FFE FFFE3FFE>
PXLC RP
5559 18319 XY F34(Kamiok)S -32 x(ande)S
XP /F34 73 197 0 0 14 30 31 16 0
<FFFE FFFE 07C0 07C0 07C0 07C0 07C0 07C0 07C0 07C0 07C0 07C0 07C0
 07C0 07C0 07C0 07C0 07C0 07C0 07C0 07C0 07C0 07C0 07C0 07C0 07C0
 07C0 07C0 07C0 FFFE FFFE>
PXLC RP
8750 18319 XY F34(I)S 15 x(I,)S 236 x(threshold)S
XP /F34 55 273 3 -1 20 29 31 24 0
<600000 7FFFC0 7FFFC0 7FFFC0 7FFF80 E00300 C00700 C00E00 C01C00
 001800 003800 007000 007000 00E000 00E000 01E000 01E000 01C000
 03C000 03C000 03C000 03C000 07C000 07C000 07C000 07C000 07C000
 07C000 07C000 07C000 038000>
PXLC RP
12001 18319 XY F34(7.)S
XP /F34 53 273 2 -1 19 29 31 24 0
<380700 3FFF00 3FFE00 3FFC00 3FF000 300000 300000 300000 300000
 300000 300000 33F000 37FC00 3F1E00 3C0F00 380F80 380780 0007C0
 0007C0 0007C0 0007C0 F807C0 F807C0 F807C0 F807C0 F80F80 600F80
 701F00 3C3E00 1FFC00 07F000>
PXLC RP
12425 18319 XY F34(5)S 236 x(Me)S
XP /F34 86 410 1 -1 31 30 32 32 0
<FFF807FE FFF807FE 0FC001F0 0FC000E0 07C000C0 07C000C0 07E001C0
 03E00180 03E00180 03F00380 01F00300 01F00300 00F80600 00F80600
 00F80600 007C0C00 007C0C00 007C0C00 003E1800 003E1800 003F1800
 001F3000 001F3000 001FF000 000FE000 000FE000 000FE000 0007C000
 0007C000 00038000 00038000 00038000>
PXLC RP
13677 18319 XY F34(V\))S 236 x(indicates)S 237 x(that)S 237 x(the)S
237 x(de\014ciency)S 238 x(of)S 236 x(electron-)S 678 y 5347 X(t)S
-15 x(yp)S 15 x(e)S 190 x(neutrinos)S 189 x(at)S 189 x(the)S 191 x
(earth)S 189 x(is)S 189 x(energy)S 190 x(dep)S 16 x(enden)S -14 x
(t,)S 189 x(if)S 189 x(the)S 190 x(rates)S 189 x(and)S 190 x(the)S
190 x(uncertain-)S 677 y 5347 X(ties)S 173 x(in)S 173 x(b)S 15 x
(oth)S 173 x(exp)S 15 x(erimen)S -15 x(ts)S 173 x(ha)S -15 x(v)S 
-16 x(e)S 174 x(b)S 15 x(een)S 174 x(correctly)S 173 x(understo)S 
15 x(o)S 15 x(d.)S 173 x(The)S 173 x(inference)S 174 x(that)S 173 x
(the)S 677 y 5347 X(de\014ciency)S 251 x(is)S 249 x(energy-dep)S 
15 x(enden)S -14 x(t)S 250 x(con\015icts)S 250 x(with)S 249 x(the)S
250 x(simplest)S 249 x(v)S -15 x(ersion)S 249 x(of)S 249 x(standard)S
678 y 5347 X(electro)S -15 x(w)S -15 x(eak)S 171 x(theory)S -45 x(.)S
171 x(F)S -46 x(ourth,)S 171 x(exp)S 16 x(erimen)S -16 x(ts)S 172 x
(are)S 171 x(b)S 15 x(eing)S 172 x(constructed)S 173 x(that)S 171 x
(ha)S -15 x(v)S -16 x(e)S 173 x(the)S 172 x(ca-)S 677 y 5347 X
(pabili)S -2 x(ties)S 182 x(to)S 182 x(determine)S 182 x(conclusiv)S
-16 x(ely)S 182 x(if)S 181 x(new)S 182 x(neutrino)S 181 x(ph)S -15 x
(ysics)S 182 x(is)S 182 x(re)S
XP /F34 113 288 2 -9 23 19 29 24 0
<07F0C0 0FF9C0 1F1DC0 3C0FC0 7C07C0 7803C0 F803C0 F003C0 F003C0
 F003C0 F003C0 F003C0 F003C0 F803C0 7803C0 7C07C0 3C0FC0 3E1FC0
 1FFBC0 07E3C0 0003C0 0003C0 0003C0 0003C0 0003C0 0003C0 0003C0
 003FFC 003FFC>
PXLC RP
20609 21706 XY F34(quired.)S
XP /F48 49 336 4 0 22 31 32 24 0
<001800 007800 03F800 FFF800 FFF800 FDF800 01F800 01F800 01F800
 01F800 01F800 01F800 01F800 01F800 01F800 01F800 01F800 01F800
 01F800 01F800 01F800 01F800 01F800 01F800 01F800 01F800 01F800
 01F800 01F800 7FFFE0 7FFFE0 7FFFE0>
PXLC RP
5347 23024 XY F48(1)S
XP /F48 46 187 4 0 11 7 8 8 0
<3C 7E FF FF FF FF 7E 3C>
PXLC RP
5683 23024 XY F48(.)S 672 x(I)S
XP /F48 110 374 2 0 27 21 22 32 0
<FF87F000 FF9FFC00 FFF8FC00 1FF07E00 1FE07E00 1FC07E00 1FC07E00
 1F807E00 1F807E00 1F807E00 1F807E00 1F807E00 1F807E00 1F807E00
 1F807E00 1F807E00 1F807E00 1F807E00 1F807E00 FFF1FFC0 FFF1FFC0
 FFF1FFC0>
PXLC RP
6793 23024 XY F48(n)S
XP /F48 116 262 1 0 17 31 32 24 0
<01C000 01C000 01C000 01C000 03C000 03C000 03C000 07C000 07C000
 0FC000 3FFF00 FFFF00 FFFF00 0FC000 0FC000 0FC000 0FC000 0FC000
 0FC000 0FC000 0FC000 0FC000 0FC000 0FC380 0FC380 0FC380 0FC380
 0FC380 0FC780 07E700 03FE00 01FC00>
PXLC RP
7148 23024 XY F48(t)S
XP /F48 114 275 2 0 20 21 22 24 0
<FF1F80 FF7FC0 FF77E0 1FE7E0 1FC7E0 1FC7E0 1F83C0 1F8000 1F8000
 1F8000 1F8000 1F8000 1F8000 1F8000 1F8000 1F8000 1F8000 1F8000
 1F8000 FFF800 FFF800 FFF800>
PXLC RP
7409 23024 XY F48(r)S
XP /F48 111 336 2 0 24 21 22 24 0
<00FE00 07FFC0 0F83E0 1F01F0 3E00F8 7E00FC 7E00FC 7E00FC FE00FE
 FE00FE FE00FE FE00FE FE00FE FE00FE FE00FE 7E00FC 7E00FC 3F01F8
 1F01F0 0F83E0 07FFC0 00FE00>
PXLC RP
7684 23024 XY F48(o)S
XP /F48 100 374 2 0 27 34 35 32 0
<0003FE00 0003FE00 0003FE00 00007E00 00007E00 00007E00 00007E00
 00007E00 00007E00 00007E00 00007E00 00007E00 00007E00 01FE7E00
 07FFFE00 1FC3FE00 3F80FE00 3F007E00 7F007E00 7E007E00 FE007E00
 FE007E00 FE007E00 FE007E00 FE007E00 FE007E00 FE007E00 FE007E00
 7E007E00 7E007E00 3F00FE00 3F01FE00 1FC3FFC0 0FFFFFC0 01FC7FC0>
PXLC RP
8039 23024 XY F48(d)S
XP /F48 117 374 2 0 27 21 22 32 0
<FF83FE00 FF83FE00 FF83FE00 1F807E00 1F807E00 1F807E00 1F807E00
 1F807E00 1F807E00 1F807E00 1F807E00 1F807E00 1F807E00 1F807E00
 1F807E00 1F807E00 1F80FE00 1F80FE00 1F81FE00 0FC3FFC0 0FFF7FC0
 03FC7FC0>
PXLC RP
8412 23024 XY F48(u)S
XP /F48 99 299 2 0 21 21 22 24 0
<01FF80 07FFE0 1FC3F0 3F83F0 3F03F0 7F03F0 7E01E0 FE0000 FE0000
 FE0000 FE0000 FE0000 FE0000 FE0000 FE0000 7F0000 7F0000 3F8070
 3F80F0 1FE1E0 07FFC0 01FF00>
PXLC RP
8786 23024 XY F48(ct)S
XP /F48 105 187 1 0 12 35 36 16 0
<1F00 3F80 7F80 7F80 7F80 7F80 3F80 1F00 0000 0000 0000 0000 0000
 0000 FF80 FF80 FF80 1F80 1F80 1F80 1F80 1F80 1F80 1F80 1F80 1F80
 1F80 1F80 1F80 1F80 1F80 1F80 1F80 FFF0 FFF0 FFF0>
PXLC RP
9346 23024 XY F48(ion)S 24305 Y 5347 X F34(There)S 136 x(is)S 136 x
(a)S 136 x(lot)S 135 x(of)S 136 x(excitemen)S -14 x(t,)S 135 x(en)S
-14 x(th)S -15 x(usiasm,)S 134 x(and)S 136 x(activit)S -16 x(y)S 
136 x(in)S 136 x(our)S 136 x(\014eld)S 136 x(to)S 15 x(da)S -16 x(y)S
-45 x(.)S 135 x(There)S 137 x(are)S 678 y 5347 X(also)S 164 x(in)S
-16 x(tense)S 167 x(argumen)S -16 x(ts)S 165 x(ab)S 15 x(out)S 165 x
(sp)S 16 x(eci\014c)S 166 x(issues)S 165 x(and)S 166 x(ab)S 14 x
(out)S 166 x(ho)S -16 x(w)S 165 x(to)S 165 x(in)S -15 x(terpret)S 
166 x(di)S
XP /F34 11 318 0 0 29 31 32 32 0
<001FC3F0 00FFFFF8 03F1FE7C 07C1FC7C 0781F87C 0F01F07C 0F00F000
 0F00F000 0F00F000 0F00F000 0F00F000 0F00F000 FFFFFF80 FFFFFF80
 0F00F000 0F00F000 0F00F000 0F00F000 0F00F000 0F00F000 0F00F000
 0F00F000 0F00F000 0F00F000 0F00F000 0F00F000 0F00F000 0F00F000
 0F00F000 0F00F000 FFC3FF00 FFC3FF00>
PXLC RP
22872 24983 XY F34(\013eren)S -14 x(t)S 677 y 5347 X(exp)S 15 x
(erimen)S -15 x(tal)S 260 x(results.)S 261 x(In)S 262 x(this)S 261 x
(situation,)S 260 x(it)S 261 x(is)S 261 x(helpful)S 261 x(to)S 261 x
(lo)S 15 x(ok)S 261 x(bac)S -15 x(kw)S -16 x(ards)S 261 x(to)S 262 x
(see)S 677 y 5347 X(where)S 155 x(w)S -16 x(e)S 155 x(came)S 155 x
(from)S 153 x(in)S 154 x(order)S 154 x(that)S 154 x(w)S -15 x(e)S 
155 x(ma)S -16 x(y)S 154 x(ha)S -15 x(v)S -16 x(e)S 155 x(a)S 154 x
(clearer)S 155 x(picture)S 154 x(of)S 154 x(what)S 154 x(w)S -15 x
(e)S 155 x(ha)S -15 x(v)S -16 x(e)S 678 y 5347 X(accomplished)S 181 x
(and,)S 181 x(p)S 15 x(erhaps,)S 182 x(where)S 182 x(w)S -15 x(e)S
182 x(ma)S -16 x(y)S 182 x(w)S -15 x(an)S -16 x(t)S 182 x(to)S 182 x
(go.)S 677 y 6165 X(It)S 153 x(is)S 153 x(exactly)S 153 x(thirt)S 
-16 x(y)S 153 x(y)S -16 x(ears)S 153 x(ago)S 153 x(this)S 152 x(mon)S
-15 x(th)S 153 x(that)S 153 x(Ra)S -16 x(y)S 153 x(and)S 152 x(I)S
153 x(\014rst)S 153 x(claimed)S 152 x(in)S 153 x(prin)S -16 x(t)S
XP /F13 /cmr8 300 399 398.506 128 [-3 -9 33 24] PXLNF RP
XP /F13 49 212 2 0 13 20 21 16 0
<0300 0F00 FF00 F700 0700 0700 0700 0700 0700 0700 0700 0700 0700
 0700 0700 0700 0700 0700 0700 7FF0 7FF0>
PXLC RP
24152 27494 XY F13(1)S 876 y 5347 X F34(that)S 201 x(the)S 202 x
(exp)S 16 x(ected)S 203 x(ev)S -15 x(en)S -15 x(t)S 202 x(rate)S 
201 x(in)S 201 x(a)S 201 x(c)S -14 x(hlori)S -2 x(ne)S 202 x(detect)S
2 x(or)S 200 x(w)S -15 x(as)S 202 x(lar)S -2 x(ge)S 202 x(enough)S
201 x(that)S 202 x(solar)S 677 y 5347 X(neutrinos)S 183 x(could)S 
184 x(b)S 15 x(e)S 184 x(observ)S -15 x(ed)S 185 x(in)S 183 x(an)S
184 x(exp)S 15 x(erimen)S -15 x(t)S 184 x(that)S 184 x(could)S 183 x
(b)S 15 x(e)S 185 x(built)S 183 x(using)S 183 x(existing)S 678 y 
5347 X(tec)S -14 x(hnolog)S -2 x(y)S -45 x(.)S 243 x(The)S 243 x
(argumen)S -16 x(t)S 244 x(that)S 243 x(detection)S 244 x(of)S 243 x
(solar)S 242 x(neutrinos)S 243 x(w)S -16 x(as)S 244 x(practical)S 
242 x(w)S -15 x(as)S 677 y 5347 X(based)S 236 x(up)S 15 x(on)S 236 x
(Ra)S -16 x(y)S
XP /F34 39 152 4 18 9 31 14 8 0
<70 F8 FC FC 7C 0C 0C 0C 1C 18 38 70 F0 60>
PXLC RP
9299 30402 XY F34('s)S 236 x(exp)S 15 x(erience)S 237 x(using)S 236 x
(a)S 236 x(c)S -15 x(hlori)S -2 x(ne)S 237 x(detector)S 237 x(to)S
236 x(try)S 236 x(to)S 235 x(observ)S -15 x(e)S 237 x(an)S -16 x
(ti-)S 678 y 5347 X(neutrinos)S 141 x(from)S 141 x(a)S 142 x
(reactor,)S 141 x(up)S 15 x(on)S 142 x(his)S 142 x(con\014dence)S 
144 x(in)S 141 x(his)S 142 x(abili)S -2 x(t)S -15 x(y)S 142 x(to)S
142 x(scale)S 142 x(up)S 143 x(the)S 142 x(smaller)S 677 y 5347 X
(detector)S 215 x(to)S 214 x(a)S 213 x(m)S -15 x(uc)S -15 x(h)S 214 x
(larger)S 213 x(si)S
XP /F34 122 242 1 0 16 19 20 16 0
<3FFF 3FFF 381F 303E 707C 60FC 60F8 61F0 03F0 03E0 07C0 0FC3 0F83
 1F03 3F07 3E06 7C06 F81E FFFE FFFE>
PXLC RP
12122 31757 XY F34(ze,)S 214 x(up)S 15 x(on)S 214 x(the)S 215 x
(\014rst)S 214 x(neutrino)S 213 x(\015ux)S 214 x(calculations)S 213 x
(with)S 214 x(a)S 678 y 5347 X(detailed)S 220 x(solar)S 219 x(mo)S
14 x(del,)S 220 x(and)S 220 x(up)S 15 x(on)S 220 x(the)S 221 x(n)S
-15 x(uclear)S 220 x(ph)S -15 x(ysics)S 220 x(calculation)S 219 x
(of)S 219 x(the)S 221 x(neutrino)S 677 y 5347 X(cross)S 180 x
(sections)S 180 x(\(including)S 178 x(the)S 181 x(crucial)S 179 x
(analo)S -2 x(g)S 180 x(state)S 180 x(con)S -15 x(tribution\).)S 
178 x(A)S -15 x(t)S 180 x(the)S 180 x(time,)S 179 x(the)S 678 y 
5347 X(only)S 229 x(motiv)S -32 x(ation)S 230 x(presen)S -15 x(ted)S
-198 y F13(1)S 198 y 255 x F34(for)S 230 x(p)S 15 x(erform)S -2 x
(ing)S 230 x(a)S 230 x(solar)S 228 x(neutrino)S 230 x(exp)S 15 x
(erimen)S -15 x(t)S 230 x(w)S -15 x(as)S 230 x(to)S
1 PP EP

1000 BP 39600 30600 PM 0 0 XY
3815 Y 5347 X F34(use)S 236 x(neutrinos)S
XP /F34 92 273 5 18 19 31 14 16 0
<180C 3C1E 381C 7038 6030 E070 C060 C060 C060 F87C FC7E FC7E 7C3E
 381C>
PXLC RP
8796 3815 XY F34(\\...)S -2 x(to)S 236 x(see)S 238 x(in)S -16 x(to)S
236 x(the)S 237 x(in)S -16 x(terior)S 235 x(of)S 236 x(a)S 236 x
(star)S 235 x(and)S 236 x(th)S -15 x(us)S 237 x(directly)S 235 x(v)S
-15 x(erify)S 235 x(the)S 678 y 5347 X(h)S -15 x(yp)S 15 x(othesis)S
219 x(of)S 218 x(n)S -15 x(uclear)S 219 x(energy)S 219 x(generation)S
218 x(in)S 219 x(stars.)S
XP /F34 34 273 2 18 16 31 14 16 0
<7038 F87C FC7E FC7E 7C3E 0C06 0C06 0C06 1C0E 180C 381C 7038 F078
 6030>
PXLC RP
17103 4493 XY F34(")S 219 x(The)S 219 x(h)S -15 x(yp)S 15 x(othesis)S
219 x(to)S 219 x(b)S 15 x(e)S 219 x(teste)S 2 x(d)S 677 y 5347 X(w)S
-16 x(as)S 244 x(that)S 243 x(stars)S 243 x(lik)S -16 x(e)S 244 x
(the)S 244 x(sun)S 244 x(shine)S 243 x(b)S -15 x(y)S 244 x(burning)S
242 x(protons)S 243 x(to)S 243 x(form)S 242 x(alpha)S 243 x
(particles,)S 678 y 5347 X(neutrinos,)S 181 x(and)S 181 x(thermal)S
181 x(energy)S
XP /F34 58 152 4 0 8 19 20 8 0
<70 F8 F8 F8 70 00 00 00 00 00 00 00 00 00 00 70 F8 F8 F8 70>
PXLC RP
12551 5848 XY F34(:)S
XP /F34 52 273 1 0 20 29 30 24 0
<000E00 000E00 001E00 003E00 003E00 007E00 007E00 00DE00 01DE00
 019E00 031E00 071E00 061E00 0C1E00 1C1E00 181E00 381E00 301E00
 601E00 E01E00 FFFFF0 FFFFF0 001E00 001E00 001E00 001E00 001E00
 001E00 01FFE0 01FFE0>
PXLC RP
10122 7203 XY F34(4)S
XP /F35 /cmmi10 329 546 545.454 128 [-2 -11 46 33] PXLNF RP
XP /F35 112 274 -2 -9 21 19 29 24 0
<03E1F8 07F7FC 06FF1E 0C7E0E 0C7C0F 1CF80F 18F80F 00F00F 00F00F
 01F01F 01F01F 01E01E 01E01E 03E03C 03E03C 03E078 03E0F0 07F1E0
 07FFC0 079F00 078000 0F8000 0F8000 0F0000 0F0000 1F0000 1F0000
 FFE000 FFE000>
PXLC RP
10395 7203 XY F35(p)S
XP /F36 /cmsy10 329 546 545.454 128 [-1 -44 50 35] PXLNF RP
XP /F36 0 425 4 10 29 11 2 32 0
<FFFFFFC0 FFFFFFC0>
PXLC RP
10821 7203 XY F36(\000)S
XP /F36 33 546 3 3 40 18 16 40 0
<0000000600 0000000700 0000000300 0000000380 00000001C0 00000000E0
 0000000070 FFFFFFFFFC FFFFFFFFFC 0000000070 00000000E0 00000001C0
 0000000380 0000000300 0000000700 0000000600>
PXLC RP
11154 7203 XY F36(!)S
XP /F13 52 212 1 0 15 20 21 16 0
<0070 00F0 00F0 01F0 03F0 0770 0670 0E70 1C70 1870 3070 7070 E070
 FFFE FFFE 0070 0070 0070 0070 03FE 03FE>
PXLC RP
11700 6977 XY F13(4)S
XP /F35 72 454 2 0 37 30 31 40 0
<00FFF9FFF0 00FFF9FFF0 000F801F00 000F001E00 000F003E00 001F003E00
 001E003C00 001E003C00 001E007C00 003E007C00 003C007800 003C007800
 003C00F800 007C00F800 007FFFF000 007FFFF000 007801F000 00F801F000
 00F001E000 00F001E000 00F003E000 01F003E000 01E003C000 01E003C000
 01E007C000 03E007C000 03C0078000 03C0078000 07C00F8000 7FFCFFF800
 FFFCFFF800>
PXLC RP
12088 7203 XY F35(H)S
XP /F35 101 254 2 0 17 19 20 16 0
<00FC 03FE 07C7 0F03 1E03 3C07 7C0E 783E 7FF8 7FE0 F800 F000 F000
 F000 F000 7003 780F 3C3E 1FFC 0FE0>
PXLC RP
12586 7203 XY F35(e)S
XP /F34 43 425 3 -5 30 26 32 32 0
<00060000 00060000 00060000 00060000 00060000 00060000 00060000
 00060000 00060000 00060000 00060000 00060000 00060000 00060000
 00060000 FFFFFFF0 FFFFFFF0 00060000 00060000 00060000 00060000
 00060000 00060000 00060000 00060000 00060000 00060000 00060000
 00060000 00060000 00060000 00060000>
PXLC RP
13143 7203 XY F34(+)S
XP /F34 50 273 2 0 19 29 30 24 0
<07F000 1FFC00 3C7E00 701F00 600F80 F80780 FC07C0 FC07C0 FC03C0
 FC03C0 3007C0 0007C0 0007C0 000F80 000F80 001F00 003E00 003C00
 007800 00F000 01E000 03C000 0780C0 0E00C0 1C00C0 3801C0 7FFF80
 FFFF80 FFFF80 FFFF80>
PXLC RP
13870 7203 XY F34(2)S F35(e)S
XP /F13 43 329 2 -4 23 19 24 24 0
<003000 003000 003000 003000 003000 003000 003000 003000 003000
 003000 003000 FFFFFC FFFFFC 003000 003000 003000 003000 003000
 003000 003000 003000 003000 003000 003000>
PXLC RP
14397 6977 XY F13(+)S 226 y 328 x F34(+)S 303 x(2)S
XP /F35 23 269 2 0 22 19 20 24 0
<0F8070 7F80F8 7F80F8 0F00F0 0F00F0 1F01F0 1F01E0 1E03E0 1E03C0
 3E0780 3E0F80 3C0F00 3C1E00 7C3C00 7C7800 79F000 7BE000 FF8000
 FE0000 F80000>
PXLC RP
16054 7203 XY F35(\027)S
XP /F14 /cmmi8 300 399 398.506 128 [-1 -8 35 24] PXLNF RP
XP /F14 101 196 1 0 13 13 14 16 0
<07F0 1FF8 3C38 7838 F070 FFF0 FFC0 E000 E000 E000 E018 7078 7FF0
 1FC0>
PXLC RP
16323 7284 XY F14(e)S -81 y 329 x F34(+)S 303 x(25)S 181 x(MeV)S
XP /F35 58 152 4 0 8 4 5 8 0
<70 F8 F8 F8 70>
PXLC RP
19461 7203 XY F35(:)S 23692 X F34(\()S
XP /F34 49 273 3 0 18 29 30 16 0
<00C0 01C0 0FC0 FFC0 F3C0 03C0 03C0 03C0 03C0 03C0 03C0 03C0 03C0
 03C0 03C0 03C0 03C0 03C0 03C0 03C0 03C0 03C0 03C0 03C0 03C0 03C0
 03C0 03C0 7FFF 7FFF>
PXLC RP
23904 7203 XY F34(1\))S
XP /F34 87 562 1 -1 44 30 32 48 0
<FFF8FFF83FF0 FFF8FFF83FF0 1F800F800F80 0F800F800700 0F800FC00300
 07C007C00600 07C007C00600 07C007C00600 03E00FE00C00 03E00FE00C00
 03E00FE00C00 03F019F01C00 01F019F01800 01F019F01800 01F830F83800
 00F830F83000 00F830F83000 00F870FC3000 007C607C6000 007C607C6000
 007CE07E6000 003EC03EC000 003EC03EC000 003EC03EC000 001F801F8000
 001F801F8000 001F801F8000 001F000F8000 000F000F0000 000F000F0000
 000E00070000 000600060000>
PXLC RP
6165 8204 XY F34(With)S 218 x(the)S 220 x(four)S 218 x(pioneering)S
218 x(solar)S 218 x(neutrino)S 218 x(exp)S 16 x(erimen)S -16 x(ts)S
220 x(that)S 219 x(are)S 219 x(curren)S -15 x(tly)S 218 x(op-)S 677 y 
5347 X(erating,)S 221 x(the)S 223 x(goal)S 222 x(of)S 222 x
(demonstrating)S 221 x(that)S 223 x(proton)S 222 x(fusion)S 222 x
(is)S 222 x(the)S 223 x(origin)S 221 x(of)S 222 x(sunshine)S 678 y 
5347 X(has)S 183 x(b)S 16 x(een)S 184 x(ac)S -15 x(hiev)S -15 x(ed.)S
184 x(Solar)S 182 x(neutrinos)S 184 x(ha)S -16 x(v)S -15 x(e)S 184 x
(b)S 15 x(een)S 185 x(observ)S -15 x(ed)S 184 x(in)S 183 x(all)S 
183 x(of)S 183 x(the)S 185 x(exp)S 15 x(erimen)S -16 x(ts)S 677 y 
5347 X(with,)S 191 x(to)S 192 x(usual)S 191 x(astronomi)S -2 x(cal)S
192 x(accuracy)S 192 x(\(a)S 192 x(factor)S 192 x(of)S 191 x(t)S 
-15 x(w)S -15 x(o)S 192 x(or)S 191 x(three\),)S 192 x(ab)S 15 x(out)S
192 x(the)S 193 x(righ)S -16 x(t)S 678 y 5347 X(n)S -15 x(um)S -16 x
(b)S 15 x(ers)S 136 x(and)S 136 x(ab)S 15 x(out)S 135 x(the)S 137 x
(righ)S -16 x(t)S 136 x(energies.)S 135 x(Moreo)S -15 x(v)S -16 x
(er,)S 136 x(the)S 136 x(fact)S 136 x(that)S 136 x(the)S 136 x
(neutrinos)S 136 x(come)S 677 y 5347 X(from)S 217 x(the)S 219 x(sun)S
219 x(w)S -15 x(as)S 218 x(established)S 219 x(directly)S 218 x(b)S
-15 x(y)S 219 x(the)S 219 x(Kamiok)S -32 x(ande)S 219 x(exp)S 16 x
(erimen)S -16 x(t)S
XP /F13 50 212 2 0 14 20 21 16 0
<1F80 7FE0 F1F0 F8F8 F878 F878 7078 0078 00F8 00F0 01F0 01E0 03C0
 0780 0E00 1C18 3818 7038 FFF0 FFF0 FFF0>
PXLC RP
22225 11393 XY F13(2)S
XP /F13 44 118 3 -6 6 3 10 8 0
<E0 F0 F0 F0 30 30 70 60 E0 C0>
PXLC RP
22437 11393 XY F13(,)S
XP /F13 51 212 1 0 15 20 21 16 0
<0FE0 3FF8 7878 783C 7C3C 783C 303C 0078 00F8 07F0 07E0 0078 003C
 001E 701E F81E F81E F83E F07C 7FF8 1FE0>
PXLC RP
22555 11393 XY F13(3)S 198 y 243 x F34(whic)S -15 x(h)S 678 y 5347 X
(sho)S -16 x(w)S -15 x(ed)S 163 x(that)S 163 x(electrons)S 163 x
(scattere)S 2 x(d)S 162 x(b)S -15 x(y)S 163 x(neutrinos)S 162 x
(recoil)S 162 x(in)S 163 x(the)S 163 x(forw)S -16 x(ard)S 162 x
(direction)S 162 x(from)S 677 y 5347 X(the)S 182 x(sun.)S 677 y 
6165 X(The)S 171 x(observ)S -30 x(ations)S 170 x(of)S 170 x(solar)S
170 x(neutrinos)S 170 x(test)S 172 x(directly)S 171 x(the)S 171 x
(theory)S 171 x(of)S 170 x(n)S -15 x(uclear)S 171 x(energy)S 678 y 
5347 X(generation)S 231 x(and)S 231 x(of)S 231 x(stellar)S 230 x(ev)S
-15 x(olution.)S
XP /F34 72 410 1 0 31 30 31 32 0
<FFFC7FFE FFFC7FFE 07C007C0 07C007C0 07C007C0 07C007C0 07C007C0
 07C007C0 07C007C0 07C007C0 07C007C0 07C007C0 07C007C0 07C007C0
 07FFFFC0 07FFFFC0 07C007C0 07C007C0 07C007C0 07C007C0 07C007C0
 07C007C0 07C007C0 07C007C0 07C007C0 07C007C0 07C007C0 07C007C0
 07C007C0 FFFC7FFE FFFC7FFE>
PXLC RP
14101 14301 XY F34(Hundreds)S 232 x(of)S 231 x(exp)S 15 x(erimen)S
-16 x(talists)S 231 x(ha)S -15 x(v)S -15 x(e)S 231 x(w)S -15 x(ork)S
-16 x(ed)S 677 y 5347 X(together)S 219 x(and)S 218 x(indep)S 16 x
(enden)S -15 x(tly)S 219 x(to)S 218 x(mak)S -16 x(e)S 220 x(these)S
219 x(measuremen)S -15 x(ts)S 219 x(and)S 219 x(man)S -16 x(y)S 219 x
(additio)S -2 x(nal)S 678 y 5347 X(h)S -15 x(undreds)S 186 x(of)S 
185 x(scien)S -15 x(tists)S 186 x(ha)S -15 x(v)S -16 x(e)S 186 x
(con)S -15 x(tributed)S 186 x(imp)S 14 x(ortan)S -16 x(tly)S 186 x
(to)S 185 x(the)S 186 x(determination)S 185 x(of)S 185 x(the)S 677 y 
5347 X(input)S 185 x(parameters)S 186 x(and)S 185 x(the)S 187 x
(theoretical)S 185 x(calculations)S 185 x(with)S 185 x(whic)S -15 x
(h)S 186 x(the)S 187 x(solar)S 184 x(neutrino)S 678 y 5347 X(observ)S
-31 x(ations)S 181 x(are)S 182 x(compared.)S 677 y 6165 X(The)S 152 x
(exp)S 15 x(erimen)S -15 x(tal)S 151 x(results)S 151 x(on)S 151 x
(solar)S 150 x(neutrinos)S 152 x(represen)S -15 x(t)S 152 x(a)S 151 x
(collectiv)S -16 x(e)S 152 x(triumph)S 151 x(for)S 678 y 5347 X(the)S
162 x(ph)S -15 x(ysics,)S 162 x(the)S 162 x(c)S -14 x(hemistry)S 
-47 x(,)S 162 x(and)S 162 x(the)S 162 x(astronom)S -16 x(y)S 162 x
(comm)S -16 x(unities)S 161 x(since)S 163 x(they)S 162 x(bring)S 
161 x(to)S 162 x(a)S 677 y 5347 X(successful)S 149 x(conclusion)S 
147 x(the)S 149 x(dev)S -15 x(elopmen)S -15 x(t)S 148 x(\(whic)S 
-15 x(h)S 148 x(spanned)S 148 x(m)S -16 x(uc)S -14 x(h)S 148 x(of)S
147 x(the)S 149 x(20th)S 147 x(cen)S -14 x(tury\))S 678 y 5347 X(of)S
181 x(a)S 181 x(theory)S 182 x(of)S 181 x(ho)S -15 x(w)S 182 x(mai)S
-2 x(n)S 182 x(sequenc)S 2 x(e)S 182 x(stars)S 182 x(shine.)S 677 y 
6165 X(Ironical)S -2 x(ly)S -46 x(,)S 246 x(most)S 246 x(of)S 245 x
(the)S 247 x(curren)S -15 x(t)S 246 x(in)S -15 x(terest)S 247 x(in)S
246 x(solar)S 245 x(neutrinos)S 245 x(is)S 246 x(fo)S 15 x(cused)S
247 x(on)S 246 x(an)S 678 y 5347 X(applicati)S -2 x(on)S 197 x(of)S
197 x(solar)S 196 x(neutrino)S 196 x(researc)S -14 x(h)S 197 x(that)S
197 x(w)S -15 x(as)S 197 x(not)S 197 x(discussed)S 198 x(at)S 197 x
(the)S 198 x(time)S 197 x(of)S 196 x(the)S 677 y 5347 X(orig)S -2 x
(inal)S 152 x(exp)S 16 x(erimen)S -16 x(tal)S 153 x(and)S 153 x
(theoretical)S 152 x(prop)S 15 x(osals.)S
XP /F34 80 372 1 0 26 30 31 32 0
<FFFFF000 FFFFFC00 07C03E00 07C00F00 07C00F80 07C00780 07C007C0
 07C007C0 07C007C0 07C007C0 07C007C0 07C00780 07C00F80 07C00F00
 07C03E00 07FFFC00 07FFF000 07C00000 07C00000 07C00000 07C00000
 07C00000 07C00000 07C00000 07C00000 07C00000 07C00000 07C00000
 07C00000 FFFE0000 FFFE0000>
PXLC RP
16761 21753 XY F34(P)S -15 x(on)S -16 x(tecorv)S -14 x(o)S -198 y 
F13(4)S 198 y 177 x F34(w)S -15 x(as)S 153 x(the)S 154 x(\014rst)S
153 x(p)S 15 x(erson)S 677 y 5347 X(to)S 132 x(emphasize)S 132 x
(that)S 132 x(one)S 133 x(can)S 132 x(use)S 133 x(solar)S 131 x
(neutrinos)S 132 x(for)S 131 x(studying)S 132 x(exp)S 15 x(erimen)S
-15 x(tall)S -2 x(y)S 133 x(asp)S 15 x(ects)S 678 y 5347 X(of)S 181 x
(the)S 182 x(w)S -15 x(eak)S 182 x(in)S -16 x(teractions)S 182 x
(that)S 182 x(are)S 181 x(not)S 182 x(accessible)S 183 x(in)S 181 x
(lab)S 15 x(orator)S -2 x(y)S 182 x(exp)S 15 x(erimen)S -15 x(ts.)S
677 y 6165 X(It)S 259 x(is)S 258 x(easy)S 259 x(to)S 258 x(see)S 
260 x(wh)S -15 x(y)S 258 x(solar)S 258 x(neutrino)S 258 x(b)S 15 x
(eams)S 258 x(are)S 259 x(of)S 258 x(suc)S -15 x(h)S 259 x(extreme)S
259 x(to)S 258 x(ph)S -15 x(ysi-)S 678 y 5347 X(cists.)S 260 x(If)S
259 x(neutrinos)S 260 x(ha)S -16 x(v)S -15 x(e)S 260 x(\014nite)S 
260 x(masses,)S 260 x(as)S 260 x(predicted)S 260 x(b)S -15 x(y)S 
260 x(theories)S 260 x(that)S 260 x(unify)S 259 x(the)S 677 y 5347 X
(strong,)S 189 x(the)S 190 x(electromagnetic,)S 189 x(and)S 190 x
(the)S 190 x(w)S -15 x(eak)S 190 x(in)S -15 x(teractions,)S 189 x
(then)S 190 x(one)S 190 x(\014gure)S 190 x(of)S 189 x(merit)S 678 y 
5347 X(for)S 226 x(the)S 227 x(sensitivit)S -16 x(y)S 227 x(of)S 
226 x(an)S 227 x(exp)S 15 x(erimen)S -15 x(t)S 227 x(is)S 226 x(the)S
227 x(prop)S 15 x(er)S 227 x(time)S 226 x(that)S 227 x(it)S 226 x
(tak)S -15 x(es)S 227 x(the)S 228 x(b)S 15 x(eam)S 677 y 5347 X
(particles)S 221 x(to)S 222 x(tra)S -16 x(v)S -15 x(erse)S 223 x
(the)S 222 x(distance)S 222 x(b)S 16 x(et)S -15 x(w)S -15 x(een)S 
223 x(the)S 222 x(p)S 15 x(oin)S -16 x(t)S 223 x(of)S 221 x(pro)S 
14 x(duction)S 222 x(and)S 222 x(the)S 222 x(de-)S 678 y 5347 X
(tectors.)S 244 x(This)S 243 x(prop)S 15 x(er)S 244 x(time)S 243 x
(is)S 244 x(prop)S 14 x(ortional)S 242 x(to)S 244 x(the)S 245 x
(ratio)S -2 x(:)S 244 x(\(P)S -15 x(ath)S
XP /F34 76 342 1 0 24 30 31 24 0
<FFFF00 FFFF00 07C000 07C000 07C000 07C000 07C000 07C000 07C000
 07C000 07C000 07C000 07C000 07C000 07C000 07C000 07C000 07C000
 07C000 07C003 07C003 07C003 07C003 07C007 07C007 07C006 07C00E
 07C01E 07C07E FFFFFE FFFFFE>
PXLC RP
20433 27173 XY F34(Length\))S
XP /F34 47 273 3 -11 18 33 45 16 0
<0003 0003 0007 0007 0006 000E 000E 000C 001C 001C 0018 0038 0038
 0030 0070 0070 0060 00E0 00E0 00C0 01C0 01C0 0180 0380 0380 0300
 0700 0700 0600 0E00 0E00 0C00 1C00 1C00 1800 3800 3800 3000 7000
 7000 6000 E000 E000 C000 C000>
PXLC RP
22319 27173 XY F34(/)S
XP /F34 69 372 1 0 28 30 31 32 0
<FFFFFF80 FFFFFF80 07C00F80 07C00380 07C00180 07C00180 07C001C0
 07C000C0 07C000C0 07C060C0 07C060C0 07C06000 07C06000 07C0E000
 07FFE000 07FFE000 07C0E000 07C06000 07C06000 07C06030 07C06030
 07C00030 07C00060 07C00060 07C00060 07C000E0 07C000E0 07C001E0
 07C007C0 FFFFFFC0 FFFFFFC0>
PXLC RP
22592 27173 XY F34(Energy)S -46 x(.)S 677 y 5347 X(F)S -46 x(or)S 
177 x(lab)S 14 x(oratory)S 176 x(exp)S 16 x(erimen)S -16 x(ts,)S 
177 x(one)S 177 x(ma)S -16 x(y)S 178 x(hop)S 15 x(e)S 177 x(to)S 
177 x(obtain)S 177 x(accurate)S 177 x(measuremen)S -15 x(ts)S 178 x
(for)S 678 y 5347 X(whic)S -16 x(h)S 240 x(this)S 239 x(ratio)S 239 x
(is:)S 238 x(\(1)S 240 x(Km\)/\(1)S
XP /F34 71 429 3 -1 32 31 33 32 0
<001FE060 007FF860 01F83CE0 03E00FE0 07C007E0 0F8003E0 1F0001E0
 3E0000E0 3E0000E0 7C0000E0 7C000060 7C000060 F8000060 F8000000
 F8000000 F8000000 F8000000 F8000000 F8000000 F800FFFC F800FFFC
 7C0003E0 7C0003E0 7C0003E0 3E0003E0 3E0003E0 1F0003E0 0F8003E0
 07C007E0 03E00FE0 01FC1EE0 007FF860 001FE000>
PXLC RP
13019 28528 XY F34(GeV\).)S 239 x(F)S -45 x(or)S 239 x(the)S 240 x
(sun,)S 239 x(the)S 240 x(corresp)S 16 x(onding)S 238 x(ratio)S 239 x
(is:)S 677 y 5347 X(\(10)S
XP /F13 56 212 1 0 15 20 21 16 0
<0FE0 1FF8 383C 601C 600C 780C 7C1C 7F38 3FF0 1FF0 1FF8 3FFC 78FE
 F03E E01E E00E E00E F01E 7C3C 3FF8 0FE0>
PXLC RP
6104 29007 XY F13(8)S 198 y 226 x F34(Km\)/\(10)S
XP /F15 /cmsy8 300 399 398.506 128 [-2 -32 39 26] PXLNF RP
XP /F15 0 329 3 7 22 8 2 24 0
<FFFFF0 FFFFF0>
PXLC RP
8663 29007 XY F15(\000)S F13(3)S 198 y 426 x F34(GeV\))S
XP /F34 61 425 3 5 30 16 12 32 0
<FFFFFFF0 FFFFFFF0 00000000 00000000 00000000 00000000 00000000
 00000000 00000000 00000000 FFFFFFF0 FFFFFFF0>
PXLC RP
11324 29205 XY F34(=)S 384 x(\(10)S -198 y F13(11)S 198 y 225 x F34
(Km\)/\(GeV\).)S 200 x(With)S 200 x(solar)S 200 x(neutrino)S 200 x
(b)S 15 x(eams,)S 201 x(one)S 677 y 5347 X(can)S 166 x(study)S 166 x
(neutrino)S 165 x(masses)S 167 x(as)S 165 x(small)S 165 x(as)S 166 x
(10)S -197 y F15(\000)S
XP /F13 54 212 1 0 15 20 21 16 0
<03F0 0FF8 1F3C 3C3C 783C 783C 7000 F080 F7F0 FFF8 FC3C F81E F01E
 F01E F01E 701E 701E 783C 3C7C 1FF8 0FE0>
PXLC RP
15360 29685 XY F13(6)S 197 y 191 x F34(eV.)S 166 x(Solar)S 164 x
(neutrino)S 166 x(exp)S 15 x(erimen)S -15 x(ts)S 166 x(ha)S -15 x(v)S
-16 x(e)S 678 y 5347 X(the)S 153 x(sensitivit)S -15 x(y)S 152 x(to)S
153 x(detec)S 2 x(t)S 153 x(neutrino)S 152 x(masses)S 153 x(in)S 
153 x(the)S 154 x(range)S 152 x(predicted)S 154 x(b)S -15 x(y)S 152 x
(represen)S -14 x(tativ)S -16 x(e)S 677 y 5347 X(grand)S 181 x
(uni\014ed)S 182 x(theories.)S 678 y 6165 X(The)S 213 x(existing)S
212 x(appli)S -2 x(cations)S 213 x(of)S 212 x(solar)S 211 x
(neutrino)S 212 x(observ)S -31 x(ations)S 212 x(to)S 212 x(the)S 
214 x(study)S 212 x(of)S 212 x(the)S 677 y 5347 X(particle)S 192 x
(ph)S -15 x(ysics)S 193 x(prop)S 15 x(erties)S 193 x(of)S 193 x
(neutrinos)S 193 x(are)S 193 x(based)S 193 x(up)S 15 x(on)S 193 x
(comparisons)S 192 x(of)S 193 x(observ)S -31 x(a-)S 678 y 5347 X
(tions)S 172 x(with)S 172 x(the)S 173 x(predictions)S 172 x(of)S 
172 x(a)S 173 x(com)S -16 x(bined)S 173 x(standard)S 172 x(mo)S 14 x
(del,)S 172 x(the)S 173 x(standard)S 172 x(mo)S 15 x(del)S 172 x(of)S
1 PP EP

1000 BP 39600 30600 PM 0 0 XY
3815 Y 5347 X F34(electro)S -15 x(w)S -15 x(eak)S 205 x(theory)S 
205 x(plus)S 205 x(the)S 206 x(standard)S 204 x(solar)S 204 x(mo)S
15 x(del.)S 204 x(The)S 206 x(standard)S 205 x(solar)S 203 x(mo)S 
15 x(del)S 205 x(is)S 678 y 5347 X(required)S 205 x(to)S 206 x
(predict)S 207 x(ho)S -16 x(w)S 207 x(man)S -16 x(y)S -46 x(,)S 206 x
(and)S 206 x(with)S 205 x(what)S 206 x(energies,)S 206 x(neutrinos)S
206 x(are)S 206 x(pro)S 15 x(duced)S 677 y 5347 X(in)S 224 x(the)S
226 x(solar)S 224 x(in)S -16 x(terior.)S 224 x(The)S 225 x(standard)S
225 x(electro)S -15 x(w)S -15 x(eak)S 225 x(mo)S 15 x(del)S
XP /F34 123 273 0 11 21 12 2 24 0
<FFFFFC FFFFFC>
PXLC RP
18376 5170 XY F34({or)S 224 x(some)S 225 x(mo)S 14 x(di\014cation)S
224 x(of)S 678 y 5347 X(the)S 218 x(standard)S 217 x(electro)S -15 x
(w)S -15 x(eak)S 218 x(mo)S 14 x(del{is)S 217 x(required)S 217 x(to)S
218 x(determine)S 218 x(what)S 217 x(happ)S 15 x(ens)S 219 x(to)S 
217 x(the)S 677 y 5347 X(neutrinos)S 209 x(as)S 210 x(they)S 211 x
(pass)S 210 x(through)S 209 x(the)S 211 x(the)S 210 x(enormous)S 
209 x(amoun)S -16 x(t)S 210 x(of)S 210 x(matter)S 209 x(in)S 210 x
(the)S 211 x(sun)S 678 y 5347 X(and)S 181 x(tra)S -15 x(v)S -16 x
(el)S 182 x(from)S 181 x(the)S 182 x(solar)S 181 x(in)S -16 x
(terior)S 181 x(to)S 181 x(detec)S 2 x(tors)S 181 x(on)S 182 x
(earth.)S
XP /F34 68 418 1 0 30 30 31 32 0
<FFFFF000 FFFFFE00 07C01F00 07C00780 07C003C0 07C001E0 07C001F0
 07C000F0 07C000F8 07C00078 07C00078 07C0007C 07C0007C 07C0007C
 07C0007C 07C0007C 07C0007C 07C0007C 07C0007C 07C0007C 07C00078
 07C00078 07C000F8 07C000F0 07C000F0 07C001E0 07C003C0 07C00780
 07C01F00 FFFFFE00 FFFFF000>
PXLC RP
6165 7880 XY F34(Do)S 139 x(neutrinos)S 139 x(c)S -15 x(hange)S 139 x
(their)S 139 x(\015a)S -15 x(v)S -15 x(or)S 138 x(from)S 138 x
(electron-t)S -15 x(yp)S 15 x(e)S 140 x(to)S 139 x(some)S 140 x
(other)S 139 x(t)S -15 x(yp)S 15 x(e)S 140 x(during)S 677 y 5347 X
(their)S
XP /F34 106 167 -3 -9 9 30 40 16 0
<0070 00F8 00F8 00F8 0070 0000 0000 0000 0000 0000 0000 0078 07F8
 07F8 00F8 0078 0078 0078 0078 0078 0078 0078 0078 0078 0078 0078
 0078 0078 0078 0078 0078 0078 0078 0078 7078 F878 F8F0 F9F0 7FE0
 3F80>
PXLC RP
6689 8557 XY F34(journey)S 218 x(from)S 218 x(the)S 220 x(sun)S 220 x
(to)S 219 x(the)S 220 x(earth)S
XP /F34 63 258 3 0 16 31 32 16 0
<0FC0 3FF0 7878 E03C F83C F83C F83C F83C 707C 0078 00F8 01E0 01C0
 03C0 0380 0380 0300 0300 0300 0300 0300 0300 0000 0000 0000 0000
 0000 0700 0F80 0F80 0F80 0700>
PXLC RP
14967 8557 XY F34(?)S 219 x(The)S 219 x(simplest)S 219 x(v)S -15 x
(ersion)S 219 x(of)S 218 x(the)S 220 x(standard)S 678 y 5347 X
(electro)S -15 x(w)S -15 x(eak)S 246 x(mo)S 15 x(del)S 246 x(sa)S 
-15 x(ys:)S 246 x(\\)S
XP /F34 78 410 1 0 31 30 31 32 0
<FFE01FFE FFE01FFE 07F001E0 07F800C0 07F800C0 06FC00C0 06FC00C0
 067E00C0 063F00C0 063F00C0 061F80C0 060FC0C0 060FC0C0 0607E0C0
 0603E0C0 0603F0C0 0601F8C0 0601F8C0 0600FCC0 06007EC0 06007EC0
 06003FC0 06001FC0 06001FC0 06000FC0 060007C0 060007C0 060003C0
 0F0003C0 FFF001C0 FFF000C0>
PXLC RP
11670 9235 XY F34(No.")S 246 x(Neutrinos)S 246 x(ha)S -15 x(v)S -16 x
(e)S 248 x(zero)S 246 x(masses)S 247 x(in)S 246 x(this)S 247 x(mo)S
14 x(del)S 247 x(and)S 677 y 5347 X(lepton)S 170 x(\015a)S -15 x(v)S
-15 x(or)S 170 x(is)S 170 x(conserv)S -15 x(ed.)S 171 x(Ho)S -15 x
(w)S -16 x(ev)S -15 x(er,)S 171 x(there)S 171 x(is)S 170 x(at)S 171 x
(least)S 171 x(one)S 171 x(b)S 15 x(eautiful)S 170 x(theory)S 170 x
(of)S 170 x(ho)S -15 x(w)S 678 y 5347 X(the)S 132 x(standard)S 132 x
(theory)S 132 x(of)S 132 x(electro)S -15 x(w)S -15 x(eak)S 132 x(in)S
-16 x(teractions)S 132 x(ma)S -16 x(y)S 133 x(b)S 15 x(e)S 132 x(mo)S
15 x(di\014ed,)S 131 x(called)S 132 x(the)S 133 x(MSW)S 677 y 5347 X
(e\013ect)S 220 x(\(after)S 218 x(its)S 219 x(disco)S -16 x(v)S -15 x
(erers)S
XP /F13 53 212 2 0 14 20 21 16 0
<7030 7FF0 7FE0 7FC0 6000 6000 6000 6000 6F80 7FE0 78F0 6070 0078
 0078 7078 F078 F078 F0F0 F1F0 7FE0 1F80>
PXLC RP
11728 11069 XY F13(5,6)S 198 y 24 x F34(\),)S 219 x(whic)S -16 x(h)S
219 x(sho)S -15 x(ws)S 219 x(that)S 218 x(if)S 218 x(at)S 219 x
(least)S 218 x(one)S 219 x(neutrino)S 218 x(has)S 219 x(a)S 678 y 
5347 X(\014nite)S 178 x(mass,)S 177 x(then)S 178 x(neutrinos)S 177 x
(of)S 178 x(the)S 178 x(electron)S 178 x(t)S -15 x(yp)S 15 x(e)S 
179 x(can)S 178 x(b)S 15 x(e)S 178 x(con)S -15 x(v)S -15 x(erted)S
178 x(to)S 178 x(neutrinos)S 177 x(of)S 677 y 5347 X(a)S 181 x
(di\013eren)S -15 x(t)S 182 x(t)S -15 x(yp)S 15 x(e)S 183 x(that)S
181 x(are)S 182 x(m)S -16 x(uc)S -14 x(h)S 182 x(mor)S -2 x(e)S 183 x
(di)S
XP /F34 14 456 0 0 36 31 32 40 0
<001FC0FE00 00FFF7FF00 03F07F8780 07C0FE0F80 0780FC0F80 0F00F80F80
 0F00F80F80 0F00780000 0F00780000 0F00780000 0F00780000 0F00780000
 FFFFFFFF80 FFFFFFFF80 0F00780780 0F00780780 0F00780780 0F00780780
 0F00780780 0F00780780 0F00780780 0F00780780 0F00780780 0F00780780
 0F00780780 0F00780780 0F00780780 0F00780780 0F00780780 0F00780780
 FFE3FF3FF8 FFE3FF3FF8>
PXLC RP
14548 12622 XY F34(\016cult)S 182 x(to)S 181 x(detec)S 2 x(t.)S
XP /F48 50 336 3 0 23 31 32 24 0
<03FC00 1FFF80 3FFFC0 7C3FE0 FC1FF0 FE0FF0 FE07F8 FE07F8 FE03F8
 7C03F8 3807F8 0007F8 0007F0 000FF0 000FE0 001FC0 001F80 003F00
 007E00 00F800 01F000 01E038 03C038 078038 0F0038 1C0078 3FFFF0
 7FFFF0 FFFFF0 FFFFF0 FFFFF0 FFFFF0>
PXLC RP
5347 13697 XY F48(2.)S 672 x(R)S
XP /F48 101 307 2 0 22 21 22 24 0
<01FE00 07FF80 1FC7E0 3F03E0 3F03F0 7E01F0 7E01F8 FE01F8 FE01F8
 FFFFF8 FFFFF8 FE0000 FE0000 FE0000 FE0000 7E0000 7F0000 3F0038
 1F8078 0FE1F0 07FFE0 00FF80>
PXLC RP
7044 13697 XY F48(e)S
XP /F48 115 265 2 0 18 21 22 24 0
<0FFF00 3FFF00 7C1F00 780F00 F00700 F00700 F80000 FF0000 FFF800
 7FFC00 7FFE00 1FFF00 0FFF80 007F80 E00F80 E00F80 F00780 F00780
 F80F00 FE1F00 FFFE00 C7F800>
PXLC RP
7351 13697 XY F48(su)S
XP /F48 108 187 1 0 12 34 35 16 0
<FF80 FF80 FF80 1F80 1F80 1F80 1F80 1F80 1F80 1F80 1F80 1F80 1F80
 1F80 1F80 1F80 1F80 1F80 1F80 1F80 1F80 1F80 1F80 1F80 1F80 1F80
 1F80 1F80 1F80 1F80 1F80 1F80 FFF0 FFF0 FFF0>
PXLC RP
7990 13697 XY F48(lts)S
XP /F48 102 205 1 0 19 34 35 24 0
<003F80 01FFC0 03F7E0 07E7E0 07E7E0 0FC7E0 0FC3C0 0FC000 0FC000
 0FC000 0FC000 0FC000 0FC000 FFFC00 FFFC00 FFFC00 0FC000 0FC000
 0FC000 0FC000 0FC000 0FC000 0FC000 0FC000 0FC000 0FC000 0FC000
 0FC000 0FC000 0FC000 0FC000 0FC000 7FFC00 7FFC00 7FFC00>
PXLC RP
8927 13697 XY F48(fro)S
XP /F48 109 560 2 0 43 21 22 48 0
<FF87F80FF000 FF9FFC3FF800 FFBC7E78FC00 1FF07FE0FE00 1FE03FC07E00
 1FC03F807E00 1FC03F807E00 1F803F007E00 1F803F007E00 1F803F007E00
 1F803F007E00 1F803F007E00 1F803F007E00 1F803F007E00 1F803F007E00
 1F803F007E00 1F803F007E00 1F803F007E00 1F803F007E00 FFF1FFE3FFC0
 FFF1FFE3FFC0 FFF1FFE3FFC0>
PXLC RP
9744 13697 XY F48(m)S 224 x(t)S
XP /F48 104 374 2 0 27 34 35 32 0
<FF800000 FF800000 FF800000 1F800000 1F800000 1F800000 1F800000
 1F800000 1F800000 1F800000 1F800000 1F800000 1F800000 1F87F000
 1F9FFC00 1FF8FC00 1FF07E00 1FE07E00 1FC07E00 1FC07E00 1F807E00
 1F807E00 1F807E00 1F807E00 1F807E00 1F807E00 1F807E00 1F807E00
 1F807E00 1F807E00 1F807E00 1F807E00 FFF1FFC0 FFF1FFC0 FFF1FFC0>
PXLC RP
10790 13697 XY F48(he)S
XP /F48 70 423 2 0 31 33 34 32 0
<FFFFFFF8 FFFFFFF8 FFFFFFF8 07F003F8 07F000F8 07F00078 07F00038
 07F0003C 07F0003C 07F0381C 07F0381C 07F0381C 07F0381C 07F07800
 07F0F800 07FFF800 07FFF800 07FFF800 07F0F800 07F07800 07F03800
 07F03800 07F03800 07F03800 07F00000 07F00000 07F00000 07F00000
 07F00000 07F00000 07F00000 FFFFE000 FFFFE000 FFFFE000>
PXLC RP
11694 13697 XY F48(F)S -56 x(our)S 224 x(O)S
XP /F48 112 374 2 -10 27 21 32 32 0
<FF9FE000 FFFFFC00 FFF0FE00 1FC07F00 1F803F80 1F803F80 1F803F80
 1F801FC0 1F801FC0 1F801FC0 1F801FC0 1F801FC0 1F801FC0 1F801FC0
 1F801FC0 1F803F80 1F803F80 1F807F00 1FC07F00 1FF1FE00 1FFFF800
 1F9FE000 1F800000 1F800000 1F800000 1F800000 1F800000 1F800000
 1F800000 FFF00000 FFF00000 FFF00000>
PXLC RP
13775 13697 XY F48(p)S 18 x(er)S
XP /F48 97 327 2 0 25 21 22 24 0
<07FE00 1FFF80 3F0FC0 3F07E0 3F07F0 3F03F0 1E03F0 0003F0 0003F0
 01FFF0 0FFFF0 3FE3F0 7F83F0 7F03F0 FF03F0 FE03F0 FE03F0 FE07F0
 FF0FF8 7F1FFF 3FFDFF 0FF0FF>
PXLC RP
14748 13697 XY F48(atin)S
XP /F48 103 336 1 -11 25 21 33 32 0
<03FE3F00 0FFFFF80 1F8FFF80 3F07E780 7E03F780 7E03F000 7E03F000
 7E03F000 7E03F000 7E03F000 7E03F000 3F07E000 1F8FC000 1FFF8000
 3BFE0000 38000000 38000000 3C000000 3FFFE000 3FFFFC00 1FFFFE00
 1FFFFF00 3FFFFF00 7C007F80 F8001F80 F8000F80 F8000F80 F8000F80
 FC001F80 7E003F00 3F80FE00 0FFFF800 01FFC000>
PXLC RP
15897 13697 XY F48(g)S 225 x(E)S
XP /F48 120 355 1 0 27 21 22 32 0
<FFF0FFE0 FFF0FFE0 FFF0FFE0 07E07800 07F0F000 03F8E000 01F9E000
 00FFC000 007F8000 007F0000 003F0000 001F8000 003FC000 007FE000
 00F7F000 01E3F000 01C1F800 03C1FC00 0780FE00 FFC1FFE0 FFC1FFE0
 FFC1FFE0>
PXLC RP
16899 13697 XY F48(xp)S 18 x(erimen)S -19 x(ts)S 14680 Y 6165 X F34
(There)S 207 x(are)S 207 x(four)S 206 x(op)S 15 x(erating)S 206 x
(solar)S 206 x(neutrino)S 206 x(exp)S 16 x(erimen)S -16 x(ts,)S 207 x
(three)S 208 x(of)S 206 x(whic)S -15 x(h)S 207 x(use)S 208 x(ra-)S
677 y 5347 X(dio)S 14 x(c)S -15 x(hemical)S 168 x(detec)S 2 x(tion)S
168 x(\(one)S 168 x(c)S -14 x(hlori)S -2 x(ne)S 169 x(and)S 169 x(t)S
-15 x(w)S -16 x(o)S 169 x(gall)S -2 x(ium)S 168 x(detectors\))S 170 x
(and)S 168 x(one)S 169 x(detector)S 677 y 5347 X(whic)S -16 x(h)S 
182 x(is)S 182 x(electronic)S 182 x(\(the)S 182 x(Kamiok)S -31 x
(ande)S 182 x(pure)S 182 x(w)S -16 x(ater)S 182 x(detec)S 2 x
(tor\).)S 678 y 6165 X(The)S 186 x(\014rst,)S 185 x(and)S 185 x(for)S
185 x(t)S -15 x(w)S -16 x(o)S 186 x(decades)S 186 x(the)S 186 x
(only)S -46 x(,)S 185 x(solar)S 184 x(neutrino)S 185 x(exp)S 16 x
(erimen)S -16 x(t)S 186 x(uses)S 186 x(a)S 186 x(ra-)S 677 y 5347 X
(dio)S 14 x(c)S -15 x(hemical)S 140 x(c)S -15 x(hlori)S -2 x(ne)S 
141 x(detector)S 141 x(to)S 139 x(observ)S -15 x(e)S 141 x
(electron-t)S -15 x(yp)S 15 x(e)S 140 x(neutrinos)S 140 x(via)S 139 x
(the)S 140 x(reaction)S
XP /F13 55 212 2 0 16 21 22 16 0
<6000 7FFE 7FFE 7FFE E01C C038 C070 00E0 00C0 01C0 0380 0380 0780
 0700 0700 0F00 0F00 0F00 0F00 0F00 0F00 0F00>
PXLC RP
24152 17191 XY F13(7)S 18744 Y 11947 X F35(\027)S
XP /F13 101 188 1 0 13 13 14 16 0
<0FC0 3FF0 7CF0 7878 FFF8 FFF8 F000 F000 F000 F000 7818 7C38 1FF0
 0FE0>
PXLC RP
12217 18826 XY F13(e)S -82 y 146 x F34(+)S -225 y 303 x F13(37)S 
225 y 25 x F34(Cl)S 151 x F36(!)S 151 x F34(e)S -225 y F15(\000)S 
225 y 147 x F34(+)S -225 y 304 x F13(37)S 225 y 24 x F34(Ar)S F35(:)S
23692 X F34(\(2\))S 19746 Y 5347 X(As)S 144 x(Ra)S -16 x(y)S 144 x
(will)S 142 x(describ)S 15 x(e)S 144 x(for)S 143 x(us)S 144 x(in)S
144 x(the)S 144 x(next)S 144 x(talk,)S 143 x(the)S -198 y 144 x F13
(37)S 198 y 25 x F34(Ar)S 143 x(atoms)S 143 x(pro)S 15 x(duced)S 
145 x(b)S -16 x(y)S 144 x(neutrino)S 677 y 5347 X(capture)S 191 x
(are)S 191 x(extracted)S 191 x(c)S -14 x(hemicall)S -2 x(y)S 191 x
(from)S 189 x(the)S 192 x(0.)S
XP /F34 54 273 2 -1 19 29 31 24 0
<00FE00 03FF00 07C780 0F0F80 1E0F80 3C0F80 3C0F80 780000 780000
 700000 F06000 F3FC00 F7FE00 F60F00 FC0780 F80780 F80380 F003C0
 F003C0 F003C0 F003C0 F003C0 7003C0 7003C0 7803C0 780780 380780
 3C0F00 1E1E00 0FFC00 03F000>
PXLC RP
16013 20423 XY F34(6)S 191 x(kilo)S -2 x(tons)S 191 x(of)S 190 x
(\015uid,)S
XP /F35 67 391 2 -1 32 31 33 32 0
<0000FE06 0007FF8E 001FC3DC 007E00FC 00F8007C 01F0007C 03E00038
 07C00038 0F800038 0F800038 1F000030 1F000030 3E000030 3E000000
 7E000000 7C000000 7C000000 7C000000 FC000000 F8000000 F8000000
 F80001C0 F8000180 F8000180 78000380 78000300 7C000700 3C000E00
 3E001C00 1F007800 0FC1F000 03FFC000 00FF0000>
PXLC RP
20567 20423 XY F35(C)S 82 y F13(2)S -82 y 25 x F35(C)S
XP /F35 108 163 2 0 12 31 32 16 0
<03E0 1FE0 1FE0 03C0 03C0 07C0 07C0 0780 0780 0F80 0F80 0F00 0F00
 1F00 1F00 1E00 1E00 3E00 3E00 3C00 3C00 7C00 7C00 7800 7800 F980
 FB80 F300 F300 FF00 7E00 3E00>
PXLC RP
21623 20423 XY F35(l)S 82 y F13(4)S -82 y 24 x F34(,)S 191 x(in)S 
190 x(whic)S -15 x(h)S 677 y 5347 X(they)S 182 x(are)S 181 x
(created)S 182 x(and)S 181 x(are)S 181 x(then)S 182 x(coun)S -15 x
(ted)S 182 x(using)S 181 x(their)S 181 x(c)S -15 x(haracteristic)S
181 x(radioa)S -2 x(ctivit)S -15 x(y)S 181 x(in)S 678 y 5347 X
(small)S -2 x(,)S 176 x(gaseous)S 176 x(prop)S 14 x(ortional)S 174 x
(coun)S -14 x(ters.)S 175 x(The)S 177 x(c)S -15 x(hlorine)S 175 x
(exp)S 16 x(erimen)S -16 x(t)S 176 x(detec)S 2 x(ts)S 176 x
(neutrinos)S 677 y 5347 X(ab)S 15 x(out)S 181 x(0.8)S 181 x(MeV)S 
182 x(with)S 182 x(a)S 181 x(sensitivit)S -15 x(y)S 181 x(that)S 
182 x(dep)S 16 x(ends)S 182 x(up)S 16 x(on)S 181 x(energy)S -45 x(,)S
181 x(but)S 182 x(without)S 678 y 6165 X(As)S 185 x(Gene)S 185 x
(Beier)S 185 x(will)S 183 x(discuss)S 185 x(more)S 184 x(fully)S 
183 x(in)S 184 x(the)S 185 x(next)S 185 x(talk,)S 184 x(the)S 185 x
(second)S 185 x(solar)S 183 x(neu-)S 677 y 5347 X(trino)S 210 x(exp)S
16 x(erimen)S -16 x(t)S 211 x(to)S 212 x(b)S 15 x(e)S 212 x(p)S 15 x
(erform)S -2 x(ed)S 212 x(is)S 211 x(based)S 212 x(up)S 15 x(on)S 
211 x(the)S 212 x(scattering)S 211 x(of)S 211 x(neutrinos)S 211 x(b)S
-15 x(y)S 678 y 5347 X(electrons,)S -198 y F13(2,3)S 25843 Y 12969 X 
F35(\027)S 157 x F34(+)S 121 x(e)S 152 x F36(!)S 152 x F35(\027)S
XP /F15 48 115 1 1 8 17 17 8 0
<0F 0F 0F 1F 1E 1E 1E 3C 3C 3C 78 78 78 70 F0 E0 60>
PXLC RP
15335 25617 XY F15(0)S 226 y 146 x F34(+)S 122 x(e)S -226 y F15(0)S
XP /F35 59 152 4 -9 9 4 14 8 0
<70 F8 FC FC 7C 0C 0C 0C 1C 18 38 70 F0 60>
PXLC RP
16524 25843 XY F35(;)S 23692 X F34(\()S
XP /F34 51 273 2 -1 19 29 31 24 0
<07F000 1FFC00 3C3F00 380F00 7C0F80 7C0F80 7E0780 7C0F80 7C0F80
 000F80 001F00 001E00 007E00 03F800 03F000 003C00 001F00 000F80
 000F80 0007C0 0007C0 7807C0 FC07C0 FC07C0 FC07C0 FC0F80 F80F80
 701F00 3C3E00 1FFC00 07F000>
PXLC RP
23904 25843 XY F34(3\))S 26844 Y 5347 X(whic)S -16 x(h)S 185 x(o)S
15 x(ccurs)S 185 x(inside)S 184 x(the)S 186 x(\014ducial)S 183 x
(mass)S 185 x(of)S 184 x(0.68)S 183 x(kilo-)S -2 x(tons)S 185 x(of)S
184 x(ultra)S 184 x(pure)S 184 x(w)S -15 x(ater.)S
XP /F34 79 425 3 -1 30 31 33 32 0
<001F8000 00FFF000 01E07800 07C03E00 0F801F00 0F000F00 1E000780
 3C0003C0 3C0003C0 7C0003E0 7C0003E0 780001E0 F80001F0 F80001F0
 F80001F0 F80001F0 F80001F0 F80001F0 F80001F0 F80001F0 F80001F0
 780001E0 7C0003E0 7C0003E0 3C0003C0 3E0007C0 1E000780 0F000F00
 0F801F00 07C03E00 01F0F800 00FFF000 001F8000>
PXLC RP
23222 26844 XY F34(Only)S 677 y 5347 X(the)S 182 x(v)S -16 x(ery)S
182 x(rare)S 181 x(8)S
XP /F34 22 273 3 25 18 26 2 16 0
<FFFF FFFF>
PXLC RP
8608 27890 XY F34(\026)S -369 y 181 x(solar)S 180 x(neutrinos)S 181 x
(are)S 181 x(detec)S 2 x(table)S 181 x(in)S 181 x(the)S 182 x(Kamio)S
-2 x(k)S -30 x(ande)S 182 x(I)S 15 x(I)S 181 x(exp)S 16 x(erimen)S
-16 x(t,)S 678 y 5347 X(for)S 200 x(whic)S -15 x(h)S 201 x(the)S 
202 x(lo)S -16 x(w)S -15 x(est)S 202 x(published)S 201 x(v)S -31 x
(alue)S 201 x(for)S 200 x(the)S 202 x(detec)S 2 x(tion)S 200 x
(threshold)S 201 x(is)S 201 x(7.5)S 200 x(MeV.)S 202 x(In)S 677 y 
5347 X(the)S 174 x(Kamiok)S -32 x(ande)S 174 x(I)S 16 x(I)S 173 x
(exp)S 16 x(erimen)S -16 x(t,)S 174 x(the)S 174 x(electrons)S 174 x
(are)S 174 x(detecte)S 2 x(d)S 174 x(b)S -16 x(y)S 174 x(the)S 175 x
(Cerenk)S -15 x(o)S -16 x(v)S 174 x(ligh)S -16 x(t)S 678 y 5347 X
(that)S 209 x(they)S 209 x(pro)S 15 x(duce)S 210 x(while)S 208 x(mo)S
-16 x(ving)S 209 x(through)S 208 x(the)S 210 x(w)S -16 x(ater.)S 
209 x(Neutrino)S 208 x(scattering)S 209 x(exp)S 16 x(er-)S 677 y 
5347 X(imen)S -16 x(ts)S 227 x(pro)S -16 x(vide)S 227 x(infor)S -2 x
(mation)S 226 x(that)S 227 x(is)S 226 x(not)S 227 x(a)S -16 x(v)S 
-30 x(ail)S -2 x(able)S 227 x(from)S 225 x(radio)S 14 x(c)S -15 x
(hemical)S 226 x(detec)S 2 x(tors,)S 678 y 5347 X(including)S 181 x
(the)S 183 x(direction)S 182 x(from)S 182 x(whic)S -16 x(h)S 183 x
(the)S 183 x(neutrinos)S 182 x(come,)S 183 x(the)S 183 x(precise)S
183 x(arriv)S -32 x(al)S 182 x(times)S 677 y 5347 X(for)S 213 x
(individual)S 213 x(ev)S -15 x(en)S -15 x(ts,)S 215 x(infor)S -2 x
(mation)S 214 x(ab)S 15 x(out)S 214 x(the)S 215 x(energy)S 215 x(sp)S
15 x(ectrum)S 215 x(of)S 214 x(the)S 216 x(neutrinos,)S 678 y 5347 X
(and)S 181 x(some)S 182 x(sensitivit)S -16 x(y)S 182 x(to)S 182 x(m)S
-16 x(uon)S 182 x(and)S 181 x(tau)S 182 x(neutrinos.)S 677 y 6165 X
(The)S 150 x(fact)S 149 x(that)S 150 x(the)S 150 x(neutrinos)S 150 x
(are)S 149 x(coming)S 149 x(from)S 148 x(the)S 150 x(sun)S 150 x(is)S
149 x(established)S 150 x(directly)S 149 x(b)S -15 x(y)S 678 y 5347 X
(the)S 213 x(Kamiok)S -31 x(ande)S 213 x(I)S 15 x(I)S 213 x(exp)S 
15 x(erimen)S -15 x(t)S 213 x(since)S 213 x(the)S 214 x(electrons)S
213 x(are)S 213 x(scattered)S 214 x(in)S 212 x(the)S 214 x(forw)S 
-17 x(ard)S
1 PP EP

1000 BP 39600 30600 PM 0 0 XY
3815 Y 5347 X F34(direction)S 196 x(in)S 196 x(reaction)S 197 x(Eq.)S
196 x(\(3\).)S 196 x(The)S 198 x(observ)S -16 x(ed)S 197 x
(directions)S 197 x(of)S 196 x(the)S 197 x(scattered)S 198 x
(electrons)S 678 y 5347 X(trace)S 182 x(out)S 182 x(the)S 182 x(p)S
15 x(osition)S 181 x(of)S 181 x(the)S 182 x(sun)S 182 x(in)S 182 x
(the)S 182 x(sky)S -45 x(.)S 677 y 6165 X(There)S 178 x(are)S 177 x
(t)S -15 x(w)S -16 x(o)S 177 x(galli)S -2 x(um)S 177 x(exp)S 16 x
(erimen)S -16 x(ts)S 178 x(in)S 177 x(progr)S -2 x(ess,)S 178 x
(GALLE)S
XP /F34 88 410 1 0 31 30 31 32 0
<7FFC7FF8 7FFC7FF8 07F01F80 03E00F00 03F00E00 01F80C00 00F81800
 00FC3800 007C3000 007E6000 003FE000 001FC000 001FC000 000F8000
 000FC000 0007E000 0007E000 000FF000 000DF000 0019F800 0038FC00
 00307C00 00607E00 00E03E00 00C03F00 01C01F80 01800F80 03800FC0
 0FC00FE0 FFF07FFE FFF07FFE>
PXLC RP
19449 5170 XY F34(X)S -198 y F13(8,)S
XP /F13 57 212 1 0 15 20 21 16 0
<0FE0 1FF0 7C78 783C F01C F01E F01E F01E F01E F03E 787E 3FFE 1FDE
 021E 001C 783C 783C 7878 78F0 3FE0 1F80>
PXLC RP
20188 4972 XY F13(9)S 198 y 201 x F34(and)S 178 x(SA)S -15 x(GE,)S
-198 y F13(1)S
XP /F13 48 212 1 0 15 20 21 16 0
<07C0 1FF0 3C78 783C 701C 701C F01E F01E F01E F01E F01E F01E F01E
 F01E F01E 701C 701C 783C 3C78 1FF0 07C0>
PXLC RP
23517 4972 XY F13(0)S
XP /F13 123 212 0 8 16 9 2 24 0
<FFFF80 FFFF80>
PXLC RP
23729 4972 XY F13({12)S 876 y 5347 X F34(that)S 195 x(pro)S -16 x
(vide)S 194 x(the)S 196 x(\014rst)S 195 x(observ)S -31 x(ational)S
193 x(inform)S -2 x(ation)S 194 x(ab)S 15 x(out)S 195 x(the)S 195 x
(lo)S -16 x(w)S 195 x(energy)S 195 x(neutrinos)S 677 y 5347 X(from)S
233 x(the)S 235 x(basic)S 235 x(proton-pr)S -2 x(oton)S 235 x
(reaction.)S 233 x(The)S 235 x(GALLEX)S 235 x(and)S 235 x(SA)S -15 x
(GE)S 235 x(exp)S 15 x(erimen)S -16 x(ts)S 678 y 5347 X(mak)S -16 x
(e)S 182 x(use)S 182 x(of)S 182 x(neutrino)S 181 x(absorption)S 180 x
(b)S -15 x(y)S 182 x(galli)S -2 x(um,)S 8557 Y 11800 X F35(\027)S 
82 y F13(e)S -82 y 147 x F34(+)S -225 y 303 x F13(71)S 225 y 25 x 
F34(Ga)S 151 x F36(!)S 152 x F34(e)S -225 y F15(\000)S 225 y 147 x 
F34(+)S -225 y 303 x F13(71)S 225 y 24 x F34(Ge)S F35(;)S 23692 X 
F34(\(4\))S 9559 Y 5347 X(whic)S -16 x(h)S 159 x(has)S 158 x(a)S 
159 x(threshold)S 158 x(of)S 158 x(only)S 157 x(0.23)S 157 x(MeV)S
159 x(for)S 158 x(the)S 159 x(detection)S 159 x(of)S 158 x
(electron-t)S -15 x(yp)S 15 x(e)S 159 x(neutri-)S 677 y 5347 X(nos.)S
163 x(This)S 164 x(lo)S -16 x(w)S 164 x(threshold)S 164 x(mak)S -16 x
(es)S 165 x(p)S 15 x(ossible)S 164 x(the)S 165 x(detection)S 165 x
(of)S 164 x(the)S 164 x(lo)S -15 x(w)S 164 x(energy)S 164 x
(neutrinos)S 678 y 5347 X(from)S 245 x(the)S 248 x(proton-proto)S 
-2 x(n)S 248 x(\(or)S 246 x F35(pp)S F34(\))S 248 x(reaction)S
XP /F34 59 152 4 -9 8 19 29 8 0
<70 F8 F8 F8 70 00 00 00 00 00 00 00 00 00 00 70 F8 F8 F8 78 18 18 18
 38 30 30 70 E0 40>
PXLC RP
15154 10914 XY F34(;)S 247 x(the)S 247 x F35(pp)S 248 x F34
(reaction)S 247 x(initiates)S 246 x(the)S 248 x(n)S -15 x(uclear)S
677 y 5347 X(fusion)S 187 x(c)S -14 x(hain)S 188 x(in)S 188 x(the)S
189 x(sun)S 189 x(b)S -15 x(y)S 188 x(pro)S 15 x(ducing)S 188 x
(neutrinos)S 189 x(with)S 188 x(a)S 188 x(maxim)S -17 x(um)S 189 x
(energy)S 189 x(of)S 188 x(only)S 678 y 5347 X(0.42)S 249 x(MeV.)S
250 x(Both)S 251 x(the)S 250 x(GALLEX)S 251 x(and)S 250 x(the)S 251 x
(SA)S -15 x(GE)S 250 x(exp)S 16 x(erimen)S -16 x(ts)S 251 x(use)S 
250 x(radio)S 14 x(c)S -14 x(hemical)S 677 y 5347 X(pro)S 14 x
(cedures)S 178 x(to)S 176 x(extract)S 177 x(and)S 177 x(coun)S -15 x
(t)S 176 x(a)S 177 x(smal)S -2 x(l)S 176 x(n)S -15 x(um)S -15 x(b)S
15 x(er)S 177 x(of)S 176 x(atoms)S 175 x(from)S 175 x(a)S 177 x
(larg)S -2 x(e)S 177 x(detec)S 2 x(tor,)S 677 y 5347 X(simi)S -2 x
(lar)S 181 x(to)S 182 x(what)S 181 x(is)S 182 x(done)S 182 x(in)S 
181 x(the)S 183 x(c)S -15 x(hlori)S -2 x(ne)S 183 x(exp)S 15 x
(erimen)S -15 x(t.)S 678 y 6165 X(Figure)S 143 x(1)S 143 x(sho)S 
-15 x(ws)S 144 x(a)S 143 x(comparison)S 142 x(b)S 15 x(et)S -14 x(w)S
-15 x(een)S 144 x(the)S 144 x(predictions)S 144 x(of)S 143 x(the)S
144 x(standard)S 143 x(mo)S 14 x(del)S -198 y F13(13)S 875 y 5347 X 
F34(and)S 213 x(the)S 214 x(four)S 212 x(op)S 15 x(erating)S 212 x
(solar)S 212 x(neutrino)S 213 x(exp)S 16 x(erimen)S -16 x(ts)S -198 y 
F13(2,3,7{1)S -2 x(2,14)S 198 y 237 x F34(.)S 213 x(The)S 214 x
(unit)S 213 x(used)S 214 x(for)S 678 y 5347 X(the)S 179 x(three)S 
179 x(radio)S 14 x(c)S -15 x(hemical)S 178 x(exp)S 16 x(erimen)S 
-16 x(ts)S 179 x(is)S 179 x(a)S
XP /F35 83 335 3 -1 27 31 33 32 0
<0007E180 001FFB80 007C3F00 00F01F00 00E00F00 01C00F00 03C00600
 03800600 03800600 03800600 07800000 07C00000 03E00000 03F80000
 03FF8000 01FFE000 00FFE000 003FF000 0003F000 0000F000 0000F000
 00007000 00007000 30007000 30007000 7000F000 7000E000 7001E000
 7801C000 FC038000 FF0F0000 EFFE0000 C3F80000>
PXLC RP
15172 15656 XY F35(S)S
XP /F35 78 439 2 0 37 30 31 40 0
<00FF807FF0 00FF807FF0 000F800F80 000FC00600 000FC00E00 001FC00E00
 001BE00C00 0019E00C00 0019E01C00 0039F01C00 0030F01800 0030F01800
 0030F83800 0070783800 0060783000 00607C3000 00603C7000 00E03C7000
 00C03C6000 00C03E6000 00C01EE000 01C01EE000 01801FC000 01800FC000
 01800FC000 03800FC000 0300078000 0700078000 0F80078000 7FF0038000
 FFF0030000>
PXLC RP
15538 15656 XY F35(N)S
XP /F35 85 373 4 -1 32 30 32 32 0
<7FFC3FF8 7FFC3FF8 07C007C0 07800300 0F800700 0F800600 0F000600
 0F000600 1F000E00 1F000C00 1E000C00 1E000C00 3E001C00 3E001800
 3C001800 3C001800 7C003800 7C003000 78003000 78003000 78007000
 F8006000 F0006000 F000E000 F000C000 F001C000 F0038000 70070000
 780F0000 3C3C0000 1FF80000 07E00000>
PXLC RP
16036 15656 XY F35(U)S 390 x F34(=)S 330 x(10)S -198 y F15(\000)S 
F13(36)S 198 y 203 x F34(in)S -16 x(teractions)S 179 x(p)S 15 x(er)S
179 x(target)S 677 y 5347 X(atom)S 174 x(p)S 15 x(er)S 175 x
(second.)S 176 x(The)S 175 x(result)S 175 x(for)S 175 x(the)S 175 x
(Kamiok)S -31 x(ande)S 175 x(w)S -15 x(ater)S 175 x(exp)S 15 x
(erimen)S -15 x(t)S 175 x(is)S 175 x(expressed,)S 678 y 5347 X(foll)S
-2 x(o)S -15 x(wing)S 135 x(the)S 136 x(exp)S 16 x(erimen)S -16 x
(talists,)S 135 x(in)S 136 x(terms)S 136 x(of)S 135 x(a)S 136 x
(ratio)S 134 x(to)S 136 x(the)S 137 x(predicted)S 136 x(ev)S -14 x
(en)S -15 x(t)S 136 x(rate.)S 135 x(The)S 677 y 5347 X(errors)S 190 x
(sho)S -15 x(wn)S 191 x(are,)S 190 x(in)S 191 x(all)S 189 x(cases,)S
192 x(e\013ect)S 2 x(iv)S -16 x(e)S 191 x(1)S
XP /F35 27 312 2 0 25 19 20 24 0
<00FFFF 03FFFF 07FFFF 0FFFFE 1F8F80 3E07C0 3C03C0 7C03C0 7803C0
 7807C0 F80780 F00780 F00780 F00F00 F00F00 F01E00 703C00 787800
 3FF000 0FC000>
PXLC RP
14975 17688 XY F35(\033)S 210 x F34(uncertain)S -15 x(ties,)S 191 x
(where)S 191 x(I)S 191 x(ha)S -15 x(v)S -15 x(e)S 191 x(com)S -15 x
(bined)S 678 y 5347 X(quadrati)S -2 x(cally)S 184 x(the)S 186 x
(quoted)S 185 x(statistical)S 184 x(and)S 185 x(systematic)S 185 x
(errors.)S 184 x(I)S 185 x(wil)S -2 x(l)S 185 x(use)S 186 x
(throughout)S 677 y 5347 X(this)S 168 x(review)S 168 x(the)S 170 x
(standard)S 168 x(solar)S 167 x(mo)S 14 x(del)S 169 x(results)S 169 x
(of)S 168 x(Bahcall)S 167 x(and)S 169 x(Pinsonneault)S -198 y F13
(13)S 198 y 192 x F34(since)S 678 y 5347 X(this)S 184 x(is)S 184 x
(the)S 185 x(only)S 183 x(standard)S 184 x(solar)S 183 x(mo)S 14 x
(del)S 184 x(published)S 185 x(so)S 184 x(far)S 183 x(to)S 184 x
(tak)S -15 x(e)S 184 x(accoun)S -14 x(t)S 184 x(of)S 184 x(helium)S
677 y 5347 X(di\013usion.)S 250 x(Ho)S -15 x(w)S -15 x(ev)S -15 x
(er,)S 252 x(accurate)S 252 x(solar)S 251 x(mo)S 14 x(dels)S 252 x
(without)S 252 x(helium)S 251 x(di\013usion)S 251 x(ha)S -15 x(v)S
-16 x(e)S 253 x(b)S 15 x(een)S 678 y 5347 X(published)S 157 x(b)S 
-15 x(y)S 157 x(man)S -16 x(y)S 157 x(other)S 157 x(authors)S 156 x
(using)S 157 x(di\013eren)S -15 x(t)S 157 x(computer)S 158 x(co)S 
15 x(des)S 158 x(and)S 157 x(are)S 156 x(in)S 157 x(go)S 15 x(o)S 
15 x(d)S 677 y 5347 X(agreemen)S -16 x(t)S 211 x(with)S 210 x(the)S
210 x(Bahcall-Pinsonneault)S 209 x(solar)S 209 x(mo)S 14 x(del)S 
211 x(without)S 209 x(helium)S 210 x(di\013usion)S 677 y 5347 X
(\(see,)S 235 x(e.g.,)S -198 y -2 x F13(15{25)S 198 y 24 x F34(\).)S
234 x(Mo)S 15 x(dels)S 234 x(that)S 234 x(include)S 235 x(b)S 15 x
(oth)S 234 x(hea)S -15 x(vy)S 234 x(elemen)S -14 x(t)S 234 x
(di\013usion)S 234 x(and)S 234 x(helium)S 678 y 5347 X(di\013usion)S
255 x(ha)S -15 x(v)S -16 x(e)S 257 x(b)S 15 x(een)S 257 x(computed)S
256 x(b)S -15 x(y)S 256 x(b)S 15 x(oth)S 256 x(Charles)S 255 x
(Pro\016tt)S 256 x(and)S 256 x(b)S -15 x(y)S 255 x(Marc)S 256 x
(Pinson-)S 677 y 5347 X(neault)S 161 x(and)S 162 x(m)S -16 x(yself)S
161 x(\(see)S 163 x(the)S 162 x(talks)S 161 x(b)S -15 x(y)S 162 x
(Pro\016tt)S 161 x(and)S 161 x(b)S -15 x(y)S 161 x(Pinsonneault)S 
161 x(in)S 162 x(this)S 161 x(v)S -15 x(olume\).)S 678 y 5347 X
(\(These)S 208 x(as-of-y)S -17 x(et)S 208 x(unpublished)S 207 x
(results)S 207 x(with)S 207 x(hea)S -15 x(vy)S 206 x(elemen)S -14 x
(t)S 207 x(di\013usion)S 206 x(raise)S 206 x(sligh)S -16 x(tly)S 
677 y 5347 X(the)S 182 x(predicted)S 183 x(ev)S -15 x(en)S -15 x(t)S
182 x(rates.\))S 814 y 6165 X(All)S 157 x(four)S 157 x(of)S 158 x
(the)S 159 x(solar)S 157 x(neutrino)S 157 x(exp)S 16 x(erimen)S -16 x
(ts)S 159 x(yield)S 157 x(v)S -30 x(alues)S 158 x(less)S 158 x(than)S
158 x(the)S 159 x(predicted)S 678 y 5347 X(v)S -31 x(alue)S 196 x
(for)S 195 x(that)S 196 x(detec)S 2 x(tor)S 195 x(and)S 196 x
(outside)S 196 x(the)S 197 x(com)S -15 x(bined)S 196 x(errors.)S 
195 x(I)S 196 x(shall)S 195 x(presen)S -15 x(t)S 197 x(later)S 195 x
(in)S 677 y 5347 X(this)S 198 x(talk)S 197 x(a)S 198 x(detailed)S 
198 x(compariso)S -2 x(n)S 199 x(b)S 15 x(et)S -15 x(w)S -15 x(een)S
199 x(the)S 199 x(the)S 199 x(theoretical)S 197 x(predictions)S 198 x
(and)S 198 x(the)S 677 y 5347 X(measured)S 210 x(rates.)S 210 x(Ho)S
-15 x(w)S -15 x(ev)S -15 x(er,)S 210 x(one)S 211 x(fact)S 210 x(is)S
211 x(apparen)S -16 x(t)S 211 x(already)S 209 x(from)S 209 x(Figure)S
210 x(1.)S 210 x(The)S 211 x(dis-)S 678 y 5347 X(crepancy)S 160 x(b)S
15 x(et)S -15 x(w)S -15 x(een)S 160 x(theory)S 160 x(and)S 159 x
(observ)S -31 x(ation)S 159 x(is)S 159 x(ab)S 15 x(out)S 159 x(a)S
159 x(factor)S 159 x(of)S 159 x(3.5)S 158 x(for)S 159 x(the)S 160 x
(c)S -15 x(hlori)S -2 x(ne)S 677 y 5347 X(exp)S 15 x(erimen)S -15 x
(t,)S 189 x(whereas)S 191 x(the)S 191 x(discrepancy)S 190 x(is)S 
190 x(only)S 190 x(a)S 190 x(factor)S 189 x(of)S 190 x(2.0)S 189 x
(for)S 189 x(the)S 191 x(Kamiok)S -31 x(ande)S 678 y 5347 X(exp)S 
15 x(erimen)S -15 x(t.)S 175 x(These)S 177 x(t)S -15 x(w)S -15 x(o)S
176 x(exp)S 15 x(erimen)S -15 x(ts)S 176 x(are)S 176 x(prima)S -2 x
(rily)S 175 x(sensitiv)S -16 x(e)S 177 x(to)S 176 x(the)S 176 x
(same)S 176 x(neutrino)S 677 y 5347 X(source,)S 185 x(the)S 185 x
(rare,)S 184 x(high-energy)S 185 x(8)S 369 y -273 x(\026)S -369 y 
185 x(solar)S 183 x(neutrinos)S 185 x(\(maxim)S -17 x(um)S 185 x
(neutrino)S 185 x(energy)S 185 x(of)S 184 x(15)S 678 y 5347 X
(MeV\).)S 190 x(Th)S -15 x(us)S 191 x(the)S 191 x(disagr)S -2 x
(eemen)S -14 x(t)S 191 x(b)S 15 x(et)S -15 x(w)S -15 x(een)S 191 x
(theory)S 191 x(and)S 190 x(exp)S 15 x(erimen)S -15 x(t)S 190 x
(seems)S 191 x(to)S 191 x(dep)S 15 x(end)S 677 y 5347 X(up)S 15 x
(on)S 241 x(the)S 241 x(threshold)S 241 x(for)S 239 x(neutrino)S 
241 x(detection,)S 241 x(b)S 15 x(eing)S 241 x(larg)S -2 x(er)S 241 x
(for)S 240 x(c)S -15 x(hlorine)S 240 x(\(0.8)S 240 x(MeV)S 678 y 
5347 X(threshold\))S 235 x(than)S 235 x(for)S 235 x(the)S 236 x
(Kamiok)S -32 x(ande)S 236 x(\(w)S -15 x(ater\))S 235 x(exp)S 16 x
(erimen)S -16 x(t)S 236 x(\(7.5)S 234 x(MeV)S 236 x(threshold\).)S
677 y 5347 X(This)S 181 x(ma)S -16 x(y)S 182 x(b)S 15 x(e)S 182 x
(the)S 183 x(most)S 181 x(signi\014can)S -16 x(t)S 182 x(fact)S 182 x
(ab)S 15 x(out)S 181 x(the)S 183 x(solar)S 180 x(neutrino)S 182 x
(problem.)S
1 PP EP

1000 BP 39600 30600 PM 0 0 XY
6587 15917 XY 0 SPB
 save 10 dict begin /Figure exch def currentpoint translate
/showpage {} def 
0 SPE
6587 15917 XY 0 SPB
 20 dict begin
72 300 div dup scale
1 setlinejoin 0 setlinecap
/Helvetica findfont 55 scalefont setfont 
/B { currentpoint stroke newpath moveto } def
/F { moveto 0 setlinecap} def
/C { CS M 1 1 3 { pop 3 1 roll 255 div } for SET_COLOUR } def
/CS { currentpoint stroke } def
/CF { currentpoint fill } def
/L { lineto } def /M { moveto } def
/P { moveto 0 1 rlineto stroke } def
/T { currentlinecap exch 1 setlinecap show setlinecap } def
errordict /nocurrentpoint { pop 0 0 M currentpoint } put
/SET_COLOUR systemdict /colorimage known
{ (setrgbcolor) } { (pop pop pop) } ifelse cvx def
 80 50 translate
gsave
CS [] 0 setdash M
CS M 1 100 mul 72 div dup setlinewidth
/P [ /moveto cvx 0 5 -1 roll .05 add dup -2 div 0 exch /rmoveto cvx /rlineto cvx /stroke cvx ] cvx def
 0 0 0 C
CS M 1 100 mul 72 div dup setlinewidth
/P [ /moveto cvx 0 5 -1 roll .05 add dup -2 div 0 exch /rmoveto cvx /rlineto cvx /stroke cvx ] cvx def
 0 0 0 C
62 218 M 62 1014 L
118 1014 L
118 218 L
62 218 L
62 219 M 118 219 L
62 221 M 118 221 L
62 223 M 118 223 L
62 225 M 118 225 L
62 227 M 118 227 L
62 229 M 118 229 L
62 231 M 118 231 L
62 233 M 118 233 L
62 235 M 118 235 L
62 237 M 118 237 L
62 239 M 118 239 L
62 241 M 118 241 L
62 243 M 118 243 L
62 245 M 118 245 L
62 247 M 118 247 L
62 249 M 118 249 L
62 250 M 118 250 L
62 252 M 118 252 L
62 254 M 118 254 L
62 256 M 118 256 L
62 258 M 118 258 L
62 260 M 118 260 L
62 262 M 118 262 L
62 264 M 118 264 L
62 266 M 118 266 L
62 268 M 118 268 L
62 270 M 118 270 L
62 272 M 118 272 L
62 274 M 118 274 L
62 276 M 118 276 L
62 278 M 118 278 L
62 280 M 118 280 L
62 282 M 118 282 L
62 284 M 118 284 L
62 286 M 118 286 L
62 288 M 118 288 L
62 290 M 118 290 L
62 291 M 118 291 L
62 293 M 118 293 L
62 295 M 118 295 L
62 297 M 118 297 L
62 299 M 118 299 L
62 301 M 118 301 L
62 303 M 118 303 L
62 305 M 118 305 L
62 307 M 118 307 L
62 309 M 118 309 L
62 311 M 118 311 L
62 313 M 118 313 L
62 315 M 118 315 L
62 317 M 118 317 L
62 319 M 118 319 L
62 321 M 118 321 L
62 323 M 118 323 L
62 325 M 118 325 L
62 327 M 118 327 L
62 329 M 118 329 L
62 331 M 118 331 L
62 333 M 118 333 L
62 334 M 118 334 L
62 336 M 118 336 L
62 338 M 118 338 L
62 340 M 118 340 L
62 342 M 118 342 L
62 344 M 118 344 L
62 346 M 118 346 L
62 348 M 118 348 L
62 350 M 118 350 L
62 352 M 118 352 L
62 354 M 118 354 L
62 356 M 118 356 L
62 358 M 118 358 L
62 360 M 118 360 L
62 362 M 118 362 L
62 364 M 118 364 L
62 366 M 118 366 L
62 368 M 118 368 L
62 370 M 118 370 L
62 372 M 118 372 L
62 374 M 118 374 L
62 375 M 118 375 L
62 377 M 118 377 L
62 379 M 118 379 L
62 381 M 118 381 L
62 383 M 118 383 L
62 385 M 118 385 L
62 387 M 118 387 L
62 389 M 118 389 L
62 391 M 118 391 L
62 393 M 118 393 L
62 395 M 118 395 L
62 397 M 118 397 L
62 399 M 118 399 L
62 401 M 118 401 L
62 403 M 118 403 L
62 405 M 118 405 L
62 407 M 118 407 L
62 409 M 118 409 L
62 411 M 118 411 L
62 413 M 118 413 L
62 415 M 118 415 L
62 416 M 118 416 L
62 418 M 118 418 L
62 420 M 118 420 L
62 422 M 118 422 L
62 424 M 118 424 L
62 426 M 118 426 L
62 428 M 118 428 L
62 430 M 118 430 L
62 432 M 118 432 L
62 434 M 118 434 L
62 436 M 118 436 L
62 438 M 118 438 L
62 440 M 118 440 L
62 442 M 118 442 L
62 444 M 118 444 L
62 446 M 118 446 L
62 448 M 118 448 L
62 450 M 118 450 L
62 452 M 118 452 L
62 454 M 118 454 L
62 456 M 118 456 L
62 458 M 118 458 L
62 459 M 118 459 L
62 461 M 118 461 L
62 463 M 118 463 L
62 465 M 118 465 L
62 467 M 118 467 L
62 469 M 118 469 L
62 471 M 118 471 L
62 473 M 118 473 L
62 475 M 118 475 L
62 477 M 118 477 L
62 479 M 118 479 L
62 481 M 118 481 L
62 483 M 118 483 L
62 485 M 118 485 L
62 487 M 118 487 L
62 489 M 118 489 L
62 491 M 118 491 L
62 493 M 118 493 L
62 495 M 118 495 L
62 497 M 118 497 L
62 499 M 118 499 L
62 500 M 118 500 L
62 502 M 118 502 L
62 504 M 118 504 L
62 506 M 118 506 L
62 508 M 118 508 L
62 510 M 118 510 L
62 512 M 118 512 L
62 514 M 118 514 L
62 516 M 118 516 L
62 518 M 118 518 L
62 520 M 118 520 L
62 522 M 118 522 L
62 524 M 118 524 L
62 526 M 118 526 L
62 528 M 118 528 L
62 530 M 118 530 L
62 532 M 118 532 L
62 534 M 118 534 L
62 536 M 118 536 L
62 538 M 118 538 L
62 540 M 118 540 L
62 541 M 118 541 L
62 543 M 118 543 L
62 545 M 118 545 L
62 547 M 118 547 L
62 549 M 118 549 L
62 551 M 118 551 L
62 553 M 118 553 L
62 555 M 118 555 L
62 557 M 118 557 L
62 559 M 118 559 L
62 561 M 118 561 L
62 563 M 118 563 L
62 565 M 118 565 L
62 567 M 118 567 L
62 569 M 118 569 L
62 571 M 118 571 L
62 573 M 118 573 L
62 575 M 118 575 L
62 577 M 118 577 L
62 579 M 118 579 L
62 581 M 118 581 L
62 583 M 118 583 L
62 584 M 118 584 L
62 586 M 118 586 L
62 588 M 118 588 L
62 590 M 118 590 L
62 592 M 118 592 L
62 594 M 118 594 L
62 596 M 118 596 L
62 598 M 118 598 L
62 600 M 118 600 L
62 602 M 118 602 L
62 604 M 118 604 L
62 606 M 118 606 L
62 608 M 118 608 L
62 610 M 118 610 L
62 612 M 118 612 L
62 614 M 118 614 L
62 616 M 118 616 L
62 618 M 118 618 L
62 620 M 118 620 L
62 622 M 118 622 L
62 624 M 118 624 L
62 625 M 118 625 L
62 627 M 118 627 L
62 629 M 118 629 L
62 631 M 118 631 L
62 633 M 118 633 L
62 635 M 118 635 L
62 637 M 118 637 L
62 639 M 118 639 L
62 641 M 118 641 L
62 643 M 118 643 L
62 645 M 118 645 L
62 647 M 118 647 L
62 649 M 118 649 L
62 651 M 118 651 L
62 653 M 118 653 L
62 655 M 118 655 L
62 657 M 118 657 L
62 659 M 118 659 L
62 661 M 118 661 L
62 663 M 118 663 L
62 665 M 118 665 L
62 666 M 118 666 L
62 668 M 118 668 L
62 670 M 118 670 L
62 672 M 118 672 L
62 674 M 118 674 L
62 676 M 118 676 L
62 678 M 118 678 L
62 680 M 118 680 L
62 682 M 118 682 L
62 684 M 118 684 L
62 686 M 118 686 L
62 688 M 118 688 L
62 690 M 118 690 L
62 692 M 118 692 L
62 694 M 118 694 L
62 696 M 118 696 L
62 698 M 118 698 L
62 700 M 118 700 L
62 702 M 118 702 L
62 704 M 118 704 L
62 706 M 118 706 L
62 708 M 118 708 L
62 709 M 118 709 L
62 711 M 118 711 L
62 713 M 118 713 L
62 715 M 118 715 L
62 717 M 118 717 L
62 719 M 118 719 L
62 721 M 118 721 L
62 723 M 118 723 L
62 725 M 118 725 L
62 727 M 118 727 L
62 729 M 118 729 L
62 731 M 118 731 L
62 733 M 118 733 L
62 735 M 118 735 L
62 737 M 118 737 L
62 739 M 118 739 L
62 741 M 118 741 L
62 743 M 118 743 L
62 745 M 118 745 L
62 747 M 118 747 L
62 749 M 118 749 L
62 750 M 118 750 L
62 752 M 118 752 L
62 754 M 118 754 L
62 756 M 118 756 L
62 758 M 118 758 L
62 760 M 118 760 L
62 762 M 118 762 L
62 764 M 118 764 L
62 766 M 118 766 L
62 768 M 118 768 L
62 770 M 118 770 L
62 772 M 118 772 L
62 774 M 118 774 L
62 776 M 118 776 L
62 778 M 118 778 L
62 780 M 118 780 L
62 782 M 118 782 L
62 784 M 118 784 L
62 786 M 118 786 L
62 788 M 118 788 L
62 790 M 118 790 L
62 791 M 118 791 L
62 793 M 118 793 L
62 795 M 118 795 L
CS M
62 797 M 118 797 L
62 799 M 118 799 L
62 801 M 118 801 L
62 803 M 118 803 L
62 805 M 118 805 L
62 807 M 118 807 L
62 809 M 118 809 L
62 811 M 118 811 L
62 813 M 118 813 L
62 815 M 118 815 L
62 817 M 118 817 L
62 819 M 118 819 L
62 821 M 118 821 L
62 823 M 118 823 L
62 825 M 118 825 L
62 827 M 118 827 L
62 829 M 118 829 L
62 831 M 118 831 L
62 833 M 118 833 L
62 834 M 118 834 L
62 836 M 118 836 L
62 838 M 118 838 L
62 840 M 118 840 L
62 842 M 118 842 L
62 844 M 118 844 L
62 846 M 118 846 L
62 848 M 118 848 L
62 850 M 118 850 L
62 852 M 118 852 L
62 854 M 118 854 L
62 856 M 118 856 L
62 858 M 118 858 L
62 860 M 118 860 L
62 862 M 118 862 L
62 864 M 118 864 L
62 866 M 118 866 L
62 868 M 118 868 L
62 870 M 118 870 L
62 872 M 118 872 L
62 874 M 118 874 L
62 875 M 118 875 L
62 877 M 118 877 L
62 879 M 118 879 L
62 881 M 118 881 L
62 883 M 118 883 L
62 885 M 118 885 L
62 887 M 118 887 L
62 889 M 118 889 L
62 891 M 118 891 L
62 893 M 118 893 L
62 895 M 118 895 L
62 897 M 118 897 L
62 899 M 118 899 L
62 901 M 118 901 L
62 903 M 118 903 L
62 905 M 118 905 L
62 907 M 118 907 L
62 909 M 118 909 L
62 911 M 118 911 L
62 913 M 118 913 L
62 915 M 118 915 L
62 916 M 118 916 L
62 918 M 118 918 L
62 920 M 118 920 L
62 922 M 118 922 L
62 924 M 118 924 L
62 926 M 118 926 L
62 928 M 118 928 L
62 930 M 118 930 L
62 932 M 118 932 L
62 934 M 118 934 L
62 936 M 118 936 L
62 938 M 118 938 L
62 940 M 118 940 L
62 942 M 118 942 L
62 944 M 118 944 L
62 946 M 118 946 L
62 948 M 118 948 L
62 950 M 118 950 L
62 952 M 118 952 L
62 954 M 118 954 L
62 956 M 118 956 L
62 958 M 118 958 L
62 959 M 118 959 L
62 961 M 118 961 L
62 963 M 118 963 L
62 965 M 118 965 L
62 967 M 118 967 L
62 969 M 118 969 L
62 971 M 118 971 L
62 973 M 118 973 L
62 975 M 118 975 L
62 977 M 118 977 L
62 979 M 118 979 L
62 981 M 118 981 L
62 983 M 118 983 L
62 985 M 118 985 L
62 987 M 118 987 L
62 989 M 118 989 L
62 991 M 118 991 L
62 993 M 118 993 L
62 995 M 118 995 L
62 997 M 118 997 L
62 999 M 118 999 L
62 1000 M 118 1000 L
62 1002 M 118 1002 L
62 1004 M 118 1004 L
62 1006 M 118 1006 L
62 1008 M 118 1008 L
62 1010 M 118 1010 L
62 1012 M 118 1012 L
 255 255 255 C
109 837 M 118 849 L
99 837 M 118 863 L
88 837 M 118 877 L
78 837 M 118 890 L
68 837 M 118 904 L
62 844 M 118 918 L
62 857 M 118 932 L
62 871 M 118 946 L
62 885 M 118 959 L
62 899 M 118 973 L
62 913 M 118 987 L
62 926 M 118 1001 L
62 940 M 117 1014 L
62 954 M 107 1014 L
62 968 M 96 1014 L
62 982 M 86 1014 L
62 996 M 76 1014 L
62 1009 M 66 1014 L
CS M 3 100 mul 72 div dup setlinewidth
/P [ /moveto cvx 0 5 -1 roll .05 add dup -2 div 0 exch /rmoveto cvx /rlineto cvx /stroke cvx ] cvx def
62 925 M 118 925 L
CS M 1 100 mul 72 div dup setlinewidth
/P [ /moveto cvx 0 5 -1 roll .05 add dup -2 div 0 exch /rmoveto cvx /rlineto cvx /stroke cvx ] cvx def
 0 0 0 C
146 925 M CS [] 0 setdash M
146 912 M 156 939 M 152 937 L
151 935 L
151 931 L
152 928 L
156 927 L
161 927 L
165 928 L
166 931 L
166 935 L
165 937 L
161 939 L
156 939 L
153 937 L
152 935 L
152 931 L
153 928 L
156 927 L
161 927 M 164 928 L
165 931 L
165 935 L
164 937 L
161 939 L
156 927 M 152 926 L
151 925 L
149 922 L
149 917 L
151 914 L
152 913 L
156 912 L
161 912 L
165 913 L
166 914 L
167 917 L
167 922 L
166 925 L
165 926 L
161 927 L
156 927 M 153 926 L
152 925 L
151 922 L
151 917 L
152 914 L
153 913 L
156 912 L
161 912 M 164 913 L
165 914 L
166 917 L
166 922 L
165 925 L
164 926 L
161 927 L
178 914 M 176 913 L
178 912 L
179 913 L
178 914 L
196 939 M 192 937 L
189 934 L
188 927 L
188 923 L
189 917 L
192 913 L
196 912 L
198 912 L
202 913 L
205 917 L
206 923 L
206 927 L
205 934 L
202 937 L
198 939 L
196 939 L
193 937 L
192 936 L
191 934 L
189 927 L
189 923 L
191 917 L
192 914 L
193 913 L
196 912 L
198 912 M 201 913 L
202 914 L
204 917 L
205 923 L
205 927 L
204 934 L
202 936 L
201 937 L
198 939 L
225 934 M 225 912 L
215 923 M 236 923 L
215 912 M 236 912 L
249 934 M 251 935 L
255 939 L
255 912 L
254 937 M 254 912 L
249 912 M 260 912 L
273 914 M 272 913 L
273 912 L
274 913 L
273 914 L
291 939 M 287 937 L
285 934 L
283 927 L
283 923 L
285 917 L
287 913 L
291 912 L
294 912 L
297 913 L
300 917 L
301 923 L
301 927 L
300 934 L
297 937 L
294 939 L
291 939 L
288 937 L
287 936 L
286 934 L
285 927 L
285 923 L
286 917 L
287 914 L
288 913 L
291 912 L
294 912 M 296 913 L
297 914 L
299 917 L
300 923 L
300 927 L
299 934 L
297 936 L
296 937 L
294 939 L
CS [] 0 setdash M
 0 0 0 C
123 218 M 123 440 L
179 440 L
179 218 L
123 218 L
 0 0 0 C
174 400 M 179 406 L
164 400 M 179 420 L
153 400 M 179 434 L
143 400 M 174 440 L
133 400 M 163 440 L
123 401 M 153 440 L
123 414 M 143 440 L
123 428 M 132 440 L
CS M 3 100 mul 72 div dup setlinewidth
/P [ /moveto cvx 0 5 -1 roll .05 add dup -2 div 0 exch /rmoveto cvx /rlineto cvx /stroke cvx ] cvx def
123 420 M 179 420 L
CS M 1 100 mul 72 div dup setlinewidth
/P [ /moveto cvx 0 5 -1 roll .05 add dup -2 div 0 exch /rmoveto cvx /rlineto cvx /stroke cvx ] cvx def
207 420 M CS [] 0 setdash M
207 406 M 212 428 M 213 427 L
212 426 L
211 427 L
211 428 L
212 431 L
213 432 L
217 433 L
222 433 L
226 432 L
227 431 L
229 428 L
229 426 L
227 423 L
223 420 L
217 418 L
214 416 L
212 414 L
211 410 L
211 406 L
222 433 M 225 432 L
226 431 L
227 428 L
CS M
227 426 L
226 423 L
222 420 L
217 418 L
211 409 M 212 410 L
214 410 L
221 408 L
225 408 L
227 409 L
229 410 L
214 410 M 221 406 L
226 406 L
227 408 L
229 410 L
229 413 L
239 409 M 237 408 L
239 406 L
240 408 L
239 409 L
250 428 M 252 427 L
250 426 L
249 427 L
249 428 L
250 431 L
252 432 L
256 433 L
261 433 L
265 432 L
266 431 L
267 428 L
267 426 L
266 423 L
262 420 L
256 418 L
253 416 L
250 414 L
249 410 L
249 406 L
261 433 M 263 432 L
265 431 L
266 428 L
266 426 L
265 423 L
261 420 L
256 418 L
249 409 M 250 410 L
253 410 L
259 408 L
263 408 L
266 409 L
267 410 L
253 410 M 259 406 L
265 406 L
266 408 L
267 410 L
267 413 L
281 433 M 277 432 L
276 429 L
276 426 L
277 423 L
281 422 L
286 422 L
290 423 L
292 426 L
292 429 L
290 432 L
286 433 L
281 433 L
279 432 L
277 429 L
277 426 L
279 423 L
281 422 L
286 422 M 289 423 L
290 426 L
290 429 L
289 432 L
286 433 L
281 422 M 277 420 L
276 419 L
275 416 L
275 411 L
276 409 L
277 408 L
281 406 L
286 406 L
290 408 L
292 409 L
293 411 L
293 416 L
292 419 L
290 420 L
286 422 L
281 422 M 279 420 L
277 419 L
276 416 L
276 411 L
277 409 L
279 408 L
281 406 L
286 406 M 289 408 L
290 409 L
292 411 L
292 416 L
290 419 L
289 420 L
286 422 L
312 428 M 312 406 L
302 418 M 322 418 L
302 406 M 322 406 L
339 433 M 335 432 L
333 428 L
332 422 L
332 418 L
333 411 L
335 408 L
339 406 L
342 406 L
346 408 L
348 411 L
350 418 L
350 422 L
348 428 L
346 432 L
342 433 L
339 433 L
337 432 L
335 431 L
334 428 L
333 422 L
333 418 L
334 411 L
335 409 L
337 408 L
339 406 L
342 406 M 344 408 L
346 409 L
347 411 L
348 418 L
348 422 L
347 428 L
346 431 L
344 432 L
342 433 L
360 409 M 358 408 L
360 406 L
361 408 L
360 409 L
371 428 M 373 427 L
371 426 L
370 427 L
370 428 L
371 431 L
373 432 L
377 433 L
382 433 L
386 432 L
387 431 L
388 428 L
388 426 L
387 423 L
383 420 L
377 418 L
374 416 L
371 414 L
370 410 L
370 406 L
382 433 M 384 432 L
386 431 L
387 428 L
387 426 L
386 423 L
382 420 L
377 418 L
370 409 M 371 410 L
374 410 L
380 408 L
384 408 L
387 409 L
388 410 L
374 410 M 380 406 L
386 406 L
387 408 L
388 410 L
388 413 L
397 428 M 398 427 L
397 426 L
396 427 L
396 428 L
397 431 L
398 432 L
402 433 L
407 433 L
411 432 L
413 429 L
413 426 L
411 423 L
407 422 L
404 422 L
407 433 M 410 432 L
411 429 L
411 426 L
410 423 L
407 422 L
410 420 L
413 418 L
414 415 L
414 411 L
413 409 L
411 408 L
407 406 L
402 406 L
398 408 L
397 409 L
396 411 L
396 413 L
397 414 L
398 413 L
397 411 L
411 419 M 413 415 L
413 411 L
411 409 L
410 408 L
407 406 L
CS [] 0 setdash M
118 175 M CS [] 0 setdash M
97 161 M 119 184 M 120 180 L
120 188 L
119 184 L
116 187 L
113 188 L
110 188 L
106 187 L
104 184 L
102 182 L
101 178 L
101 171 L
102 167 L
104 165 L
106 162 L
110 161 L
113 161 L
116 162 L
119 165 L
120 167 L
110 188 M 107 187 L
105 184 L
104 182 L
102 178 L
102 171 L
104 167 L
105 165 L
107 162 L
110 161 L
131 188 M 131 161 L
132 188 M 132 161 L
127 188 M 132 188 L
127 161 M 136 161 L
CS [] 0 setdash M
 0 0 0 C
901 218 M 901 1014 L
956 1014 L
956 218 L
901 218 L
901 219 M 956 219 L
901 221 M 956 221 L
901 223 M 956 223 L
901 225 M 956 225 L
901 227 M 956 227 L
901 229 M 956 229 L
901 231 M 956 231 L
901 233 M 956 233 L
901 235 M 956 235 L
901 237 M 956 237 L
901 239 M 956 239 L
901 241 M 956 241 L
901 243 M 956 243 L
901 245 M 956 245 L
901 247 M 956 247 L
901 249 M 956 249 L
901 250 M 956 250 L
901 252 M 956 252 L
901 254 M 956 254 L
901 256 M 956 256 L
901 258 M 956 258 L
901 260 M 956 260 L
901 262 M 956 262 L
901 264 M 956 264 L
901 266 M 956 266 L
901 268 M 956 268 L
901 270 M 956 270 L
901 272 M 956 272 L
901 274 M 956 274 L
901 276 M 956 276 L
901 278 M 956 278 L
901 280 M 956 280 L
901 282 M 956 282 L
901 284 M 956 284 L
901 286 M 956 286 L
901 288 M 956 288 L
901 290 M 956 290 L
901 291 M 956 291 L
901 293 M 956 293 L
901 295 M 956 295 L
CS M
901 297 M 956 297 L
901 299 M 956 299 L
901 301 M 956 301 L
901 303 M 956 303 L
901 305 M 956 305 L
901 307 M 956 307 L
901 309 M 956 309 L
901 311 M 956 311 L
901 313 M 956 313 L
901 315 M 956 315 L
901 317 M 956 317 L
901 319 M 956 319 L
901 321 M 956 321 L
901 323 M 956 323 L
901 325 M 956 325 L
901 327 M 956 327 L
901 329 M 956 329 L
901 331 M 956 331 L
901 333 M 956 333 L
901 334 M 956 334 L
901 336 M 956 336 L
901 338 M 956 338 L
901 340 M 956 340 L
901 342 M 956 342 L
901 344 M 956 344 L
901 346 M 956 346 L
901 348 M 956 348 L
901 350 M 956 350 L
901 352 M 956 352 L
901 354 M 956 354 L
901 356 M 956 356 L
901 358 M 956 358 L
901 360 M 956 360 L
901 362 M 956 362 L
901 364 M 956 364 L
901 366 M 956 366 L
901 368 M 956 368 L
901 370 M 956 370 L
901 372 M 956 372 L
901 374 M 956 374 L
901 375 M 956 375 L
901 377 M 956 377 L
901 379 M 956 379 L
901 381 M 956 381 L
901 383 M 956 383 L
901 385 M 956 385 L
901 387 M 956 387 L
901 389 M 956 389 L
901 391 M 956 391 L
901 393 M 956 393 L
901 395 M 956 395 L
901 397 M 956 397 L
901 399 M 956 399 L
901 401 M 956 401 L
901 403 M 956 403 L
901 405 M 956 405 L
901 407 M 956 407 L
901 409 M 956 409 L
901 411 M 956 411 L
901 413 M 956 413 L
901 415 M 956 415 L
901 416 M 956 416 L
901 418 M 956 418 L
901 420 M 956 420 L
901 422 M 956 422 L
901 424 M 956 424 L
901 426 M 956 426 L
901 428 M 956 428 L
901 430 M 956 430 L
901 432 M 956 432 L
901 434 M 956 434 L
901 436 M 956 436 L
901 438 M 956 438 L
901 440 M 956 440 L
901 442 M 956 442 L
901 444 M 956 444 L
901 446 M 956 446 L
901 448 M 956 448 L
901 450 M 956 450 L
901 452 M 956 452 L
901 454 M 956 454 L
901 456 M 956 456 L
901 458 M 956 458 L
901 459 M 956 459 L
901 461 M 956 461 L
901 463 M 956 463 L
901 465 M 956 465 L
901 467 M 956 467 L
901 469 M 956 469 L
901 471 M 956 471 L
901 473 M 956 473 L
901 475 M 956 475 L
901 477 M 956 477 L
901 479 M 956 479 L
901 481 M 956 481 L
901 483 M 956 483 L
901 485 M 956 485 L
901 487 M 956 487 L
901 489 M 956 489 L
901 491 M 956 491 L
901 493 M 956 493 L
901 495 M 956 495 L
901 497 M 956 497 L
901 499 M 956 499 L
901 500 M 956 500 L
901 502 M 956 502 L
901 504 M 956 504 L
901 506 M 956 506 L
901 508 M 956 508 L
901 510 M 956 510 L
901 512 M 956 512 L
901 514 M 956 514 L
901 516 M 956 516 L
901 518 M 956 518 L
901 520 M 956 520 L
901 522 M 956 522 L
901 524 M 956 524 L
901 526 M 956 526 L
901 528 M 956 528 L
901 530 M 956 530 L
901 532 M 956 532 L
901 534 M 956 534 L
901 536 M 956 536 L
901 538 M 956 538 L
901 540 M 956 540 L
901 541 M 956 541 L
901 543 M 956 543 L
901 545 M 956 545 L
901 547 M 956 547 L
901 549 M 956 549 L
901 551 M 956 551 L
901 553 M 956 553 L
901 555 M 956 555 L
901 557 M 956 557 L
901 559 M 956 559 L
901 561 M 956 561 L
901 563 M 956 563 L
901 565 M 956 565 L
901 567 M 956 567 L
901 569 M 956 569 L
901 571 M 956 571 L
901 573 M 956 573 L
901 575 M 956 575 L
901 577 M 956 577 L
901 579 M 956 579 L
901 581 M 956 581 L
901 583 M 956 583 L
901 584 M 956 584 L
901 586 M 956 586 L
901 588 M 956 588 L
901 590 M 956 590 L
901 592 M 956 592 L
901 594 M 956 594 L
901 596 M 956 596 L
901 598 M 956 598 L
901 600 M 956 600 L
901 602 M 956 602 L
901 604 M 956 604 L
901 606 M 956 606 L
901 608 M 956 608 L
901 610 M 956 610 L
901 612 M 956 612 L
901 614 M 956 614 L
901 616 M 956 616 L
901 618 M 956 618 L
901 620 M 956 620 L
901 621 M 956 621 L
901 623 M 956 623 L
901 625 M 956 625 L
901 627 M 956 627 L
901 629 M 956 629 L
901 631 M 956 631 L
901 633 M 956 633 L
901 635 M 956 635 L
901 637 M 956 637 L
901 639 M 956 639 L
901 641 M 956 641 L
901 643 M 956 643 L
901 645 M 956 645 L
901 647 M 956 647 L
901 649 M 956 649 L
901 651 M 956 651 L
901 653 M 956 653 L
901 655 M 956 655 L
901 657 M 956 657 L
901 659 M 956 659 L
901 661 M 956 661 L
901 662 M 956 662 L
901 664 M 956 664 L
901 666 M 956 666 L
901 668 M 956 668 L
901 670 M 956 670 L
901 672 M 956 672 L
901 674 M 956 674 L
901 676 M 956 676 L
901 678 M 956 678 L
901 680 M 956 680 L
901 682 M 956 682 L
901 684 M 956 684 L
901 686 M 956 686 L
901 688 M 956 688 L
901 690 M 956 690 L
901 692 M 956 692 L
901 694 M 956 694 L
901 696 M 956 696 L
901 698 M 956 698 L
901 700 M 956 700 L
901 702 M 956 702 L
901 704 M 956 704 L
901 705 M 956 705 L
901 707 M 956 707 L
901 709 M 956 709 L
901 711 M 956 711 L
901 713 M 956 713 L
901 715 M 956 715 L
901 717 M 956 717 L
901 719 M 956 719 L
901 721 M 956 721 L
901 723 M 956 723 L
901 725 M 956 725 L
901 727 M 956 727 L
901 729 M 956 729 L
901 731 M 956 731 L
901 733 M 956 733 L
901 735 M 956 735 L
901 737 M 956 737 L
901 739 M 956 739 L
901 741 M 956 741 L
901 743 M 956 743 L
901 745 M 956 745 L
901 746 M 956 746 L
901 748 M 956 748 L
901 750 M 956 750 L
901 752 M 956 752 L
901 754 M 956 754 L
901 756 M 956 756 L
901 758 M 956 758 L
901 760 M 956 760 L
901 762 M 956 762 L
901 764 M 956 764 L
901 766 M 956 766 L
901 768 M 956 768 L
901 770 M 956 770 L
901 772 M 956 772 L
901 774 M 956 774 L
901 776 M 956 776 L
901 778 M 956 778 L
901 780 M 956 780 L
901 782 M 956 782 L
901 784 M 956 784 L
901 786 M 956 786 L
901 787 M 956 787 L
901 789 M 956 789 L
901 791 M 956 791 L
901 793 M 956 793 L
901 795 M 956 795 L
901 797 M 956 797 L
901 799 M 956 799 L
901 801 M 956 801 L
901 803 M 956 803 L
901 805 M 956 805 L
901 807 M 956 807 L
901 809 M 956 809 L
901 811 M 956 811 L
901 813 M 956 813 L
901 815 M 956 815 L
901 817 M 956 817 L
901 819 M 956 819 L
901 821 M 956 821 L
901 823 M 956 823 L
901 825 M 956 825 L
901 827 M 956 827 L
901 829 M 956 829 L
901 830 M 956 830 L
901 832 M 956 832 L
901 834 M 956 834 L
901 836 M 956 836 L
901 838 M 956 838 L
901 840 M 956 840 L
901 842 M 956 842 L
901 844 M 956 844 L
901 846 M 956 846 L
901 848 M 956 848 L
901 850 M 956 850 L
901 852 M 956 852 L
901 854 M 956 854 L
901 856 M 956 856 L
901 858 M 956 858 L
901 860 M 956 860 L
901 862 M 956 862 L
901 864 M 956 864 L
901 866 M 956 866 L
901 868 M 956 868 L
901 870 M 956 870 L
901 871 M 956 871 L
901 873 M 956 873 L
901 875 M 956 875 L
901 877 M 956 877 L
901 879 M 956 879 L
901 881 M 956 881 L
CS M
901 883 M 956 883 L
901 885 M 956 885 L
901 887 M 956 887 L
901 889 M 956 889 L
901 891 M 956 891 L
901 893 M 956 893 L
901 895 M 956 895 L
901 897 M 956 897 L
901 899 M 956 899 L
901 901 M 956 901 L
901 903 M 956 903 L
901 905 M 956 905 L
901 907 M 956 907 L
901 909 M 956 909 L
901 911 M 956 911 L
901 912 M 956 912 L
901 914 M 956 914 L
901 916 M 956 916 L
901 918 M 956 918 L
901 920 M 956 920 L
901 922 M 956 922 L
901 924 M 956 924 L
901 926 M 956 926 L
901 928 M 956 928 L
901 930 M 956 930 L
901 932 M 956 932 L
901 934 M 956 934 L
901 936 M 956 936 L
901 938 M 956 938 L
901 940 M 956 940 L
901 942 M 956 942 L
901 944 M 956 944 L
901 946 M 956 946 L
901 948 M 956 948 L
901 950 M 956 950 L
901 952 M 956 952 L
901 954 M 956 954 L
901 955 M 956 955 L
901 957 M 956 957 L
901 959 M 956 959 L
901 961 M 956 961 L
901 963 M 956 963 L
901 965 M 956 965 L
901 967 M 956 967 L
901 969 M 956 969 L
901 971 M 956 971 L
901 973 M 956 973 L
901 975 M 956 975 L
901 977 M 956 977 L
901 979 M 956 979 L
901 981 M 956 981 L
901 983 M 956 983 L
901 985 M 956 985 L
901 987 M 956 987 L
901 989 M 956 989 L
901 991 M 956 991 L
901 993 M 956 993 L
901 995 M 956 995 L
901 996 M 956 996 L
901 998 M 956 998 L
901 1000 M 956 1000 L
901 1002 M 956 1002 L
901 1004 M 956 1004 L
901 1006 M 956 1006 L
901 1008 M 956 1008 L
901 1010 M 956 1010 L
901 1012 M 956 1012 L
 255 255 255 C
955 934 M 956 936 L
944 934 M 956 950 L
934 934 M 956 963 L
924 934 M 956 977 L
913 934 M 956 991 L
903 934 M 956 1005 L
901 944 M 953 1014 L
901 958 M 942 1014 L
901 972 M 932 1014 L
901 986 M 922 1014 L
901 1000 M 911 1014 L
901 1013 M 901 1014 L
CS M 3 100 mul 72 div dup setlinewidth
/P [ /moveto cvx 0 5 -1 roll .05 add dup -2 div 0 exch /rmoveto cvx /rlineto cvx /stroke cvx ] cvx def
901 974 M 956 974 L
 0 0 0 C
CS M 1 100 mul 72 div dup setlinewidth
/P [ /moveto cvx 0 5 -1 roll .05 add dup -2 div 0 exch /rmoveto cvx /rlineto cvx /stroke cvx ] cvx def
979 974 M CS [] 0 setdash M
979 960 M 986 982 M 989 983 L
993 987 L
993 960 L
991 986 M 991 960 L
986 960 M 998 960 L
1009 982 M 1011 981 L
1009 979 L
1008 981 L
1008 982 L
1009 984 L
1011 986 L
1015 987 L
1020 987 L
1024 986 L
1025 983 L
1025 979 L
1024 977 L
1020 975 L
1016 975 L
1020 987 M 1022 986 L
1024 983 L
1024 979 L
1022 977 L
1020 975 L
1022 974 L
1025 972 L
1026 969 L
1026 965 L
1025 963 L
1024 961 L
1020 960 L
1015 960 L
1011 961 L
1009 963 L
1008 965 L
1008 966 L
1009 968 L
1011 966 L
1009 965 L
1024 973 M 1025 969 L
1025 965 L
1024 963 L
1022 961 L
1020 960 L
1035 982 M 1036 981 L
1035 979 L
1034 981 L
1034 982 L
1035 984 L
1036 986 L
1040 987 L
1045 987 L
1049 986 L
1051 984 L
1052 982 L
1052 979 L
1051 977 L
1047 974 L
1040 972 L
1038 970 L
1035 968 L
1034 964 L
1034 960 L
1045 987 M 1048 986 L
1049 984 L
1051 982 L
1051 979 L
1049 977 L
1045 974 L
1040 972 L
1034 963 M 1035 964 L
1038 964 L
1044 961 L
1048 961 L
1051 963 L
1052 964 L
1038 964 M 1044 960 L
1049 960 L
1051 961 L
1052 964 L
1052 966 L
1071 982 M 1071 960 L
1061 972 M 1082 972 L
1061 960 M 1082 960 L
1091 987 M 1091 979 L
1091 982 M 1092 984 L
1094 987 L
1097 987 L
1103 983 L
1106 983 L
1107 984 L
1109 987 L
1092 984 M 1094 986 L
1097 986 L
1103 983 L
1109 987 M 1109 983 L
1107 979 L
1102 973 L
1101 970 L
1100 966 L
1100 960 L
1107 979 M 1101 973 L
1100 970 L
1098 966 L
1098 960 L
CS [] 0 setdash M
840 218 M 840 745 L
895 745 L
895 218 L
840 218 L
888 527 M 895 536 L
878 527 M 895 550 L
868 527 M 895 564 L
858 527 M 895 578 L
847 527 M 895 592 L
840 531 M 895 605 L
840 545 M 895 619 L
840 559 M 895 633 L
840 573 M 895 647 L
840 586 M 895 661 L
840 600 M 895 675 L
840 614 M 895 688 L
840 628 M 895 702 L
840 642 M 895 716 L
840 655 M 895 730 L
840 669 M 895 744 L
840 683 M 886 745 L
840 697 M 876 745 L
840 711 M 865 745 L
840 725 M 855 745 L
840 738 M 845 745 L
CS M 3 100 mul 72 div dup setlinewidth
/P [ /moveto cvx 0 5 -1 roll .05 add dup -2 div 0 exch /rmoveto cvx /rlineto cvx /stroke cvx ] cvx def
840 636 M 895 636 L
CS M 1 100 mul 72 div dup setlinewidth
/P [ /moveto cvx 0 5 -1 roll .05 add dup -2 div 0 exch /rmoveto cvx /rlineto cvx /stroke cvx ] cvx def
812 636 M CS [] 0 setdash M
678 622 M 682 650 M 682 642 L
682 644 M 683 647 L
686 650 L
688 650 L
695 646 L
697 646 L
698 647 L
700 650 L
683 647 M 686 648 L
688 648 L
695 646 L
700 650 M 700 646 L
698 642 L
693 635 L
692 633 L
691 629 L
691 622 L
698 642 M 692 635 L
691 633 L
689 629 L
689 622 L
709 644 M 710 643 L
709 642 L
707 643 L
707 644 L
709 647 L
710 648 L
714 650 L
719 650 L
723 648 L
724 646 L
724 642 L
723 639 L
719 638 L
715 638 L
719 650 M 722 648 L
723 646 L
723 642 L
722 639 L
719 638 L
722 637 L
724 634 L
726 631 L
726 628 L
724 625 L
723 624 L
719 622 L
714 622 L
710 624 L
709 625 L
707 628 L
707 629 L
709 630 L
710 629 L
709 628 L
723 635 M 724 631 L
724 628 L
723 625 L
722 624 L
719 622 L
745 644 M 745 622 L
734 634 M 755 634 L
734 622 M 755 622 L
768 644 M 771 646 L
774 650 L
774 622 L
773 648 M 773 622 L
768 622 M 780 622 L
807 640 M 805 637 L
803 634 L
799 633 L
798 633 L
794 634 L
791 637 L
790 640 L
790 642 L
791 646 L
794 648 L
798 650 L
800 650 L
804 648 L
807 646 L
808 642 L
808 634 L
807 629 L
805 626 L
803 624 L
799 622 L
795 622 L
CS M
792 624 L
791 626 L
791 628 L
792 629 L
794 628 L
792 626 L
798 633 M 795 634 L
792 637 L
791 640 L
791 642 L
792 646 L
795 648 L
798 650 L
800 650 M 803 648 L
805 646 L
807 642 L
807 634 L
805 629 L
804 626 L
802 624 L
799 622 L
CS [] 0 setdash M
962 218 M 962 737 L
1017 737 L
1017 218 L
962 218 L
1007 603 M 1017 617 L
997 603 M 1017 631 L
987 603 M 1017 645 L
976 603 M 1017 659 L
966 603 M 1017 672 L
962 612 M 1017 686 L
962 626 M 1017 700 L
962 639 M 1017 714 L
962 653 M 1017 728 L
962 667 M 1014 737 L
962 681 M 1004 737 L
962 695 M 994 737 L
962 708 M 983 737 L
962 722 M 973 737 L
962 736 M 963 737 L
CS M 3 100 mul 72 div dup setlinewidth
/P [ /moveto cvx 0 5 -1 roll .05 add dup -2 div 0 exch /rmoveto cvx /rlineto cvx /stroke cvx ] cvx def
962 670 M 1017 670 L
CS M 1 100 mul 72 div dup setlinewidth
/P [ /moveto cvx 0 5 -1 roll .05 add dup -2 div 0 exch /rmoveto cvx /rlineto cvx /stroke cvx ] cvx def
1040 670 M CS [] 0 setdash M
1040 657 M 1044 684 M 1044 676 L
1044 679 M 1045 681 L
1047 684 L
1050 684 L
1056 680 L
1059 680 L
1060 681 L
1062 684 L
1045 681 M 1047 683 L
1050 683 L
1056 680 L
1062 684 M 1062 680 L
1060 676 L
1055 670 L
1054 667 L
1052 663 L
1052 657 L
1060 676 M 1054 670 L
1052 667 L
1051 663 L
1051 657 L
1086 675 M 1085 671 L
1082 668 L
1078 667 L
1077 667 L
1073 668 L
1071 671 L
1069 675 L
1069 676 L
1071 680 L
1073 683 L
1077 684 L
1080 684 L
1083 683 L
1086 680 L
1087 676 L
1087 668 L
1086 663 L
1085 661 L
1082 658 L
1078 657 L
1074 657 L
1072 658 L
1071 661 L
1071 662 L
1072 663 L
1073 662 L
1072 661 L
1077 667 M 1074 668 L
1072 671 L
1071 675 L
1071 676 L
1072 680 L
1074 683 L
1077 684 L
1080 684 M 1082 683 L
1085 680 L
1086 676 L
1086 668 L
1085 663 L
1083 661 L
1081 658 L
1078 657 L
1107 679 M 1107 657 L
1096 668 M 1117 668 L
1096 657 M 1117 657 L
1130 679 M 1132 680 L
1136 684 L
1136 657 L
1135 683 M 1135 657 L
1130 657 M 1141 657 L
1156 679 M 1158 680 L
1162 684 L
1162 657 L
1161 683 M 1161 657 L
1156 657 M 1167 657 L
1180 659 M 1179 658 L
1180 657 L
1181 658 L
1180 659 L
1190 684 M 1190 676 L
1190 679 M 1192 681 L
1194 684 L
1197 684 L
1203 680 L
1206 680 L
1207 681 L
1208 684 L
1192 681 M 1194 683 L
1197 683 L
1203 680 L
1208 684 M 1208 680 L
1207 676 L
1202 670 L
1201 667 L
1199 663 L
1199 657 L
1207 676 M 1201 670 L
1199 667 L
1198 663 L
1198 657 L
CS [] 0 setdash M
929 175 M CS [] 0 setdash M
901 161 M 923 184 M 924 180 L
924 188 L
923 184 L
920 187 L
916 188 L
914 188 L
910 187 L
907 184 L
906 182 L
905 178 L
905 171 L
906 167 L
907 165 L
910 162 L
914 161 L
916 161 L
920 162 L
923 165 L
914 188 M 911 187 L
909 184 L
907 182 L
906 178 L
906 171 L
907 167 L
909 165 L
911 162 L
914 161 L
923 171 M 923 161 L
924 171 M 924 161 L
919 171 M 928 171 L
937 176 M 937 175 L
936 175 L
936 176 L
937 178 L
939 179 L
945 179 L
947 178 L
948 176 L
950 174 L
950 165 L
951 162 L
952 161 L
948 176 M 948 165 L
950 162 L
952 161 L
954 161 L
948 174 M 947 173 L
939 171 L
936 170 L
934 167 L
934 165 L
936 162 L
939 161 L
943 161 L
946 162 L
948 165 L
939 171 M 937 170 L
936 167 L
936 165 L
937 162 L
939 161 L
CS [] 0 setdash M
 0 0 0 C
451 218 M 451 1016 L
507 1016 L
507 218 L
451 218 L
451 219 M 507 219 L
451 221 M 507 221 L
451 223 M 507 223 L
451 225 M 507 225 L
451 227 M 507 227 L
451 229 M 507 229 L
451 231 M 507 231 L
451 233 M 507 233 L
451 235 M 507 235 L
451 237 M 507 237 L
451 239 M 507 239 L
451 241 M 507 241 L
451 243 M 507 243 L
451 245 M 507 245 L
451 247 M 507 247 L
451 249 M 507 249 L
451 250 M 507 250 L
451 252 M 507 252 L
451 254 M 507 254 L
451 256 M 507 256 L
451 258 M 507 258 L
451 260 M 507 260 L
451 262 M 507 262 L
451 264 M 507 264 L
451 266 M 507 266 L
451 268 M 507 268 L
451 270 M 507 270 L
451 272 M 507 272 L
451 274 M 507 274 L
451 276 M 507 276 L
451 278 M 507 278 L
451 280 M 507 280 L
451 282 M 507 282 L
451 284 M 507 284 L
451 286 M 507 286 L
451 288 M 507 288 L
451 290 M 507 290 L
451 291 M 507 291 L
451 293 M 507 293 L
451 295 M 507 295 L
451 297 M 507 297 L
451 299 M 507 299 L
451 301 M 507 301 L
451 303 M 507 303 L
451 305 M 507 305 L
451 307 M 507 307 L
451 309 M 507 309 L
451 311 M 507 311 L
451 313 M 507 313 L
451 315 M 507 315 L
451 317 M 507 317 L
451 319 M 507 319 L
451 321 M 507 321 L
451 323 M 507 323 L
451 325 M 507 325 L
451 327 M 507 327 L
451 329 M 507 329 L
451 331 M 507 331 L
451 333 M 507 333 L
451 334 M 507 334 L
451 336 M 507 336 L
451 338 M 507 338 L
451 340 M 507 340 L
451 342 M 507 342 L
451 344 M 507 344 L
451 346 M 507 346 L
451 348 M 507 348 L
451 350 M 507 350 L
451 352 M 507 352 L
451 354 M 507 354 L
451 356 M 507 356 L
451 358 M 507 358 L
451 360 M 507 360 L
451 362 M 507 362 L
451 364 M 507 364 L
451 366 M 507 366 L
451 368 M 507 368 L
451 370 M 507 370 L
451 372 M 507 372 L
451 374 M 507 374 L
451 375 M 507 375 L
451 377 M 507 377 L
451 379 M 507 379 L
451 381 M 507 381 L
451 383 M 507 383 L
451 385 M 507 385 L
451 387 M 507 387 L
451 389 M 507 389 L
451 391 M 507 391 L
451 393 M 507 393 L
451 395 M 507 395 L
451 397 M 507 397 L
451 399 M 507 399 L
CS M
451 401 M 507 401 L
451 403 M 507 403 L
451 405 M 507 405 L
451 407 M 507 407 L
451 409 M 507 409 L
451 411 M 507 411 L
451 413 M 507 413 L
451 415 M 507 415 L
451 416 M 507 416 L
451 418 M 507 418 L
451 420 M 507 420 L
451 422 M 507 422 L
451 424 M 507 424 L
451 426 M 507 426 L
451 428 M 507 428 L
451 430 M 507 430 L
451 432 M 507 432 L
451 434 M 507 434 L
451 436 M 507 436 L
451 438 M 507 438 L
451 440 M 507 440 L
451 442 M 507 442 L
451 444 M 507 444 L
451 446 M 507 446 L
451 448 M 507 448 L
451 450 M 507 450 L
451 452 M 507 452 L
451 454 M 507 454 L
451 456 M 507 456 L
451 458 M 507 458 L
451 459 M 507 459 L
451 461 M 507 461 L
451 463 M 507 463 L
451 465 M 507 465 L
451 467 M 507 467 L
451 469 M 507 469 L
451 471 M 507 471 L
451 473 M 507 473 L
451 475 M 507 475 L
451 477 M 507 477 L
451 479 M 507 479 L
451 481 M 507 481 L
451 483 M 507 483 L
451 485 M 507 485 L
451 487 M 507 487 L
451 489 M 507 489 L
451 491 M 507 491 L
451 493 M 507 493 L
451 495 M 507 495 L
451 497 M 507 497 L
451 499 M 507 499 L
451 500 M 507 500 L
451 502 M 507 502 L
451 504 M 507 504 L
451 506 M 507 506 L
451 508 M 507 508 L
451 510 M 507 510 L
451 512 M 507 512 L
451 514 M 507 514 L
451 516 M 507 516 L
451 518 M 507 518 L
451 520 M 507 520 L
451 522 M 507 522 L
451 524 M 507 524 L
451 526 M 507 526 L
451 528 M 507 528 L
451 530 M 507 530 L
451 532 M 507 532 L
451 534 M 507 534 L
451 536 M 507 536 L
451 537 M 507 537 L
451 539 M 507 539 L
451 541 M 507 541 L
451 543 M 507 543 L
451 545 M 507 545 L
451 547 M 507 547 L
451 549 M 507 549 L
451 551 M 507 551 L
451 553 M 507 553 L
451 555 M 507 555 L
451 557 M 507 557 L
451 559 M 507 559 L
451 561 M 507 561 L
451 563 M 507 563 L
451 565 M 507 565 L
451 567 M 507 567 L
451 569 M 507 569 L
451 571 M 507 571 L
451 573 M 507 573 L
451 575 M 507 575 L
451 577 M 507 577 L
451 579 M 507 579 L
451 580 M 507 580 L
451 582 M 507 582 L
451 584 M 507 584 L
451 586 M 507 586 L
451 588 M 507 588 L
451 590 M 507 590 L
451 592 M 507 592 L
451 594 M 507 594 L
451 596 M 507 596 L
451 598 M 507 598 L
451 600 M 507 600 L
451 602 M 507 602 L
451 604 M 507 604 L
451 606 M 507 606 L
451 608 M 507 608 L
451 610 M 507 610 L
451 612 M 507 612 L
451 614 M 507 614 L
451 616 M 507 616 L
451 618 M 507 618 L
451 620 M 507 620 L
451 621 M 507 621 L
451 623 M 507 623 L
451 625 M 507 625 L
451 627 M 507 627 L
451 629 M 507 629 L
451 631 M 507 631 L
451 633 M 507 633 L
451 635 M 507 635 L
451 637 M 507 637 L
451 639 M 507 639 L
451 641 M 507 641 L
451 643 M 507 643 L
451 645 M 507 645 L
451 647 M 507 647 L
451 649 M 507 649 L
451 651 M 507 651 L
451 653 M 507 653 L
451 655 M 507 655 L
451 657 M 507 657 L
451 659 M 507 659 L
451 661 M 507 661 L
451 662 M 507 662 L
451 664 M 507 664 L
451 666 M 507 666 L
451 668 M 507 668 L
451 670 M 507 670 L
451 672 M 507 672 L
451 674 M 507 674 L
451 676 M 507 676 L
451 678 M 507 678 L
451 680 M 507 680 L
451 682 M 507 682 L
451 684 M 507 684 L
451 686 M 507 686 L
451 688 M 507 688 L
451 690 M 507 690 L
451 692 M 507 692 L
451 694 M 507 694 L
451 696 M 507 696 L
451 698 M 507 698 L
451 700 M 507 700 L
451 702 M 507 702 L
451 704 M 507 704 L
451 705 M 507 705 L
451 707 M 507 707 L
451 709 M 507 709 L
451 711 M 507 711 L
451 713 M 507 713 L
451 715 M 507 715 L
451 717 M 507 717 L
451 719 M 507 719 L
451 721 M 507 721 L
451 723 M 507 723 L
451 725 M 507 725 L
451 727 M 507 727 L
451 729 M 507 729 L
451 731 M 507 731 L
451 733 M 507 733 L
451 735 M 507 735 L
451 737 M 507 737 L
451 739 M 507 739 L
451 741 M 507 741 L
451 743 M 507 743 L
451 745 M 507 745 L
451 746 M 507 746 L
451 748 M 507 748 L
451 750 M 507 750 L
451 752 M 507 752 L
451 754 M 507 754 L
451 756 M 507 756 L
451 758 M 507 758 L
451 760 M 507 760 L
451 762 M 507 762 L
451 764 M 507 764 L
451 766 M 507 766 L
451 768 M 507 768 L
451 770 M 507 770 L
451 772 M 507 772 L
451 774 M 507 774 L
451 776 M 507 776 L
451 778 M 507 778 L
451 780 M 507 780 L
451 782 M 507 782 L
451 784 M 507 784 L
451 786 M 507 786 L
451 787 M 507 787 L
451 789 M 507 789 L
451 791 M 507 791 L
451 793 M 507 793 L
451 795 M 507 795 L
451 797 M 507 797 L
451 799 M 507 799 L
451 801 M 507 801 L
451 803 M 507 803 L
451 805 M 507 805 L
451 807 M 507 807 L
451 809 M 507 809 L
451 811 M 507 811 L
451 813 M 507 813 L
451 815 M 507 815 L
451 817 M 507 817 L
451 819 M 507 819 L
451 821 M 507 821 L
451 823 M 507 823 L
451 825 M 507 825 L
451 827 M 507 827 L
451 829 M 507 829 L
451 830 M 507 830 L
451 832 M 507 832 L
451 834 M 507 834 L
451 836 M 507 836 L
451 838 M 507 838 L
451 840 M 507 840 L
451 842 M 507 842 L
451 844 M 507 844 L
451 846 M 507 846 L
451 848 M 507 848 L
451 850 M 507 850 L
451 852 M 507 852 L
451 854 M 507 854 L
451 856 M 507 856 L
451 858 M 507 858 L
451 860 M 507 860 L
451 862 M 507 862 L
451 864 M 507 864 L
451 866 M 507 866 L
451 868 M 507 868 L
451 870 M 507 870 L
451 871 M 507 871 L
451 873 M 507 873 L
451 875 M 507 875 L
451 877 M 507 877 L
451 879 M 507 879 L
451 881 M 507 881 L
451 883 M 507 883 L
451 885 M 507 885 L
451 887 M 507 887 L
451 889 M 507 889 L
451 891 M 507 891 L
451 893 M 507 893 L
451 895 M 507 895 L
451 897 M 507 897 L
451 899 M 507 899 L
451 901 M 507 901 L
451 903 M 507 903 L
451 905 M 507 905 L
451 907 M 507 907 L
451 909 M 507 909 L
451 911 M 507 911 L
451 912 M 507 912 L
451 914 M 507 914 L
451 916 M 507 916 L
451 918 M 507 918 L
451 920 M 507 920 L
451 922 M 507 922 L
451 924 M 507 924 L
451 926 M 507 926 L
451 928 M 507 928 L
451 930 M 507 930 L
451 932 M 507 932 L
451 934 M 507 934 L
451 936 M 507 936 L
451 938 M 507 938 L
451 940 M 507 940 L
451 942 M 507 942 L
451 944 M 507 944 L
451 946 M 507 946 L
451 948 M 507 948 L
451 950 M 507 950 L
451 952 M 507 952 L
451 954 M 507 954 L
451 955 M 507 955 L
451 957 M 507 957 L
451 959 M 507 959 L
451 961 M 507 961 L
451 963 M 507 963 L
451 965 M 507 965 L
451 967 M 507 967 L
451 969 M 507 969 L
451 971 M 507 971 L
451 973 M 507 973 L
451 975 M 507 975 L
451 977 M 507 977 L
451 979 M 507 979 L
451 981 M 507 981 L
451 983 M 507 983 L
451 985 M 507 985 L
CS M
451 987 M 507 987 L
451 989 M 507 989 L
451 991 M 507 991 L
451 993 M 507 993 L
451 995 M 507 995 L
451 996 M 507 996 L
451 998 M 507 998 L
451 1000 M 507 1000 L
451 1002 M 507 1002 L
451 1004 M 507 1004 L
451 1006 M 507 1006 L
451 1008 M 507 1008 L
451 1010 M 507 1010 L
451 1012 M 507 1012 L
451 1014 M 507 1014 L
451 1016 M 507 1016 L
 255 255 255 C
499 820 M 507 831 L
488 820 M 507 845 L
478 820 M 507 858 L
468 820 M 507 872 L
457 820 M 507 886 L
451 825 M 507 900 L
451 839 M 507 914 L
451 853 M 507 927 L
451 867 M 507 941 L
451 881 M 507 955 L
451 895 M 507 969 L
451 908 M 507 983 L
451 922 M 507 996 L
451 936 M 507 1010 L
451 950 M 501 1016 L
451 964 M 490 1016 L
451 977 M 480 1016 L
451 991 M 470 1016 L
451 1005 M 459 1016 L
CS M 3 100 mul 72 div dup setlinewidth
/P [ /moveto cvx 0 5 -1 roll .05 add dup -2 div 0 exch /rmoveto cvx /rlineto cvx /stroke cvx ] cvx def
451 918 M 507 918 L
CS M 1 100 mul 72 div dup setlinewidth
/P [ /moveto cvx 0 5 -1 roll .05 add dup -2 div 0 exch /rmoveto cvx /rlineto cvx /stroke cvx ] cvx def
 0 0 0 C
534 918 M CS [] 0 setdash M
534 905 M 542 927 M 545 928 L
548 932 L
548 905 L
547 930 M 547 905 L
542 905 M 554 905 L
566 907 M 565 906 L
566 905 L
568 906 L
566 907 L
584 932 M 581 930 L
578 927 L
577 920 L
577 916 L
578 910 L
581 906 L
584 905 L
587 905 L
591 906 L
593 910 L
595 916 L
595 920 L
593 927 L
591 930 L
587 932 L
584 932 L
582 930 L
581 929 L
579 927 L
578 920 L
578 916 L
579 910 L
581 907 L
582 906 L
584 905 L
587 905 M 590 906 L
591 907 L
592 910 L
593 916 L
593 920 L
592 927 L
591 929 L
590 930 L
587 932 L
610 932 M 606 930 L
604 927 L
602 920 L
602 916 L
604 910 L
606 906 L
610 905 L
613 905 L
617 906 L
619 910 L
620 916 L
620 920 L
619 927 L
617 930 L
613 932 L
610 932 L
608 930 L
606 929 L
605 927 L
604 920 L
604 916 L
605 910 L
606 907 L
608 906 L
610 905 L
613 905 M 615 906 L
617 907 L
618 910 L
619 916 L
619 920 L
618 927 L
617 929 L
615 930 L
613 932 L
640 927 M 640 905 L
629 916 M 650 916 L
629 905 M 650 905 L
667 932 M 663 930 L
660 927 L
659 920 L
659 916 L
660 910 L
663 906 L
667 905 L
669 905 L
673 906 L
676 910 L
677 916 L
677 920 L
676 927 L
673 930 L
669 932 L
667 932 L
664 930 L
663 929 L
662 927 L
660 920 L
660 916 L
662 910 L
663 907 L
664 906 L
667 905 L
669 905 M 672 906 L
673 907 L
675 910 L
676 916 L
676 920 L
675 927 L
673 929 L
672 930 L
669 932 L
687 907 M 686 906 L
687 905 L
689 906 L
687 907 L
702 927 M 704 928 L
708 932 L
708 905 L
707 930 M 707 905 L
702 905 M 713 905 L
735 929 M 735 905 L
736 932 M 736 905 L
736 932 M 722 912 L
743 912 L
731 905 M 740 905 L
CS [] 0 setdash M
512 218 M 512 617 L
568 617 L
568 218 L
512 218 L
563 505 M 568 512 L
552 505 M 568 526 L
542 505 M 568 540 L
532 505 M 568 554 L
521 505 M 568 567 L
512 507 M 568 581 L
512 521 M 568 595 L
512 534 M 568 609 L
512 548 M 564 617 L
512 562 M 553 617 L
512 576 M 543 617 L
512 590 M 533 617 L
512 603 M 522 617 L
512 617 M 512 617 L
CS M 3 100 mul 72 div dup setlinewidth
/P [ /moveto cvx 0 5 -1 roll .05 add dup -2 div 0 exch /rmoveto cvx /rlineto cvx /stroke cvx ] cvx def
512 561 M 568 561 L
CS M 1 100 mul 72 div dup setlinewidth
/P [ /moveto cvx 0 5 -1 roll .05 add dup -2 div 0 exch /rmoveto cvx /rlineto cvx /stroke cvx ] cvx def
590 561 M CS [] 0 setdash M
590 548 M 601 575 M 598 574 L
595 570 L
594 563 L
594 559 L
595 553 L
598 549 L
601 548 L
604 548 L
608 549 L
610 553 L
612 559 L
612 563 L
610 570 L
608 574 L
604 575 L
601 575 L
599 574 L
598 572 L
596 570 L
595 563 L
595 559 L
596 553 L
598 550 L
599 549 L
601 548 L
604 548 M 607 549 L
608 550 L
609 553 L
610 559 L
610 563 L
609 570 L
608 572 L
607 574 L
604 575 L
622 550 M 621 549 L
622 548 L
623 549 L
622 550 L
644 572 M 644 548 L
645 575 M 645 548 L
645 575 M 631 555 L
652 555 L
640 548 M 649 548 L
675 566 M 674 562 L
671 559 L
667 558 L
666 558 L
662 559 L
659 562 L
658 566 L
658 567 L
659 571 L
662 574 L
666 575 L
668 575 L
672 574 L
675 571 L
676 567 L
676 559 L
675 554 L
674 552 L
671 549 L
667 548 L
663 548 L
661 549 L
659 552 L
659 553 L
661 554 L
662 553 L
661 552 L
666 558 M 663 559 L
661 562 L
659 566 L
659 567 L
661 571 L
663 574 L
666 575 L
668 575 M 671 574 L
674 571 L
675 567 L
675 559 L
674 554 L
672 552 L
670 549 L
667 548 L
695 570 M 695 548 L
685 559 M 706 559 L
685 548 M 706 548 L
722 575 M 719 574 L
716 570 L
715 563 L
715 559 L
716 553 L
719 549 L
722 548 L
725 548 L
729 549 L
731 553 L
733 559 L
733 563 L
731 570 L
729 574 L
725 575 L
722 575 L
720 574 L
719 572 L
717 570 L
716 563 L
716 559 L
717 553 L
719 550 L
720 549 L
722 548 L
725 548 M 728 549 L
729 550 L
730 553 L
731 559 L
CS M
731 563 L
730 570 L
729 572 L
728 574 L
725 575 L
743 550 M 742 549 L
743 548 L
744 549 L
743 550 L
761 575 M 757 574 L
755 570 L
753 563 L
753 559 L
755 553 L
757 549 L
761 548 L
764 548 L
767 549 L
770 553 L
771 559 L
771 563 L
770 570 L
767 574 L
764 575 L
761 575 L
758 574 L
757 572 L
756 570 L
755 563 L
755 559 L
756 553 L
757 550 L
758 549 L
761 548 L
764 548 M 766 549 L
767 550 L
769 553 L
770 559 L
770 563 L
769 570 L
767 572 L
766 574 L
764 575 L
785 575 M 782 574 L
780 571 L
780 567 L
782 564 L
785 563 L
791 563 L
795 564 L
796 567 L
796 571 L
795 574 L
791 575 L
785 575 L
783 574 L
782 571 L
782 567 L
783 564 L
785 563 L
791 563 M 793 564 L
795 567 L
795 571 L
793 574 L
791 575 L
785 563 M 782 562 L
780 561 L
779 558 L
779 553 L
780 550 L
782 549 L
785 548 L
791 548 L
795 549 L
796 550 L
797 553 L
797 558 L
796 561 L
795 562 L
791 563 L
785 563 M 783 562 L
782 561 L
780 558 L
780 553 L
782 550 L
783 549 L
785 548 L
791 548 M 793 549 L
795 550 L
796 553 L
796 558 L
795 561 L
793 562 L
791 563 L
CS [] 0 setdash M
507 175 M CS [] 0 setdash M
469 161 M 476 188 M 476 161 L
477 188 M 477 161 L
492 188 M 492 161 L
494 188 M 494 161 L
472 188 M 481 188 L
488 188 M 497 188 L
477 175 M 492 175 L
472 161 M 481 161 L
488 161 M 497 161 L
503 162 M 504 161 L
503 160 L
502 161 L
502 162 L
503 163 L
504 164 L
506 165 L
509 165 L
512 164 L
512 163 L
513 162 L
513 160 L
512 159 L
510 157 L
506 156 L
505 155 L
503 153 L
502 151 L
502 149 L
509 165 M 511 164 L
512 163 L
512 162 L
512 160 L
512 159 L
509 157 L
506 156 L
502 150 M 503 151 L
505 151 L
509 149 L
511 149 L
512 150 L
513 151 L
505 151 M 509 149 L
512 149 L
512 149 L
513 151 L
513 152 L
528 188 M 525 187 L
522 184 L
521 182 L
519 176 L
519 173 L
521 167 L
522 165 L
525 162 L
528 161 L
531 161 L
535 162 L
537 165 L
539 167 L
540 173 L
540 176 L
539 182 L
537 184 L
535 187 L
531 188 L
528 188 L
526 187 L
523 184 L
522 182 L
521 176 L
521 173 L
522 167 L
523 165 L
526 162 L
528 161 L
531 161 M 534 162 L
536 165 L
537 167 L
539 173 L
539 176 L
537 182 L
536 184 L
534 187 L
531 188 L
CS [] 0 setdash M
221 96 M 251 96 L
221 98 M 251 98 L
221 100 M 251 100 L
221 102 M 251 102 L
221 104 M 251 104 L
221 106 M 251 106 L
221 108 M 251 108 L
221 110 M 251 110 L
221 112 M 251 112 L
221 114 M 251 114 L
221 116 M 251 116 L
221 118 M 251 118 L
221 120 M 251 120 L
221 121 M 251 121 L
282 106 M CS [] 0 setdash M
282 93 M 293 120 M 293 93 L
294 120 M 294 93 L
285 120 M 284 112 L
284 120 L
304 120 L
304 112 L
302 120 L
289 93 M 298 93 L
312 120 M 312 93 L
314 120 M 314 93 L
314 107 M 316 110 L
320 111 L
323 111 L
327 110 L
328 107 L
328 93 L
323 111 M 325 110 L
327 107 L
327 93 L
309 120 M 314 120 L
309 93 M 318 93 L
323 93 M 332 93 L
339 103 M 355 103 L
355 106 L
354 108 L
352 110 L
350 111 L
346 111 L
342 110 L
339 107 L
338 103 L
338 101 L
339 97 L
342 94 L
346 93 L
349 93 L
352 94 L
355 97 L
354 103 M 354 107 L
352 110 L
346 111 M 343 110 L
341 107 L
339 103 L
339 101 L
341 97 L
343 94 L
346 93 L
370 111 M 367 110 L
364 107 L
363 103 L
363 101 L
364 97 L
367 94 L
370 93 L
373 93 L
377 94 L
379 97 L
381 101 L
381 103 L
379 107 L
377 110 L
373 111 L
370 111 L
368 110 L
365 107 L
364 103 L
364 101 L
365 97 L
368 94 L
370 93 L
373 93 M 375 94 L
378 97 L
379 101 L
379 103 L
378 107 L
375 110 L
373 111 L
391 111 M 391 93 L
392 111 M 392 93 L
392 103 M 394 107 L
396 110 L
399 111 L
403 111 L
404 110 L
404 108 L
403 107 L
401 108 L
403 110 L
387 111 M 392 111 L
387 93 M 396 93 L
412 111 M 419 93 L
413 111 M 419 95 L
427 111 M 419 93 L
417 88 L
414 85 L
412 84 L
410 84 L
409 85 L
410 86 L
412 85 L
409 111 M 417 111 L
422 111 M 430 111 L
CS [] 0 setdash M
770 96 M 801 96 L
801 122 L
770 122 L
770 96 L
831 106 M CS [] 0 setdash M
831 93 M 838 120 M 838 93 L
839 120 M 839 93 L
CS M
847 112 M 847 102 L
834 120 M 855 120 L
855 112 L
853 120 L
839 107 M 847 107 L
834 93 M 855 93 L
855 101 L
853 93 L
864 111 M 878 93 L
865 111 M 879 93 L
879 111 M 864 93 L
861 111 M 869 111 L
874 111 M 882 111 L
861 93 M 869 93 L
874 93 M 882 93 L
891 111 M 891 84 L
892 111 M 892 84 L
892 107 M 895 110 L
897 111 L
900 111 L
904 110 L
906 107 L
907 103 L
907 101 L
906 97 L
904 94 L
900 93 L
897 93 L
895 94 L
892 97 L
900 111 M 902 110 L
905 107 L
906 103 L
906 101 L
905 97 L
902 94 L
900 93 L
887 111 M 892 111 L
887 84 M 896 84 L
916 103 M 932 103 L
932 106 L
931 108 L
929 110 L
927 111 L
923 111 L
919 110 L
916 107 L
915 103 L
915 101 L
916 97 L
919 94 L
923 93 L
925 93 L
929 94 L
932 97 L
931 103 M 931 107 L
929 110 L
923 111 M 920 110 L
918 107 L
916 103 L
916 101 L
918 97 L
920 94 L
923 93 L
942 111 M 942 93 L
943 111 M 943 93 L
943 103 M 945 107 L
947 110 L
950 111 L
954 111 L
955 110 L
955 108 L
954 107 L
952 108 L
954 110 L
938 111 M 943 111 L
938 93 M 947 93 L
964 120 M 963 119 L
964 117 L
965 119 L
964 120 L
964 111 M 964 93 L
965 111 M 965 93 L
960 111 M 965 111 L
960 93 M 969 93 L
978 111 M 978 93 L
979 111 M 979 93 L
979 107 M 982 110 L
986 111 L
988 111 L
992 110 L
994 107 L
994 93 L
988 111 M 991 110 L
992 107 L
992 93 L
994 107 M 996 110 L
1000 111 L
1003 111 L
1007 110 L
1008 107 L
1008 93 L
1003 111 M 1005 110 L
1007 107 L
1007 93 L
974 111 M 979 111 L
974 93 M 983 93 L
988 93 M 997 93 L
1003 93 M 1012 93 L
1019 103 M 1035 103 L
1035 106 L
1034 108 L
1032 110 L
1030 111 L
1026 111 L
1022 110 L
1019 107 L
1018 103 L
1018 101 L
1019 97 L
1022 94 L
1026 93 L
1028 93 L
1032 94 L
1035 97 L
1034 103 M 1034 107 L
1032 110 L
1026 111 M 1023 110 L
1021 107 L
1019 103 L
1019 101 L
1021 97 L
1023 94 L
1026 93 L
1045 111 M 1045 93 L
1046 111 M 1046 93 L
1046 107 M 1049 110 L
1053 111 L
1055 111 L
1059 110 L
1061 107 L
1061 93 L
1055 111 M 1058 110 L
1059 107 L
1059 93 L
1041 111 M 1046 111 L
1041 93 M 1050 93 L
1055 93 M 1064 93 L
1073 120 M 1073 98 L
1075 94 L
1077 93 L
1080 93 L
1082 94 L
1084 97 L
1075 120 M 1075 98 L
1076 94 L
1077 93 L
1070 111 M 1080 111 L
CS [] 0 setdash M
stroke
grestore
showpage
end
0 SPE
6587 15917 XY 0 SPB
 clear Figure end restore 
0 SPE
XP /F25 /cmr10 300 498 498.132 128 [-3 -11 40 30] PXLF RP
7309 15943 XY F25(Figure)S 178 x(1.)S 178 x(Comparison)S 178 x(of)S
178 x(measure)S 2 x(d)S 177 x(rates)S
XP /F8 /cmr7 300 349 348.692 128 [-2 -8 31 21] PXLNF RP
XP /F8 50 199 2 0 13 18 19 16 0
<1F80 7FE0 E3E0 E0F0 E0F0 E070 00F0 00F0 00E0 01C0 0380 0700 0E00
 1C30 3830 7070 FFE0 FFE0 FFE0>
PXLC RP
16105 15762 XY F8(2)S
XP /F8 44 113 2 -6 5 3 10 8 0
<E0 F0 F0 F0 10 10 30 20 60 40>
PXLC RP
16303 15762 XY F8(,)S
XP /F8 51 199 1 0 14 18 19 16 0
<1FC0 7FF0 70F8 7078 7078 0078 0078 01F0 0FC0 00F0 0078 003C 003C
 E03C E03C E07C F0F8 7FF0 1FC0>
PXLC RP
16416 15762 XY F8(3,)S
XP /F8 55 199 2 0 15 19 20 16 0
<6000 7FFC 7FFC 7FFC E038 C070 C0E0 01C0 0180 0380 0300 0700 0700
 0600 0E00 0E00 0E00 0E00 0E00 0E00>
PXLC RP
16727 15762 XY F8(7)S
XP /F8 123 199 0 8 15 8 1 16 0
<FFFF>
PXLC RP
16926 15762 XY F8({)S
XP /F8 49 199 3 0 13 18 19 16 0
<0600 1E00 FE00 EE00 0E00 0E00 0E00 0E00 0E00 0E00 0E00 0E00 0E00
 0E00 0E00 0E00 0E00 FFE0 FFE0>
PXLC RP
17125 15762 XY F8(12,)S -2 x(1)S
XP /F8 52 199 1 0 14 18 19 16 0
<0060 00E0 01E0 03E0 06E0 04E0 0CE0 18E0 30E0 60E0 C0E0 FFFC FFFC
 00E0 00E0 00E0 00E0 07FC 07FC>
PXLC RP
17833 15762 XY F8(4)S 181 y 202 x F25(and)S 177 x(standard-mo)S 14 x
(del)S 598 y 7309 X(predictions)S -181 y F8(13)S 181 y 191 x F25
(for)S 166 x(four)S 166 x(solar-)S 2 x(neutrino)S 165 x(exp)S 14 x
(erime)S 2 x(n)S -14 x(ts.)S
XP /F48 51 336 2 0 24 31 32 24 0
<01FE00 07FFC0 1FFFE0 1F0FF0 3F07F8 3F07F8 3F83F8 3F03F8 3F03F8
 0E07F8 0007F0 0007F0 000FE0 001FC0 01FF00 01FF00 000FE0 0003F8
 0003FC 0001FC 0001FE 0001FE 7C01FE FE01FE FE01FE FE01FE FE03FC
 FE03FC 7E07F8 3FFFF0 1FFFC0 03FF00>
PXLC RP
5347 19676 XY F48(3.)S 672 x(Theoreti)S -2 x(cal)S 224 x(Neutri)S 
-2 x(no)S 224 x(Fluxes)S 20658 Y 5347 X F34(T)S -46 x(able)S 204 x
(1)S 205 x(sho)S -16 x(ws)S 205 x(the)S 205 x(solar)S 203 x
(neutrino)S 204 x(\015uxes)S 205 x(computed)S 205 x(with)S 204 x
(the)S 205 x(aid)S 204 x(of)S 204 x(the)S 205 x(standard)S 678 y 
5347 X(solar)S 216 x(mo)S 15 x(del.)S 217 x(The)S 218 x F35(pp)S 
218 x F34(neutrino)S 217 x(\015ux)S 218 x(is)S 217 x(predicted)S 
218 x(to)S 218 x(b)S 15 x(e)S 218 x(the)S 218 x(largest)S 217 x
(\015ux)S 218 x(b)S -15 x(y)S 217 x(an)S 218 x(or-)S 677 y 5347 X
(der)S 200 x(of)S 200 x(magnitude,)S 200 x(but)S 200 x(is)S 200 x
(not)S 201 x(observ)S -31 x(able)S 201 x(in)S 200 x(the)S 201 x(c)S
-15 x(hlorine)S 200 x(and)S 200 x(in)S 200 x(the)S 201 x(Kamiok)S 
-31 x(ande)S 678 y 5347 X(exp)S 15 x(erimen)S -15 x(ts.)S 214 x
(Only)S 214 x(the)S 216 x(gall)S -2 x(ium)S 214 x(exp)S 16 x(erimen)S
-16 x(ts)S 215 x(ha)S -15 x(v)S -16 x(e)S 215 x(a)S 215 x(lo)S -16 x
(w)S 215 x(enough)S 214 x(threshold)S 215 x(to)S 214 x(b)S 15 x(e)S
677 y 5347 X(sensitiv)S -16 x(e)S 173 x(to)S 173 x(the)S 173 x F35
(pp)S 174 x F34(neutrinos.)S 171 x(The)S 173 x(second)S 174 x(most)S
172 x(abundan)S -15 x(t)S 172 x(neutrino)S 172 x(source)S 173 x(is)S
-198 y 173 x F13(7)S 198 y 24 x F34(Be,)S 678 y 5347 X(whic)S -16 x
(h)S 184 x(pro)S 15 x(duces)S 184 x(t)S -15 x(w)S -15 x(o)S 184 x
(lines.)S 182 x(The)S -198 y 184 x F13(7)S 198 y 25 x F34(Be)S 184 x
(neutrinos)S 184 x(are)S 183 x(exp)S 16 x(ected)S 185 x(to)S 183 x
(con)S -15 x(tribute)S 184 x(a)S 183 x(small)S 677 y 5347 X(amoun)S
-16 x(t)S 147 x(to)S 146 x(the)S 148 x(capture)S 147 x(rate)S 147 x
(in)S 147 x(the)S 147 x(c)S -15 x(hlorine)S 146 x(exp)S 16 x(erimen)S
-16 x(t)S 147 x(\(15)S
XP /F34 37 456 3 -3 33 33 37 32 0
<0F800030 1FC00070 3DE000F0 78F801E0 707E0FC0 F07FFFC0 F033FB80
 F0300700 F0300F00 F0300E00 F0301C00 F0303C00 F0703800 70607000
 78E0F000 3DE0E000 1FC1C000 0F83C000 000381F0 000703F8 000707B8
 000E0F1C 001E0E0C 001C1E0E 00381E06 00781E06 00701E06 00E01E06
 01E01E06 01C01E06 03801E06 07801E0E 07000E0C 0E000F1C 1E0007B8
 1C0003F8 180001F0>
PXLC RP
18933 24723 XY F34(
(standard)S 678 y 5347 X(mo)S 14 x(del)S 181 x(prediction\))S 180 x
(and)S 181 x(a)S 180 x(somewhat)S 181 x(lar)S -2 x(ger)S 181 x
(fraction)S 180 x(\(25
(rate\))S 181 x(to)S 180 x(the)S 677 y 5347 X(gall)S -2 x(ium)S 181 x
(exp)S 16 x(erimen)S -16 x(t,)S 181 x(but)S 182 x(are)S 182 x(b)S 
15 x(elo)S -15 x(w)S 182 x(threshold)S 181 x(in)S 181 x(the)S 183 x
(Kamiok)S -32 x(ande)S 182 x(exp)S 16 x(erimen)S -16 x(t.)S 784 y 
6165 X(The)S 234 x(most)S 233 x(easily)S 232 x(detec)S 2 x(ted)S 
234 x(neutrinos)S 233 x(are)S 233 x(the)S 234 x(v)S -15 x(ery)S 233 x
(rare,)S 233 x(but)S 233 x(higher-energy)S -46 x(,)S 233 x(8)S 368 y 
-273 x(\026)S 309 y 5347 X(neutrinos.)S 160 x(They)S 162 x(are)S 
161 x(predicted)S 162 x(to)S 161 x(b)S 16 x(e)S 161 x(four)S 161 x
(orders)S 161 x(of)S 161 x(magni)S -2 x(tude)S 162 x(rarer)S 161 x
(than)S 161 x(the)S 162 x(lo)S -16 x(w-)S 678 y 5347 X(energy)S 225 x 
F35(pp)S 225 x F34(neutrinos,)S 224 x(but)S 225 x(b)S 16 x(ecause)S
226 x(the)S 225 x(8)S 368 y -273 x(\026)S -368 y 225 x(neutrinos)S
224 x(ha)S -15 x(v)S -15 x(e)S 225 x(relativ)S -16 x(ely)S 224 x
(high)S 224 x(energies)S 677 y 5347 X(they)S 236 x(dominate)S 236 x
(the)S 237 x(predicted)S 237 x(capture)S 237 x(rate)S 236 x(for)S 
235 x(the)S 237 x(c)S -15 x(hlorine)S 235 x(exp)S 16 x(erimen)S -16 x
(t)S 237 x(\(almo)S -2 x(st)S 678 y 5347 X(80
(the)S 206 x(total)S 203 x(predicted)S 206 x(rate\))S 204 x(and)S 
205 x(are)S 204 x(the)S 205 x(only)S 204 x(neutrino)S 204 x(source)S
205 x(to)S 205 x(whic)S -16 x(h)S 205 x(the)S 677 y 5347 X(Kamio)S
-2 x(k)S -30 x(ande)S 182 x(exp)S 16 x(erimen)S -16 x(t)S 182 x(is)S
181 x(sensitiv)S -15 x(e.)S 677 y 6165 X(T)S -46 x(able)S 189 x(1)S
189 x(sho)S -16 x(ws)S 189 x(the)S 190 x(most)S 188 x(imp)S 15 x
(ortan)S -16 x(t)S 189 x(neutrino)S 188 x(\015uxes)S 189 x(and)S 
189 x(the)S 190 x(e\013ectiv)S -15 x(e)S 190 x(1)S F35(\033)S 208 x 
F34(error)S 678 y 5347 X(bars)S 218 x(that)S 219 x(ha)S -16 x(v)S 
-15 x(e)S 219 x(b)S 15 x(een)S 220 x(calculated)S 219 x(with)S 218 x
(the)S 219 x(standard)S 219 x(solar)S 217 x(mo)S 14 x(del.)S 219 x
(The)S 219 x(size)S 219 x(of)S 218 x(the)S 677 y 5347 X(uncertain)S
-16 x(ties)S 229 x(is)S 228 x(of)S 228 x(critical)S 227 x(imp)S 14 x
(ortance.)S 228 x(I)S 228 x(ha)S -15 x(v)S -15 x(e)S 229 x
(therefore)S 228 x(dev)S -15 x(oted)S 229 x(a)S 228 x(full)S 227 x
(c)S -15 x(hapter,)S 678 y 5347 X(Chapter)S 190 x(7,)S 189 x(in)S 
190 x(m)S -16 x(y)S 190 x(b)S 16 x(o)S 14 x(ok)S
XP /F37 /cmti10 329 546 545.454 128 [-2 -12 49 33] PXLNF RP
XP /F37 78 406 3 0 37 30 31 40 0
<01FF00FFE0 01FF00FFE0 001F001F00 001F800C00 001F801C00 003F801C00
 0037C01800 0033C01800 0033C03800 0073E03800 0061E03000 0061E03000
 0061F07000 00E0F07000 00C0F06000 00C0F86000 00C078E000 01C078E000
 018078C000 01807CC000 01803DC000 03803DC000 03003F8000 03001F8000
 03001F8000 07001F8000 06000F0000 0E000F0000 1F000F0000 FFE0070000
 FFE0060000>
PXLC RP
11011 32959 XY F37(N)S
XP /F37 101 251 4 0 20 19 20 24 0
<00FE00 03FF00 07C780 0F0380 1E0180 3C0380 7C0700 781F00 7FFC00
 7FF000 F80000 F00000 F00000 F00000 700000 700300 780700 3C3E00
 1FFC00 0FE000>
PXLC RP
11417 32959 XY F37(e)S
XP /F37 117 293 3 0 24 19 20 24 0
<0F80E0 1FC1F0 39E1F0 71E1E0 63E1E0 63C3E0 E3C3C0 07C3C0 0783C0
 0787C0 0F8780 0F0780 0F0780 0F0F9C 0F0F18 0F0F18 0F0F38 0F3F30
 07FFF0 03E3E0>
PXLC RP
11668 32959 XY F37(u)S
XP /F37 116 181 4 0 15 27 28 16 0
<01C0 03E0 03E0 03C0 03C0 07C0 07C0 0780 FFF0 FFF0 0F80 0F00 0F00
 1F00 1F00 1E00 1E00 3E00 3E00 3C00 3C00 7C70 7C60 7860 78E0 79C0
 3F80 1F00>
PXLC RP
11960 32959 XY F37(t)S
XP /F37 114 230 3 0 20 19 20 24 0
<1F1F00 3FBF80 33F1C0 73E3C0 63C3C0 63C3C0 E7C380 078000 078000
 078000 0F8000 0F0000 0F0000 0F0000 1F0000 1E0000 1E0000 1E0000
 3E0000 1C0000>
PXLC RP
12142 32959 XY F37(r)S
XP /F37 105 167 3 0 14 30 31 16 0
<00F0 00F0 00F0 00E0 0000 0000 0000 0000 0000 0000 0000 0F80 1FC0
 3BC0 73C0 63C0 67C0 E780 0780 0F80 0F00 0F00 1F00 1E00 1E70 3E60
 3C60 3CE0 3DC0 3F80 1F00>
PXLC RP
12372 32959 XY F37(i)S
XP /F37 110 307 3 0 26 19 20 24 0
<1F0FC0 3F9FF0 33F8F0 73F078 63E078 63C0F8 E7C0F0 0780F0 0780F0
 0781F0 0F81E0 0F01E0 0F03E0 0F03C7 1F03C6 1E07C6 1E078E 1E079C
 3E03F8 1C01F0>
PXLC RP
12539 32959 XY F37(n)S
XP /F37 111 279 4 0 22 19 20 24 0
<007E00 03FF80 07C780 0F03C0 1E01C0 3C01E0 3C01E0 7801E0 7801E0
 7803E0 F803E0 F003C0 F003C0 F00780 F00780 700F00 781E00 3C7C00
 3FF800 0FC000>
PXLC RP
12846 32959 XY F37(o)S
XP /F37 65 406 2 0 30 31 32 32 0
<00000700 00000700 00000F00 00000F00 00001F00 00003F00 00003F00
 00006F00 00006F00 0000CF00 0000CF00 00018F00 00038F00 00030F00
 00060F00 00060F00 000C0F80 000C0780 00180780 00180780 003FFF80
 007FFF80 00600780 00C00780 00C00780 01800780 01800780 03000780
 07000780 1F8007C0 7FC07FF8 FFC07FF8>
PXLC RP
13327 32959 XY F37(A)S
XP /F37 115 223 3 0 18 19 20 16 0
<00FC 03FE 0787 0F0F 0E0F 0E0F 1F00 1FF0 1FF8 0FFC 07FE 03FE 003E
 701E F01C F01C F03C F078 7FF0 1FC0>
PXLC RP
13733 32959 XY F37(str)S -28 x(o)S
XP /F37 112 279 0 -9 22 19 29 24 0
<03E1F0 07F7F8 067F3C 0E7C1C 0C781E 0CF81E 1CF81E 00F01E 00F01E
 01F03E 01F03E 01E03C 01E03C 03E078 03E078 03C0F0 03E1E0 07F3C0
 07FF80 07BE00 078000 0F8000 0F8000 0F0000 0F0000 1F0000 1F0000
 FFE000 FFE000>
PXLC RP
14618 32959 XY F37(p)S
XP /F37 104 279 2 0 23 31 32 24 0
<01F000 0FF000 0FF000 01E000 01E000 01E000 03E000 03C000 03C000
 03C000 07C000 078000 07BF00 07FFC0 0FE3C0 0FC1E0 0F81E0 0F03E0
 1F03C0 1E03C0 1E03C0 1E07C0 3E0780 3C0780 3C0F80 3C0F1C 7C0F18
 781F18 781E38 781E70 F80FE0 7007C0>
PXLC RP
14897 32959 XY F37(h)S
XP /F37 121 265 3 -9 23 19 29 24 0
<0F8070 1FC0F8 39E0F8 71E0F0 63E0F0 63C1F0 E3C1E0 07C1E0 0781E0
 0783E0 0F83C0 0F03C0 0F03C0 0F07C0 0F0780 0F0780 0F0F80 0F1F80
 07FF00 03FF00 000F00 001F00 101E00 783E00 783C00 787800 71F000
 7FE000 1F8000>
PXLC RP
15175 32959 XY F37(ysi)S
XP /F37 99 251 4 0 20 19 20 24 0
<007E00 03FF00 07C380 0F0780 1E0780 3C0780 3C0200 780000 780000
 780000 F80000 F00000 F00000 F00000 F00000 700300 780700 3C3E00
 3FFC00 0FE000>
PXLC RP
15830 32959 XY F37(cs)S 191 x F34(to)S 190 x(the)S 190 x(estimation)S
189 x(of)S 190 x(the)S 191 x(errors)S 189 x(in)S 677 y 5347 X(eac)S
-15 x(h)S 179 x(neutrino)S 178 x(\015ux.)S 178 x(F)S -45 x(or)S 178 x
(a)S 179 x(recen)S -15 x(t)S 179 x(detailed)S 179 x(calculation)S 
177 x(of)S 178 x(the)S 180 x(errors)S 178 x(and)S 178 x(a)S 179 x
(compar-)S
1 PP EP

1000 BP 39600 30600 PM 0 0 XY
4389 Y 11668 X F34(T)S -46 x(able)S 182 x(1.)S 181 x(Neutrino)S 181 x
(Fluxes)S -198 y F13(13)S 720 y 9349 X 11040 24 R 773 y 9648 X F34
(Source)S 16907 X(Flux)S 522 y 9349 X 11040 24 R 170 y 9349 X 
11040 24 R 773 y 14825 X(\(10)S -198 y F13(10)S 198 y 206 x F34(cm)S
-198 y F15(\000)S F13(2)S 198 y 25 x F34(s)S -198 y F15(\000)S F13
(1)S 198 y 25 x F34(\))S 677 y 10042 X(p-p)S 14825 X(6)S F35(:)S F34
(0)S 181 x(\(1)S
XP /F36 6 425 3 -1 30 30 32 32 0
<00060000 00060000 00060000 00060000 00060000 00060000 00060000
 00060000 00060000 00060000 00060000 00060000 00060000 00060000
 00060000 FFFFFFF0 FFFFFFF0 00060000 00060000 00060000 00060000
 00060000 00060000 00060000 00060000 00060000 00060000 00060000
 00060000 00060000 FFFFFFF0 FFFFFFF0>
PXLC RP
16310 7826 XY F36(\006)S 121 x F34(0)S F35(:)S F34(007\))S 678 y 
10004 X(p)S 16 x(ep)S 14825 X(1)S F35(:)S F34(4)S
XP /F36 2 425 6 -1 27 22 24 24 0
<C0000C E0001C F0003C 780078 3C00F0 1E01E0 0F03C0 078780 03CF00
 01FE00 00FC00 007800 007800 00FC00 01FE00 03CF00 078780 0F03C0
 1E01E0 3C00F0 780078 F0003C E0001C C0000C>
PXLC RP
15643 8504 XY F36(\002)S 122 x F34(10)S -198 y F15(\000)S F13(2)S 
198 y 206 x F34(\(1)S 121 x F36(\006)S 121 x F34(0)S F35(:)S F34
(012\))S 677 y 10012 X(hep)S 14825 X(1)S F35(:)S F34(2)S 120 x F36
(\002)S 122 x F34(10)S -198 y F15(\000)S F13(7)S 678 y 10004 X(7)S
198 y 24 x F34(Be)S 14825 X(4)S F35(:)S
XP /F34 57 273 2 -1 19 29 31 24 0
<03F000 0FFC00 1E1E00 3C0F00 780700 780780 F00780 F00380 F00380
 F003C0 F003C0 F003C0 F003C0 F003C0 7007C0 7807C0 780FC0 3C1BC0
 1FFBC0 0FF3C0 0183C0 000380 000780 000780 7C0700 7C0F00 7C1E00
 7C3C00 787800 3FF000 0FC000>
PXLC RP
15249 9859 XY F34(9)S 121 x F36(\002)S 122 x F34(10)S -198 y F15
(\000)S F13(1)S 198 y 206 x F34(\(1)S 121 x F36(\006)S 121 x F34(0)S
F35(:)S F34(06\))S 479 y 10125 X F13(8)S 198 y 24 x F34(B)S 14825 X
(5)S F35(:)S F34(7)S 120 x F36(\002)S 122 x F34(10)S -198 y F15
(\000)S F13(4)S 198 y 206 x F34(\(1)S 121 x F36(\006)S 121 x F34(0)S
F35(:)S F34(14\))S 480 y 10008 X F13(13)S 197 y 24 x F34(N)S 14825 X
(5)S 121 x F36(\002)S 121 x F34(10)S -197 y F15(\000)S F13(2)S 197 y 
207 x F34(\(1)S 121 x F36(\006)S 121 x F34(0)S F35(:)S F34(17\))S 
480 y 10000 X F13(15)S 198 y 24 x F34(O)S 14825 X(4)S 121 x F36
(\002)S 121 x F34(10)S -198 y F15(\000)S F13(2)S 198 y 207 x F34
(\(1)S 121 x F36(\006)S 121 x F34(0)S F35(:)S F34(19\))S 522 y 9349 X 
11040 24 R 14505 Y 5347 X(ison)S 181 x(with)S 181 x(the)S 183 x
(uncertain)S -15 x(ties)S 182 x(estimated)S 181 x(b)S -15 x(y)S 182 x
(di\013eren)S -15 x(t)S 182 x(authors,)S 181 x(see)S 183 x(Ref..)S
-198 y -2 x F13(13)S 876 y 6165 X F34(The)S 266 x F35(pp)S 267 x F34
(neutrinos)S 265 x(are)S 266 x(calculated)S 266 x(with)S 265 x(a)S
266 x(precision)S 265 x(that)S 266 x(is)S 265 x(b)S 16 x(etter)S 
266 x(than)S 266 x(1
229 x(abundan)S -15 x(t)S 229 x(neutrinos,)S 229 x(the)S -198 y 231 x 
F13(7)S 198 y 24 x F34(Be)S 231 x(neutrinos,)S 229 x(are)S 229 x
(calculated)S 230 x(with)S 229 x(an)S 678 y 5347 X(uncertain)S -16 x
(t)S -15 x(y)S 167 x(of)S 165 x F36(\006)S F34(6
167 x(rare)S 165 x(8)S 368 y -273 x(\026)S -368 y 167 x(neutrinos)S
166 x(are)S 166 x(calculated)S 166 x(with)S 166 x(the)S 167 x(least)S
166 x(accuracy)S -45 x(,)S 677 y 5347 X F36(\006)S F34(14
XP /F34 85 410 1 -1 31 30 32 32 0
<FFFE1FFE FFFE1FFE 07C001E0 07C000C0 07C000C0 07C000C0 07C000C0
 07C000C0 07C000C0 07C000C0 07C000C0 07C000C0 07C000C0 07C000C0
 07C000C0 07C000C0 07C000C0 07C000C0 07C000C0 07C000C0 07C000C0
 07C000C0 07C000C0 07C000C0 03C001C0 03E00180 03E00380 01F00700
 00F80F00 007C1E00 003FF800 0007F000>
PXLC RP
7104 17215 XY F34(Unfortunately)S -47 x(,)S 181 x(the)S 182 x
(easier)S 181 x(solar)S 180 x(neutrinos)S 181 x(are)S 181 x(to)S 
181 x(detect,)S 182 x(the)S 182 x(more)S 181 x(di\016cult)S 677 y 
5347 X(they)S 182 x(are)S 182 x(to)S 181 x(calculate.)S
XP /F48 52 336 2 0 24 31 32 24 0
<0000E0 0001E0 0003E0 0007E0 000FE0 000FE0 001FE0 003FE0 0077E0
 00E7E0 00E7E0 01C7E0 0387E0 0707E0 0E07E0 0E07E0 1C07E0 3807E0
 7007E0 E007E0 FFFFFE FFFFFE FFFFFE 0007E0 0007E0 0007E0 0007E0
 0007E0 0007E0 00FFFE 00FFFE 00FFFE>
PXLC RP
5347 18967 XY F48(4.)S
XP /F48 67 486 3 0 35 33 34 40 0
<0007FE0180 003FFFC380 00FFFFEF80 03FF81FF80 07FC007F80 0FF0003F80
 1FE0001F80 1FC0000F80 3FC0000780 7F80000780 7F80000780 7F80000380
 FF00000380 FF00000000 FF00000000 FF00000000 FF00000000 FF00000000
 FF00000000 FF00000000 FF00000000 FF00000000 7F80000380 7F80000380
 7F80000380 3FC0000780 1FC0000700 1FE0000F00 0FF0001E00 07FC003C00
 03FF81F800 00FFFFF000 003FFFC000 0007FF0000>
PXLC RP
6542 18967 XY F48(Comparison)S 225 x(of)S 226 x(the)S 225 x(Chlori)S
-2 x(ne)S 225 x(and)S 225 x(the)S 224 x(Electron)S
XP /F48 45 224 1 9 14 13 5 16 0
<FFFC FFFC FFFC FFFC FFFC>
PXLC RP
19918 18967 XY F48(-Scatter)S -2 x(ing)S 225 x(Ex-)S 722 y 5347 X(p)S
18 x(erimen)S -20 x(ts)S
XP /F48 119 486 1 0 37 21 22 40 0
<FFE3FF8FF8 FFE3FF8FF8 FFE3FF8FF8 1FC07C03C0 0FC07E0380 0FC0FE0380
 07E0FE0700 07E1FF0700 07E1DF0700 03F1DF0E00 03F3DF8E00 03FB8F9E00
 01FB8FDC00 01FF07DC00 01FF07FC00 00FF07F800 00FE03F800 007E03F000
 007E03F000 007C01F000 003C01E000 003C01E000>
PXLC RP
8480 19689 XY F48(with)S 223 x(Theor)S
XP /F48 121 355 1 -10 27 21 32 32 0
<FFF03FE0 FFF03FE0 FFF03FE0 0FC00E00 0FE01E00 07E01C00 07E01C00
 03F03800 03F03800 03F87800 01F87000 01FCF000 00FCE000 00FCE000
 007FC000 007FC000 007FC000 003F8000 003F8000 001F0000 001F0000
 000E0000 000E0000 001C0000 001C0000 783C0000 FC380000 FC780000
 FC700000 79E00000 7FC00000 1F000000>
PXLC RP
11771 19689 XY F48(y)S 20672 Y 5347 X F34(W)S -46 x(e)S 279 x(will)S
278 x(no)S -15 x(w)S 279 x(compare)S 278 x(the)S 280 x(results)S 
279 x(of)S 279 x(the)S 280 x(c)S -15 x(hlori)S -2 x(ne)S 280 x(and)S
279 x(the)S 280 x(electron-scattering)S 678 y 5347 X(\(Kamiok)S -32 x
(ande\))S 124 x(exp)S 15 x(erimen)S -16 x(ts)S 124 x(with)S 122 x
(the)S 124 x(theoretical)S 123 x(exp)S 15 x(ectations)S 124 x(for)S
122 x(eac)S -15 x(h)S 123 x(exp)S 16 x(erimen)S -16 x(t.)S 677 y 
5347 X(The)S 202 x(predicted)S 202 x(ev)S -15 x(en)S -15 x(t)S 202 x
(rate,)S 201 x(8)S 134 x F36(\006)S 135 x F34(1)S 201 x(SNU,)S 201 x
(for)S 201 x(the)S 202 x(c)S -15 x(hlorine)S 201 x(exp)S 15 x
(erimen)S -15 x(t)S 201 x(is)S 202 x(domi)S -2 x(nated)S 677 y 5347 X
(b)S -15 x(y)S 183 x(the)S 184 x(6.2)S 183 x(SNU)S 183 x(from)S 182 x
(the)S 184 x(rare)S 183 x(8)S 369 y -273 x(\026)S -369 y 184 x
(neutrinos.)S 182 x(The)S 184 x(next)S 184 x(most)S 183 x(imp)S 15 x
(ortan)S -16 x(t)S 183 x(source,)S 184 x(ac-)S 678 y 5347 X(cording)S
212 x(to)S 212 x(the)S 214 x(standard)S 212 x(mo)S 15 x(del,)S 212 x
(for)S 212 x(this)S 213 x(exp)S 15 x(erimen)S -15 x(t)S 213 x(is)S
212 x(the)S 214 x(electron-capture)S 213 x(line)S 677 y 5347 X(from)S
-198 y 211 x F13(7)S 198 y 25 x F34(Be,)S 214 x(whic)S -16 x(h)S 
214 x(is)S 212 x(predicted)S 214 x(to)S 213 x(pro)S 15 x(duce)S 214 x
(a)S 213 x(1.2)S 212 x(SNU)S 213 x(capture)S 214 x(rate.)S 212 x
(The)S 214 x F35(pep)S 214 x F34(and)S 678 y 5347 X(CNO)S 182 x
(neutrinos)S 181 x(are)S 182 x(exp)S 15 x(ecte)S 2 x(d)S 182 x(to)S
181 x(pro)S 15 x(duce)S 183 x(together)S 182 x(a)S 181 x(rate)S 182 x
(of)S 181 x(0.6)S 181 x(SNU.)S 677 y 6165 X(The)S 182 x(exp)S 16 x
(erimen)S -16 x(tal)S 181 x(rate)S 182 x(is)S -198 y F13(7)S 198 y 
206 x F34(2)S F35(:)S F34(28)S 120 x F36(\006)S 121 x F34(0)S F35(:)S
F34(23)S 181 x(SNU.)S 678 y 6165 X(In)S 147 x(order)S 146 x(to)S 
146 x(assess)S 148 x(the)S 147 x(signi\014cance)S 147 x(of)S 146 x
(the)S 147 x(disagreemen)S -16 x(t)S 147 x(b)S 15 x(et)S -14 x(w)S
-15 x(een)S 147 x(theory)S 147 x(and)S 146 x(ob-)S 677 y 5347 X
(serv)S -31 x(ation)S 178 x(for)S 177 x(the)S 178 x(solar)S 177 x
(neutrino)S 177 x(exp)S 16 x(erimen)S -16 x(ts,)S 178 x(I)S 178 x
(ha)S -15 x(v)S -16 x(e)S 179 x(p)S 15 x(erformed)S 177 x(a)S 178 x
(series)S 178 x(of)S 178 x(Mon)S -16 x(te)S 678 y 5347 X(Carlo)S 
241 x(calculations.)S 242 x(The)S 244 x(results)S 243 x(are)S 243 x
(sho)S -15 x(wn)S 243 x(in)S 243 x(Figure)S 242 x(2,)S 243 x(whic)S
-16 x(h)S 244 x(w)S -15 x(as)S 243 x(constructed)S 677 y 5347 X
(using)S 180 x(the)S 181 x(results)S 180 x(from)S 179 x(a)S 180 x
(thousand)S 180 x(implemen)S -16 x(tations)S 180 x(of)S 180 x(the)S
181 x(standard)S 180 x(solar)S 179 x(mo)S 14 x(del.)S 678 y 5347 X
(F)S -46 x(or)S 206 x(eac)S -15 x(h)S 206 x(mo)S 14 x(del,)S 206 x
(all)S 204 x(of)S 206 x(the)S 206 x(imp)S 14 x(ortan)S -16 x(t)S 
206 x(input)S 206 x(parameters)S 205 x(\(including)S 205 x(n)S -15 x
(uclear)S 206 x(reac-)S 677 y 5347 X(tion)S 151 x(rates)S 152 x(and)S
152 x(c)S -15 x(hemical)S 151 x(comp)S 15 x(osition\))S 151 x(w)S 
-15 x(ere)S 152 x(c)S -15 x(hosen)S 153 x(from)S 150 x(normal)S 150 x
(distributions)S 151 x(that)S 677 y 5347 X(had)S 228 x(means)S 228 x
(and)S 228 x(standard)S 228 x(distributio)S -2 x(ns)S 229 x(equal)S
228 x(to)S 228 x(the)S 229 x(exp)S 16 x(erimen)S -16 x
(tally-determi)S -2 x(ned)S 678 y 5347 X(v)S -31 x(alues.)S 222 x(F)S
-46 x(or)S 222 x(a)S 222 x(eac)S -15 x(h)S 222 x(solar)S 221 x(mo)S
15 x(del,)S 221 x(ev)S -15 x(ery)S 222 x(parameter)S 222 x(w)S -15 x
(as)S 222 x(c)S -15 x(hosen)S 222 x(from)S 221 x(its)S 222 x(o)S 
-16 x(wn)S 223 x(nor-)S 677 y 5347 X(mal)S 191 x(distributi)S -2 x
(on)S 192 x(and)S 192 x(the)S 193 x(solar)S 191 x(calculatio)S -2 x
(ns)S 193 x(w)S -16 x(ere)S 193 x(iterated)S 192 x(to)S 192 x(matc)S
-15 x(h)S 192 x(the)S 193 x(observ)S -16 x(ed)S 678 y 5347 X(c)S 
-15 x(haracteristics)S 151 x(of)S 150 x(the)S 152 x(presen)S -15 x
(t-da)S -15 x(y)S 151 x(sun.)S 151 x(This)S 151 x(pro)S 14 x(cedure)S
152 x(is)S 151 x(required)S 151 x(in)S 150 x(order)S 151 x(to)S 151 x
(tak)S -16 x(e)S 677 y 5347 X(accoun)S -15 x(t)S 155 x(of)S 155 x
(the)S 155 x(strong)S 155 x(e\013ect)S 2 x(s)S 155 x(of)S 154 x(b)S
16 x(oundary)S 154 x(conditions)S 154 x(and)S 155 x(the)S 156 x
(coupling)S 154 x(of)S 155 x(di\013eren)S -15 x(t)S 678 y 5347 X
(calculated)S 177 x(neutrino)S 176 x(\015uxes)S 178 x(that)S 177 x
(exists)S 177 x(among)S 176 x(the)S 178 x(solutions)S 176 x(of)S 
176 x(the)S 178 x(coupled)S 177 x(partial)S
1 PP EP

1000 BP 39600 30600 PM 0 0 XY
4787 15917 XY 0 SPB
 save 10 dict begin /Figure exch def currentpoint translate
/showpage {} def 
0 SPE
4787 15917 XY 0 SPB
save 50 dict begin /psplot exch def
/StartPSPlot
   {newpath 0 0 moveto 0 setlinewidth 0 setgray 0 setlinecap
    1 setlinejoin 72 300 div dup scale}def
/pending {false} def
/finish {pending {currentpoint stroke moveto /pending false def} if} def
/r {finish newpath moveto} def
/d {lineto /pending true def} def
/l {finish 4 2 roll moveto lineto currentpoint stroke moveto} def
/p {finish newpath moveto currentpoint lineto currentpoint stroke moveto} def
/e {finish gsave showpage grestore newpath 0 0 moveto} def
/lw {finish setlinewidth} def
/lt0 {finish [] 0 setdash} def
/lt1 {finish [3 5] 0 setdash} def
/lt2 {finish [20 10] 0 setdash} def
/lt3 {finish [60 10] 0 setdash} def
/lt4 {finish [3 10 20 10] 0 setdash} def
/lt5 {finish [3 10 60 10] 0 setdash} def
/lt6 {finish [20 10 60 10] 0 setdash} def
/EndPSPlot {clear psplot end restore}def
StartPSPlot
gsave 
90 rotate 4 72 mul .55 -72 mul moveto /Times-Roman findfont
20 scalefont setfont 0.3 setgray (astro-ph/9404070   28 Apr 94) show grestore
   1 lw lt0  200  200 r 1600  200 d  200  200 r  200  238 d  225  200 r
  225  219 d  251  200 r  251  219 d  276  200 r  276  219 d  302  200 r
  302  219 d  327  200 r  327  238 d  353  200 r  353  219 d  378  200 r
  378  219 d  404  200 r  404  219 d  429  200 r  429  219 d  455  200 r
  455  238 d  480  200 r  480  219 d  505  200 r  505  219 d  531  200 r
  531  219 d  556  200 r  556  219 d  582  200 r  582  238 d  607  200 r
  607  219 d  633  200 r  633  219 d  658  200 r  658  219 d  684  200 r
  684  219 d  709  200 r  709  238 d  735  200 r  735  219 d  760  200 r
  760  219 d  785  200 r  785  219 d  811  200 r  811  219 d  836  200 r
  836  238 d  862  200 r  862  219 d  887  200 r  887  219 d  913  200 r
  913  219 d  938  200 r  938  219 d  964  200 r  964  238 d  989  200 r
  989  219 d 1015  200 r 1015  219 d 1040  200 r 1040  219 d 1065  200 r
 1065  219 d 1091  200 r 1091  238 d 1116  200 r 1116  219 d 1142  200 r
 1142  219 d 1167  200 r 1167  219 d 1193  200 r 1193  219 d 1218  200 r
 1218  238 d 1244  200 r 1244  219 d 1269  200 r 1269  219 d 1295  200 r
 1295  219 d 1320  200 r 1320  219 d 1345  200 r 1345  238 d 1371  200 r
 1371  219 d 1396  200 r 1396  219 d 1422  200 r 1422  219 d 1447  200 r
 1447  219 d 1473  200 r 1473  238 d 1498  200 r 1498  219 d 1524  200 r
 1524  219 d 1549  200 r 1549  219 d 1575  200 r 1575  219 d 1600  200 r
 1600  238 d  199  176 r  195  175 d  193  172 d  192  166 d  192  162 d
  193  156 d  195  152 d  199  151 d  201  151 d  205  152 d  207  156 d
  208  162 d  208  166 d  207  172 d  205  175 d  201  176 d  199  176 d
  196  175 d  195  174 d  194  172 d  193  166 d  193  162 d  194  156 d
  195  154 d  196  152 d  199  151 d  201  151 r  204  152 d  205  154 d
  206  156 d  207  162 d  207  166 d  206  172 d  205  174 d  204  175 d
  201  176 d  322  172 r  325  173 d  328  176 d  328  151 d  327  175 r
  327  151 d  322  151 r  333  151 d  448  172 r  449  170 d  448  169 d
  447  170 d  447  172 d  448  174 d  449  175 d  453  176 d  457  176 d
  461  175 d  462  174 d  463  172 d  463  169 d  462  167 d  459  164 d
  453  162 d  450  161 d  448  158 d  447  155 d  447  151 d  457  176 r
  460  175 d  461  174 d  462  172 d  462  169 d  461  167 d  457  164 d
  453  162 d  447  154 r  448  155 d  450  155 d  456  152 d  460  152 d
  462  154 d  463  155 d  450  155 r  456  151 d  461  151 d  462  152 d
  463  155 d  463  157 d  575  173 r  576  172 d  575  170 d  574  172 d
  574  173 d  576  175 d  580  176 d  584  176 d  588  175 d  589  173 d
  589  169 d  588  167 d  584  166 d  581  166 d  584  176 r  587  175 d
  588  173 d  588  169 d  587  167 d  584  166 d  587  164 d  589  162 d
  590  160 d  590  156 d  589  154 d  588  152 d  584  151 d  580  151 d
  576  152 d  575  154 d  574  156 d  574  157 d  575  158 d  576  157 d
  575  156 d  588  163 r  589  160 d  589  156 d  588  154 d  587  152 d
  584  151 d  711  174 r  711  151 d  713  176 r  713  151 d  713  176 r
  699  158 d  719  158 d  708  151 r  716  151 d  830  176 r  828  164 d
  830  166 d  834  167 d  837  167 d  841  166 d  843  163 d  844  160 d
  844  158 d  843  155 d  841  152 d  837  151 d  834  151 d  830  152 d
  829  154 d  828  156 d  828  157 d  829  158 d  830  157 d  829  156 d
  837  167 r  840  166 d  842  163 d  843  160 d  843  158 d  842  155 d
  840  152 d  837  151 d  830  176 r  842  176 d  830  175 r  836  175 d
  842  176 d  970  173 r  969  172 d  970  170 d  971  172 d  971  173 d
  970  175 d  968  176 d  964  176 d  960  175 d  958  173 d  957  170 d
  956  166 d  956  158 d  957  155 d  959  152 d  963  151 d  965  151 d
  969  152 d  971  155 d  972  158 d  972  160 d  971  163 d  969  166 d
  965  167 d  964  167 d  960  166 d  958  163 d  957  160 d  964  176 r
  962  175 d  959  173 d  958  170 d  957  166 d  957  158 d  958  155 d
  960  152 d  963  151 d  965  151 r  968  152 d  970  155 d  971  158 d
  971  160 d  970  163 d  968  166 d  965  167 d 1083  176 r 1083  169 d
 1083  172 r 1084  174 d 1086  176 d 1089  176 d 1095  173 d 1097  173 d
 1098  174 d 1099  176 d 1084  174 r 1086  175 d 1089  175 d 1095  173 d
 1099  176 r 1099  173 d 1098  169 d 1093  163 d 1092  161 d 1091  157 d
 1091  151 d 1098  169 r 1092  163 d 1091  161 d 1090  157 d 1090  151 d
 1216  176 r 1212  175 d 1211  173 d 1211  169 d 1212  167 d 1216  166 d
 1220  166 d 1224  167 d 1225  169 d 1225  173 d 1224  175 d 1220  176 d
 1216  176 d 1213  175 d 1212  173 d 1212  169 d 1213  167 d 1216  166 d
 1220  166 r 1223  167 d 1224  169 d 1224  173 d 1223  175 d 1220  176 d
 1216  166 r 1212  164 d 1211  163 d 1210  161 d 1210  156 d 1211  154 d
 1212  152 d 1216  151 d 1220  151 d 1224  152 d 1225  154 d 1226  156 d
 1226  161 d 1225  163 d 1224  164 d 1220  166 d 1216  166 r 1213  164 d
 1212  163 d 1211  161 d 1211  156 d 1212  154 d 1213  152 d 1216  151 d
 1220  151 r 1223  152 d 1224  154 d 1225  156 d 1225  161 d 1224  163 d
 1223  164 d 1220  166 d 1352  168 r 1351  164 d 1349  162 d 1345  161 d
 1344  161 d 1340  162 d 1338  164 d 1337  168 d 1337  169 d 1338  173 d
 1340  175 d 1344  176 d 1346  176 d 1350  175 d 1352  173 d 1353  169 d
 1353  162 d 1352  157 d 1351  155 d 1349  152 d 1345  151 d 1341  151 d
 1339  152 d 1338  155 d 1338  156 d 1339  157 d 1340  156 d 1339  155 d
 1344  161 r 1341  162 d 1339  164 d 1338  168 d 1338  169 d 1339  173 d
 1341  175 d 1344  176 d 1346  176 r 1349  175 d 1351  173 d 1352  169 d
 1352  162 d 1351  157 d 1350  155 d 1347  152 d 1345  151 d 1456  172 r
 1459  173 d 1462  176 d 1462  151 d 1461  175 r 1461  151 d 1456  151 r
 1467  151 d 1484  176 r 1480  175 d 1478  172 d 1477  166 d 1477  162 d
 1478  156 d 1480  152 d 1484  151 d 1486  151 d 1490  152 d 1492  156 d
 1493  162 d 1493  166 d 1492  172 d 1490  175 d 1486  176 d 1484  176 d
 1481  175 d 1480  174 d 1479  172 d 1478  166 d 1478  162 d 1479  156 d
 1480  154 d 1481  152 d 1484  151 d 1486  151 r 1489  152 d 1490  154 d
 1491  156 d 1492  162 d 1492  166 d 1491  172 d 1490  174 d 1489  175 d
 1486  176 d 1583  172 r 1586  173 d 1589  176 d 1589  151 d 1588  175 r
 1588  151 d 1583  151 r 1594  151 d 1607  172 r 1610  173 d 1613  176 d
 1613  151 d 1612  175 r 1612  151 d 1607  151 r 1618  151 d  200 1100 r
 1600 1100 d  200 1100 r  200 1062 d  225 1100 r  225 1081 d  251 1100 r
  251 1081 d  276 1100 r  276 1081 d  302 1100 r  302 1081 d  327 1100 r
  327 1062 d  353 1100 r  353 1081 d  378 1100 r  378 1081 d  404 1100 r
  404 1081 d  429 1100 r  429 1081 d  455 1100 r  455 1062 d  480 1100 r
  480 1081 d  505 1100 r  505 1081 d  531 1100 r  531 1081 d  556 1100 r
  556 1081 d  582 1100 r  582 1062 d  607 1100 r  607 1081 d  633 1100 r
  633 1081 d  658 1100 r  658 1081 d  684 1100 r  684 1081 d  709 1100 r
  709 1062 d  735 1100 r  735 1081 d  760 1100 r  760 1081 d  785 1100 r
  785 1081 d  811 1100 r  811 1081 d  836 1100 r  836 1062 d  862 1100 r
  862 1081 d  887 1100 r  887 1081 d  913 1100 r  913 1081 d  938 1100 r
  938 1081 d  964 1100 r  964 1062 d  989 1100 r  989 1081 d 1015 1100 r
 1015 1081 d 1040 1100 r 1040 1081 d 1065 1100 r 1065 1081 d 1091 1100 r
 1091 1062 d 1116 1100 r 1116 1081 d 1142 1100 r 1142 1081 d 1167 1100 r
 1167 1081 d 1193 1100 r 1193 1081 d 1218 1100 r 1218 1062 d 1244 1100 r
 1244 1081 d 1269 1100 r 1269 1081 d 1295 1100 r 1295 1081 d 1320 1100 r
 1320 1081 d 1345 1100 r 1345 1062 d 1371 1100 r 1371 1081 d 1396 1100 r
 1396 1081 d 1422 1100 r 1422 1081 d 1447 1100 r 1447 1081 d 1473 1100 r
 1473 1062 d 1498 1100 r 1498 1081 d 1524 1100 r 1524 1081 d 1549 1100 r
 1549 1081 d 1575 1100 r 1575 1081 d 1600 1100 r 1600 1062 d  200  200 r
  200 1100 d  200  200 r  238  200 d  200  256 r  219  256 d  200  313 r
  219  313 d  200  369 r  219  369 d  200  425 r  219  425 d  200  481 r
  238  481 d  200  538 r  219  538 d  200  594 r  219  594 d  200  650 r
  219  650 d  200  706 r  219  706 d  200  762 r  238  762 d  200  819 r
  219  819 d  200  875 r  219  875 d  200  931 r  219  931 d  200  987 r
  219  987 d  200 1044 r  238 1044 d  200 1100 r  219 1100 d  168  214 r
  164  213 d  162  210 d  161  204 d  161  200 d  162  194 d  164  190 d
  168  189 d  170  189 d  174  190 d  176  194 d  177  200 d  177  204 d
  176  210 d  174  213 d  170  214 d  168  214 d  165  213 d  164  212 d
  163  210 d  162  204 d  162  200 d  163  194 d  164  192 d  165  190 d
  168  189 d  170  189 r  173  190 d  174  192 d  175  194 d  176  200 d
  176  204 d  175  210 d  174  212 d  173  213 d  170  214 d  139  495 r
  137  483 d  139  485 d  143  486 d  146  486 d  150  485 d  152  482 d
  153  479 d  153  477 d  152  474 d  150  471 d  146  470 d  143  470 d
  139  471 d  138  473 d  137  475 d  137  476 d  138  477 d  139  476 d
  138  475 d  146  486 r  149  485 d  151  482 d  152  479 d  152  477 d
  151  474 d  149  471 d  146  470 d  139  495 r  151  495 d  139  494 r
  145  494 d  151  495 d  168  495 r  164  494 d  162  491 d  161  485 d
  161  481 d  162  475 d  164  471 d  168  470 d  170  470 d  174  471 d
  176  475 d  177  481 d  177  485 d  176  491 d  174  494 d  170  495 d
  168  495 d  165  494 d  164  493 d  163  491 d  162  485 d  162  481 d
  163  475 d  164  473 d  165  471 d  168  470 d  170  470 r  173  471 d
  174  473 d  175  475 d  176  481 d  176  485 d  175  491 d  174  493 d
  173  494 d  170  495 d  116  772 r  119  773 d  122  776 d  122  751 d
  121  775 r  121  751 d  116  751 r  127  751 d  144  776 r  140  775 d
  138  772 d  137  766 d  137  762 d  138  756 d  140  752 d  144  751 d
  146  751 d  150  752 d  152  756 d  153  762 d  153  766 d  152  772 d
  150  775 d  146  776 d  144  776 d  141  775 d  140  774 d  139  772 d
  138  766 d  138  762 d  139  756 d  140  754 d  141  752 d  144  751 d
  146  751 r  149  752 d  150  754 d  151  756 d  152  762 d  152  766 d
  151  772 d  150  774 d  149  775 d  146  776 d  168  776 r  164  775 d
  162  772 d  161  766 d  161  762 d  162  756 d  164  752 d  168  751 d
  170  751 d  174  752 d  176  756 d  177  762 d  177  766 d  176  772 d
  174  775 d  170  776 d  168  776 d  165  775 d  164  774 d  163  772 d
  162  766 d  162  762 d  163  756 d  164  754 d  165  752 d  168  751 d
  170  751 r  173  752 d  174  754 d  175  756 d  176  762 d  176  766 d
  175  772 d  174  774 d  173  775 d  170  776 d  116 1054 r  119 1055 d
  122 1058 d  122 1033 d  121 1057 r  121 1033 d  116 1033 r  127 1033 d
  139 1058 r  137 1046 d  139 1048 d  143 1049 d  146 1049 d  150 1048 d
  152 1045 d  153 1042 d  153 1040 d  152 1037 d  150 1034 d  146 1033 d
  143 1033 d  139 1034 d  138 1036 d  137 1038 d  137 1039 d  138 1040 d
  139 1039 d  138 1038 d  146 1049 r  149 1048 d  151 1045 d  152 1042 d
  152 1040 d  151 1037 d  149 1034 d  146 1033 d  139 1058 r  151 1058 d
  139 1057 r  145 1057 d  151 1058 d  168 1058 r  164 1057 d  162 1054 d
  161 1048 d  161 1044 d  162 1038 d  164 1034 d  168 1033 d  170 1033 d
  174 1034 d  176 1038 d  177 1044 d  177 1048 d  176 1054 d  174 1057 d
  170 1058 d  168 1058 d  165 1057 d  164 1056 d  163 1054 d  162 1048 d
  162 1044 d  163 1038 d  164 1036 d  165 1034 d  168 1033 d  170 1033 r
  173 1034 d  174 1036 d  175 1038 d  176 1044 d  176 1048 d  175 1054 d
  174 1056 d  173 1057 d  170 1058 d 1600  200 r 1600 1100 d 1600  200 r
 1562  200 d 1600  256 r 1581  256 d 1600  313 r 1581  313 d 1600  369 r
 1581  369 d 1600  425 r 1581  425 d 1600  481 r 1562  481 d 1600  538 r
 1581  538 d 1600  594 r 1581  594 d 1600  650 r 1581  650 d 1600  706 r
 1581  706 d 1600  762 r 1562  762 d 1600  819 r 1581  819 d 1600  875 r
 1581  875 d 1600  931 r 1581  931 d 1600  987 r 1581  987 d 1600 1044 r
 1562 1044 d 1600 1100 r 1581 1100 d 1600  200 r 1575  200 d 1575  206 d
 1549  206 d 1549  211 d 1524  211 d 1524  211 p 1498  211 d 1498  234 d
 1473  234 d 1473  217 d 1447  217 d 1447  313 d 1422  313 d 1422  273 d
 1396  273 d 1396  391 d 1371  391 d 1371  386 d 1345  386 d 1345  408 d
 1320  408 d 1320  583 d 1295  583 d 1295  639 d 1269  639 d 1269  656 d
 1244  656 d 1244  757 d 1218  757 d 1218  650 d 1193  650 d 1193  650 p
 1167  650 d 1167  611 d 1142  611 d 1142  689 d 1116  689 d 1116  571 d
 1091  571 d 1091  476 d 1065  476 d 1065  358 d 1040  358 d 1040  358 p
 1015  358 d 1015  296 d  989  296 d  989  251 d  964  251 d  964  234 d
  938  234 d  938  200 d  913  200 d  913  200 p  887  200 d  887  200 p
  862  200 d  862  200 p  836  200 d  836  200 p  811  200 d  811  200 p
  785  200 d  785  200 p  760  200 d  760  200 p  735  200 d  735  200 p
  709  200 d  709  200 p  684  200 d  684  200 p  658  200 d  658  200 p
  633  200 d  633  200 p  607  200 d  607  200 p  582  200 d  582  200 p
  556  200 d  556  200 p  531  200 d  531  200 p  505  200 d  505  200 p
  480  200 d  480  200 p  455  200 d  455  200 p  429  200 d  429  200 p
  404  200 d  404  200 p  378  200 d  378  200 p  353  200 d  353  200 p
  327  200 d  327  200 p  302  200 d  302  200 p  276  200 d  276  200 p
  251  200 d  251  200 p  225  200 d  225  200 p  200  200 d  495  308 r
  485  308 d  495  309 r  485  309 d  495  310 r  485  310 d  495  311 r
  485  311 d  495  312 r  485  312 d  495  313 r  485  313 d  495  314 r
  485  314 d  495  315 r  485  315 d  495  316 r  485  316 d  495  317 r
  485  317 d  495  318 r  485  318 d  490  313 r  519  313 d  519  320 r
  519  306 d  519  320 d  490  313 r  461  313 d  461  320 r  461  306 d
  461  320 d 1025  999 r 1027  996 d 1027 1002 d 1025  999 d 1023 1001 d
 1019 1002 d 1017 1002 d 1013 1001 d 1011  999 d 1010  996 d 1009  993 d
 1009  987 d 1010  983 d 1011  981 d 1013  978 d 1017  977 d 1019  977 d
 1023  978 d 1025  981 d 1027  983 d 1017 1002 r 1015 1001 d 1012  999 d
 1011  996 d 1010  993 d 1010  987 d 1011  983 d 1012  981 d 1015  978 d
 1017  977 d 1036 1002 r 1036  977 d 1037 1002 r 1037  977 d 1037  990 r
 1040  993 d 1043  994 d 1046  994 d 1049  993 d 1051  990 d 1051  977 d
 1046  994 r 1048  993 d 1049  990 d 1049  977 d 1033 1002 r 1037 1002 d
 1033  977 r 1041  977 d 1046  977 r 1054  977 d 1063 1002 r 1063  977 d
 1064 1002 r 1064  977 d 1059 1002 r 1064 1002 d 1059  977 r 1067  977 d
 1081  994 r 1077  993 d 1075  990 d 1073  987 d 1073  984 d 1075  981 d
 1077  978 d 1081  977 d 1083  977 d 1087  978 d 1089  981 d 1090  984 d
 1090  987 d 1089  990 d 1087  993 d 1083  994 d 1081  994 d 1078  993 d
 1076  990 d 1075  987 d 1075  984 d 1076  981 d 1078  978 d 1081  977 d
 1083  977 r 1085  978 d 1088  981 d 1089  984 d 1089  987 d 1088  990 d
 1085  993 d 1083  994 d 1100  994 r 1100  977 d 1101  994 r 1101  977 d
 1101  987 r 1102  990 d 1105  993 d 1107  994 d 1111  994 d 1112  993 d
 1112  992 d 1111  990 d 1109  992 d 1111  993 d 1096  994 r 1101  994 d
 1096  977 r 1105  977 d 1120 1002 r 1119 1001 d 1120 1000 d 1121 1001 d
 1120 1002 d 1120  994 r 1120  977 d 1121  994 r 1121  977 d 1117  994 r
 1121  994 d 1117  977 r 1125  977 d 1133  994 r 1133  977 d 1135  994 r
 1135  977 d 1135  990 r 1137  993 d 1141  994 d 1143  994 d 1147  993 d
 1148  990 d 1148  977 d 1143  994 r 1145  993 d 1147  990 d 1147  977 d
 1130  994 r 1135  994 d 1130  977 r 1138  977 d 1143  977 r 1151  977 d
 1159  987 r 1173  987 d 1173  989 d 1172  992 d 1171  993 d 1168  994 d
 1165  994 d 1161  993 d 1159  990 d 1157  987 d 1157  984 d 1159  981 d
 1161  978 d 1165  977 d 1167  977 d 1171  978 d 1173  981 d 1172  987 r
 1172  990 d 1171  993 d 1165  994 r 1162  993 d 1160  990 d 1159  987 d
 1159  984 d 1160  981 d 1162  978 d 1165  977 d  878  885 r  880  886 d
  884  889 d  884  864 d  883  888 r  883  864 d  878  864 r  889  864 d
  905  889 r  902  888 d  899  885 d  898  879 d  898  875 d  899  869 d
  902  865 d  905  864 d  908  864 d  911  865 d  914  869 d  915  875 d
  915  879 d  914  885 d  911  888 d  908  889 d  905  889 d  903  888 d
  902  887 d  901  885 d  899  879 d  899  875 d  901  869 d  902  867 d
  903  865 d  905  864 d  908  864 r  910  865 d  911  867 d  913  869 d
  914  875 d  914  879 d  913  885 d  911  887 d  910  888 d  908  889 d
  929  889 r  926  888 d  923  885 d  922  879 d  922  875 d  923  869 d
  926  865 d  929  864 d  932  864 d  935  865 d  938  869 d  939  875 d
  939  879 d  938  885 d  935  888 d  932  889 d  929  889 d  927  888 d
  926  887 d  925  885 d  923  879 d  923  875 d  925  869 d  926  867 d
  927  865 d  929  864 d  932  864 r  934  865 d  935  867 d  937  869 d
  938  875 d  938  879 d  937  885 d  935  887 d  934  888 d  932  889 d
  953  889 r  950  888 d  947  885 d  946  879 d  946  875 d  947  869 d
  950  865 d  953  864 d  956  864 d  959  865 d  962  869 d  963  875 d
  963  879 d  962  885 d  959  888 d  956  889 d  953  889 d  951  888 d
  950  887 d  949  885 d  947  879 d  947  875 d  949  869 d  950  867 d
  951  865 d  953  864 d  956  864 r  958  865 d  959  867 d  961  869 d
  962  875 d  962  879 d  961  885 d  959  887 d  958  888 d  956  889 d
 1007  887 r 1009  889 d 1009  885 d 1007  887 d 1006  888 d 1003  889 d
  998  889 d  994  888 d  992  886 d  992  883 d  993  881 d  994  880 d
  997  879 d 1004  876 d 1006  875 d 1009  873 d  992  883 r  994  881 d
  997  880 d 1004  877 d 1006  876 d 1007  875 d 1009  873 d 1009  868 d
 1006  865 d 1003  864 d  998  864 d  994  865 d  993  867 d  992  869 d
  992  864 d  993  867 d 1018  889 r 1018  869 d 1019  865 d 1022  864 d
 1024  864 d 1027  865 d 1028  868 d 1019  889 r 1019  869 d 1021  865 d
 1022  864 d 1015  881 r 1024  881 d 1036  879 r 1036  877 d 1035  877 d
 1035  879 d 1036  880 d 1039  881 d 1043  881 d 1046  880 d 1047  879 d
 1048  876 d 1048  868 d 1049  865 d 1051  864 d 1047  879 r 1047  868 d
 1048  865 d 1051  864 d 1052  864 d 1047  876 r 1046  875 d 1039  874 d
 1035  873 d 1034  870 d 1034  868 d 1035  865 d 1039  864 d 1042  864 d
 1045  865 d 1047  868 d 1039  874 r 1036  873 d 1035  870 d 1035  868 d
 1036  865 d 1039  864 d 1060  881 r 1060  864 d 1061  881 r 1061  864 d
 1061  877 r 1064  880 d 1067  881 d 1070  881 d 1073  880 d 1075  877 d
 1075  864 d 1070  881 r 1072  880 d 1073  877 d 1073  864 d 1057  881 r
 1061  881 d 1057  864 r 1065  864 d 1070  864 r 1078  864 d 1099  889 r
 1099  864 d 1100  889 r 1100  864 d 1099  877 r 1096  880 d 1094  881 d
 1091  881 d 1088  880 d 1085  877 d 1084  874 d 1084  871 d 1085  868 d
 1088  865 d 1091  864 d 1094  864 d 1096  865 d 1099  868 d 1091  881 r
 1089  880 d 1087  877 d 1085  874 d 1085  871 d 1087  868 d 1089  865 d
 1091  864 d 1095  889 r 1100  889 d 1099  864 r 1103  864 d 1112  879 r
 1112  877 d 1111  877 d 1111  879 d 1112  880 d 1114  881 d 1119  881 d
 1121  880 d 1123  879 d 1124  876 d 1124  868 d 1125  865 d 1126  864 d
 1123  879 r 1123  868 d 1124  865 d 1126  864 d 1127  864 d 1123  876 r
 1121  875 d 1114  874 d 1111  873 d 1109  870 d 1109  868 d 1111  865 d
 1114  864 d 1118  864 d 1120  865 d 1123  868 d 1114  874 r 1112  873 d
 1111  870 d 1111  868 d 1112  865 d 1114  864 d 1136  881 r 1136  864 d
 1137  881 r 1137  864 d 1137  874 r 1138  877 d 1141  880 d 1143  881 d
 1147  881 d 1148  880 d 1148  879 d 1147  877 d 1145  879 d 1147  880 d
 1132  881 r 1137  881 d 1132  864 r 1141  864 d 1168  889 r 1168  864 d
 1169  889 r 1169  864 d 1168  877 r 1166  880 d 1163  881 d 1161  881 d
 1157  880 d 1155  877 d 1154  874 d 1154  871 d 1155  868 d 1157  865 d
 1161  864 d 1163  864 d 1166  865 d 1168  868 d 1161  881 r 1159  880 d
 1156  877 d 1155  874 d 1155  871 d 1156  868 d 1159  865 d 1161  864 d
 1165  889 r 1169  889 d 1168  864 r 1173  864 d 1203  889 r 1203  864 d
 1204  889 r 1211  868 d 1203  889 r 1211  864 d 1220  889 r 1211  864 d
 1220  889 r 1220  864 d 1221  889 r 1221  864 d 1199  889 r 1204  889 d
 1220  889 r 1225  889 d 1199  864 r 1207  864 d 1216  864 r 1225  864 d
 1238  881 r 1234  880 d 1232  877 d 1231  874 d 1231  871 d 1232  868 d
 1234  865 d 1238  864 d 1240  864 d 1244  865 d 1246  868 d 1247  871 d
 1247  874 d 1246  877 d 1244  880 d 1240  881 d 1238  881 d 1235  880 d
 1233  877 d 1232  874 d 1232  871 d 1233  868 d 1235  865 d 1238  864 d
 1240  864 r 1243  865 d 1245  868 d 1246  871 d 1246  874 d 1245  877 d
 1243  880 d 1240  881 d 1269  889 r 1269  864 d 1270  889 r 1270  864 d
 1269  877 r 1267  880 d 1264  881 d 1262  881 d 1258  880 d 1256  877 d
 1255  874 d 1255  871 d 1256  868 d 1258  865 d 1262  864 d 1264  864 d
 1267  865 d 1269  868 d 1262  881 r 1259  880 d 1257  877 d 1256  874 d
 1256  871 d 1257  868 d 1259  865 d 1262  864 d 1265  889 r 1270  889 d
 1269  864 r 1274  864 d 1281  874 r 1295  874 d 1295  876 d 1294  879 d
 1293  880 d 1291  881 d 1287  881 d 1283  880 d 1281  877 d 1280  874 d
 1280  871 d 1281  868 d 1283  865 d 1287  864 d 1289  864 d 1293  865 d
 1295  868 d 1294  874 r 1294  877 d 1293  880 d 1287  881 r 1285  880 d
 1282  877 d 1281  874 d 1281  871 d 1282  868 d 1285  865 d 1287  864 d
 1305  889 r 1305  864 d 1306  889 r 1306  864 d 1301  889 r 1306  889 d
 1301  864 r 1310  864 d 1328  880 r 1329  881 d 1329  877 d 1328  880 d
 1324  881 d 1321  881 d 1317  880 d 1316  879 d 1316  877 d 1317  875 d
 1319  874 d 1325  871 d 1328  870 d 1329  869 d 1316  877 r 1317  876 d
 1319  875 d 1325  873 d 1328  871 d 1329  869 d 1329  867 d 1328  865 d
 1324  864 d 1321  864 d 1317  865 d 1316  868 d 1316  864 d 1317  865 d
  490  763 r  490  352 d  487  345 r  490  341 d  492  345 d  484  348 r
  490  342 d  496  348 d  490  363 r  490  342 d  479  799 r  482  801 d
  476  801 d  479  799 d  477  797 d  476  793 d  476  791 d  477  787 d
  479  785 d  482  784 d  485  783 d  491  783 d  495  784 d  497  785 d
  500  787 d  501  791 d  501  793 d  500  797 d  497  799 d  495  801 d
  476  791 r  477  789 d  479  786 d  482  785 d  485  784 d  491  784 d
  495  785 d  497  786 d  500  789 d  501  791 d  476  810 r  501  810 d
  476  811 r  501  811 d  476  807 r  476  811 d  501  807 r  501  815 d
  476  845 r  501  845 d  476  846 r  501  846 d  483  853 r  492  853 d
  476  841 r  476  861 d  482  861 d  476  859 d  488  846 r  488  853 d
  501  841 r  501  861 d  495  861 d  501  859 d  484  870 r  501  883 d
  484  871 r  501  885 d  484  885 r  501  870 d  484  867 r  484  875 d
  484  881 r  484  888 d  501  867 r  501  874 d  501  880 r  501  888 d
  484  897 r  509  897 d  484  898 r  509  898 d  488  898 r  485  900 d
  484  903 d  484  905 d  485  909 d  488  911 d  491  912 d  494  912 d
  497  911 d  500  909 d  501  905 d  501  903 d  500  900 d  497  898 d
  484  905 r  485  907 d  488  910 d  491  911 d  494  911 d  497  910 d
  500  907 d  501  905 d  484  893 r  484  898 d  509  893 r  509  901 d
   57  435 r   85  435 d   57  436 r   82  452 d   60  436 r   85  452 d
   57  452 r   85  452 d   57  431 r   57  436 d   57  448 r   57  456 d
   85  431 r   85  439 d   67  465 r   81  465 d   83  466 d   85  470 d
   85  473 d   83  477 d   81  479 d   67  466 r   81  466 d   83  467 d
   85  470 d   67  479 r   85  479 d   67  480 r   85  480 d   67  461 r
   67  466 d   67  475 r   67  480 d   85  479 r   85  484 d   67  493 r
   85  493 d   67  495 r   85  495 d   70  495 r   68  497 d   67  501 d
   67  504 d   68  508 d   70  509 d   85  509 d   67  504 r   68  506 d
   70  508 d   85  508 d   70  509 r   68  512 d   67  516 d   67  518 d
   68  522 d   70  523 d   85  523 d   67  518 r   68  521 d   70  522 d
   85  522 d   67  490 r   67  495 d   85  490 r   85  499 d   85  504 r
   85  513 d   85  518 r   85  527 d   57  536 r   85  536 d   57  538 r
   85  538 d   70  538 r   68  540 d   67  543 d   67  545 d   68  549 d
   70  552 d   74  553 d   77  553 d   81  552 d   83  549 d   85  545 d
   85  543 d   83  540 d   81  538 d   67  545 r   68  548 d   70  551 d
   74  552 d   77  552 d   81  551 d   83  548 d   85  545 d   57  532 r
   57  538 d   74  562 r   74  578 d   72  578 d   69  577 d   68  575 d
   67  573 d   67  569 d   68  565 d   70  562 d   74  561 d   77  561 d
   81  562 d   83  565 d   85  569 d   85  571 d   83  575 d   81  578 d
   74  577 r   70  577 d   68  575 d   67  569 r   68  566 d   70  564 d
   74  562 d   77  562 d   81  564 d   83  566 d   85  569 d   67  588 r
   85  588 d   67  590 r   85  590 d   74  590 r   70  591 d   68  594 d
   67  596 d   67  600 d   68  601 d   69  601 d   70  600 d   69  599 d
   68  600 d   67  584 r   67  590 d   85  584 r   85  594 d   67  639 r
   68  635 d   70  633 d   74  631 d   77  631 d   81  633 d   83  635 d
   85  639 d   85  642 d   83  646 d   81  648 d   77  649 d   74  649 d
   70  648 d   68  646 d   67  642 d   67  639 d   68  636 d   70  634 d
   74  633 d   77  633 d   81  634 d   83  636 d   85  639 d   85  642 r
   83  644 d   81  647 d   77  648 d   74  648 d   70  647 d   68  644 d
   67  642 d   59  666 r   60  665 d   61  666 d   60  668 d   59  668 d
   57  666 d   57  664 d   59  661 d   61  660 d   85  660 d   57  664 r
   59  662 d   61  661 d   85  661 d   67  656 r   67  665 d   85  656 r
   85  665 d   57  699 r   85  699 d   57  700 r   81  708 d   57  699 r
   85  708 d   57  717 r   85  708 d   57  717 r   85  717 d   57  718 r
   85  718 d   57  695 r   57  700 d   57  717 r   57  722 d   85  695 r
   85  703 d   85  713 r   85  722 d   67  737 r   68  733 d   70  730 d
   74  729 d   77  729 d   81  730 d   83  733 d   85  737 d   85  739 d
   83  743 d   81  746 d   77  747 d   74  747 d   70  746 d   68  743 d
   67  739 d   67  737 d   68  734 d   70  731 d   74  730 d   77  730 d
   81  731 d   83  734 d   85  737 d   85  739 r   83  742 d   81  744 d
   77  746 d   74  746 d   70  744 d   68  742 d   67  739 d   57  770 r
   85  770 d   57  772 r   85  772 d   70  770 r   68  768 d   67  765 d
   67  763 d   68  759 d   70  756 d   74  755 d   77  755 d   81  756 d
   83  759 d   85  763 d   85  765 d   83  768 d   81  770 d   67  763 r
   68  760 d   70  757 d   74  756 d   77  756 d   81  757 d   83  760 d
   85  763 d   57  766 r   57  772 d   85  770 r   85  776 d   74  783 r
   74  799 d   72  799 d   69  798 d   68  796 d   67  794 d   67  790 d
   68  786 d   70  783 d   74  782 d   77  782 d   81  783 d   83  786 d
   85  790 d   85  792 d   83  796 d   81  799 d   74  798 r   70  798 d
   68  796 d   67  790 r   68  787 d   70  785 d   74  783 d   77  783 d
   81  785 d   83  787 d   85  790 d   57  809 r   85  809 d   57  811 r
   85  811 d   57  805 r   57  811 d   85  805 r   85  815 d   68  834 r
   67  835 d   70  835 d   68  834 d   67  830 d   67  826 d   68  822 d
   69  821 d   70  821 d   73  822 d   74  825 d   77  831 d   78  834 d
   80  835 d   70  821 r   72  822 d   73  825 d   76  831 d   77  834 d
   80  835 d   82  835 d   83  834 d   85  830 d   85  826 d   83  822 d
   81  821 d   85  821 d   83  822 d  741  129 r  742  126 d  742  133 d
  741  129 d  739  131 d  735  133 d  732  133 d  728  131 d  726  129 d
  724  126 d  723  122 d  723  116 d  724  112 d  726  109 d  728  107 d
  732  105 d  735  105 d  739  107 d  741  109 d  742  112 d  732  133 r
  729  131 d  727  129 d  726  126 d  724  122 d  724  116 d  726  112 d
  727  109 d  729  107 d  732  105 d  753  133 r  753  105 d  754  133 r
  754  105 d  749  133 r  754  133 d  749  105 r  758  105 d  805  130 r
  806  133 d  806  127 d  805  130 d  804  131 d  800  133 d  794  133 d
  791  131 d  788  129 d  788  126 d  789  124 d  791  122 d  793  121 d
  801  118 d  804  117 d  806  114 d  788  126 r  791  124 d  793  122 d
  801  120 d  804  118 d  805  117 d  806  114 d  806  109 d  804  107 d
  800  105 d  794  105 d  791  107 d  789  108 d  788  111 d  788  105 d
  789  108 d  817  133 r  817  105 d  818  133 r  833  108 d  818  130 r
  833  105 d  833  133 r  833  105 d  813  133 r  818  133 d  830  133 r
  837  133 d  813  105 r  820  105 d  846  133 r  846  113 d  848  109 d
  850  107 d  854  105 d  857  105 d  861  107 d  863  109 d  865  113 d
  865  133 d  848  133 r  848  113 d  849  109 d  852  107 d  854  105 d
  843  133 r  852  133 d  861  133 r  869  133 d  876  124 r  875  122 d
  876  121 d  878  122 d  876  124 d  876  108 r  875  107 d  876  105 d
  878  107 d  876  108 d  940  133 r  940  105 d  941  133 r  941  105 d
  932  133 r  931  126 d  931  133 d  950  133 d  950  126 d  949  133 d
  936  105 r  945  105 d  960  133 r  960  105 d  961  133 r  961  105 d
  961  120 r  963  122 d  967  124 d  970  124 d  974  122 d  975  120 d
  975  105 d  970  124 r  973  122 d  974  120 d  974  105 d  956  133 r
  961  133 d  956  105 r  965  105 d  970  105 r  979  105 d  987  116 r
 1002  116 d 1002  118 d 1001  121 d 1000  122 d  997  124 d  993  124 d
  989  122 d  987  120 d  986  116 d  986  113 d  987  109 d  989  107 d
  993  105 d  996  105 d 1000  107 d 1002  109 d 1001  116 r 1001  120 d
 1000  122 d  993  124 r  991  122 d  988  120 d  987  116 d  987  113 d
  988  109 d  991  107 d  993  105 d 1018  124 r 1014  122 d 1012  120 d
 1010  116 d 1010  113 d 1012  109 d 1014  107 d 1018  105 d 1021  105 d
 1025  107 d 1027  109 d 1028  113 d 1028  116 d 1027  120 d 1025  122 d
 1021  124 d 1018  124 d 1015  122 d 1013  120 d 1012  116 d 1012  113 d
 1013  109 d 1015  107 d 1018  105 d 1021  105 r 1023  107 d 1026  109 d
 1027  113 d 1027  116 d 1026  120 d 1023  122 d 1021  124 d 1039  124 r
 1039  105 d 1040  124 r 1040  105 d 1040  116 r 1041  120 d 1044  122 d
 1047  124 d 1051  124 d 1052  122 d 1052  121 d 1051  120 d 1049  121 d
 1051  122 d 1035  124 r 1040  124 d 1035  105 r 1044  105 d 1060  124 r
 1067  105 d 1061  124 r 1067  108 d 1075  124 r 1067  105 d 1065  100 d
 1062   98 d 1060   96 d 1058   96 d 1057   98 d 1058   99 d 1060   98 d
 1056  124 r 1065  124 d 1071  124 r 1079  124 d
e
EndPSPlot
0 SPE
4787 15917 XY 0 SPB
 clear Figure end restore 
0 SPE 5509 15723 XY F25(Figure)S 248 x(2.)S 495 x(1000)S 247 x
(solar)S 248 x(mo)S 14 x(dels)S 249 x(vs)S 247 x(exp)S 14 x(erimen)S
-13 x(ts.)S -181 y F8(2)S
XP /F8 54 199 1 0 14 18 19 16 0
<03F0 0FF8 1E38 3838 7000 7000 E000 E7E0 FC70 F038 F01C E01C E01C
 E01C 701C 703C 3878 3FF0 0FC0>
PXLC RP
15983 15542 XY F8(6)S 181 y 272 x F25(The)S 247 x(n)S -14 x(um)S 
-14 x(b)S 14 x(er)S 248 x(of)S 247 x(precise)S 2 x(ly)S 247 x(calc)S
2 x(ulated)S 598 y 5509 X(solar)S 159 x(mo)S 14 x(dels)S 160 x(that)S
158 x(predict)S 159 x(di\013eren)S -14 x(t)S 159 x(solar)S 159 x
(neutrino)S 159 x(ev)S -14 x(en)S -14 x(t)S 159 x(rates)S 159 x(are)S
159 x(sho)S -13 x(wn)S 158 x(for)S 159 x(the)S 158 x(c)S -13 x
(hlorine)S 598 y 5509 X(exp)S 13 x(erime)S 2 x(n)S -14 x(t.)S -181 y 
F8(7)S 181 y 183 x F25(Eac)S -13 x(h)S 158 x(input)S 158 x
(parameter)S 160 x(in)S 159 x(eac)S -13 x(h)S 158 x(solar)S 160 x
(mo)S 14 x(del)S 160 x(w)S -14 x(as)S 159 x(dra)S -14 x(wn)S 158 x
(indep)S 14 x(enden)S -14 x(tly)S 159 x(from)S 597 y 5509 X(a)S 166 x
(normal)S 166 x(distribution)S 166 x(ha)S -14 x(ving)S 167 x(the)S
165 x(mean)S 167 x(and)S 165 x(the)S 166 x(standard)S 166 x
(deviation)S 166 x(appropriate)S 167 x(to)S 165 x(that)S 598 y 5509 X
(parameter.)S 167 x(The)S 166 x(exp)S 14 x(erimen)S -13 x(tal)S 166 x
(error)S 167 x(bar)S 166 x(includes)S 167 x(only)S 166 x(statistic)S
2 x(al)S 166 x(errors)S 167 x(\(1)S
XP /F26 /cmmi10 300 498 498.132 128 [-1 -10 42 30] PXLNF RP
XP /F26 27 285 1 0 22 17 18 24 0
<01FFFC 07FFFC 0FFFFC 1F0F00 3C0700 380700 700380 700780 F00700
 E00700 E00700 E00F00 E00E00 E01C00 F03C00 78F800 3FE000 1F8000>
PXLC RP
21280 18114 XY F26(\033)S 18 x F25(\).)S 20396 Y 5347 X F34
(di\013eren)S -15 x(tial)S 180 x(equations)S 182 x(of)S 181 x
(stellar)S 181 x(ev)S -15 x(olution.)S 783 y 6165 X(None)S 218 x(of)S
217 x(the)S 219 x(1,000)S 216 x(solar)S 217 x(mo)S 14 x(dels)S 218 x
(represen)S -14 x(ted)S 218 x(in)S 218 x(Figure)S 217 x(2)S 218 x
(has)S 218 x(a)S 217 x(neutrino)S 218 x(\015ux)S 678 y 5347 X(that)S
182 x(is)S 181 x(in)S 181 x(agreemen)S -15 x(t)S 182 x(with)S 182 x
(the)S 182 x(observ)S -15 x(ed)S 182 x(rate.)S 677 y 6165 X(Figure)S
291 x(3)S 292 x(sho)S -15 x(ws)S 292 x(a)S 292 x(simila)S -2 x(r)S
292 x(comparison)S 291 x(for)S 291 x(the)S 293 x(neutrino-electron)S
292 x(scattering)S 678 y 5347 X(\(Kamiok)S -32 x(ande)S 264 x(I)S 
15 x(I\))S 264 x(exp)S 16 x(erimen)S -16 x(t)S 264 x(and)S 264 x
(the)S 264 x(1000)S 263 x(solar)S 263 x(mo)S 14 x(dels.)S 264 x(The)S
264 x(Kamiok)S -31 x(ande)S 264 x(I)S 15 x(I)S 677 y 5347 X(exp)S 
15 x(erimen)S -15 x(t)S 253 x(is)S 253 x(only)S 252 x(sensitiv)S 
-15 x(e)S 254 x(to)S 253 x(the)S 254 x(high-energy)S 252 x(side)S 
253 x(of)S 253 x(the)S 254 x(8)S 368 y -273 x(\026)S -368 y 253 x
(neutrino)S 253 x(energy)S 678 y 5347 X(sp)S 15 x(ectrum.)S 227 x
(Although)S 226 x(for)S 227 x(the)S 228 x(Kamio)S -2 x(k)S -30 x
(ande)S 227 x(I)S 15 x(I)S 228 x(exp)S 15 x(erimen)S -15 x(t)S 227 x
(none)S 227 x(of)S 227 x(the)S 228 x(1000)S 226 x(solar)S 677 y 
5347 X(mo)S 14 x(dels)S 206 x(are)S 206 x(consisten)S -14 x(t)S 206 x
(with)S 205 x(the)S 207 x(observ)S -15 x(ed)S 206 x(v)S -30 x(alue,)S
205 x(the)S 206 x(discrepancy)S 207 x(is)S 205 x(only)S 206 x(a)S 
205 x(factor)S 677 y 5347 X(of)S 175 x(t)S -15 x(w)S -15 x(o)S -198 y 
F13(3,14)S 198 y 199 x F34(in)S 176 x(this)S 175 x(case)S 177 x
(\(compared)S 175 x(to)S 176 x(the)S 177 x(factor)S 175 x(of)S 175 x
(3.5)S 175 x(for)S 175 x(the)S 177 x(c)S -15 x(hlori)S -2 x(ne)S 
177 x(exp)S 15 x(erimen)S -15 x(t)S 678 y 5347 X(whic)S -16 x(h)S 
182 x(has)S 182 x(an)S 182 x(energy)S 182 x(threshold)S 181 x(an)S
182 x(order-of)S -2 x(-magnitude)S 181 x(lo)S -16 x(w)S -15 x(er\).)S
814 y 6165 X(Can)S 248 x(one)S 248 x(understand)S 248 x(wh)S -15 x
(y)S 248 x(the)S 248 x(Mon)S -15 x(te)S 248 x(Carlo)S 247 x(sim)S 
-16 x(ulations)S 247 x(pro)S 14 x(duce)S 249 x(suc)S -14 x(h)S 248 x
(w)S -16 x(ell-)S 677 y 5347 X(de\014ned)S 185 x(theoretical)S 185 x
(predictions?)S
XP /F34 89 410 0 0 32 30 31 40 0
<FFFC03FF80 FFFC03FF80 07E000F800 03E0007000 03F0006000 01F000C000
 01F801C000 00F8018000 00FC038000 007E030000 007E060000 003F060000
 001F0C0000 001F9C0000 000F980000 000FF80000 0007F00000 0007E00000
 0003E00000 0003E00000 0003E00000 0003E00000 0003E00000 0003E00000
 0003E00000 0003E00000 0003E00000 0003E00000 0003E00000 007FFF0000
 007FFF0000>
PXLC RP
12983 28090 XY F34(Y)S -46 x(es,)S 185 x(there)S 185 x(are)S 185 x
(\014v)S -15 x(e)S 185 x(reasons,)S 184 x(whic)S -16 x(h)S 185 x(I)S
185 x(list)S 184 x(b)S 15 x(elo)S -15 x(w)S 184 x(in)S 678 y 5347 X
(what)S 172 x(I)S 172 x(judge)S 172 x(to)S 172 x(b)S 16 x(e)S 172 x
(the)S 173 x(relativ)S -16 x(e)S 172 x(order)S 172 x(of)S 172 x(imp)S
14 x(ortance.)S 172 x(1\))S
XP /F37 84 391 8 0 35 30 31 32 0
<1FFFFFF0 1FFFFFF0 1F0781E0 3C0780E0 38078060 300F8060 700F0060
 600F00E0 600F00C0 E01F00C0 C01E00C0 001E0000 001E0000 003E0000
 003C0000 003C0000 003C0000 007C0000 00780000 00780000 00780000
 00F80000 00F00000 00F00000 00F00000 01F00000 01E00000 01E00000
 03E00000 FFFF0000 FFFF0000>
PXLC RP
18475 28768 XY F37(The)S
XP /F37 108 139 3 0 13 31 32 16 0
<03E0 1FE0 1FE0 03C0 03C0 07C0 07C0 0780 0780 0F80 0F80 0F00 0F00
 1F00 1F00 1E00 1E00 3E00 3E00 3C00 3C00 7C00 7C00 7800 7800 F9C0
 F980 F180 F380 F300 7F00 3E00>
PXLC RP
19581 28768 XY F37(lu)S
XP /F37 109 447 3 0 37 19 20 40 0
<1F0FC1F800 3F9FF3FE00 33F8F71E00 73F07E0F00 63E07C0F00 63C0F81F00
 E7C0F81E00 0780F01E00 0780F01E00 0781F03E00 0F81F03C00 0F01E03C00
 0F01E07C00 0F03E078E0 1F03E078C0 1E03C0F8C0 1E03C0F1C0 1E07C0F380
 3E07C07F00 1C03803E00>
PXLC RP
20014 28768 XY F37(minosity)S
XP /F37 98 251 4 0 20 31 32 24 0
<07C000 3FC000 3FC000 078000 078000 0F8000 0F8000 0F0000 0F0000
 1F0000 1F0000 1E0000 1EFC00 3FFE00 3F8F00 3F0700 3E0780 7C0780
 7C0780 780780 780780 780F80 F80F80 F00F00 F00F00 F01E00 F01E00
 F03C00 707800 78F000 3FE000 1F8000>
PXLC RP
22235 28768 XY F37(b)S -28 x(oun)S
XP /F37 100 279 4 0 24 31 32 24 0
<0000F8 0007F8 0007F8 0000F0 0000F0 0001F0 0001F0 0001E0 0001E0
 0003E0 0003E0 0003C0 00FBC0 03FFC0 079FC0 0F0F80 1E0780 3C0F80
 3C0F80 780F00 780F00 781F00 F81F00 F01E00 F01E00 F03E70 F03E60
 F03C60 707CE0 79FCC0 3FDFC0 1F0F80>
PXLC RP
23336 28768 XY F37(d)S
XP /F37 97 279 4 0 23 19 20 24 0
<00FB80 03FFC0 079FC0 0F0F80 1E0780 3C0F80 3C0F00 780F00 780F00
 781F00 F81E00 F01E00 F01E00 F03E70 F03C60 F03C60 707CE0 79FCC0
 3FDFC0 1F0F80>
PXLC RP
23615 28768 XY F37(ary)S 677 y 5347 X(c)S -28 x(ondition)S 166 x F34
(requires)S 167 x(that)S 166 x(the)S 168 x(computed)S 167 x(photon)S
167 x(lumi)S -2 x(nosit)S -15 x(y)S 166 x(of)S 167 x(the)S 167 x
(presen)S -14 x(t-da)S -16 x(y)S 167 x(solar)S 678 y 5347 X(mo)S 
14 x(del)S 191 x(equals)S 191 x(the)S 192 x(measured)S 191 x(solar)S
190 x(luminosi)S -2 x(t)S -15 x(y)S -45 x(,)S
XP /F35 76 372 2 0 27 30 31 32 0
<00FFFC00 00FFFC00 000F8000 000F0000 000F0000 001F0000 001E0000
 001E0000 001E0000 003E0000 003C0000 003C0000 003C0000 007C0000
 00780000 00780000 00780000 00F80000 00F00000 00F000C0 00F001C0
 01F00180 01E00180 01E00380 01E00300 03E00700 03C00F00 03C01E00
 07C07E00 7FFFFE00 FFFFFC00>
PXLC RP
15939 30123 XY F35(L)S
XP /F15 12 329 2 -4 23 19 24 24 0
<01FE00 07FF80 0F87C0 1E01E0 3C00F0 380070 700038 700038 E0001C
 E0781C C0FC0C C0FC0C C0FC0C C0FC0C E0781C E0001C 700038 700038
 380070 3C00F0 1E01E0 0F87C0 07FF80 01FE00>
PXLC RP
16310 30204 XY F15(\014)S -81 y 26 x F34(,)S 190 x(whic)S -15 x(h)S
191 x(is)S 191 x(kno)S -15 x(wn)S 191 x(exp)S 16 x(erimen)S -16 x
(tally)S 677 y 5347 X(to)S 135 x(an)S 136 x(accuracy)S 136 x(of)S 
136 x(ab)S 14 x(out)S 136 x(t)S -15 x(w)S -15 x(o)S 136 x(parts)S 
135 x(in)S 136 x(a)S 135 x(thousand.)S 136 x(If)S 135 x(one)S 136 x
(o)S -15 x(v)S -16 x(ersimpli\014es)S 135 x(the)S 137 x(problem)S 
677 y 5347 X(of)S 208 x(stellar)S 208 x(ev)S -15 x(olution)S 208 x
(and)S 209 x(represen)S -15 x(ts)S 209 x(the)S 210 x(output)S 209 x
(of)S 208 x(a)S 209 x(solar)S 208 x(mo)S 14 x(del)S 209 x(in)S 209 x
(terms)S 208 x(of)S 209 x(just)S 678 y 5347 X(the)S 271 x(cen)S -14 x
(tral)S 270 x(temp)S 15 x(erature,)S
XP /F35 84 319 1 0 30 30 31 32 0
<1FFFFFFC 1FFFFFFC 1F03C078 3C03C038 3803C018 3007C018 70078018
 60078038 60078030 E00F8030 C00F0030 000F0000 000F0000 001F0000
 001E0000 001E0000 001E0000 003E0000 003C0000 003C0000 003C0000
 007C0000 00780000 00780000 00780000 00F80000 00F00000 00F00000
 01F00000 7FFFC000 7FFFC000>
PXLC RP
11618 32155 XY F35(T)S
XP /F14 99 183 1 0 13 13 14 16 0
<03F0 0FF8 3E78 7878 7070 E000 E000 E000 E000 E000 E018 F078 7FF0
 1FC0>
PXLC RP
11937 32237 XY F14(c)S -82 y 296 x F34(\(as)S 271 x(is)S 270 x(done)S
271 x(in)S 271 x(sev)S -15 x(eral)S 270 x(recen)S -14 x(t)S 271 x
(pap)S 15 x(ers)S 271 x(b)S -15 x(y)S 271 x(di\013eren)S -15 x(t)S
677 y 5347 X(authors\),)S 195 x(then)S -198 y F13(27)S 198 y 222 x 
F34(the)S 197 x(\015ux)S 197 x(of)S 196 x(the)S 197 x(most-sensitiv)S
-16 x(e)S 197 x(neutrino)S 196 x(branc)S -15 x(h,)S 196 x(the)S 198 x
(8)S 369 y -273 x(\026)S -369 y 196 x(neutrinos)S 678 y 5347 X(is)S
XP /F35 30 325 2 -9 25 31 41 24 0
<0000C0 0000C0 0000C0 0001C0 0001C0 000180 000180 000380 000380
 000300 000300 000700 007FF0 01FFF8 07C63C 0F8E1E 1F0E1E 3E0C0F
 3C0C0F 7C1C0F 781C0F 78181F F8181F F0381E F0383E F0303C F0307C
 7870F8 7871F0 3C63E0 1FFF80 0FFE00 00E000 00C000 00C000 01C000
 01C000 018000 018000 038000 038000>
PXLC RP
5931 33510 XY F35(\036)S F34(\()S -198 y F13(8)S 198 y 25 x F34(B\))S
XP /F36 47 425 3 0 30 19 20 32 0
<07E001F0 1FF807F0 3FFC0F80 387E1E00 703F3C00 601FB800 E00FF000
 C007F000 C007E000 C003E000 C001F000 C001F800 C003F800 E003FC00
 60077E00 700F3F00 301E1F80 3C7C0FF0 1FF807F0 07E001F0>
PXLC RP
7514 33510 XY F36(/)S 211 x F35(T)S -198 y 75 x F13(18)S 333 y -499 x 
F14(c)S -135 y 558 x F34(and)S 217 x(the)S 218 x(luminosi)S -2 x(t)S
-15 x(y)S 218 x F35(L)S 82 y F15(\014)S -82 y 236 x F36(/)S 211 x 
F35(T)S -198 y 75 x F13(4)S 333 y -287 x F14(c)S -135 y 129 x F34(.)S
217 x(One)S 218 x(concludes)S 218 x(b)S -15 x(y)S 217 x(this)S 218 x
(argum)S -2 x(en)S -14 x(t)S
1 PP EP

1000 BP 39600 30600 PM 0 0 XY
4787 15917 XY 0 SPB
 save 10 dict begin /Figure exch def currentpoint translate
/showpage {} def 
0 SPE
4787 15917 XY 0 SPB
save 50 dict begin /psplot exch def
/StartPSPlot
   {newpath 0 0 moveto 0 setlinewidth 0 setgray 0 setlinecap
    1 setlinejoin 72 300 div dup scale}def
/pending {false} def
/finish {pending {currentpoint stroke moveto /pending false def} if} def
/r {finish newpath moveto} def
/d {lineto /pending true def} def
/l {finish 4 2 roll moveto lineto currentpoint stroke moveto} def
/p {finish newpath moveto currentpoint lineto currentpoint stroke moveto} def
/e {finish gsave showpage grestore newpath 0 0 moveto} def
/lw {finish setlinewidth} def
/lt0 {finish [] 0 setdash} def
/lt1 {finish [3 5] 0 setdash} def
/lt2 {finish [20 10] 0 setdash} def
/lt3 {finish [60 10] 0 setdash} def
/lt4 {finish [3 10 20 10] 0 setdash} def
/lt5 {finish [3 10 60 10] 0 setdash} def
/lt6 {finish [20 10 60 10] 0 setdash} def
/EndPSPlot {clear psplot end restore}def
StartPSPlot
   1 lw lt0  200  200 r 1600  200 d  200  200 r  200  238 d  270  200 r
  270  219 d  340  200 r  340  219 d  410  200 r  410  219 d  480  200 r
  480  219 d  550  200 r  550  238 d  620  200 r  620  219 d  690  200 r
  690  219 d  760  200 r  760  219 d  830  200 r  830  219 d  900  200 r
  900  238 d  970  200 r  970  219 d 1040  200 r 1040  219 d 1110  200 r
 1110  219 d 1180  200 r 1180  219 d 1250  200 r 1250  238 d 1320  200 r
 1320  219 d 1390  200 r 1390  219 d 1460  200 r 1460  219 d 1530  200 r
 1530  219 d 1600  200 r 1600  238 d  199  176 r  195  175 d  193  172 d
  192  166 d  192  162 d  193  156 d  195  152 d  199  151 d  201  151 d
  205  152 d  207  156 d  208  162 d  208  166 d  207  172 d  205  175 d
  201  176 d  199  176 d  196  175 d  195  174 d  194  172 d  193  166 d
  193  162 d  194  156 d  195  154 d  196  152 d  199  151 d  201  151 r
  204  152 d  205  154 d  206  156 d  207  162 d  207  166 d  206  172 d
  205  174 d  204  175 d  201  176 d  538  154 r  537  152 d  538  151 d
  539  152 d  538  154 d  549  176 r  546  164 d  549  166 d  552  167 d
  556  167 d  559  166 d  562  163 d  563  160 d  563  158 d  562  155 d
  559  152 d  556  151 d  552  151 d  549  152 d  547  154 d  546  156 d
  546  157 d  547  158 d  549  157 d  547  156 d  556  167 r  558  166 d
  561  163 d  562  160 d  562  158 d  561  155 d  558  152 d  556  151 d
  549  176 r  561  176 d  549  175 r  555  175 d  561  176 d  895  172 r
  898  173 d  901  176 d  901  151 d  900  175 r  900  151 d  895  151 r
  906  151 d 1228  172 r 1231  173 d 1234  176 d 1234  151 d 1233  175 r
 1233  151 d 1228  151 r 1239  151 d 1250  154 r 1249  152 d 1250  151 d
 1251  152 d 1250  154 d 1261  176 r 1258  164 d 1261  166 d 1264  167 d
 1268  167 d 1271  166 d 1274  163 d 1275  160 d 1275  158 d 1274  155 d
 1271  152 d 1268  151 d 1264  151 d 1261  152 d 1259  154 d 1258  156 d
 1258  157 d 1259  158 d 1261  157 d 1259  156 d 1268  167 r 1270  166 d
 1273  163 d 1274  160 d 1274  158 d 1273  155 d 1270  152 d 1268  151 d
 1261  176 r 1273  176 d 1261  175 r 1267  175 d 1273  176 d 1593  172 r
 1594  170 d 1593  169 d 1592  170 d 1592  172 d 1593  174 d 1594  175 d
 1598  176 d 1602  176 d 1606  175 d 1607  174 d 1608  172 d 1608  169 d
 1607  167 d 1604  164 d 1598  162 d 1595  161 d 1593  158 d 1592  155 d
 1592  151 d 1602  176 r 1605  175 d 1606  174 d 1607  172 d 1607  169 d
 1606  167 d 1602  164 d 1598  162 d 1592  154 r 1593  155 d 1595  155 d
 1601  152 d 1605  152 d 1607  154 d 1608  155 d 1595  155 r 1601  151 d
 1606  151 d 1607  152 d 1608  155 d 1608  157 d  200 1100 r 1600 1100 d
  200 1100 r  200 1062 d  270 1100 r  270 1081 d  340 1100 r  340 1081 d
  410 1100 r  410 1081 d  480 1100 r  480 1081 d  550 1100 r  550 1062 d
  620 1100 r  620 1081 d  690 1100 r  690 1081 d  760 1100 r  760 1081 d
  830 1100 r  830 1081 d  900 1100 r  900 1062 d  970 1100 r  970 1081 d
 1040 1100 r 1040 1081 d 1110 1100 r 1110 1081 d 1180 1100 r 1180 1081 d
 1250 1100 r 1250 1062 d 1320 1100 r 1320 1081 d 1390 1100 r 1390 1081 d
 1460 1100 r 1460 1081 d 1530 1100 r 1530 1081 d 1600 1100 r 1600 1062 d
  200  200 r  200 1100 d  200  200 r  238  200 d  200  230 r  219  230 d
  200  260 r  219  260 d  200  290 r  219  290 d  200  320 r  219  320 d
  200  350 r  238  350 d  200  380 r  219  380 d  200  410 r  219  410 d
  200  440 r  219  440 d  200  470 r  219  470 d  200  500 r  238  500 d
  200  530 r  219  530 d  200  560 r  219  560 d  200  590 r  219  590 d
  200  620 r  219  620 d  200  650 r  238  650 d  200  680 r  219  680 d
  200  710 r  219  710 d  200  740 r  219  740 d  200  770 r  219  770 d
  200  800 r  238  800 d  200  830 r  219  830 d  200  860 r  219  860 d
  200  890 r  219  890 d  200  920 r  219  920 d  200  950 r  238  950 d
  200  980 r  219  980 d  200 1010 r  219 1010 d  200 1040 r  219 1040 d
  200 1070 r  219 1070 d  200 1100 r  238 1100 d  168  214 r  164  213 d
  162  210 d  161  204 d  161  200 d  162  194 d  164  190 d  168  189 d
  170  189 d  174  190 d  176  194 d  177  200 d  177  204 d  176  210 d
  174  213 d  170  214 d  168  214 d  165  213 d  164  212 d  163  210 d
  162  204 d  162  200 d  163  194 d  164  192 d  165  190 d  168  189 d
  170  189 r  173  190 d  174  192 d  175  194 d  176  200 d  176  204 d
  175  210 d  174  212 d  173  213 d  170  214 d  139  364 r  137  352 d
  139  354 d  143  355 d  146  355 d  150  354 d  152  351 d  153  348 d
  153  346 d  152  343 d  150  340 d  146  339 d  143  339 d  139  340 d
  138  342 d  137  344 d  137  345 d  138  346 d  139  345 d  138  344 d
  146  355 r  149  354 d  151  351 d  152  348 d  152  346 d  151  343 d
  149  340 d  146  339 d  139  364 r  151  364 d  139  363 r  145  363 d
  151  364 d  168  364 r  164  363 d  162  360 d  161  354 d  161  350 d
  162  344 d  164  340 d  168  339 d  170  339 d  174  340 d  176  344 d
  177  350 d  177  354 d  176  360 d  174  363 d  170  364 d  168  364 d
  165  363 d  164  362 d  163  360 d  162  354 d  162  350 d  163  344 d
  164  342 d  165  340 d  168  339 d  170  339 r  173  340 d  174  342 d
  175  344 d  176  350 d  176  354 d  175  360 d  174  362 d  173  363 d
  170  364 d  116  510 r  119  511 d  122  514 d  122  489 d  121  513 r
  121  489 d  116  489 r  127  489 d  144  514 r  140  513 d  138  510 d
  137  504 d  137  500 d  138  494 d  140  490 d  144  489 d  146  489 d
  150  490 d  152  494 d  153  500 d  153  504 d  152  510 d  150  513 d
  146  514 d  144  514 d  141  513 d  140  512 d  139  510 d  138  504 d
  138  500 d  139  494 d  140  492 d  141  490 d  144  489 d  146  489 r
  149  490 d  150  492 d  151  494 d  152  500 d  152  504 d  151  510 d
  150  512 d  149  513 d  146  514 d  168  514 r  164  513 d  162  510 d
  161  504 d  161  500 d  162  494 d  164  490 d  168  489 d  170  489 d
  174  490 d  176  494 d  177  500 d  177  504 d  176  510 d  174  513 d
  170  514 d  168  514 d  165  513 d  164  512 d  163  510 d  162  504 d
  162  500 d  163  494 d  164  492 d  165  490 d  168  489 d  170  489 r
  173  490 d  174  492 d  175  494 d  176  500 d  176  504 d  175  510 d
  174  512 d  173  513 d  170  514 d  116  660 r  119  661 d  122  664 d
  122  639 d  121  663 r  121  639 d  116  639 r  127  639 d  139  664 r
  137  652 d  139  654 d  143  655 d  146  655 d  150  654 d  152  651 d
  153  648 d  153  646 d  152  643 d  150  640 d  146  639 d  143  639 d
  139  640 d  138  642 d  137  644 d  137  645 d  138  646 d  139  645 d
  138  644 d  146  655 r  149  654 d  151  651 d  152  648 d  152  646 d
  151  643 d  149  640 d  146  639 d  139  664 r  151  664 d  139  663 r
  145  663 d  151  664 d  168  664 r  164  663 d  162  660 d  161  654 d
  161  650 d  162  644 d  164  640 d  168  639 d  170  639 d  174  640 d
  176  644 d  177  650 d  177  654 d  176  660 d  174  663 d  170  664 d
  168  664 d  165  663 d  164  662 d  163  660 d  162  654 d  162  650 d
  163  644 d  164  642 d  165  640 d  168  639 d  170  639 r  173  640 d
  174  642 d  175  644 d  176  650 d  176  654 d  175  660 d  174  662 d
  173  663 d  170  664 d  114  810 r  115  808 d  114  807 d  113  808 d
  113  810 d  114  812 d  115  813 d  119  814 d  123  814 d  127  813 d
  128  812 d  129  810 d  129  807 d  128  805 d  125  802 d  119  800 d
  116  799 d  114  796 d  113  793 d  113  789 d  123  814 r  126  813 d
  127  812 d  128  810 d  128  807 d  127  805 d  123  802 d  119  800 d
  113  792 r  114  793 d  116  793 d  122  790 d  126  790 d  128  792 d
  129  793 d  116  793 r  122  789 d  127  789 d  128  790 d  129  793 d
  129  795 d  144  814 r  140  813 d  138  810 d  137  804 d  137  800 d
  138  794 d  140  790 d  144  789 d  146  789 d  150  790 d  152  794 d
  153  800 d  153  804 d  152  810 d  150  813 d  146  814 d  144  814 d
  141  813 d  140  812 d  139  810 d  138  804 d  138  800 d  139  794 d
  140  792 d  141  790 d  144  789 d  146  789 r  149  790 d  150  792 d
  151  794 d  152  800 d  152  804 d  151  810 d  150  812 d  149  813 d
  146  814 d  168  814 r  164  813 d  162  810 d  161  804 d  161  800 d
  162  794 d  164  790 d  168  789 d  170  789 d  174  790 d  176  794 d
  177  800 d  177  804 d  176  810 d  174  813 d  170  814 d  168  814 d
  165  813 d  164  812 d  163  810 d  162  804 d  162  800 d  163  794 d
  164  792 d  165  790 d  168  789 d  170  789 r  173  790 d  174  792 d
  175  794 d  176  800 d  176  804 d  175  810 d  174  812 d  173  813 d
  170  814 d  114  960 r  115  958 d  114  957 d  113  958 d  113  960 d
  114  962 d  115  963 d  119  964 d  123  964 d  127  963 d  128  962 d
  129  960 d  129  957 d  128  955 d  125  952 d  119  950 d  116  949 d
  114  946 d  113  943 d  113  939 d  123  964 r  126  963 d  127  962 d
  128  960 d  128  957 d  127  955 d  123  952 d  119  950 d  113  942 r
  114  943 d  116  943 d  122  940 d  126  940 d  128  942 d  129  943 d
  116  943 r  122  939 d  127  939 d  128  940 d  129  943 d  129  945 d
  139  964 r  137  952 d  139  954 d  143  955 d  146  955 d  150  954 d
  152  951 d  153  948 d  153  946 d  152  943 d  150  940 d  146  939 d
  143  939 d  139  940 d  138  942 d  137  944 d  137  945 d  138  946 d
  139  945 d  138  944 d  146  955 r  149  954 d  151  951 d  152  948 d
  152  946 d  151  943 d  149  940 d  146  939 d  139  964 r  151  964 d
  139  963 r  145  963 d  151  964 d  168  964 r  164  963 d  162  960 d
  161  954 d  161  950 d  162  944 d  164  940 d  168  939 d  170  939 d
  174  940 d  176  944 d  177  950 d  177  954 d  176  960 d  174  963 d
  170  964 d  168  964 d  165  963 d  164  962 d  163  960 d  162  954 d
  162  950 d  163  944 d  164  942 d  165  940 d  168  939 d  170  939 r
  173  940 d  174  942 d  175  944 d  176  950 d  176  954 d  175  960 d
  174  962 d  173  963 d  170  964 d  114 1111 r  115 1110 d  114 1108 d
  113 1110 d  113 1111 d  115 1113 d  119 1114 d  123 1114 d  127 1113 d
  128 1111 d  128 1107 d  127 1105 d  123 1104 d  120 1104 d  123 1114 r
  126 1113 d  127 1111 d  127 1107 d  126 1105 d  123 1104 d  126 1102 d
  128 1100 d  129 1098 d  129 1094 d  128 1092 d  127 1090 d  123 1089 d
  119 1089 d  115 1090 d  114 1092 d  113 1094 d  113 1095 d  114 1096 d
  115 1095 d  114 1094 d  127 1101 r  128 1098 d  128 1094 d  127 1092 d
  126 1090 d  123 1089 d  144 1114 r  140 1113 d  138 1110 d  137 1104 d
  137 1100 d  138 1094 d  140 1090 d  144 1089 d  146 1089 d  150 1090 d
  152 1094 d  153 1100 d  153 1104 d  152 1110 d  150 1113 d  146 1114 d
  144 1114 d  141 1113 d  140 1112 d  139 1110 d  138 1104 d  138 1100 d
  139 1094 d  140 1092 d  141 1090 d  144 1089 d  146 1089 r  149 1090 d
  150 1092 d  151 1094 d  152 1100 d  152 1104 d  151 1110 d  150 1112 d
  149 1113 d  146 1114 d  168 1114 r  164 1113 d  162 1110 d  161 1104 d
  161 1100 d  162 1094 d  164 1090 d  168 1089 d  170 1089 d  174 1090 d
  176 1094 d  177 1100 d  177 1104 d  176 1110 d  174 1113 d  170 1114 d
  168 1114 d  165 1113 d  164 1112 d  163 1110 d  162 1104 d  162 1100 d
  163 1094 d  164 1092 d  165 1090 d  168 1089 d  170 1089 r  173 1090 d
  174 1092 d  175 1094 d  176 1100 d  176 1104 d  175 1110 d  174 1112 d
  173 1113 d  170 1114 d 1600  200 r 1600 1100 d 1600  200 r 1562  200 d
 1600  230 r 1581  230 d 1600  260 r 1581  260 d 1600  290 r 1581  290 d
 1600  320 r 1581  320 d 1600  350 r 1562  350 d 1600  380 r 1581  380 d
 1600  410 r 1581  410 d 1600  440 r 1581  440 d 1600  470 r 1581  470 d
 1600  500 r 1562  500 d 1600  530 r 1581  530 d 1600  560 r 1581  560 d
 1600  590 r 1581  590 d 1600  620 r 1581  620 d 1600  650 r 1562  650 d
 1600  680 r 1581  680 d 1600  710 r 1581  710 d 1600  740 r 1581  740 d
 1600  770 r 1581  770 d 1600  800 r 1562  800 d 1600  830 r 1581  830 d
 1600  860 r 1581  860 d 1600  890 r 1581  890 d 1600  920 r 1581  920 d
 1600  950 r 1562  950 d 1600  980 r 1581  980 d 1600 1010 r 1581 1010 d
 1600 1040 r 1581 1040 d 1600 1070 r 1581 1070 d 1600 1100 r 1562 1100 d
 1600  200 r 1565  200 d 1565  200 p 1530  200 d 1530  200 p 1495  200 d
 1495  200 p 1460  200 d 1460  200 p 1425  200 d 1425  200 p 1390  200 d
 1390  200 p 1355  200 d 1355  200 p 1320  200 d 1320  200 p 1285  200 d
 1285  200 p 1250  200 d 1250  200 p 1215  200 d 1215  200 p 1180  200 d
 1180  206 d 1145  206 d 1145  218 d 1110  218 d 1110  239 d 1075  239 d
 1075  299 d 1040  299 d 1040  368 d 1005  368 d 1005  473 d  970  473 d
  970  635 d  935  635 d  935  674 d  900  674 d  900  611 d  865  611 d
  865  647 d  830  647 d  830  509 d  795  509 d  795  368 d  760  368 d
  760  320 d  725  320 d  725  227 d  690  227 d  690  206 d  655  206 d
  655  200 d  620  200 d  620  200 p  585  200 d  585  200 p  550  200 d
  550  200 p  515  200 d  515  200 p  480  200 d  480  200 p  445  200 d
  445  200 p  410  200 d  410  200 p  375  200 d  375  200 p  340  200 d
  340  200 p  305  200 d  305  200 p  270  200 d  270  200 p  235  200 d
  235  200 p  200  200 d  541  315 r  531  315 d  541  316 r  531  316 d
  541  317 r  531  317 d  541  318 r  531  318 d  541  319 r  531  319 d
  541  320 r  531  320 d  541  321 r  531  321 d  541  322 r  531  322 d
  541  323 r  531  323 d  541  324 r  531  324 d  541  325 r  531  325 d
  536  320 r  571  320 d  571  327 r  571  313 d  571  327 d  536  320 r
  501  320 d  501  327 r  501  313 d  501  327 d 1015  978 r 1013  977 d
 1012  976 d 1012  974 d 1013  972 d 1015  971 d 1018  971 d 1020  972 d
 1021  974 d 1021  976 d 1020  977 d 1018  978 d 1015  978 d 1014  977 d
 1013  976 d 1013  974 d 1014  972 d 1015  971 d 1018  971 r 1020  972 d
 1020  974 d 1020  976 d 1020  977 d 1018  978 d 1015  971 r 1013  971 d
 1012  970 d 1012  968 d 1012  966 d 1012  964 d 1013  963 d 1015  963 d
 1018  963 d 1020  963 d 1021  964 d 1022  966 d 1022  968 d 1021  970 d
 1020  971 d 1018  971 d 1015  971 r 1014  971 d 1013  970 d 1012  968 d
 1012  966 d 1013  964 d 1014  963 d 1015  963 d 1018  963 r 1020  963 d
 1020  964 d 1021  966 d 1021  968 d 1020  970 d 1020  971 d 1018  971 d
 1030  964 r 1030  939 d 1031  964 r 1031  939 d 1026  964 r 1041  964 d
 1044  963 d 1046  962 d 1047  960 d 1047  957 d 1046  955 d 1044  954 d
 1041  952 d 1041  964 r 1043  963 d 1044  962 d 1046  960 d 1046  957 d
 1044  955 d 1043  954 d 1041  952 d 1031  952 r 1041  952 d 1044  951 d
 1046  950 d 1047  948 d 1047  944 d 1046  942 d 1044  940 d 1041  939 d
 1026  939 d 1041  952 r 1043  951 d 1044  950 d 1046  948 d 1046  944 d
 1044  942 d 1043  940 d 1041  939 d 1078  964 r 1078  939 d 1079  964 r
 1079  939 d 1086  957 r 1086  948 d 1074  964 r 1094  964 d 1094  958 d
 1092  964 d 1079  952 r 1086  952 d 1074  939 r 1083  939 d 1102  964 r
 1102  939 d 1103  964 r 1103  939 d 1098  964 r 1103  964 d 1098  939 r
 1107  939 d 1115  956 r 1115  943 d 1116  940 d 1120  939 d 1122  939 d
 1126  940 d 1128  943 d 1116  956 r 1116  943 d 1118  940 d 1120  939 d
 1128  956 r 1128  939 d 1130  956 r 1130  939 d 1112  956 r 1116  956 d
 1125  956 r 1130  956 d 1128  939 r 1133  939 d 1142  956 r 1155  939 d
 1143  956 r 1156  939 d 1156  956 r 1142  939 d 1138  956 r 1146  956 d
 1152  956 r 1160  956 d 1138  939 r 1145  939 d 1151  939 r 1160  939 d
  851  870 r  854  871 d  857  874 d  857  849 d  856  873 r  856  849 d
  851  849 r  862  849 d  879  874 r  875  873 d  873  870 d  872  864 d
  872  860 d  873  854 d  875  850 d  879  849 d  881  849 d  885  850 d
  887  854 d  888  860 d  888  864 d  887  870 d  885  873 d  881  874 d
  879  874 d  876  873 d  875  872 d  874  870 d  873  864 d  873  860 d
  874  854 d  875  852 d  876  850 d  879  849 d  881  849 r  884  850 d
  885  852 d  886  854 d  887  860 d  887  864 d  886  870 d  885  872 d
  884  873 d  881  874 d  903  874 r  899  873 d  897  870 d  896  864 d
  896  860 d  897  854 d  899  850 d  903  849 d  905  849 d  909  850 d
  911  854 d  912  860 d  912  864 d  911  870 d  909  873 d  905  874 d
  903  874 d  900  873 d  899  872 d  898  870 d  897  864 d  897  860 d
  898  854 d  899  852 d  900  850 d  903  849 d  905  849 r  908  850 d
  909  852 d  910  854 d  911  860 d  911  864 d  910  870 d  909  872 d
  908  873 d  905  874 d  927  874 r  923  873 d  921  870 d  920  864 d
  920  860 d  921  854 d  923  850 d  927  849 d  929  849 d  933  850 d
  935  854 d  936  860 d  936  864 d  935  870 d  933  873 d  929  874 d
  927  874 d  924  873 d  923  872 d  922  870 d  921  864 d  921  860 d
  922  854 d  923  852 d  924  850 d  927  849 d  929  849 r  932  850 d
  933  852 d  934  854 d  935  860 d  935  864 d  934  870 d  933  872 d
  932  873 d  929  874 d  981  872 r  982  874 d  982  870 d  981  872 d
  980  873 d  976  874 d  971  874 d  968  873 d  965  871 d  965  868 d
  966  866 d  968  865 d  970  864 d  977  861 d  980  860 d  982  858 d
  965  868 r  968  866 d  970  865 d  977  862 d  980  861 d  981  860 d
  982  858 d  982  853 d  980  850 d  976  849 d  971  849 d  968  850 d
  966  852 d  965  854 d  965  849 d  966  852 d  992  874 r  992  854 d
  993  850 d  995  849 d  998  849 d 1000  850 d 1001  853 d  993  874 r
  993  854 d  994  850 d  995  849 d  988  866 r  998  866 d 1010  864 r
 1010  862 d 1008  862 d 1008  864 d 1010  865 d 1012  866 d 1017  866 d
 1019  865 d 1020  864 d 1022  861 d 1022  853 d 1023  850 d 1024  849 d
 1020  864 r 1020  853 d 1022  850 d 1024  849 d 1025  849 d 1020  861 r
 1019  860 d 1012  859 d 1008  858 d 1007  855 d 1007  853 d 1008  850 d
 1012  849 d 1016  849 d 1018  850 d 1020  853 d 1012  859 r 1010  858 d
 1008  855 d 1008  853 d 1010  850 d 1012  849 d 1034  866 r 1034  849 d
 1035  866 r 1035  849 d 1035  862 r 1037  865 d 1041  866 d 1043  866 d
 1047  865 d 1048  862 d 1048  849 d 1043  866 r 1046  865 d 1047  862 d
 1047  849 d 1030  866 r 1035  866 d 1030  849 r 1038  849 d 1043  849 r
 1052  849 d 1072  874 r 1072  849 d 1073  874 r 1073  849 d 1072  862 r
 1070  865 d 1067  866 d 1065  866 d 1061  865 d 1059  862 d 1058  859 d
 1058  856 d 1059  853 d 1061  850 d 1065  849 d 1067  849 d 1070  850 d
 1072  853 d 1065  866 r 1062  865 d 1060  862 d 1059  859 d 1059  856 d
 1060  853 d 1062  850 d 1065  849 d 1068  874 r 1073  874 d 1072  849 r
 1077  849 d 1085  864 r 1085  862 d 1084  862 d 1084  864 d 1085  865 d
 1088  866 d 1092  866 d 1095  865 d 1096  864 d 1097  861 d 1097  853 d
 1098  850 d 1100  849 d 1096  864 r 1096  853 d 1097  850 d 1100  849 d
 1101  849 d 1096  861 r 1095  860 d 1088  859 d 1084  858 d 1083  855 d
 1083  853 d 1084  850 d 1088  849 d 1091  849 d 1094  850 d 1096  853 d
 1088  859 r 1085  858 d 1084  855 d 1084  853 d 1085  850 d 1088  849 d
 1109  866 r 1109  849 d 1110  866 r 1110  849 d 1110  859 r 1112  862 d
 1114  865 d 1116  866 d 1120  866 d 1121  865 d 1121  864 d 1120  862 d
 1119  864 d 1120  865 d 1106  866 r 1110  866 d 1106  849 r 1114  849 d
 1142  874 r 1142  849 d 1143  874 r 1143  849 d 1142  862 r 1139  865 d
 1137  866 d 1134  866 d 1131  865 d 1128  862 d 1127  859 d 1127  856 d
 1128  853 d 1131  850 d 1134  849 d 1137  849 d 1139  850 d 1142  853 d
 1134  866 r 1132  865 d 1130  862 d 1128  859 d 1128  856 d 1130  853 d
 1132  850 d 1134  849 d 1138  874 r 1143  874 d 1142  849 r 1146  849 d
 1176  874 r 1176  849 d 1178  874 r 1185  853 d 1176  874 r 1185  849 d
 1193  874 r 1185  849 d 1193  874 r 1193  849 d 1194  874 r 1194  849 d
 1173  874 r 1178  874 d 1193  874 r 1198  874 d 1173  849 r 1180  849 d
 1190  849 r 1198  849 d 1211  866 r 1208  865 d 1205  862 d 1204  859 d
 1204  856 d 1205  853 d 1208  850 d 1211  849 d 1214  849 d 1217  850 d
 1220  853 d 1221  856 d 1221  859 d 1220  862 d 1217  865 d 1214  866 d
 1211  866 d 1209  865 d 1206  862 d 1205  859 d 1205  856 d 1206  853 d
 1209  850 d 1211  849 d 1214  849 r 1216  850 d 1218  853 d 1220  856 d
 1220  859 d 1218  862 d 1216  865 d 1214  866 d 1242  874 r 1242  849 d
 1244  874 r 1244  849 d 1242  862 r 1240  865 d 1238  866 d 1235  866 d
 1232  865 d 1229  862 d 1228  859 d 1228  856 d 1229  853 d 1232  850 d
 1235  849 d 1238  849 d 1240  850 d 1242  853 d 1235  866 r 1233  865 d
 1230  862 d 1229  859 d 1229  856 d 1230  853 d 1233  850 d 1235  849 d
 1239  874 r 1244  874 d 1242  849 r 1247  849 d 1254  859 r 1269  859 d
 1269  861 d 1268  864 d 1266  865 d 1264  866 d 1260  866 d 1257  865 d
 1254  862 d 1253  859 d 1253  856 d 1254  853 d 1257  850 d 1260  849 d
 1263  849 d 1266  850 d 1269  853 d 1268  859 r 1268  862 d 1266  865 d
 1260  866 r 1258  865 d 1256  862 d 1254  859 d 1254  856 d 1256  853 d
 1258  850 d 1260  849 d 1278  874 r 1278  849 d 1280  874 r 1280  849 d
 1275  874 r 1280  874 d 1275  849 r 1283  849 d 1301  865 r 1302  866 d
 1302  862 d 1301  865 d 1298  866 d 1294  866 d 1290  865 d 1289  864 d
 1289  862 d 1290  860 d 1293  859 d 1299  856 d 1301  855 d 1302  854 d
 1289  862 r 1290  861 d 1293  860 d 1299  858 d 1301  856 d 1302  854 d
 1302  852 d 1301  850 d 1298  849 d 1294  849 d 1290  850 d 1289  853 d
 1289  849 d 1290  850 d  536  599 r  536  350 d  533  343 r  536  339 d
  538  343 d  530  346 r  536  340 d  542  346 d  536  361 r  536  340 d
  522  626 r  547  626 d  522  627 r  547  627 d  522  643 r  537  627 d
  532  633 r  547  643 d  532  632 r  547  642 d  522  622 r  522  631 d
  522  639 r  522  646 d  547  622 r  547  631 d  547  638 r  547  646 d
  532  654 r  534  654 d  534  652 d  532  652 d  531  654 d  530  656 d
  530  661 d  531  663 d  532  664 d  535  666 d  543  666 d  546  667 d
  547  668 d  532  664 r  543  664 d  546  666 d  547  668 d  547  669 d
  535  664 r  536  663 d  537  656 d  538  652 d  541  651 d  543  651 d
  546  652 d  547  656 d  547  660 d  546  662 d  543  664 d  537  656 r
  538  654 d  541  652 d  543  652 d  546  654 d  547  656 d  530  678 r
  547  678 d  530  679 r  547  679 d  534  679 r  531  681 d  530  685 d
  530  687 d  531  691 d  534  692 d  547  692 d  530  687 r  531  690 d
  534  691 d  547  691 d  534  692 r  531  694 d  530  698 d  530  700 d
  531  704 d  534  705 d  547  705 d  530  700 r  531  703 d  534  704 d
  547  704 d  530  674 r  530  679 d  547  674 r  547  682 d  547  687 r
  547  696 d  547  700 r  547  709 d  522  717 r  523  716 d  524  717 d
  523  718 d  522  717 d  530  717 r  547  717 d  530  718 r  547  718 d
  530  714 r  530  718 d  547  714 r  547  722 d  530  735 r  531  732 d
  534  729 d  537  728 d  540  728 d  543  729 d  546  732 d  547  735 d
  547  738 d  546  741 d  543  744 d  540  745 d  537  745 d  534  744 d
  531  741 d  530  738 d  530  735 d  531  733 d  534  730 d  537  729 d
  540  729 d  543  730 d  546  733 d  547  735 d  547  738 r  546  740 d
  543  742 d  540  744 d  537  744 d  534  742 d  531  740 d  530  738 d
  522  754 r  547  754 d  522  756 r  547  756 d  530  768 r  542  756 d
  537  762 r  547  769 d  537  760 r  547  768 d  522  751 r  522  756 d
  530  764 r  530  771 d  547  751 r  547  759 d  547  764 r  547  772 d
  532  781 r  534  781 d  534  780 d  532  780 d  531  781 d  530  783 d
  530  788 d  531  790 d  532  792 d  535  793 d  543  793 d  546  794 d
  547  795 d  532  792 r  543  792 d  546  793 d  547  795 d  547  796 d
  535  792 r  536  790 d  537  783 d  538  780 d  541  778 d  543  778 d
  546  780 d  547  783 d  547  787 d  546  789 d  543  792 d  537  783 r
  538  781 d  541  780 d  543  780 d  546  781 d  547  783 d  530  805 r
  547  805 d  530  806 r  547  806 d  534  806 r  531  808 d  530  812 d
  530  814 d  531  818 d  534  819 d  547  819 d  530  814 r  531  817 d
  534  818 d  547  818 d  530  801 r  530  806 d  547  801 r  547  810 d
  547  814 r  547  823 d  522  843 r  547  843 d  522  844 r  547  844 d
  534  843 r  531  841 d  530  838 d  530  836 d  531  832 d  534  830 d
  537  829 d  540  829 d  543  830 d  546  832 d  547  836 d  547  838 d
  546  841 d  543  843 d  530  836 r  531  834 d  534  831 d  537  830 d
  540  830 d  543  831 d  546  834 d  547  836 d  522  840 r  522  844 d
  547  843 r  547  848 d  537  855 r  537  870 d  535  870 d  532  868 d
  531  867 d  530  865 d  530  861 d  531  858 d  534  855 d  537  854 d
  540  854 d  543  855 d  546  858 d  547  861 d  547  864 d  546  867 d
  543  870 d  537  868 r  534  868 d  531  867 d  530  861 r  531  859 d
  534  856 d  537  855 d  540  855 d  543  856 d  546  859 d  547  861 d
   44  463 r   72  463 d   44  464 r   69  480 d   47  464 r   72  480 d
   44  480 r   72  480 d   44  459 r   44  464 d   44  476 r   44  484 d
   72  459 r   72  467 d   54  493 r   68  493 d   70  494 d   72  498 d
   72  501 d   70  505 d   68  507 d   54  494 r   68  494 d   70  495 d
   72  498 d   54  507 r   72  507 d   54  508 r   72  508 d   54  489 r
   54  494 d   54  503 r   54  508 d   72  507 r   72  512 d   54  521 r
   72  521 d   54  523 r   72  523 d   57  523 r   55  525 d   54  529 d
   54  532 d   55  536 d   57  537 d   72  537 d   54  532 r   55  534 d
   57  536 d   72  536 d   57  537 r   55  540 d   54  543 d   54  546 d
   55  550 d   57  551 d   72  551 d   54  546 r   55  549 d   57  550 d
   72  550 d   54  517 r   54  523 d   72  517 r   72  527 d   72  532 r
   72  541 d   72  546 r   72  555 d   44  564 r   72  564 d   44  566 r
   72  566 d   57  566 r   55  568 d   54  571 d   54  573 d   55  577 d
   57  580 d   61  581 d   64  581 d   68  580 d   70  577 d   72  573 d
   72  571 d   70  568 d   68  566 d   54  573 r   55  576 d   57  579 d
   61  580 d   64  580 d   68  579 d   70  576 d   72  573 d   44  560 r
   44  566 d   61  590 r   61  606 d   59  606 d   56  605 d   55  603 d
   54  601 d   54  597 d   55  593 d   57  590 d   61  589 d   64  589 d
   68  590 d   70  593 d   72  597 d   72  599 d   70  603 d   68  606 d
   61  605 r   57  605 d   55  603 d   54  597 r   55  594 d   57  592 d
   61  590 d   64  590 d   68  592 d   70  594 d   72  597 d   54  616 r
   72  616 d   54  618 r   72  618 d   61  618 r   57  619 d   55  622 d
   54  624 d   54  628 d   55  629 d   56  629 d   57  628 d   56  627 d
   55  628 d   54  612 r   54  618 d   72  612 r   72  622 d   54  667 r
   55  663 d   57  661 d   61  659 d   64  659 d   68  661 d   70  663 d
   72  667 d   72  670 d   70  674 d   68  676 d   64  677 d   61  677 d
   57  676 d   55  674 d   54  670 d   54  667 d   55  664 d   57  662 d
   61  661 d   64  661 d   68  662 d   70  664 d   72  667 d   72  670 r
   70  672 d   68  675 d   64  676 d   61  676 d   57  675 d   55  672 d
   54  670 d   46  694 r   47  693 d   48  694 d   47  696 d   46  696 d
   44  694 d   44  692 d   46  689 d   48  688 d   72  688 d   44  692 r
   46  690 d   48  689 d   72  689 d   54  684 r   54  693 d   72  684 r
   72  693 d   44  727 r   72  727 d   44  728 r   68  736 d   44  727 r
   72  736 d   44  745 r   72  736 d   44  745 r   72  745 d   44  746 r
   72  746 d   44  723 r   44  728 d   44  745 r   44  750 d   72  723 r
   72  731 d   72  741 r   72  750 d   54  765 r   55  761 d   57  758 d
   61  757 d   64  757 d   68  758 d   70  761 d   72  765 d   72  767 d
   70  771 d   68  774 d   64  775 d   61  775 d   57  774 d   55  771 d
   54  767 d   54  765 d   55  762 d   57  759 d   61  758 d   64  758 d
   68  759 d   70  762 d   72  765 d   72  767 r   70  770 d   68  772 d
   64  774 d   61  774 d   57  772 d   55  770 d   54  767 d   44  798 r
   72  798 d   44  800 r   72  800 d   57  798 r   55  796 d   54  793 d
   54  791 d   55  787 d   57  784 d   61  783 d   64  783 d   68  784 d
   70  787 d   72  791 d   72  793 d   70  796 d   68  798 d   54  791 r
   55  788 d   57  785 d   61  784 d   64  784 d   68  785 d   70  788 d
   72  791 d   44  794 r   44  800 d   72  798 r   72  804 d   61  811 r
   61  827 d   59  827 d   56  826 d   55  824 d   54  822 d   54  818 d
   55  814 d   57  811 d   61  810 d   64  810 d   68  811 d   70  814 d
   72  818 d   72  820 d   70  824 d   68  827 d   61  826 r   57  826 d
   55  824 d   54  818 r   55  815 d   57  813 d   61  811 d   64  811 d
   68  813 d   70  815 d   72  818 d   44  837 r   72  837 d   44  839 r
   72  839 d   44  833 r   44  839 d   72  833 r   72  843 d   55  862 r
   54  863 d   57  863 d   55  862 d   54  858 d   54  854 d   55  850 d
   56  849 d   57  849 d   60  850 d   61  853 d   64  859 d   65  862 d
   67  863 d   57  849 r   59  850 d   60  853 d   63  859 d   64  862 d
   67  863 d   69  863 d   70  862 d   72  858 d   72  854 d   70  850 d
   68  849 d   72  849 d   70  850 d  609  140 r  607  139 d  606  138 d
  606  135 d  607  134 d  609  133 d  612  133 d  615  134 d  615  135 d
  615  138 d  615  139 d  612  140 d  609  140 d  608  139 d  607  138 d
  607  135 d  608  134 d  609  133 d  612  133 r  614  134 d  615  135 d
  615  138 d  614  139 d  612  140 d  609  133 r  607  132 d  606  132 d
  605  130 d  605  127 d  606  125 d  607  125 d  609  124 d  612  124 d
  615  125 d  615  125 p  616  127 d  616  130 d  615  132 d  615  132 p
  612  133 d  609  133 r  608  132 d  607  132 d  606  130 d  606  127 d
  607  125 d  608  125 d  609  124 d  612  124 r  614  125 d  615  125 d
  615  127 d  615  130 d  615  132 d  614  132 d  612  133 d  625  126 r
  625   98 d  626  126 r  626   98 d  621  126 r  637  126 d  641  124 d
  642  123 d  643  120 d  643  118 d  642  115 d  641  114 d  637  113 d
  637  126 r  639  124 d  641  123 d  642  120 d  642  118 d  641  115 d
  639  114 d  637  113 d  626  113 r  637  113 d  641  111 d  642  110 d
  643  107 d  643  104 d  642  101 d  641  100 d  637   98 d  621   98 d
  637  113 r  639  111 d  641  110 d  642  107 d  642  104 d  641  101 d
  639  100 d  637   98 d  677  126 r  677   98 d  678  126 r  678   98 d
  686  118 r  686  107 d  673  126 r  694  126 d  694  119 d  693  126 d
  678  113 r  686  113 d  673   98 r  682   98 d  703  126 r  703   98 d
  704  126 r  704   98 d  699  126 r  704  126 d  699   98 r  708   98 d
  717  117 r  717  102 d  719  100 d  723   98 d  725   98 d  729  100 d
  732  102 d  719  117 r  719  102 d  720  100 d  723   98 d  732  117 r
  732   98 d  733  117 r  733   98 d  713  117 r  719  117 d  728  117 r
  733  117 d  732   98 r  737   98 d  746  117 r  760   98 d  747  117 r
  762   98 d  762  117 r  746   98 d  742  117 r  751  117 d  758  117 r
  765  117 d  742   98 r  750   98 d  756   98 r  765   98 d  794  131 r
  771   89 d  811  131 r  808  128 d  806  124 d  803  119 d  802  113 d
  802  107 d  803  101 d  806   96 d  808   92 d  811   89 d  808  128 r
  806  123 d  804  119 d  803  113 d  803  107 d  804  101 d  806   97 d
  808   92 d  828  126 r  819   98 d  828  126 r  837   98 d  828  122 r
  836   98 d  821  105 r  834  105 d  815   98 r  823   98 d  832   98 r
  841   98 d  846  117 r  854   98 d  847  117 r  854  101 d  862  117 r
  854   98 d  842  117 r  851  117 d  858  117 r  866  117 d  872  109 r
  888  109 d  888  111 d  886  114 d  885  115 d  882  117 d  879  117 d
  875  115 d  872  113 d  871  109 d  871  106 d  872  102 d  875  100 d
  879   98 d  881   98 d  885  100 d  888  102 d  886  109 r  886  113 d
  885  115 d  879  117 r  876  115 d  873  113 d  872  109 d  872  106 d
  873  102 d  876  100 d  879   98 d  898  117 r  898   98 d  899  117 r
  899   98 d  899  109 r  901  113 d  903  115 d  906  117 d  910  117 d
  911  115 d  911  114 d  910  113 d  908  114 d  910  115 d  894  117 r
  899  117 d  894   98 r  903   98 d  920  114 r  920  113 d  919  113 d
  919  114 d  920  115 d  923  117 d  928  117 d  931  115 d  932  114 d
  933  111 d  933  102 d  934  100 d  936   98 d  932  114 r  932  102 d
  933  100 d  936   98 d  937   98 d  932  111 r  931  110 d  923  109 d
  919  107 d  918  105 d  918  102 d  919  100 d  923   98 d  927   98 d
  929  100 d  932  102 d  923  109 r  920  107 d  919  105 d  919  102 d
  920  100 d  923   98 d  951  117 r  949  115 d  947  114 d  946  111 d
  946  109 d  947  106 d  949  105 d  951  104 d  954  104 d  957  105 d
  958  106 d  959  109 d  959  111 d  958  114 d  957  115 d  954  117 d
  951  117 d  949  115 r  947  113 d  947  107 d  949  105 d  957  105 r
  958  107 d  958  113 d  957  115 d  958  114 r  959  115 d  962  117 d
  962  115 d  959  115 d  947  106 r  946  105 d  945  102 d  945  101 d
  946   98 d  950   97 d  957   97 d  960   96 d  962   94 d  945  101 r
  946  100 d  950   98 d  957   98 d  960   97 d  962   94 d  962   93 d
  960   91 d  957   89 d  949   89 d  945   91 d  944   93 d  944   94 d
  945   97 d  949   98 d  971  109 r  986  109 d  986  111 d  985  114 d
  984  115 d  981  117 d  977  117 d  973  115 d  971  113 d  970  109 d
  970  106 d  971  102 d  973  100 d  977   98 d  980   98 d  984  100 d
  986  102 d  985  109 r  985  113 d  984  115 d  977  117 r  975  115 d
  972  113 d  971  109 d  971  106 d  972  102 d  975  100 d  977   98 d
 1020  140 r 1018  139 d 1017  138 d 1017  135 d 1018  134 d 1020  133 d
 1023  133 d 1025  134 d 1026  135 d 1026  138 d 1025  139 d 1023  140 d
 1020  140 d 1018  139 d 1018  138 d 1018  135 d 1018  134 d 1020  133 d
 1023  133 r 1025  134 d 1025  135 d 1025  138 d 1025  139 d 1023  140 d
 1020  133 r 1018  132 d 1017  132 d 1016  130 d 1016  127 d 1017  125 d
 1018  125 d 1020  124 d 1023  124 d 1025  125 d 1026  125 d 1027  127 d
 1027  130 d 1026  132 d 1025  132 d 1023  133 d 1020  133 r 1018  132 d
 1018  132 p 1017  130 d 1017  127 d 1018  125 d 1018  125 p 1020  124 d
 1023  124 r 1025  125 d 1025  125 p 1026  127 d 1026  130 d 1025  132 d
 1025  132 p 1023  133 d 1036  126 r 1036   98 d 1037  126 r 1037   98 d
 1032  126 r 1048  126 d 1051  124 d 1053  123 d 1054  120 d 1054  118 d
 1053  115 d 1051  114 d 1048  113 d 1048  126 r 1050  124 d 1051  123 d
 1053  120 d 1053  118 d 1051  115 d 1050  114 d 1048  113 d 1037  113 r
 1048  113 d 1051  111 d 1053  110 d 1054  107 d 1054  104 d 1053  101 d
 1051  100 d 1048   98 d 1032   98 d 1048  113 r 1050  111 d 1051  110 d
 1053  107 d 1053  104 d 1051  101 d 1050  100 d 1048   98 d 1088  126 r
 1088   98 d 1089  126 r 1089   98 d 1097  118 r 1097  107 d 1084  126 r
 1105  126 d 1105  119 d 1103  126 d 1089  113 r 1097  113 d 1084   98 r
 1093   98 d 1114  126 r 1114   98 d 1115  126 r 1115   98 d 1110  126 r
 1115  126 d 1110   98 r 1119   98 d 1128  117 r 1128  102 d 1130  100 d
 1133   98 d 1136   98 d 1140  100 d 1143  102 d 1130  117 r 1130  102 d
 1131  100 d 1133   98 d 1143  117 r 1143   98 d 1144  117 r 1144   98 d
 1124  117 r 1130  117 d 1139  117 r 1144  117 d 1143   98 r 1148   98 d
 1157  117 r 1171   98 d 1158  117 r 1172   98 d 1172  117 r 1157   98 d
 1153  117 r 1162  117 d 1168  117 r 1176  117 d 1153   98 r 1161   98 d
 1167   98 r 1176   98 d 1183  131 r 1185  128 d 1188  124 d 1191  119 d
 1192  113 d 1192  107 d 1191  101 d 1188   96 d 1185   92 d 1183   89 d
 1185  128 r 1188  123 d 1189  119 d 1191  113 d 1191  107 d 1189  101 d
 1188   97 d 1185   92 d
e
EndPSPlot
0 SPE
4787 15917 XY 0 SPB
 clear Figure end restore 
0 SPE 5509 15724 XY F25(Figure)S 150 x(3.)S 298 x(1000)S 150 x
(solar)S 150 x(mo)S 14 x(dels)S 150 x(vs)S 149 x(exp)S 14 x(erime)S
2 x(n)S -14 x(ts.)S -180 y F8(26)S 180 y 173 x F25(The)S 149 x(n)S
-14 x(um)S -14 x(b)S 14 x(er)S 149 x(of)S 150 x(precise)S 2 x(ly)S
149 x(calcula)S 2 x(ted)S 149 x(solar)S 598 y 5509 X(mo)S 14 x(dels)S
175 x(that)S 174 x(predict)S 175 x(di\013eren)S -13 x(t)S 174 x
(solar)S 176 x(neutrino)S 174 x(ev)S -14 x(en)S -13 x(t)S 174 x
(rates)S 175 x(are)S 175 x(sho)S -13 x(wn)S 174 x(for)S 175 x(the)S
174 x(Kamiok)S -26 x(ande)S 598 y 5509 X(exp)S 13 x(erime)S 2 x(n)S
-14 x(t.)S -181 y F8(2,3)S 181 y 223 x F25(The)S 199 x(solar)S 200 x
(mo)S 14 x(dels)S 200 x(from)S 200 x(whic)S -13 x(h)S 198 x(the)S 
200 x(\015uxes)S 199 x(w)S -14 x(ere)S 200 x(deriv)S -13 x(ed)S 199 x
(satisfy)S 200 x(the)S 199 x(equa-)S 598 y 5509 X(tions)S 168 x(of)S
169 x(stella)S 2 x(r)S 168 x(ev)S -13 x(olution)S 168 x(including)S
169 x(the)S 168 x(b)S 14 x(oundary)S 168 x(conditions)S 169 x(that)S
167 x(the)S 169 x(mo)S 14 x(del)S 169 x(luminosit)S -13 x(y)S -42 x
(,)S 597 y 5509 X(c)S -14 x(hemical)S 140 x(comp)S 14 x(osition,)S
140 x(and)S 138 x(e\013ectiv)S -13 x(e)S 138 x(temp)S 14 x(erature)S
139 x(at)S 138 x(the)S 138 x(curren)S -13 x(t)S 138 x(solar)S 140 x
(age)S 138 x(b)S 14 x(e)S 139 x(equal)S 139 x(to)S 138 x(the)S 598 y 
5509 X(observ)S -14 x(ed)S 201 x(v)S -28 x(alues.)S 202 x(Eac)S -14 x
(h)S 200 x(input)S 199 x(parameter)S 201 x(in)S 201 x(eac)S -13 x(h)S
199 x(solar)S 202 x(mo)S 14 x(del)S 200 x(w)S -14 x(as)S 201 x(dra)S
-14 x(wn)S 200 x(indep)S 13 x(enden)S -13 x(tly)S 598 y 5509 X(from)S
158 x(a)S 158 x(normal)S 158 x(distribution)S 158 x(ha)S -14 x(ving)S
158 x(the)S 158 x(mean)S 158 x(and)S 157 x(the)S 158 x(standard)S 
158 x(deviation)S 158 x(appropriate)S 158 x(to)S 598 y 5509 X(that)S
165 x(parameter.)S 167 x(The)S 166 x(exp)S 14 x(erimen)S -13 x(tal)S
166 x(error)S 167 x(bar)S 166 x(includes)S 167 x(only)S 166 x
(statistic)S 2 x(al)S 166 x(errors)S 168 x(\(1)S F26(\033)S 17 x F25
(\).)S 22099 Y 5347 X F34(that)S 182 x(the)S 182 x(uncertain)S -15 x
(t)S -15 x(y)S 181 x(in)S 182 x(the)S 182 x(8)S 368 y -273 x(\026)S
-368 y 182 x(neutrino)S 181 x(\015ux)S 182 x(is)S 181 x(v)S -15 x
(ery)S 182 x(small)S -2 x(,)S
XP /F67 /cmex10 300 498 498.132 128 [-1 -123 60 31] PXLNF RP
XP /F67 32 394 10 -122 30 1 124 24 0
<000018 000038 000070 0000E0 0001C0 0003C0 000380 000780 000F00
 000E00 001E00 001C00 003C00 003800 007800 007800 00F000 00F000
 01E000 01E000 03E000 03C000 03C000 07C000 078000 078000 0F8000
 0F8000 0F0000 0F0000 1F0000 1F0000 1E0000 1E0000 3E0000 3E0000
 3E0000 3E0000 3C0000 3C0000 7C0000 7C0000 7C0000 7C0000 7C0000
 7C0000 7C0000 7C0000 F80000 F80000 F80000 F80000 F80000 F80000
 F80000 F80000 F80000 F80000 F80000 F80000 F80000 F80000 F80000
 F80000 F80000 F80000 F80000 F80000 F80000 F80000 F80000 F80000
 F80000 F80000 F80000 F80000 7C0000 7C0000 7C0000 7C0000 7C0000
 7C0000 7C0000 7C0000 3C0000 3C0000 3E0000 3E0000 3E0000 3E0000
 1E0000 1E0000 1F0000 1F0000 0F0000 0F0000 0F8000 0F8000 078000
 078000 07C000 03C000 03C000 03E000 01E000 01E000 00F000 00F000
 007800 007800 003800 003C00 001C00 001E00 000E00 000F00 000780
 000380 0003C0 0001C0 0000E0 000070 000038 000018>
PXLC RP
9845 22677 XY F67( )S
XP /F34 1 456 2 0 34 31 32 40 0
<0001C00000 0001C00000 0003E00000 0003E00000 0007F00000 0007F00000
 000DF80000 000CF80000 0018FC0000 00187C0000 00387E0000 00303E0000
 00703F0000 00601F0000 00E01F8000 00C00F8000 01C00FC000 018007C000
 038007E000 030007E000 070003F000 060003F000 0E0001F800 0C0001F800
 0C0000F800 180000FC00 1800007C00 3000007E00 3FFFFFFE00 7FFFFFFF00
 7FFFFFFF00 FFFFFFFF80>
PXLC RP
10299 23172 XY F34(\001)S F35(\036)S F34(\()S -198 y F13(8)S 198 y 
25 x F34(B\))S 242 y 10299 X 1824 24 R 501 y 10527 X F35(\036)S F34
(\()S -158 y F13(8)S 158 y 24 x F34(B\))S
XP /F67 33 394 1 -122 21 1 124 24 0
<C00000 E00000 700000 380000 1C0000 1E0000 0E0000 0F0000 078000
 038000 03C000 01C000 01E000 00E000 00F000 00F000 007800 007800
 003C00 003C00 003E00 001E00 001E00 001F00 000F00 000F00 000F80
 000F80 000780 000780 0007C0 0007C0 0003C0 0003C0 0003E0 0003E0
 0003E0 0003E0 0001E0 0001E0 0001F0 0001F0 0001F0 0001F0 0001F0
 0001F0 0001F0 0001F0 0000F8 0000F8 0000F8 0000F8 0000F8 0000F8
 0000F8 0000F8 0000F8 0000F8 0000F8 0000F8 0000F8 0000F8 0000F8
 0000F8 0000F8 0000F8 0000F8 0000F8 0000F8 0000F8 0000F8 0000F8
 0000F8 0000F8 0000F8 0000F8 0001F0 0001F0 0001F0 0001F0 0001F0
 0001F0 0001F0 0001F0 0001E0 0001E0 0003E0 0003E0 0003E0 0003E0
 0003C0 0003C0 0007C0 0007C0 000780 000780 000F80 000F80 000F00
 000F00 001F00 001E00 001E00 003E00 003C00 003C00 007800 007800
 00F000 00F000 00E000 01E000 01C000 03C000 038000 078000 0F0000
 0E0000 1E0000 1C0000 380000 700000 E00000 C00000>
PXLC RP
12186 22677 XY F67(!)S 864 y 334 x F34(=)S
XP /F67 18 367 9 -97 28 1 99 24 0
<000070 0000F0 0001E0 0001C0 000380 000700 000E00 001E00 001C00
 003800 007800 007000 00F000 00E000 01E000 01C000 03C000 038000
 078000 078000 070000 0F0000 0F0000 0E0000 1E0000 1E0000 1E0000
 3C0000 3C0000 3C0000 3C0000 380000 780000 780000 780000 780000
 780000 780000 780000 F00000 F00000 F00000 F00000 F00000 F00000
 F00000 F00000 F00000 F00000 F00000 F00000 F00000 F00000 F00000
 F00000 F00000 F00000 F00000 F00000 F00000 780000 780000 780000
 780000 780000 780000 780000 380000 3C0000 3C0000 3C0000 3C0000
 1E0000 1E0000 1E0000 0E0000 0F0000 0F0000 070000 078000 078000
 038000 03C000 01C000 01E000 00E000 00F000 007000 007800 003800
 001C00 001E00 000E00 000700 000380 0001C0 0001E0 0000F0 000070>
PXLC RP
13671 22826 XY F67(\022)S 346 y 60 x F34(18)S 242 y -546 x 
540 24 R 501 y -404 x(4)S
XP /F67 19 367 1 -97 20 1 99 24 0
<E00000 F00000 780000 380000 1C0000 0E0000 070000 078000 038000
 01C000 01E000 00E000 00F000 007000 007800 003800 003C00 001C00
 001E00 001E00 000E00 000F00 000F00 000700 000780 000780 000780
 0003C0 0003C0 0003C0 0003C0 0001C0 0001E0 0001E0 0001E0 0001E0
 0001E0 0001E0 0001E0 0000F0 0000F0 0000F0 0000F0 0000F0 0000F0
 0000F0 0000F0 0000F0 0000F0 0000F0 0000F0 0000F0 0000F0 0000F0
 0000F0 0000F0 0000F0 0000F0 0000F0 0000F0 0001E0 0001E0 0001E0
 0001E0 0001E0 0001E0 0001E0 0001C0 0003C0 0003C0 0003C0 0003C0
 000780 000780 000780 000700 000F00 000F00 000E00 001E00 001E00
 001C00 003C00 003800 007800 007000 00F000 00E000 01E000 01C000
 038000 078000 070000 0E0000 1C0000 380000 780000 F00000 E00000>
PXLC RP
14703 22826 XY F67(\023)S 715 y 121 x F36(\002)S -715 y 121 x F67
(\022)S 346 y 60 x F34(\001)S F35(L)S 81 y F15(\014)S 161 y 16163 X 
1176 24 R 501 y 16390 X F35(L)S 82 y F15(\014)S 22826 Y 312 x F67
(\023)S
XP /F35 60 425 4 -3 29 24 28 32 0
<000000C0 000003C0 00000FC0 00003F00 0000FC00 0003F000 000FC000
 001F0000 007C0000 01F80000 07E00000 1F800000 7E000000 F8000000
 F8000000 7E000000 1F800000 07E00000 01F80000 007C0000 001F0000
 000FC000 0003F000 0000FC00 00003F00 00000FC0 000003C0 000000C0>
PXLC RP
18102 23541 XY F35(<)S 334 x F34(0)S F35(:)S F34(02)S F35(:)S 23692 X 
F34(\(5\))S 24983 Y 5347 X(This)S 202 x(argumen)S -16 x(t)S 203 x
(suggests)S 204 x(that)S 202 x(the)S 204 x(uncertain)S -15 x(t)S 
-15 x(y)S 202 x(in)S 203 x(the)S 203 x(8)S 368 y -273 x(\026)S -368 y 
203 x(neutrino)S 202 x(\015ux)S 203 x(is)S 203 x(less)S 203 x(than)S
677 y 5347 X(2
(uncertain)S -16 x(t)S -15 x(y)S 198 x(I)S 198 x(estimate)S 198 x
(is)S 197 x(v)S -15 x(ery)S 197 x(m)S -15 x(uc)S -15 x(h)S 198 x
(larger,)S 196 x(14
5347 X(for)S 145 x(the)S 146 x(discrepancy)S 147 x(b)S 15 x(et)S 
-15 x(w)S -15 x(een)S 147 x(Eq.)S 146 x(\(5\))S 146 x(and)S 146 x
(the)S 146 x(uncertain)S -15 x(t)S -15 x(y)S 146 x(obtained)S 146 x
(from)S 144 x(a)S 146 x(detailed)S 678 y 5347 X(analysi)S -2 x(s)S
-198 y F13(13)S 198 y 233 x F34(is)S 208 x(that)S 208 x(the)S 208 x
(represen)S -14 x(tation)S 207 x(of)S 208 x(a)S 208 x(solar)S 206 x
(mo)S 15 x(del)S 208 x(in)S 208 x(terms)S 208 x(of)S 207 x(just)S 
208 x(a)S 208 x(cen)S -14 x(tral)S 677 y 5347 X(temp)S 15 x(erature)S
264 x(is)S 264 x(a)S 264 x(gross)S 263 x(o)S -15 x(v)S -16 x
(ersimpli\014cati)S -2 x(on.)S 264 x(\(The)S 264 x(computed)S 265 x
(neutrino)S 263 x(\015ux)S 264 x(is)S 264 x(an)S 678 y 5347 X(in)S
-16 x(tegration)S 213 x(of)S 212 x(the)S 214 x(lo)S 15 x(cal)S 213 x
(pro)S 14 x(duction)S 214 x(rate)S 213 x(o)S -15 x(v)S -16 x(er)S 
214 x(the)S 214 x(temp)S 15 x(erature-densit)S -15 x(y)S 214 x
(pro\014l)S -2 x(e)S 214 x(of)S 677 y 5347 X(the)S 147 x(mo)S 14 x
(del)S 147 x(sun)S 146 x(and)S 147 x(also)S 145 x(dep)S 16 x(ends,)S
146 x(for)S 146 x(example,)S 145 x(in)S 146 x(di\013eren)S -14 x(t)S
146 x(w)S -15 x(a)S -15 x(ys)S 146 x(up)S 15 x(on)S 146 x(the)S 148 x
(di\013eren)S -15 x(t)S 678 y 5347 X(input)S 225 x(n)S -15 x(uclear)S
225 x(cross)S 226 x(sections.\))S 226 x(Nev)S -15 x(ertheless,)S 
226 x(y)S -15 x(ou)S 225 x(can)S 226 x(see)S 226 x(b)S -15 x(y)S 
226 x(this)S 225 x(argumen)S -16 x(t)S 226 x(that)S 677 y 5347 X
(the)S 216 x(luminosi)S -2 x(t)S -15 x(y)S 216 x(b)S 15 x(oundary)S
215 x(condition)S 216 x(pro)S -16 x(vides)S 216 x(a)S 216 x(sev)S 
-15 x(ere)S 216 x(constrain)S -15 x(t)S 216 x(on)S 216 x(the)S 216 x
(allo)S -17 x(w)S -15 x(ed)S 678 y 5347 X(v)S -31 x(alues)S 146 x
(of)S 145 x(the)S 146 x(neutrino)S 145 x(\015uxes.)S 146 x(2\))S 
146 x F37(The)S 162 x(pr)S -29 x(e)S -28 x(cision)S 162 x(o)S
XP /F37 102 167 -2 -9 19 31 41 24 0
<0000F8 0003FC 0003BC 0007BC 0007BC 000F00 000F00 000F00 000F00
 001F00 001E00 001E00 03FFF0 03FFF0 003E00 003C00 003C00 003C00
 003C00 007C00 007800 007800 007800 007800 007800 00F800 00F000
 00F000 00F000 00F000 01E000 01E000 01E000 01E000 01C000 03C000
 73C000 7B8000 FF0000 7F0000 7C0000>
PXLC RP
16470 31080 XY F37(f)S 162 x(the)S 162 x(input)S 162 x(p)S -28 x(ar)S
-28 x(ameters)S 145 x F34(has)S 146 x(greatly)S 677 y 5347 X(impr)S
-2 x(o)S -15 x(v)S -15 x(ed)S 222 x(o)S -15 x(v)S -15 x(er)S 222 x
(the)S 222 x(y)S -15 x(ears)S 222 x(as)S 222 x(man)S -16 x(y)S 222 x
(individua)S -2 x(ls)S 222 x(and)S 222 x(groups)S 221 x(\(ph)S -15 x
(ysicists,)S 222 x(c)S -15 x(hemists,)S 678 y 5347 X(and)S 224 x
(astronomers\))S 223 x(ha)S -15 x(v)S -15 x(e)S 224 x(remeasured)S
225 x(and)S 224 x(recalculated)S 224 x(the)S 225 x(quan)S -15 x
(tities)S 224 x(required)S 224 x(to)S 677 y 5347 X(determine)S 192 x
(the)S 194 x(solar)S 191 x(mo)S 14 x(del)S 193 x(neutrino)S 192 x
(\015uxes.)S 192 x(The)S 193 x(recen)S -14 x(tly-ev)S -30 x(aluated)S
192 x(uncertain)S -15 x(ties)S 678 y 5347 X(are)S 162 x(relativ)S 
-16 x(ely)S 163 x(small)S -2 x(,)S 163 x(in)S 162 x(large)S 162 x
(part,)S 162 x(b)S 16 x(ecause)S 164 x(of)S 162 x(this)S 163 x
(success)S 2 x(ful)S 162 x(comm)S -16 x(unit)S -16 x(y)S 163 x
(e\013ort.)S 163 x(3\))S
1 PP EP

1000 BP 39600 30600 PM 0 0 XY
XP /F37 72 406 3 0 37 30 31 40 0
<01FFF3FFE0 01FFF3FFE0 001F003E00 001E003C00 001E007C00 003E007C00
 003C007800 003C007800 003C00F800 007C00F800 007800F000 007800F000
 007801F000 00F801F000 00FFFFE000 00FFFFE000 00F003E000 01F003E000
 01E003C000 01E003C000 01E007C000 03E007C000 03C0078000 03C0078000
 03C00F8000 07C00F8000 07800F0000 07800F0000 0F801F0000 FFF9FFF000
 FFF9FFF000>
PXLC RP
5347 3815 XY F37(Helioseismolo)S
XP /F37 103 251 2 -9 22 19 29 24 0
<003E70 00FFF8 01E3F8 03C1F0 0781F0 0F01F0 0F01F0 1E01E0 1E01E0
 3E03E0 3E03E0 3C03C0 3C03C0 3C07C0 3C07C0 3C0F80 1C1F80 1E3F80
 0FFF80 07EF00 000F00 001F00 001F00 001E00 703E00 787C00 F8F800
 FFF000 3FC000>
PXLC RP
8568 3815 XY F37(gists)S 193 x F34(ha)S -16 x(v)S -15 x(e)S 193 x
(measured)S 192 x(the)S 193 x(frequencies)S 193 x(of)S 192 x
(thousands)S 192 x(of)S 192 x(solar)S 191 x(pressure)S 678 y 5347 X
(mo)S 14 x(des)S 271 x(to)S 270 x(an)S 270 x(accuracy)S 270 x(of)S
270 x(b)S 15 x(etter)S 271 x(than)S 270 x(one)S 270 x(part)S 270 x
(in)S 269 x(a)S 270 x(thousand.)S 270 x(The)S 270 x(standard)S 677 y 
5347 X(solar)S 190 x(mo)S 15 x(des)S 193 x(used)S 192 x(to)S 192 x
(calculate)S 192 x(solar)S 191 x(neutrino)S 191 x(\015uxes)S 193 x
(repro)S 14 x(duce)S 193 x(the)S 193 x(measured)S 192 x F35(p)S 128 x 
F36(\000)S
XP /F35 109 480 1 0 38 19 20 40 0
<1F07F03F00 3F8FFCFFC0 37FC3DE3C0 63F83F83C0 63E01F01E0 E3E03E03E0
 C7C03E03C0 07803C03C0 07803C03C0 07807C07C0 0F807C0780 0F00780780
 0F00780F80 0F00F80F1C 1F00F80F18 1E00F01F18 1E00F01E38 1E01F01E70
 3E01F00FE0 1C00E007C0>
PXLC RP
5347 5848 XY F35(m)S
XP /F35 111 264 2 0 20 19 20 24 0
<007E00 03FF80 07C780 0F03C0 1E01C0 3C01E0 3C01E0 7801E0 7801E0
 7803E0 F803E0 F003C0 F003C0 F00780 F00780 700F00 781E00 3C7C00
 3FF800 0FC000>
PXLC RP
5826 5848 XY F35(o)S
XP /F35 100 284 2 0 23 31 32 24 0
<00007C 0003FC 0003FC 000078 000078 0000F8 0000F8 0000F0 0000F0
 0001F0 0001F0 0001E0 00F9E0 03FFE0 078FE0 0F07C0 1E07C0 3C07C0
 3C07C0 780780 780780 780F80 F80F80 F00F00 F00F00 F01F18 F01F38
 F03E30 707E30 78FF60 3FEFE0 1F87C0>
PXLC RP
6090 5848 XY F35(de)S 162 x F34(eigenfrequencies)S 164 x(to)S 162 x
(t)S -15 x(ypically)S 161 x(one)S 163 x(part)S 162 x(in)S 162 x(a)S
162 x(thousand,)S 162 x(establishing)S 162 x(the)S 163 x(basic)S 
677 y 5347 X(correctness)S 212 x(of)S 211 x(the)S 211 x(solar)S 210 x
(mo)S 15 x(del)S 211 x(to)S 211 x(a)S 211 x(depth)S 212 x(of)S 210 x
(at)S 211 x(least)S 211 x(half)S 211 x(the)S 211 x(solar)S 210 x
(radius.)S 210 x(One)S 678 y 5347 X(no)S 226 x(longer)S 225 x(has)S
226 x(the)S 227 x(freedom)S 226 x(to)S 226 x(sp)S 15 x(eculate)S 
227 x(ab)S 15 x(out)S 226 x(radicall)S -2 x(y)S 227 x(di\013eren)S
-15 x(t)S 226 x(p)S 15 x(ossible)S 226 x(solar)S 677 y 5347 X(mo)S
14 x(dels)S 226 x(b)S 15 x(ecause)S 227 x(of)S 225 x(the)S 226 x
(man)S -16 x(y)S 225 x(precisely)S 226 x(measured)S 225 x
(helioseismol)S -2 x(ogical)S 225 x(frequencies.)S 677 y 5347 X(4\))S
204 x F37(The)S 215 x(sun)S 216 x(is)S 216 x(in)S 215 x(a)S 216 x
(simple)S 216 x(state)S 216 x(of)S 216 x(stel)S 28 x(lar)S 216 x(e)S
XP /F37 118 251 3 0 21 19 20 24 0
<0F81C0 1FC3E0 39E3E0 71E3E0 63E1E0 63C0E0 E3C0C0 07C0C0 0780C0
 0781C0 0F8180 0F0180 0F0180 0F0300 0F0300 0F0700 0F0E00 0F9C00
 07F800 03F000>
PXLC RP
15651 8557 XY F37(volution)S F34(,)S 203 x(the)S 205 x(main)S 203 x
(sequence)S 2 x(,)S 204 x(and)S 204 x(w)S -16 x(e)S 678 y 5347 X
(kno)S -16 x(w)S 221 x(more)S 220 x(ab)S 15 x(out)S 220 x(it)S 221 x
(exp)S 15 x(erimen)S -16 x(tally)S 220 x(than)S 221 x(ab)S 14 x(out)S
221 x(an)S -15 x(y)S 220 x(other)S 221 x(star.)S 220 x(The)S 221 x
(ph)S -15 x(ysics)S 220 x(of)S 677 y 5347 X(the)S 215 x(in)S -15 x
(terior)S 214 x(of)S 214 x(the)S 216 x(sun)S 216 x(is)S 214 x
(relativ)S -16 x(ely)S 215 x(simple;)S 214 x(for)S 214 x(example,)S
214 x(detailed)S 215 x(corrections)S 215 x(to)S 678 y 5347 X(the)S
165 x(equation)S 164 x(of)S 164 x(state)S 165 x(are)S 164 x(only)S
164 x(of)S 164 x(order)S 164 x(of)S 164 x(a)S 165 x(few)S 164 x(p)S
15 x(ercen)S -14 x(t.)S 164 x(5\))S 165 x F37(Ther)S -28 x(e)S 179 x
(ar)S -28 x(e)S 179 x(many)S 179 x(input)S 677 y 5347 X(p)S -28 x
(ar)S -28 x(ameters)S F34(,)S 185 x(including)S 185 x(the)S 187 x
(cross)S 186 x(sections)S 187 x(for)S 185 x(all)S 185 x(of)S 185 x
(the)S 187 x(relev)S -30 x(an)S -16 x(t)S 187 x(n)S -15 x(uclear)S
185 x(reactions,)S 678 y 5347 X(the)S 133 x(solar)S 132 x(lumino)S
-2 x(sit)S -15 x(y)S -45 x(,)S 132 x(and)S 133 x(the)S 133 x
(surface)S 134 x(hea)S -15 x(vy)S 132 x(elemen)S -14 x(t)S 133 x
(abundances.)S 133 x(In)S 133 x(an)S -16 x(y)S 133 x(particular)S 
677 y 5347 X(time)S 171 x(p)S 15 x(erio)S 14 x(d,)S 171 x(the)S 172 x
(impro)S -17 x(v)S -15 x(emen)S -15 x(ts)S 172 x(in)S 171 x(some)S
171 x(of)S 171 x(these)S 172 x(parameters)S 171 x(cause)S 172 x(the)S
172 x(calculated)S 678 y 5347 X(neutrino)S 154 x(ev)S -15 x(en)S 
-15 x(t)S 155 x(rates)S 154 x(to)S 154 x(increase)S 155 x(and)S 154 x
(the)S 155 x(impro)S -17 x(v)S -15 x(emen)S -15 x(ts)S 155 x(in)S 
154 x(other)S 154 x(parameters)S 154 x(cause)S 677 y 5347 X(the)S 
139 x(calculated)S 138 x(neutrino)S 138 x(ev)S -15 x(en)S -14 x(t)S
138 x(rates)S 139 x(to)S 138 x(decrease)S 2 x(.)S 138 x(On)S 138 x
(the)S 140 x(a)S -16 x(v)S -15 x(erage,)S 138 x(the)S 139 x(b)S 15 x
(est-estimate)S 678 y 5347 X(for)S 160 x(the)S 162 x(c)S -14 x
(hlori)S -2 x(ne)S 162 x(exp)S 15 x(erimen)S -15 x(t)S 162 x(has)S
161 x(remained)S 161 x(within)S 160 x(a)S 162 x(narro)S -17 x(w)S 
162 x(range)S 161 x(o)S -16 x(v)S -15 x(er)S 162 x(the)S 162 x(past)S
161 x(25)S 677 y 5347 X(y)S -16 x(ears)S 163 x(\(see)S 163 x(Figure)S
162 x(1.2)S 161 x(of)S 161 x(Ref.)S -198 y F13(27)S 198 y 24 x F34
(\).)S 162 x(If)S 161 x(w)S -15 x(e)S 163 x(consider)S 162 x(all)S
161 x(of)S 162 x(m)S -16 x(y)S 162 x(published)S 162 x(calculations)S
161 x(in)S 678 y 5347 X(whic)S -16 x(h)S 191 x(a)S 189 x(full)S 189 x
(ev)S -30 x(aluatio)S -2 x(n)S 191 x(w)S -16 x(as)S 190 x(made)S 
190 x(including)S 189 x(an)S 190 x(estimated)S 190 x(theoretical)S
189 x(error,)S 189 x(then)S 677 y 5347 X(the)S 149 x(range)S 149 x
(o)S -15 x(v)S -15 x(er)S 149 x(the)S 150 x(last)S 148 x(quarter)S
149 x(cen)S -14 x(tury)S 149 x(has)S 149 x(b)S 15 x(een)S 150 x(b)S
15 x(et)S -14 x(w)S -16 x(een)S 150 x(5)S F35(:)S F34(8)S 149 x(SNU)S
149 x(and)S 149 x(10)S F35(:)S F34(5)S 148 x(SNU,)S 677 y 5347 X
(the)S 182 x(midp)S 14 x(oin)S -16 x(t)S 182 x(of)S 182 x(whic)S 
-16 x(h)S 182 x(is)S 182 x(within)S 181 x(0.3)S 181 x(SNU)S 182 x
(of)S 181 x(the)S 182 x(curren)S -15 x(t)S 182 x(b)S 15 x(est)S 183 x
(estimate.)S
XP /F48 53 336 3 0 23 31 32 24 0
<300060 3E03E0 3FFFE0 3FFFC0 3FFF80 3FFF00 3FFE00 3FF800 380000
 380000 380000 380000 39FE00 3FFF80 3F0FC0 3C07E0 3803F0 0003F0
 0003F8 0003F8 0003F8 7803F8 FC03F8 FC03F8 FC03F8 FC03F0 FC07F0
 7807E0 7E1FE0 3FFFC0 1FFF00 07F800>
PXLC RP
5347 18439 XY F48(5.)S
XP /F48 68 515 2 0 38 33 34 40 0
<FFFFFF8000 FFFFFFF000 FFFFFFFC00 07F003FE00 07F0007F00 07F0003F80
 07F0001FC0 07F0000FE0 07F0000FE0 07F0000FF0 07F00007F0 07F00007F0
 07F00007F0 07F00007F8 07F00007F8 07F00007F8 07F00007F8 07F00007F8
 07F00007F8 07F00007F8 07F00007F8 07F00007F8 07F00007F0 07F00007F0
 07F00007F0 07F0000FE0 07F0000FE0 07F0001FC0 07F0003F80 07F0007F00
 07F003FE00 FFFFFFFC00 FFFFFFF000 FFFFFF8000>
PXLC RP
6542 18439 XY F48(Direct)S 224 x(Comparison)S 223 x(of)S 225 x
(Chlorin)S -2 x(e)S 224 x(and)S 722 y 5347 X(Electron-S)S -2 x
(catterin)S -2 x(g)S 225 x(Exp)S 18 x(erimen)S -19 x(ts)S 20144 Y 
5347 X F34(The)S 155 x(c)S -15 x(hlorine)S 154 x(and)S 155 x(the)S
156 x(Kamio)S -2 x(k)S -30 x(ande)S 155 x(exp)S 16 x(erimen)S -16 x
(ts)S 155 x(are)S 155 x(sensitiv)S -15 x(e,)S 155 x(to)S 155 x(a)S
154 x(large)S 154 x(exten)S -14 x(t,)S 155 x(to)S 678 y 5347 X(the)S
129 x(same)S 129 x(neutrino)S 128 x(source,)S 129 x(the)S 130 x
(rare)S 128 x(8)S 368 y -273 x(\026)S -368 y 128 x(neutrinos.)S 129 x
(The)S 129 x(Kamiok)S -32 x(ande)S 129 x(exp)S 16 x(erimen)S -16 x
(t)S 129 x(mea-)S 677 y 5347 X(sures)S 177 x(only)S 175 x(8)S 368 y 
-273 x(\026)S -368 y 177 x(neutrinos.)S 176 x(F)S -46 x(or)S 176 x
(the)S 177 x(c)S -14 x(hlori)S -2 x(ne)S 177 x(exp)S 16 x(erimen)S
-16 x(t,)S 176 x(ab)S 15 x(out)S 177 x(78
177 x(standard-)S 677 y 5347 X(mo)S 14 x(del)S 151 x(calculated)S 
151 x(rate)S 152 x(is)S 151 x(from)S 149 x(the)S 152 x(same)S 151 x
(source.)S 151 x(The)S 152 x(c)S -15 x(hlorine)S 150 x(and)S 151 x
(the)S 152 x(Kamiok)S -31 x(ande)S 678 y 5347 X(exp)S 15 x(erimen)S
-15 x(ts)S 152 x(di\013er)S 152 x(in)S 151 x(that)S 152 x(the)S 153 x
(threshold)S 152 x(for)S 151 x(c)S -15 x(hlorine)S 151 x(\(0.8)S 
151 x(MeV\))S 153 x(is)S 152 x(ab)S 15 x(out)S 152 x(an)S 151 x
(order-)S 677 y 5347 X(of-m)S -2 x(agnitude)S 182 x(larger)S 180 x
(than)S 182 x(for)S 181 x(Kamiok)S -32 x(ande)S 182 x(\(7.5)S 181 x
(MeV\).)S 678 y 6165 X(W)S -46 x(e)S 140 x(will)S 137 x(compare)S 
139 x(directly)S 139 x(the)S 140 x(results)S 139 x(for)S 138 x
(these)S 141 x(t)S -15 x(w)S -16 x(o)S 139 x(exp)S 16 x(erimen)S 
-16 x(ts)S 140 x(using)S 138 x(a)S 139 x(lemma,)S 677 y 5347 X(pro)S
-16 x(v)S -15 x(ed)S 208 x(in)S 208 x(Ref.,)S -198 y F13(2)S -2 x(8)S
198 y 233 x F34(that)S 208 x(states)S 209 x(that)S 209 x(the)S 208 x
(shap)S 15 x(e)S 209 x(of)S 208 x(the)S 208 x(8)S 369 y -273 x(\026)S
-369 y 208 x(neutrino)S 208 x(sp)S 15 x(ectrum)S 209 x(that)S 208 x
(is)S 678 y 5347 X(pro)S 14 x(duced)S 173 x(in)S 172 x(the)S 173 x
(cen)S -14 x(ter)S 172 x(of)S 171 x(the)S 173 x(sun)S 172 x(is)S 
172 x(the)S 173 x(same,)S 172 x(to)S 172 x(an)S 171 x(accuracy)S 
173 x(of)S 171 x(one)S 173 x(part)S 171 x(in)S 172 x(10)S -198 y F13
(5)S 198 y 24 x F34(,)S 677 y 5347 X(as)S 181 x(the)S 182 x(shap)S
16 x(e)S 182 x(of)S 181 x(the)S 182 x(kno)S -16 x(wn)S 182 x(sp)S 
15 x(ectrum)S 182 x(that)S 182 x(is)S 181 x(pro)S 15 x(duced)S 182 x
(in)S 182 x(terrestrial)S 180 x(lab)S 15 x(orato)S -2 x(ries.)S 678 y 
5347 X(The)S 194 x(larg)S -2 x(est)S 195 x(impri)S -2 x(n)S -15 x
(ts)S 194 x(of)S 193 x(the)S 195 x(solar)S 192 x(en)S -14 x(viro)S
-2 x(nmen)S -15 x(t)S 194 x(are)S 194 x(caused)S 194 x(b)S -15 x(y)S
194 x(Doppler)S 193 x(shifts)S 193 x(and)S 677 y 5347 X(b)S -15 x(y)S
229 x(the)S 230 x(gra)S -17 x(vitational)S 228 x(redshift,)S 228 x
(but)S 229 x(b)S 16 x(oth)S 229 x(of)S 228 x(these)S 231 x
(e\013ects)S 231 x(are)S 229 x(negligi)S -2 x(bly)S 229 x(small)S 
228 x(for)S 678 y 5347 X(our)S 199 x(purp)S 14 x(oses.)S 200 x
(Therefore,)S 199 x(the)S 200 x(shap)S 15 x(e)S 200 x(of)S 199 x
(the)S 200 x(neutrino)S 199 x(sp)S 15 x(ectrum)S 200 x(m)S -16 x
(ust)S 200 x(b)S 15 x(e)S 200 x(the)S 200 x(same)S 677 y 5347 X(in)S
166 x(a)S 166 x(terrestrial)S 166 x(lab)S 14 x(oratory)S 165 x(and)S
167 x(in)S 166 x(the)S 167 x(cen)S -14 x(ter)S 167 x(of)S 166 x(the)S
167 x(sun)S 167 x(unless)S 167 x(ph)S -15 x(ysics)S 166 x(b)S 16 x
(ey)S -15 x(ond)S 166 x(the)S 677 y 5347 X(standard)S 252 x(electro)S
-15 x(w)S -15 x(eak)S 253 x(mo)S 14 x(del)S 253 x(causes)S 254 x
(energy-dep)S 16 x(enden)S -14 x(t)S 253 x(c)S -15 x(hanges)S 253 x
(in)S 253 x(the)S 253 x(neutrino)S 678 y 5347 X(sp)S 15 x(ectrum.)S
677 y 6165 X(W)S -46 x(e)S 181 x(kno)S -15 x(w)S 180 x(from)S 179 x
(the)S 182 x(Kamiok)S -32 x(ande)S 181 x(exp)S 15 x(erimen)S -15 x
(t)S 181 x(ho)S -16 x(w)S 181 x(man)S -16 x(y)S 181 x(8)S 369 y -273 x
(\026)S -369 y 180 x(neutrinos)S 180 x(reac)S -15 x(h)S 181 x(the)S
678 y 5347 X(earth)S 234 x(with)S 235 x(energies)S 235 x(ab)S 15 x
(out)S 235 x(7.5)S 233 x(MeV.)S 235 x(If)S 235 x(standard)S 234 x
(electro)S -15 x(w)S -15 x(eak)S 235 x(theory)S 234 x(is)S 235 x
(correct,)S 677 y 5347 X(then)S 199 x(w)S -15 x(e)S 199 x(can)S 198 x
(extend)S 200 x(the)S 199 x(lab)S 15 x(orato)S -2 x(ry)S 199 x(8)S
369 y -273 x(\026)S -369 y 198 x(sp)S 15 x(ectrum,)S 199 x(norm)S 
-2 x(alized)S 199 x(b)S -15 x(y)S 198 x(the)S 199 x(Kamiok)S -31 x
(ande)S 678 y 5347 X(results,)S 190 x(do)S -15 x(wn)S 191 x(to)S 
191 x(0.8)S 190 x(MeV,)S 191 x(the)S 191 x(threshold)S 191 x(for)S
190 x(the)S 192 x(c)S -15 x(hlori)S -2 x(ne)S 192 x(exp)S 15 x
(erimen)S -15 x(t.)S 190 x(This)S 191 x(leads)S 677 y 5347 X(to)S 
222 x(a)S 221 x(minim)S -17 x(um)S 222 x(predicted)S 223 x(rate)S 
222 x(for)S 221 x(the)S 222 x(c)S -14 x(hlori)S -2 x(ne)S 223 x(exp)S
15 x(erimen)S -15 x(t)S 222 x(based)S 222 x(on)S 222 x(scaling)S 
221 x(the)S
1 PP EP

1000 BP 39600 30600 PM 0 0 XY
3815 Y 5347 X F34(Kamio)S -2 x(k)S -30 x(ande)S 249 x(results)S 249 x
(do)S -15 x(wn)S 248 x(to)S 249 x(the)S 250 x(c)S -15 x(hlori)S -2 x
(ne)S 250 x(threshold)S 248 x(and)S 249 x(on)S 248 x(ignoring)S 247 x
(all)S 248 x(other)S 678 y 5347 X(neutrino)S 181 x(sources)S 182 x
(exce)S 2 x(pt)S 182 x(the)S 182 x(rare)S 181 x(8)S 368 y -273 x
(\026)S -368 y 182 x(neutrinos.)S 181 x(This)S 181 x(minim)S -17 x
(um)S 182 x(v)S -31 x(alue)S 182 x(is)S 6476 Y 6285 X(Cl)S 181 x
(Rate)S 182 x(\()S -225 y F13(8)S 225 y 25 x F34(B)S 182 x(only\))S
333 x(=)S -714 y 333 x F67(\022)S 345 y 92 x F34(Rate)S 182 x
(Observ)S -15 x(ed)S 242 y 12085 X 3588 24 R 501 y 12085 X(Rate)S 
181 x(Predicted)S 5762 Y 61 x F67(\023)S
XP /F13 75 329 1 0 24 22 23 24 0
<FFC1FE FFC1FE 1E00F0 1E01E0 1E03C0 1E0780 1E0F00 1E1E00 1E3C00
 1E7C00 1EFC00 1FFE00 1FDF00 1F9F00 1F0F80 1E07C0 1E07C0 1E03E0
 1E01F0 1E01F0 1E00F8 FFC3FF FFC3FF>
PXLC RP
16104 6962 XY F13(K)S
XP /F13 97 212 1 0 16 13 14 16 0
<3FC0 7FF0 78F8 7838 3038 0FF8 3F38 7C38 F838 F03B F03B F8FB 7FFE
 3F1C>
PXLC RP
16432 6962 XY F13(a)S
XP /F13 109 353 0 0 28 13 14 32 0
<FCFC3F00 FFFEFF80 1F0FC3C0 1E0781C0 1C0701C0 1C0701C0 1C0701C0
 1C0701C0 1C0701C0 1C0701C0 1C0701C0 1C0701C0 FF9FE7F8 FF9FE7F8>
PXLC RP
16644 6962 XY F13(m)S
XP /F13 105 118 0 0 7 22 23 8 0
<3C 3C 3C 3C 00 00 00 00 00 7C 7C 1C 1C 1C 1C 1C 1C 1C 1C 1C 1C FF FF>
PXLC RP
16997 6962 XY F13(i)S
XP /F13 111 212 1 0 15 13 14 16 0
<07C0 1FF0 3838 701C 701C F01E F01E F01E F01E F01E 701C 3838 1FF0
 07C0>
PXLC RP
17115 6962 XY F13(o)S
XP /F13 107 223 0 0 17 22 23 24 0
<FC0000 FC0000 1C0000 1C0000 1C0000 1C0000 1C0000 1C0000 1C0000
 1C7F80 1C7F80 1C7C00 1CF000 1DE000 1FC000 1FE000 1EF000 1CF800
 1C7800 1C3C00 1C1E00 FF3FC0 FF3FC0>
PXLC RP
17326 6962 XY F13(k)S -23 x(a)S
XP /F13 110 235 0 0 18 13 14 24 0
<FCFC00 FFFE00 1F0F00 1E0700 1C0700 1C0700 1C0700 1C0700 1C0700
 1C0700 1C0700 1C0700 FF9FE0 FF9FE0>
PXLC RP
17738 6962 XY F13(n)S
XP /F13 100 235 1 0 18 22 23 24 0
<007E00 007E00 000E00 000E00 000E00 000E00 000E00 000E00 000E00
 0FEE00 3FFE00 7C3E00 780E00 F00E00 F00E00 F00E00 F00E00 F00E00
 F00E00 781E00 7C3E00 3FFFC0 0FCFC0>
PXLC RP
17973 6962 XY F13(de)S
XP /F67 16 297 8 -72 22 1 74 16 0
<000E 001E 001C 0038 0070 00F0 00E0 01C0 03C0 0380 0780 0700 0F00
 0F00 0E00 1E00 1E00 1C00 3C00 3C00 3C00 3800 7800 7800 7800 7800
 7800 7000 F000 F000 F000 F000 F000 F000 F000 F000 F000 F000 F000
 F000 F000 F000 F000 F000 F000 F000 7000 7800 7800 7800 7800 7800
 3800 3C00 3C00 3C00 1C00 1E00 1E00 0E00 0F00 0F00 0700 0780 0380
 03C0 01C0 00E0 00F0 0070 0038 001C 001E 000E>
PXLC RP
18512 5911 XY F67(\020)S 340 y F13(8)S 225 y 26 x F34(B)S
XP /F67 17 297 1 -72 15 1 74 16 0
<E000 F000 7000 3800 1C00 1E00 0E00 0700 0780 0380 03C0 01C0 01E0
 01E0 00E0 00F0 00F0 0070 0078 0078 0078 0038 003C 003C 003C 003C
 003C 001C 001E 001E 001E 001E 001E 001E 001E 001E 001E 001E 001E
 001E 001E 001E 001E 001E 001E 001E 001C 003C 003C 003C 003C 003C
 0038 0078 0078 0078 0070 00F0 00F0 00E0 01E0 01E0 01C0 03C0 0380
 0780 0700 0E00 1E00 1C00 3800 7000 F000 E000>
PXLC RP
19433 5911 XY F67(\021)S 565 y 122 x F36(\002)S 121 x F34(6)S F35(:)S
F34(2)S 181 x F35(S)S 32 x(N)S 60 x(U;)S 23692 X F34(\(6\))S 7755 Y 
5347 X(or)S 9110 Y 9416 X(Cl)S 181 x(Rate)S 182 x(\()S -225 y F13(8)S
225 y 25 x F34(B)S 182 x(only\))S
XP /F36 21 425 4 -7 29 28 36 32 0
<E0000000 F8000000 7E000000 1F800000 07E00000 01F80000 007C0000
 001F0000 000FC000 0003F000 0000FC00 00003F00 00000FC0 000003C0
 00000FC0 00003F00 0000FC00 0003F000 000FC000 001F0000 007C0000
 01F80000 07E00000 1F800000 7E000000 F8000000 E0000000 00000000
 00000000 00000000 00000000 00000000 00000000 00000000 FFFFFFC0
 FFFFFFC0>
PXLC RP
14031 9110 XY F36(\025)S 334 x F34(3)S F35(:)S F34(1)S 181 x F35(S)S
32 x(N)S 59 x(U)S
XP /F35 62 425 4 -3 29 24 28 32 0
<C0000000 F0000000 FC000000 3F000000 0FC00000 03F00000 00FC0000
 003E0000 000F8000 0007E000 0001F800 00007E00 00001F80 000007C0
 000007C0 00001F80 00007E00 0001F800 0007E000 000F8000 003E0000
 00FC0000 03F00000 0FC00000 3F000000 FC000000 F0000000 C0000000>
PXLC RP
17297 9110 XY F35(>)S 333 x F34(2)S F35(:)S F34(2)S 181 x F35(S)S 
32 x(N)S 60 x(U:)S 23692 X F34(\(7\))S 10091 Y 5347 X(In)S 252 x
(Eq.)S 251 x(\(6\),)S 251 x(6.2)S 251 x(SNU)S 252 x(is)S 252 x(the)S
252 x(capture)S 253 x(rate)S 252 x(for)S 250 x(c)S -14 x(hlori)S 
-2 x(ne)S 252 x(that)S 252 x(is)S 252 x(predicted)S 252 x(b)S -15 x
(y)S 252 x(the)S 678 y 5347 X(standard)S 211 x(mo)S 14 x(del)S 212 x
(for)S 211 x(just)S 212 x(the)S 212 x(8)S 368 y -273 x(\026)S -368 y 
211 x(neutrinos.)S 211 x(The)S 212 x(result)S 212 x(sho)S -15 x(wn)S
211 x(in)S 212 x(Eq.)S 211 x(\(7\))S 212 x(indicates)S 677 y 5347 X
(that)S 186 x(the)S 186 x(\015ux)S 186 x(of)S 186 x(just)S 186 x(8)S
368 y -273 x(\026)S -368 y 185 x(neutrinos)S 186 x(that)S 186 x(are)S
186 x(seen)S 187 x(in)S 185 x(the)S 187 x(Kamiok)S -32 x(ande)S 187 x
(I)S 15 x(I)S 186 x(exp)S 15 x(erimen)S -15 x(t)S 677 y 5347 X(is)S
156 x(b)S -16 x(y)S 157 x(itself)S 155 x(su\016cien)S -15 x(t)S 157 x
(to)S 156 x(yield)S 155 x(a)S 156 x(capture)S 157 x(rate)S 156 x(in)S
156 x(excess)S 158 x(of)S 155 x(the)S 157 x(c)S -15 x(hlorine)S 155 x
(exp)S 16 x(erimen)S -16 x(tal)S 678 y 5347 X(v)S -31 x(alue)S 241 x
(of)S 240 x(2)S F35(:)S F34(28)S 159 x F36(\006)S 161 x F34(0)S F35
(:)S F34(23)S 239 x(SNU.)S 241 x(The)S 241 x(additio)S -2 x(nal)S 
240 x(neutrinos)S 241 x(from)S 239 x(other,)S 240 x(more)S 240 x
(reliably)S 677 y 5347 X(calculated)S 181 x(branc)S -15 x(hes)S 183 x
(of)S 181 x(the)S 182 x F35(pp)S 183 x F34(fusion)S 181 x(c)S -15 x
(hain,)S 181 x(further)S 181 x(increase)S 182 x(the)S 183 x
(discrepancy)S -46 x(.)S 678 y 6165 X(What)S 152 x(is)S 153 x(the)S
154 x(most)S 152 x(serious)S 153 x(mistak)S -16 x(e)S 153 x(that)S
153 x(w)S -15 x(e)S 153 x(could)S 153 x(ha)S -15 x(v)S -16 x(e)S 
154 x(made)S 152 x(in)S 153 x(the)S 154 x(solar)S 151 x(mo)S 15 x
(del)S 677 y 5347 X(calculatio)S -2 x(ns?)S 157 x(The)S 157 x(most)S
157 x(crucial)S 156 x(error)S 156 x(w)S -15 x(ould)S 156 x(ha)S -16 x
(v)S -15 x(e)S 157 x(b)S 15 x(een)S 158 x(to)S 157 x(ha)S -16 x(v)S
-15 x(e)S 157 x(calculated)S 157 x(wrongly)S 678 y 5347 X(the)S 214 x
(8)S 368 y -273 x(\026)S -368 y 214 x(neutrino)S 214 x(\015ux)S 214 x
(since)S 215 x(only)S 213 x(8)S 368 y -273 x(\026)S -368 y 214 x
(neutrinos)S 214 x(are)S 214 x(observ)S -15 x(ed)S 214 x(in)S 214 x
(the)S 215 x(Kamiok)S -32 x(ande)S 215 x(ex-)S 677 y 5347 X(p)S 15 x
(erimen)S -16 x(t)S 243 x(and)S 242 x(8)S 369 y -273 x(\026)S -369 y 
241 x(neutrinos)S 242 x(also)S 242 x(accoun)S -15 x(t)S 242 x(for)S
241 x(nearly)S 242 x(80
(exp)S 16 x(ected)S 243 x(in)S 678 y 5347 X(the)S 175 x(c)S -14 x
(hlori)S -2 x(ne)S 176 x(exp)S 15 x(erimen)S -15 x(t.)S 174 x(Supp)S
16 x(ose)S 175 x(that)S 175 x(this)S 175 x(\015ux)S 175 x(w)S -15 x
(as)S 175 x(calculated)S 175 x(wrongl)S -2 x(y)S -45 x(,)S 175 x(p)S
15 x(erhaps)S 677 y 5347 X(b)S 15 x(ecause)S 189 x(all)S 186 x(of)S
187 x(the)S 189 x(lab)S 14 x(oratory)S 186 x(n)S -15 x(uclear)S 188 x
(ph)S -16 x(ysics)S 188 x(measuremen)S -15 x(ts)S 188 x(of)S 187 x
(the)S 188 x(reaction)S 188 x(that)S 678 y 5347 X(pro)S 14 x(duces)S
210 x(8)S 368 y -273 x(\026)S -368 y 209 x(ha)S -16 x(v)S -15 x(e)S
209 x(b)S 15 x(een)S 210 x(seriously)S 208 x(in)S 208 x(error.)S 
208 x(W)S -46 x(ould)S 208 x(it)S 209 x(then)S 209 x(b)S 15 x(e)S 
209 x(p)S 16 x(ossible)S 208 x(to)S 209 x(reconcile)S 677 y 5347 X
(the)S 182 x(c)S -15 x(hlorine)S 181 x(and)S 182 x(the)S 182 x
(Kamiok)S -32 x(ande)S 182 x(exp)S 16 x(erimen)S -16 x(ts?)S 678 y 
6165 X(The)S 173 x(answ)S -15 x(er)S 173 x(to)S 172 x(this)S 173 x
(question)S 173 x(is)S 172 x(giv)S -16 x(en)S 173 x(in)S 173 x
(Figure)S 172 x(4)S 172 x(and)S 173 x(is)S 172 x(\\No".)S 172 x(F)S
-45 x(or)S 172 x(eac)S -15 x(h)S 173 x(of)S 172 x(the)S 677 y 5347 X
(1000)S 224 x(solar)S 224 x(mo)S 15 x(dels)S 225 x(discussed)S 226 x
(earlier,)S 224 x(I)S 225 x(ha)S -15 x(v)S -15 x(e)S 226 x(replaced)S
225 x(the)S 226 x(calculated)S 225 x(8)S 368 y -273 x(\026)S -368 y 
226 x(\015ux)S 225 x(b)S -15 x(y)S 225 x(a)S 677 y 5347 X(v)S -31 x
(alue)S 124 x(dra)S -16 x(wn)S 123 x(from)S 122 x(a)S 124 x(norma)S
-2 x(l)S 124 x(distributi)S -2 x(on)S 124 x(with)S 123 x(the)S 124 x
(mean)S 123 x(and)S 124 x(the)S 124 x(standard)S 123 x(deviation)S
678 y 5347 X(determined)S 164 x(b)S -15 x(y)S 164 x(the)S 165 x
(Kamiok)S -32 x(ande)S 165 x(exp)S 15 x(erimen)S -15 x(t.)S 163 x
(This)S 164 x(assumption)S 163 x(reduces)S 166 x F37(ad)S 178 x(ho)S
-28 x(c)S 164 x F34(the)S 677 y 5347 X(mean)S 202 x(rate)S 201 x(b)S
-15 x(y)S 202 x(ab)S 15 x(out)S 202 x(3.1)S 201 x(SNU,)S 202 x(as)S
202 x(indicated)S 202 x(b)S -15 x(y)S 202 x(Eq.)S 201 x(\(7\).)S 
202 x(The)S 202 x(resulting)S 202 x(histogr)S -2 x(am)S 678 y 5347 X
(is)S 257 x(no)S -16 x(w)S 258 x(cen)S -14 x(tered)S 258 x(just)S 
257 x(b)S 15 x(elo)S -15 x(w)S 257 x(5)S 257 x(SNU,)S 257 x(instead)S
258 x(of)S 257 x(at)S 257 x(8)S 257 x(SNU,)S 257 x(as)S 258 x(in)S
257 x(the)S 258 x(unfudged)S 677 y 5347 X(orig)S -2 x(inal)S 224 x
(calculations)S 224 x(\(see)S 226 x(Figure)S 224 x(2\).)S 225 x(In)S
225 x(additio)S -2 x(n,)S 225 x(the)S 225 x(width)S 225 x(of)S 224 x
(the)S 226 x(histogram)S 223 x(is)S 678 y 5347 X(m)S -16 x(uc)S -15 x
(h)S 219 x(narro)S -16 x(w)S -15 x(er)S 219 x(than)S 219 x(in)S 218 x
(the)S 220 x(actual)S 218 x(calculations)S 218 x(b)S 15 x(ecause)S
221 x(the)S 219 x(con)S -15 x(tribution)S 218 x(of)S 218 x(the)S 
677 y 5347 X(8)S 369 y -273 x(\026)S -369 y 227 x(neutrinos)S 227 x
(is)S 227 x(reduced)S 229 x(and)S 227 x(8)S 369 y -273 x(\026)S -369 y 
227 x(neutrinos)S 227 x(are)S 227 x(the)S 228 x(most)S 227 x
(uncertain)S 228 x(of)S 227 x(all)S 226 x(the)S 228 x(solar)S 678 y 
5347 X(neutrino)S 181 x(sources.)S 677 y 6165 X(Ev)S -15 x(en)S 145 x
(in)S 145 x(the)S 146 x(w)S -15 x(orst)S 145 x(case)S 146 x
(scenario)S 144 x(sho)S -15 x(wn)S 145 x(in)S 145 x(Figure)S 145 x
(4,)S 144 x(in)S 145 x(whic)S -16 x(h)S 146 x(the)S 145 x(normali)S
-2 x(zation)S 678 y 5347 X(of)S 135 x(the)S 137 x(8)S 368 y -273 x
(\026)S -368 y 136 x(neutrino)S 136 x(\015ux)S 136 x(is)S 136 x
(arti\014cial)S -2 x(ly)S 136 x(adjusted)S 136 x(to)S 136 x(equal)S
136 x(the)S 137 x(measured)S 136 x(Kamiok)S -32 x(ande)S 137 x(I)S
15 x(I)S 677 y 5347 X(v)S -31 x(alue,)S 155 x(the)S 156 x
(calculated)S 155 x(rate)S 156 x(for)S 154 x(the)S 156 x(c)S -14 x
(hlori)S -2 x(ne)S 156 x(exp)S 15 x(erimen)S -15 x(t)S 155 x(is)S 
156 x(man)S -16 x(y)S 155 x(exp)S 16 x(erimen)S -16 x(tal)S 155 x
(stan-)S 678 y 5347 X(dard)S 134 x(deviations)S 133 x(larger)S 134 x
(than)S 134 x(the)S 135 x(observ)S -15 x(ed)S 135 x(rate.)S 134 x
(Hans)S 135 x(Bethe)S 136 x(and)S 134 x(I)S 134 x(ha)S -15 x(v)S 
-15 x(e)S 135 x(concluded)S -198 y F13(26)S 875 y 5347 X F34(on)S 
136 x(the)S 136 x(basis)S 136 x(of)S 136 x(Figure)S 135 x(4)S 136 x
(that)S 137 x(either)S 136 x(new)S 137 x(ph)S -15 x(ysics)S 136 x
(\(b)S 15 x(ey)S -15 x(ond)S 136 x(the)S 137 x(standard)S 136 x
(electro)S -15 x(w)S -15 x(eak)S 677 y 5347 X(mo)S 14 x(del\))S 165 x
(is)S 165 x(required)S 165 x(to)S 165 x(c)S -15 x(hange)S 165 x(the)S
166 x(shap)S 15 x(e)S 165 x(of)S 165 x(the)S 166 x(8)S 369 y -273 x
(\026)S -369 y 165 x(neutrino)S 164 x(energy)S 165 x(sp)S 16 x
(ectrum)S 165 x(or)S 165 x(one)S 678 y 5347 X(of)S 181 x(the)S 182 x
(t)S -15 x(w)S -15 x(o)S 182 x(exp)S 15 x(erimen)S -15 x(ts)S 182 x
(\(c)S -15 x(hlori)S -2 x(ne)S 183 x(and)S 181 x(Kamiok)S -31 x
(ande)S 182 x(I)S 15 x(I\))S 182 x(is)S 181 x(wrong.)S
XP /F48 54 336 2 0 24 31 32 24 0
<003FC0 01FFE0 03FFF0 07F0F8 0FC1F8 1F81F8 3F01F8 3F01F8 7F00F0
 7E0000 7E0000 FE0800 FE7FC0 FEFFF0 FFC1F8 FF81F8 FF00FC FF00FE
 FE00FE FE00FE FE00FE FE00FE FE00FE 7E00FE 7E00FE 7E00FE 3F00FC
 3F01FC 1FC3F8 0FFFF0 07FFC0 00FF00>
PXLC RP
5347 31452 XY F48(6.)S 672 x(The)S 224 x(Galli)S -2 x(um)S 225 x
(Exp)S 18 x(erimen)S -19 x(ts:)S 223 x(F)S -56 x(urth)S -2 x(er)S 
224 x(E)S
XP /F48 118 355 1 0 27 21 22 32 0
<FFF03FE0 FFF03FE0 FFF03FE0 0FC00E00 0FE01E00 07E01C00 07E01C00
 03F03800 03F03800 03F87800 01F87000 01FCF000 00FCE000 00FCE000
 007FC000 007FC000 007FC000 003F8000 003F8000 001F0000 001F0000
 000E0000>
PXLC RP
17502 31452 XY F48(vidence)S 32435 Y 5347 X F34(More)S 189 x(than)S
189 x(half)S 189 x(\(54
189 x(of)S 189 x(the)S 190 x(predicted)S 190 x(standard)S 189 x(mo)S
15 x(del)S 189 x(ev)S -15 x(en)S -14 x(t)S 189 x(rate,)S 677 y 5347 X
(132)S -229 y F13(+7)S 378 y -542 x F15(\000)S F13(6)S -149 y 188 x 
F34(SNU,)S 162 x(in)S 163 x(the)S 163 x(gall)S -2 x(ium)S 162 x(exp)S
16 x(erimen)S -16 x(ts)S 163 x(comes)S 163 x(from)S 162 x(the)S 163 x
(lo)S -16 x(w-energy)S 163 x F35(pp)S 164 x F34(neutrinos.)S 678 y 
5347 X(The)S 184 x(standard)S 184 x(\015ux)S 183 x(of)S 184 x(these)S
185 x(neutrinos)S 184 x(can)S 184 x(b)S 15 x(e)S 184 x(calculated)S
184 x(with)S 184 x(precision)S 183 x(\(accuracy)S
1 PP EP

1000 BP 39600 30600 PM 0 0 XY
4787 15917 XY 0 SPB
 save 10 dict begin /Figure exch def currentpoint translate
/showpage {} def 
0 SPE
4787 15917 XY 0 SPB
save 50 dict begin /psplot exch def
/StartPSPlot
   {newpath 0 0 moveto 0 setlinewidth 0 setgray 0 setlinecap
    1 setlinejoin 72 300 div dup scale}def
/pending {false} def
/finish {pending {currentpoint stroke moveto /pending false def} if} def
/r {finish newpath moveto} def
/d {lineto /pending true def} def
/l {finish 4 2 roll moveto lineto currentpoint stroke moveto} def
/p {finish newpath moveto currentpoint lineto currentpoint stroke moveto} def
/e {finish gsave showpage grestore newpath 0 0 moveto} def
/lw {finish setlinewidth} def
/lt0 {finish [] 0 setdash} def
/lt1 {finish [3 5] 0 setdash} def
/lt2 {finish [20 10] 0 setdash} def
/lt3 {finish [60 10] 0 setdash} def
/lt4 {finish [3 10 20 10] 0 setdash} def
/lt5 {finish [3 10 60 10] 0 setdash} def
/lt6 {finish [20 10 60 10] 0 setdash} def
/EndPSPlot {clear psplot end restore}def
StartPSPlot
   1 lw lt0  200  200 r 1600  200 d  200  200 r  200  238 d  225  200 r
  225  219 d  251  200 r  251  219 d  276  200 r  276  219 d  302  200 r
  302  219 d  327  200 r  327  238 d  353  200 r  353  219 d  378  200 r
  378  219 d  404  200 r  404  219 d  429  200 r  429  219 d  455  200 r
  455  238 d  480  200 r  480  219 d  505  200 r  505  219 d  531  200 r
  531  219 d  556  200 r  556  219 d  582  200 r  582  238 d  607  200 r
  607  219 d  633  200 r  633  219 d  658  200 r  658  219 d  684  200 r
  684  219 d  709  200 r  709  238 d  735  200 r  735  219 d  760  200 r
  760  219 d  785  200 r  785  219 d  811  200 r  811  219 d  836  200 r
  836  238 d  862  200 r  862  219 d  887  200 r  887  219 d  913  200 r
  913  219 d  938  200 r  938  219 d  964  200 r  964  238 d  989  200 r
  989  219 d 1015  200 r 1015  219 d 1040  200 r 1040  219 d 1065  200 r
 1065  219 d 1091  200 r 1091  238 d 1116  200 r 1116  219 d 1142  200 r
 1142  219 d 1167  200 r 1167  219 d 1193  200 r 1193  219 d 1218  200 r
 1218  238 d 1244  200 r 1244  219 d 1269  200 r 1269  219 d 1295  200 r
 1295  219 d 1320  200 r 1320  219 d 1345  200 r 1345  238 d 1371  200 r
 1371  219 d 1396  200 r 1396  219 d 1422  200 r 1422  219 d 1447  200 r
 1447  219 d 1473  200 r 1473  238 d 1498  200 r 1498  219 d 1524  200 r
 1524  219 d 1549  200 r 1549  219 d 1575  200 r 1575  219 d 1600  200 r
 1600  238 d  199  176 r  195  175 d  193  172 d  192  166 d  192  162 d
  193  156 d  195  152 d  199  151 d  201  151 d  205  152 d  207  156 d
  208  162 d  208  166 d  207  172 d  205  175 d  201  176 d  199  176 d
  196  175 d  195  174 d  194  172 d  193  166 d  193  162 d  194  156 d
  195  154 d  196  152 d  199  151 d  201  151 r  204  152 d  205  154 d
  206  156 d  207  162 d  207  166 d  206  172 d  205  174 d  204  175 d
  201  176 d  322  172 r  325  173 d  328  176 d  328  151 d  327  175 r
  327  151 d  322  151 r  333  151 d  448  172 r  449  170 d  448  169 d
  447  170 d  447  172 d  448  174 d  449  175 d  453  176 d  457  176 d
  461  175 d  462  174 d  463  172 d  463  169 d  462  167 d  459  164 d
  453  162 d  450  161 d  448  158 d  447  155 d  447  151 d  457  176 r
  460  175 d  461  174 d  462  172 d  462  169 d  461  167 d  457  164 d
  453  162 d  447  154 r  448  155 d  450  155 d  456  152 d  460  152 d
  462  154 d  463  155 d  450  155 r  456  151 d  461  151 d  462  152 d
  463  155 d  463  157 d  575  173 r  576  172 d  575  170 d  574  172 d
  574  173 d  576  175 d  580  176 d  584  176 d  588  175 d  589  173 d
  589  169 d  588  167 d  584  166 d  581  166 d  584  176 r  587  175 d
  588  173 d  588  169 d  587  167 d  584  166 d  587  164 d  589  162 d
  590  160 d  590  156 d  589  154 d  588  152 d  584  151 d  580  151 d
  576  152 d  575  154 d  574  156 d  574  157 d  575  158 d  576  157 d
  575  156 d  588  163 r  589  160 d  589  156 d  588  154 d  587  152 d
  584  151 d  711  174 r  711  151 d  713  176 r  713  151 d  713  176 r
  699  158 d  719  158 d  708  151 r  716  151 d  830  176 r  828  164 d
  830  166 d  834  167 d  837  167 d  841  166 d  843  163 d  844  160 d
  844  158 d  843  155 d  841  152 d  837  151 d  834  151 d  830  152 d
  829  154 d  828  156 d  828  157 d  829  158 d  830  157 d  829  156 d
  837  167 r  840  166 d  842  163 d  843  160 d  843  158 d  842  155 d
  840  152 d  837  151 d  830  176 r  842  176 d  830  175 r  836  175 d
  842  176 d  970  173 r  969  172 d  970  170 d  971  172 d  971  173 d
  970  175 d  968  176 d  964  176 d  960  175 d  958  173 d  957  170 d
  956  166 d  956  158 d  957  155 d  959  152 d  963  151 d  965  151 d
  969  152 d  971  155 d  972  158 d  972  160 d  971  163 d  969  166 d
  965  167 d  964  167 d  960  166 d  958  163 d  957  160 d  964  176 r
  962  175 d  959  173 d  958  170 d  957  166 d  957  158 d  958  155 d
  960  152 d  963  151 d  965  151 r  968  152 d  970  155 d  971  158 d
  971  160 d  970  163 d  968  166 d  965  167 d 1083  176 r 1083  169 d
 1083  172 r 1084  174 d 1086  176 d 1089  176 d 1095  173 d 1097  173 d
 1098  174 d 1099  176 d 1084  174 r 1086  175 d 1089  175 d 1095  173 d
 1099  176 r 1099  173 d 1098  169 d 1093  163 d 1092  161 d 1091  157 d
 1091  151 d 1098  169 r 1092  163 d 1091  161 d 1090  157 d 1090  151 d
 1216  176 r 1212  175 d 1211  173 d 1211  169 d 1212  167 d 1216  166 d
 1220  166 d 1224  167 d 1225  169 d 1225  173 d 1224  175 d 1220  176 d
 1216  176 d 1213  175 d 1212  173 d 1212  169 d 1213  167 d 1216  166 d
 1220  166 r 1223  167 d 1224  169 d 1224  173 d 1223  175 d 1220  176 d
 1216  166 r 1212  164 d 1211  163 d 1210  161 d 1210  156 d 1211  154 d
 1212  152 d 1216  151 d 1220  151 d 1224  152 d 1225  154 d 1226  156 d
 1226  161 d 1225  163 d 1224  164 d 1220  166 d 1216  166 r 1213  164 d
 1212  163 d 1211  161 d 1211  156 d 1212  154 d 1213  152 d 1216  151 d
 1220  151 r 1223  152 d 1224  154 d 1225  156 d 1225  161 d 1224  163 d
 1223  164 d 1220  166 d 1352  168 r 1351  164 d 1349  162 d 1345  161 d
 1344  161 d 1340  162 d 1338  164 d 1337  168 d 1337  169 d 1338  173 d
 1340  175 d 1344  176 d 1346  176 d 1350  175 d 1352  173 d 1353  169 d
 1353  162 d 1352  157 d 1351  155 d 1349  152 d 1345  151 d 1341  151 d
 1339  152 d 1338  155 d 1338  156 d 1339  157 d 1340  156 d 1339  155 d
 1344  161 r 1341  162 d 1339  164 d 1338  168 d 1338  169 d 1339  173 d
 1341  175 d 1344  176 d 1346  176 r 1349  175 d 1351  173 d 1352  169 d
 1352  162 d 1351  157 d 1350  155 d 1347  152 d 1345  151 d 1456  172 r
 1459  173 d 1462  176 d 1462  151 d 1461  175 r 1461  151 d 1456  151 r
 1467  151 d 1484  176 r 1480  175 d 1478  172 d 1477  166 d 1477  162 d
 1478  156 d 1480  152 d 1484  151 d 1486  151 d 1490  152 d 1492  156 d
 1493  162 d 1493  166 d 1492  172 d 1490  175 d 1486  176 d 1484  176 d
 1481  175 d 1480  174 d 1479  172 d 1478  166 d 1478  162 d 1479  156 d
 1480  154 d 1481  152 d 1484  151 d 1486  151 r 1489  152 d 1490  154 d
 1491  156 d 1492  162 d 1492  166 d 1491  172 d 1490  174 d 1489  175 d
 1486  176 d 1583  172 r 1586  173 d 1589  176 d 1589  151 d 1588  175 r
 1588  151 d 1583  151 r 1594  151 d 1607  172 r 1610  173 d 1613  176 d
 1613  151 d 1612  175 r 1612  151 d 1607  151 r 1618  151 d  200 1100 r
 1600 1100 d  200 1100 r  200 1062 d  225 1100 r  225 1081 d  251 1100 r
  251 1081 d  276 1100 r  276 1081 d  302 1100 r  302 1081 d  327 1100 r
  327 1062 d  353 1100 r  353 1081 d  378 1100 r  378 1081 d  404 1100 r
  404 1081 d  429 1100 r  429 1081 d  455 1100 r  455 1062 d  480 1100 r
  480 1081 d  505 1100 r  505 1081 d  531 1100 r  531 1081 d  556 1100 r
  556 1081 d  582 1100 r  582 1062 d  607 1100 r  607 1081 d  633 1100 r
  633 1081 d  658 1100 r  658 1081 d  684 1100 r  684 1081 d  709 1100 r
  709 1062 d  735 1100 r  735 1081 d  760 1100 r  760 1081 d  785 1100 r
  785 1081 d  811 1100 r  811 1081 d  836 1100 r  836 1062 d  862 1100 r
  862 1081 d  887 1100 r  887 1081 d  913 1100 r  913 1081 d  938 1100 r
  938 1081 d  964 1100 r  964 1062 d  989 1100 r  989 1081 d 1015 1100 r
 1015 1081 d 1040 1100 r 1040 1081 d 1065 1100 r 1065 1081 d 1091 1100 r
 1091 1062 d 1116 1100 r 1116 1081 d 1142 1100 r 1142 1081 d 1167 1100 r
 1167 1081 d 1193 1100 r 1193 1081 d 1218 1100 r 1218 1062 d 1244 1100 r
 1244 1081 d 1269 1100 r 1269 1081 d 1295 1100 r 1295 1081 d 1320 1100 r
 1320 1081 d 1345 1100 r 1345 1062 d 1371 1100 r 1371 1081 d 1396 1100 r
 1396 1081 d 1422 1100 r 1422 1081 d 1447 1100 r 1447 1081 d 1473 1100 r
 1473 1062 d 1498 1100 r 1498 1081 d 1524 1100 r 1524 1081 d 1549 1100 r
 1549 1081 d 1575 1100 r 1575 1081 d 1600 1100 r 1600 1062 d  200  200 r
  200 1100 d  200  200 r  238  200 d  200  221 r  219  221 d  200  243 r
  219  243 d  200  264 r  219  264 d  200  286 r  219  286 d  200  307 r
  238  307 d  200  329 r  219  329 d  200  350 r  219  350 d  200  371 r
  219  371 d  200  393 r  219  393 d  200  414 r  238  414 d  200  436 r
  219  436 d  200  457 r  219  457 d  200  479 r  219  479 d  200  500 r
  219  500 d  200  521 r  238  521 d  200  543 r  219  543 d  200  564 r
  219  564 d  200  586 r  219  586 d  200  607 r  219  607 d  200  629 r
  238  629 d  200  650 r  219  650 d  200  671 r  219  671 d  200  693 r
  219  693 d  200  714 r  219  714 d  200  736 r  238  736 d  200  757 r
  219  757 d  200  779 r  219  779 d  200  800 r  219  800 d  200  821 r
  219  821 d  200  843 r  238  843 d  200  864 r  219  864 d  200  886 r
  219  886 d  200  907 r  219  907 d  200  929 r  219  929 d  200  950 r
  238  950 d  200  971 r  219  971 d  200  993 r  219  993 d  200 1014 r
  219 1014 d  200 1036 r  219 1036 d  200 1057 r  238 1057 d  200 1079 r
  219 1079 d  200 1100 r  219 1100 d  168  214 r  164  213 d  162  210 d
  161  204 d  161  200 d  162  194 d  164  190 d  168  189 d  170  189 d
  174  190 d  176  194 d  177  200 d  177  204 d  176  210 d  174  213 d
  170  214 d  168  214 d  165  213 d  164  212 d  163  210 d  162  204 d
  162  200 d  163  194 d  164  192 d  165  190 d  168  189 d  170  189 r
  173  190 d  174  192 d  175  194 d  176  200 d  176  204 d  175  210 d
  174  212 d  173  213 d  170  214 d  139  321 r  137  309 d  139  311 d
  143  312 d  146  312 d  150  311 d  152  308 d  153  305 d  153  303 d
  152  300 d  150  297 d  146  296 d  143  296 d  139  297 d  138  299 d
  137  301 d  137  302 d  138  303 d  139  302 d  138  301 d  146  312 r
  149  311 d  151  308 d  152  305 d  152  303 d  151  300 d  149  297 d
  146  296 d  139  321 r  151  321 d  139  320 r  145  320 d  151  321 d
  168  321 r  164  320 d  162  317 d  161  311 d  161  307 d  162  301 d
  164  297 d  168  296 d  170  296 d  174  297 d  176  301 d  177  307 d
  177  311 d  176  317 d  174  320 d  170  321 d  168  321 d  165  320 d
  164  319 d  163  317 d  162  311 d  162  307 d  163  301 d  164  299 d
  165  297 d  168  296 d  170  296 r  173  297 d  174  299 d  175  301 d
  176  307 d  176  311 d  175  317 d  174  319 d  173  320 d  170  321 d
  116  424 r  119  425 d  122  428 d  122  403 d  121  427 r  121  403 d
  116  403 r  127  403 d  144  428 r  140  427 d  138  424 d  137  418 d
  137  414 d  138  408 d  140  404 d  144  403 d  146  403 d  150  404 d
  152  408 d  153  414 d  153  418 d  152  424 d  150  427 d  146  428 d
  144  428 d  141  427 d  140  426 d  139  424 d  138  418 d  138  414 d
  139  408 d  140  406 d  141  404 d  144  403 d  146  403 r  149  404 d
  150  406 d  151  408 d  152  414 d  152  418 d  151  424 d  150  426 d
  149  427 d  146  428 d  168  428 r  164  427 d  162  424 d  161  418 d
  161  414 d  162  408 d  164  404 d  168  403 d  170  403 d  174  404 d
  176  408 d  177  414 d  177  418 d  176  424 d  174  427 d  170  428 d
  168  428 d  165  427 d  164  426 d  163  424 d  162  418 d  162  414 d
  163  408 d  164  406 d  165  404 d  168  403 d  170  403 r  173  404 d
  174  406 d  175  408 d  176  414 d  176  418 d  175  424 d  174  426 d
  173  427 d  170  428 d  116  531 r  119  532 d  122  535 d  122  510 d
  121  534 r  121  510 d  116  510 r  127  510 d  139  535 r  137  523 d
  139  525 d  143  526 d  146  526 d  150  525 d  152  522 d  153  519 d
  153  517 d  152  514 d  150  511 d  146  510 d  143  510 d  139  511 d
  138  513 d  137  515 d  137  516 d  138  517 d  139  516 d  138  515 d
  146  526 r  149  525 d  151  522 d  152  519 d  152  517 d  151  514 d
  149  511 d  146  510 d  139  535 r  151  535 d  139  534 r  145  534 d
  151  535 d  168  535 r  164  534 d  162  531 d  161  525 d  161  521 d
  162  515 d  164  511 d  168  510 d  170  510 d  174  511 d  176  515 d
  177  521 d  177  525 d  176  531 d  174  534 d  170  535 d  168  535 d
  165  534 d  164  533 d  163  531 d  162  525 d  162  521 d  163  515 d
  164  513 d  165  511 d  168  510 d  170  510 r  173  511 d  174  513 d
  175  515 d  176  521 d  176  525 d  175  531 d  174  533 d  173  534 d
  170  535 d  114  639 r  115  637 d  114  636 d  113  637 d  113  639 d
  114  641 d  115  642 d  119  643 d  123  643 d  127  642 d  128  641 d
  129  639 d  129  636 d  128  634 d  125  631 d  119  629 d  116  628 d
  114  625 d  113  622 d  113  618 d  123  643 r  126  642 d  127  641 d
  128  639 d  128  636 d  127  634 d  123  631 d  119  629 d  113  621 r
  114  622 d  116  622 d  122  619 d  126  619 d  128  621 d  129  622 d
  116  622 r  122  618 d  127  618 d  128  619 d  129  622 d  129  624 d
  144  643 r  140  642 d  138  639 d  137  633 d  137  629 d  138  623 d
  140  619 d  144  618 d  146  618 d  150  619 d  152  623 d  153  629 d
  153  633 d  152  639 d  150  642 d  146  643 d  144  643 d  141  642 d
  140  641 d  139  639 d  138  633 d  138  629 d  139  623 d  140  621 d
  141  619 d  144  618 d  146  618 r  149  619 d  150  621 d  151  623 d
  152  629 d  152  633 d  151  639 d  150  641 d  149  642 d  146  643 d
  168  643 r  164  642 d  162  639 d  161  633 d  161  629 d  162  623 d
  164  619 d  168  618 d  170  618 d  174  619 d  176  623 d  177  629 d
  177  633 d  176  639 d  174  642 d  170  643 d  168  643 d  165  642 d
  164  641 d  163  639 d  162  633 d  162  629 d  163  623 d  164  621 d
  165  619 d  168  618 d  170  618 r  173  619 d  174  621 d  175  623 d
  176  629 d  176  633 d  175  639 d  174  641 d  173  642 d  170  643 d
  114  746 r  115  744 d  114  743 d  113  744 d  113  746 d  114  748 d
  115  749 d  119  750 d  123  750 d  127  749 d  128  748 d  129  746 d
  129  743 d  128  741 d  125  738 d  119  736 d  116  735 d  114  732 d
  113  729 d  113  725 d  123  750 r  126  749 d  127  748 d  128  746 d
  128  743 d  127  741 d  123  738 d  119  736 d  113  728 r  114  729 d
  116  729 d  122  726 d  126  726 d  128  728 d  129  729 d  116  729 r
  122  725 d  127  725 d  128  726 d  129  729 d  129  731 d  139  750 r
  137  738 d  139  740 d  143  741 d  146  741 d  150  740 d  152  737 d
  153  734 d  153  732 d  152  729 d  150  726 d  146  725 d  143  725 d
  139  726 d  138  728 d  137  730 d  137  731 d  138  732 d  139  731 d
  138  730 d  146  741 r  149  740 d  151  737 d  152  734 d  152  732 d
  151  729 d  149  726 d  146  725 d  139  750 r  151  750 d  139  749 r
  145  749 d  151  750 d  168  750 r  164  749 d  162  746 d  161  740 d
  161  736 d  162  730 d  164  726 d  168  725 d  170  725 d  174  726 d
  176  730 d  177  736 d  177  740 d  176  746 d  174  749 d  170  750 d
  168  750 d  165  749 d  164  748 d  163  746 d  162  740 d  162  736 d
  163  730 d  164  728 d  165  726 d  168  725 d  170  725 r  173  726 d
  174  728 d  175  730 d  176  736 d  176  740 d  175  746 d  174  748 d
  173  749 d  170  750 d  114  854 r  115  853 d  114  851 d  113  853 d
  113  854 d  115  856 d  119  857 d  123  857 d  127  856 d  128  854 d
  128  850 d  127  848 d  123  847 d  120  847 d  123  857 r  126  856 d
  127  854 d  127  850 d  126  848 d  123  847 d  126  845 d  128  843 d
  129  841 d  129  837 d  128  835 d  127  833 d  123  832 d  119  832 d
  115  833 d  114  835 d  113  837 d  113  838 d  114  839 d  115  838 d
  114  837 d  127  844 r  128  841 d  128  837 d  127  835 d  126  833 d
  123  832 d  144  857 r  140  856 d  138  853 d  137  847 d  137  843 d
  138  837 d  140  833 d  144  832 d  146  832 d  150  833 d  152  837 d
  153  843 d  153  847 d  152  853 d  150  856 d  146  857 d  144  857 d
  141  856 d  140  855 d  139  853 d  138  847 d  138  843 d  139  837 d
  140  835 d  141  833 d  144  832 d  146  832 r  149  833 d  150  835 d
  151  837 d  152  843 d  152  847 d  151  853 d  150  855 d  149  856 d
  146  857 d  168  857 r  164  856 d  162  853 d  161  847 d  161  843 d
  162  837 d  164  833 d  168  832 d  170  832 d  174  833 d  176  837 d
  177  843 d  177  847 d  176  853 d  174  856 d  170  857 d  168  857 d
  165  856 d  164  855 d  163  853 d  162  847 d  162  843 d  163  837 d
  164  835 d  165  833 d  168  832 d  170  832 r  173  833 d  174  835 d
  175  837 d  176  843 d  176  847 d  175  853 d  174  855 d  173  856 d
  170  857 d  114  961 r  115  960 d  114  958 d  113  960 d  113  961 d
  115  963 d  119  964 d  123  964 d  127  963 d  128  961 d  128  957 d
  127  955 d  123  954 d  120  954 d  123  964 r  126  963 d  127  961 d
  127  957 d  126  955 d  123  954 d  126  952 d  128  950 d  129  948 d
  129  944 d  128  942 d  127  940 d  123  939 d  119  939 d  115  940 d
  114  942 d  113  944 d  113  945 d  114  946 d  115  945 d  114  944 d
  127  951 r  128  948 d  128  944 d  127  942 d  126  940 d  123  939 d
  139  964 r  137  952 d  139  954 d  143  955 d  146  955 d  150  954 d
  152  951 d  153  948 d  153  946 d  152  943 d  150  940 d  146  939 d
  143  939 d  139  940 d  138  942 d  137  944 d  137  945 d  138  946 d
  139  945 d  138  944 d  146  955 r  149  954 d  151  951 d  152  948 d
  152  946 d  151  943 d  149  940 d  146  939 d  139  964 r  151  964 d
  139  963 r  145  963 d  151  964 d  168  964 r  164  963 d  162  960 d
  161  954 d  161  950 d  162  944 d  164  940 d  168  939 d  170  939 d
  174  940 d  176  944 d  177  950 d  177  954 d  176  960 d  174  963 d
  170  964 d  168  964 d  165  963 d  164  962 d  163  960 d  162  954 d
  162  950 d  163  944 d  164  942 d  165  940 d  168  939 d  170  939 r
  173  940 d  174  942 d  175  944 d  176  950 d  176  954 d  175  960 d
  174  962 d  173  963 d  170  964 d  123 1069 r  123 1046 d  125 1071 r
  125 1046 d  125 1071 r  111 1053 d  131 1053 d  120 1046 r  128 1046 d
  144 1071 r  140 1070 d  138 1067 d  137 1061 d  137 1057 d  138 1051 d
  140 1047 d  144 1046 d  146 1046 d  150 1047 d  152 1051 d  153 1057 d
  153 1061 d  152 1067 d  150 1070 d  146 1071 d  144 1071 d  141 1070 d
  140 1069 d  139 1067 d  138 1061 d  138 1057 d  139 1051 d  140 1049 d
  141 1047 d  144 1046 d  146 1046 r  149 1047 d  150 1049 d  151 1051 d
  152 1057 d  152 1061 d  151 1067 d  150 1069 d  149 1070 d  146 1071 d
  168 1071 r  164 1070 d  162 1067 d  161 1061 d  161 1057 d  162 1051 d
  164 1047 d  168 1046 d  170 1046 d  174 1047 d  176 1051 d  177 1057 d
  177 1061 d  176 1067 d  174 1070 d  170 1071 d  168 1071 d  165 1070 d
  164 1069 d  163 1067 d  162 1061 d  162 1057 d  163 1051 d  164 1049 d
  165 1047 d  168 1046 d  170 1046 r  173 1047 d  174 1049 d  175 1051 d
  176 1057 d  176 1061 d  175 1067 d  174 1069 d  173 1070 d  170 1071 d
 1600  200 r 1600 1100 d 1600  200 r 1562  200 d 1600  221 r 1581  221 d
 1600  243 r 1581  243 d 1600  264 r 1581  264 d 1600  286 r 1581  286 d
 1600  307 r 1562  307 d 1600  329 r 1581  329 d 1600  350 r 1581  350 d
 1600  371 r 1581  371 d 1600  393 r 1581  393 d 1600  414 r 1562  414 d
 1600  436 r 1581  436 d 1600  457 r 1581  457 d 1600  479 r 1581  479 d
 1600  500 r 1581  500 d 1600  521 r 1562  521 d 1600  543 r 1581  543 d
 1600  564 r 1581  564 d 1600  586 r 1581  586 d 1600  607 r 1581  607 d
 1600  629 r 1562  629 d 1600  650 r 1581  650 d 1600  671 r 1581  671 d
 1600  693 r 1581  693 d 1600  714 r 1581  714 d 1600  736 r 1562  736 d
 1600  757 r 1581  757 d 1600  779 r 1581  779 d 1600  800 r 1581  800 d
 1600  821 r 1581  821 d 1600  843 r 1562  843 d 1600  864 r 1581  864 d
 1600  886 r 1581  886 d 1600  907 r 1581  907 d 1600  929 r 1581  929 d
 1600  950 r 1562  950 d 1600  971 r 1581  971 d 1600  993 r 1581  993 d
 1600 1014 r 1581 1014 d 1600 1036 r 1581 1036 d 1600 1057 r 1562 1057 d
 1600 1079 r 1581 1079 d 1600 1100 r 1581 1100 d 1600  200 r 1575  200 d
 1575  200 p 1549  200 d 1549  200 p 1524  200 d 1524  200 p 1498  200 d
 1498  200 p 1473  200 d 1473  200 p 1447  200 d 1447  200 p 1422  200 d
 1422  200 p 1396  200 d 1396  200 p 1371  200 d 1371  200 p 1345  200 d
 1345  200 p 1320  200 d 1320  200 p 1295  200 d 1295  200 p 1269  200 d
 1269  200 p 1244  200 d 1244  200 p 1218  200 d 1218  200 p 1193  200 d
 1193  200 p 1167  200 d 1167  200 p 1142  200 d 1142  200 p 1116  200 d
 1116  200 p 1091  200 d 1091  200 p 1065  200 d 1065  200 p 1040  200 d
 1040  200 p 1015  200 d 1015  200 p  989  200 d  989  200 p  964  200 d
  964  202 d  938  202 d  938  200 d  913  200 d  913  215 d  887  215 d
  887  277 d  862  277 d  862  481 d  836  481 d  836  699 d  811  699 d
  811  809 d  785  809 d  785  646 d  760  646 d  760  380 d  735  380 d
  735  234 d  709  234 d  709  200 d  684  200 d  684  200 p  658  200 d
  658  200 p  633  200 d  633  200 p  607  200 d  607  200 p  582  200 d
  582  200 p  556  200 d  556  200 p  531  200 d  531  200 p  505  200 d
  505  200 p  480  200 d  480  200 p  455  200 d  455  200 p  429  200 d
  429  200 p  404  200 d  404  200 p  378  200 d  378  200 p  353  200 d
  353  200 p  327  200 d  327  200 p  302  200 d  302  200 p  276  200 d
  276  200 p  251  200 d  251  200 p  225  200 d  225  200 p  200  200 d
  495  302 r  485  302 d  495  303 r  485  303 d  495  304 r  485  304 d
  495  305 r  485  305 d  495  306 r  485  306 d  495  307 r  485  307 d
  495  308 r  485  308 d  495  309 r  485  309 d  495  310 r  485  310 d
  495  311 r  485  311 d  495  312 r  485  312 d  490  307 r  519  307 d
  519  314 r  519  300 d  519  314 d  490  307 r  461  307 d  461  314 r
  461  300 d  461  314 d  942 1004 r  944 1001 d  944 1007 d  942 1004 d
  940 1006 d  936 1007 d  934 1007 d  930 1006 d  928 1004 d  927 1001 d
  926  998 d  926  992 d  927  988 d  928  986 d  930  983 d  934  982 d
  936  982 d  940  983 d  942  986 d  944  988 d  934 1007 r  932 1006 d
  929 1004 d  928 1001 d  927  998 d  927  992 d  928  988 d  929  986 d
  932  983 d  934  982 d  953 1007 r  953  982 d  954 1007 r  954  982 d
  954  995 r  957  998 d  960  999 d  963  999 d  966  998 d  968  995 d
  968  982 d  963  999 r  965  998 d  966  995 d  966  982 d  950 1007 r
  954 1007 d  950  982 r  958  982 d  963  982 r  971  982 d  980 1007 r
  980  982 d  981 1007 r  981  982 d  976 1007 r  981 1007 d  976  982 r
  984  982 d  998  999 r  994  998 d  992  995 d  990  992 d  990  989 d
  992  986 d  994  983 d  998  982 d 1000  982 d 1004  983 d 1006  986 d
 1007  989 d 1007  992 d 1006  995 d 1004  998 d 1000  999 d  998  999 d
  995  998 d  993  995 d  992  992 d  992  989 d  993  986 d  995  983 d
  998  982 d 1000  982 r 1002  983 d 1005  986 d 1006  989 d 1006  992 d
 1005  995 d 1002  998 d 1000  999 d 1017  999 r 1017  982 d 1018  999 r
 1018  982 d 1018  992 r 1019  995 d 1022  998 d 1024  999 d 1028  999 d
 1029  998 d 1029  997 d 1028  995 d 1026  997 d 1028  998 d 1013  999 r
 1018  999 d 1013  982 r 1022  982 d 1037 1007 r 1036 1006 d 1037 1005 d
 1038 1006 d 1037 1007 d 1037  999 r 1037  982 d 1038  999 r 1038  982 d
 1034  999 r 1038  999 d 1034  982 r 1042  982 d 1050  999 r 1050  982 d
 1052  999 r 1052  982 d 1052  995 r 1054  998 d 1058  999 d 1060  999 d
 1064  998 d 1065  995 d 1065  982 d 1060  999 r 1062  998 d 1064  995 d
 1064  982 d 1047  999 r 1052  999 d 1047  982 r 1055  982 d 1060  982 r
 1068  982 d 1076  992 r 1090  992 d 1090  994 d 1089  997 d 1088  998 d
 1085  999 d 1082  999 d 1078  998 d 1076  995 d 1074  992 d 1074  989 d
 1076  986 d 1078  983 d 1082  982 d 1084  982 d 1088  983 d 1090  986 d
 1089  992 r 1089  995 d 1088  998 d 1082  999 r 1079  998 d 1077  995 d
 1076  992 d 1076  989 d 1077  986 d 1079  983 d 1082  982 d  584  921 r
  575  896 d  584  921 r  592  896 d  584  918 r  591  896 d  578  902 r
  590  902 d  572  896 r  579  896 d  587  896 r  596  896 d  611  912 r
  612  913 d  612  909 d  611  912 d  608  913 d  604  913 d  600  912 d
  599  911 d  599  909 d  600  907 d  603  906 d  609  903 d  611  902 d
  612  901 d  599  909 r  600  908 d  603  907 d  609  905 d  611  903 d
  612  901 d  612  899 d  611  897 d  608  896 d  604  896 d  600  897 d
  599  900 d  599  896 d  600  897 d  632  912 r  633  913 d  633  909 d
  632  912 d  628  913 d  624  913 d  621  912 d  620  911 d  620  909 d
  621  907 d  623  906 d  629  903 d  632  902 d  633  901 d  620  909 r
  621  908 d  623  907 d  629  905 d  632  903 d  633  901 d  633  899 d
  632  897 d  628  896 d  624  896 d  621  897 d  620  900 d  620  896 d
  621  897 d  642  913 r  642  900 d  644  897 d  647  896 d  650  896 d
  653  897 d  656  900 d  644  913 r  644  900 d  645  897 d  647  896 d
  656  913 r  656  896 d  657  913 r  657  896 d  639  913 r  644  913 d
  652  913 r  657  913 d  656  896 r  660  896 d  669  913 r  669  896 d
  670  913 r  670  896 d  670  909 r  672  912 d  676  913 d  678  913 d
  682  912 d  683  909 d  683  896 d  678  913 r  681  912 d  682  909 d
  682  896 d  683  909 r  686  912 d  689  913 d  692  913 d  695  912 d
  696  909 d  696  896 d  692  913 r  694  912 d  695  909 d  695  896 d
  665  913 r  670  913 d  665  896 r  674  896 d  678  896 r  687  896 d
  692  896 r  700  896 d  707  906 r  722  906 d  722  908 d  720  911 d
  719  912 d  717  913 d  713  913 d  710  912 d  707  909 d  706  906 d
  706  903 d  707  900 d  710  897 d  713  896 d  716  896 d  719  897 d
  722  900 d  720  906 r  720  909 d  719  912 d  713  913 r  711  912 d
  708  909 d  707  906 d  707  903 d  708  900 d  711  897 d  713  896 d
  730  913 r  729  912 d  730  911 d  731  912 d  730  913 d  730  899 r
  729  897 d  730  896 d  731  897 d  730  899 d  770  926 r  767  924 d
  765  920 d  762  915 d  761  909 d  761  905 d  762  899 d  765  894 d
  767  890 d  770  888 d  767  924 r  765  919 d  764  915 d  762  909 d
  762  905 d  764  899 d  765  895 d  767  890 d  784  921 r  780  920 d
  778  917 d  777  911 d  777  907 d  778  901 d  780  897 d  784  896 d
  786  896 d  790  897 d  792  901 d  794  907 d  794  911 d  792  917 d
  790  920 d  786  921 d  784  921 d  782  920 d  780  919 d  779  917 d
  778  911 d  778  907 d  779  901 d  780  899 d  782  897 d  784  896 d
  786  896 r  789  897 d  790  899 d  791  901 d  792  907 d  792  911 d
  791  917 d  790  919 d  789  920 d  786  921 d  802  899 r  801  897 d
  802  896 d  803  897 d  802  899 d  821  919 r  821  896 d  822  921 r
  822  896 d  822  921 r  809  903 d  828  903 d  818  896 r  826  896 d
  840  921 r  837  920 d  836  918 d  836  914 d  837  912 d  840  911 d
  845  911 d  849  912 d  850  914 d  850  918 d  849  920 d  845  921 d
  840  921 d  838  920 d  837  918 d  837  914 d  838  912 d  840  911 d
  845  911 r  848  912 d  849  914 d  849  918 d  848  920 d  845  921 d
  840  911 r  837  909 d  836  908 d  834  906 d  834  901 d  836  899 d
  837  897 d  840  896 d  845  896 d  849  897 d  850  899 d  851  901 d
  851  906 d  850  908 d  849  909 d  845  911 d  840  911 r  838  909 d
  837  908 d  836  906 d  836  901 d  837  899 d  838  897 d  840  896 d
  845  896 r  848  897 d  849  899 d  850  901 d  850  906 d  849  908 d
  848  909 d  845  911 d  890  917 r  890  896 d  880  907 r  899  907 d
  880  896 r  899  896 d  935  921 r  932  920 d  929  917 d  928  911 d
  928  907 d  929  901 d  932  897 d  935  896 d  938  896 d  941  897 d
  944  901 d  945  907 d  945  911 d  944  917 d  941  920 d  938  921 d
  935  921 d  933  920 d  932  919 d  930  917 d  929  911 d  929  907 d
  930  901 d  932  899 d  933  897 d  935  896 d  938  896 r  940  897 d
  941  899 d  942  901 d  944  907 d  944  911 d  942  917 d  941  919 d
  940  920 d  938  921 d  953  899 r  952  897 d  953  896 d  954  897 d
  953  899 d  969  921 r  965  920 d  963  917 d  962  911 d  962  907 d
  963  901 d  965  897 d  969  896 d  971  896 d  975  897 d  977  901 d
  978  907 d  978  911 d  977  917 d  975  920 d  971  921 d  969  921 d
  966  920 d  965  919 d  964  917 d  963  911 d  963  907 d  964  901 d
  965  899 d  966  897 d  969  896 d  971  896 r  974  897 d  975  899 d
  976  901 d  977  907 d  977  911 d  976  917 d  975  919 d  974  920 d
  971  921 d  992  921 r  988  920 d  987  918 d  987  914 d  988  912 d
  992  911 d  996  911 d 1000  912 d 1001  914 d 1001  918 d 1000  920 d
  996  921 d  992  921 d  989  920 d  988  918 d  988  914 d  989  912 d
  992  911 d  996  911 r  999  912 d 1000  914 d 1000  918 d  999  920 d
  996  921 d  992  911 r  988  909 d  987  908 d  986  906 d  986  901 d
  987  899 d  988  897 d  992  896 d  996  896 d 1000  897 d 1001  899 d
 1002  901 d 1002  906 d 1001  908 d 1000  909 d  996  911 d  992  911 r
  989  909 d  988  908 d  987  906 d  987  901 d  988  899 d  989  897 d
  992  896 d  996  896 r  999  897 d 1000  899 d 1001  901 d 1001  906 d
 1000  908 d  999  909 d  996  911 d 1010  926 r 1012  924 d 1014  920 d
 1017  915 d 1018  909 d 1018  905 d 1017  899 d 1014  894 d 1012  890 d
 1010  888 d 1012  924 r 1014  919 d 1016  915 d 1017  909 d 1017  905 d
 1016  899 d 1014  895 d 1012  890 d 1055  913 r 1052  912 d 1049  909 d
 1048  906 d 1048  903 d 1049  900 d 1052  897 d 1055  896 d 1058  896 d
 1061  897 d 1064  900 d 1065  903 d 1065  906 d 1064  909 d 1061  912 d
 1058  913 d 1055  913 d 1053  912 d 1050  909 d 1049  906 d 1049  903 d
 1050  900 d 1053  897 d 1055  896 d 1058  896 r 1060  897 d 1062  900 d
 1064  903 d 1064  906 d 1062  909 d 1060  912 d 1058  913 d 1080  920 r
 1079  919 d 1080  918 d 1082  919 d 1082  920 d 1080  921 d 1078  921 d
 1076  920 d 1074  918 d 1074  896 d 1078  921 r 1077  920 d 1076  918 d
 1076  896 d 1071  913 r 1079  913 d 1071  896 r 1079  896 d 1124  919 r
 1125  921 d 1125  917 d 1124  919 d 1122  920 d 1119  921 d 1114  921 d
 1110  920 d 1108  918 d 1108  915 d 1109  913 d 1110  912 d 1113  911 d
 1120  908 d 1122  907 d 1125  905 d 1108  915 r 1110  913 d 1113  912 d
 1120  909 d 1122  908 d 1124  907 d 1125  905 d 1125  900 d 1122  897 d
 1119  896 d 1114  896 d 1110  897 d 1109  899 d 1108  901 d 1108  896 d
 1109  899 d 1134  921 r 1134  901 d 1136  897 d 1138  896 d 1140  896 d
 1143  897 d 1144  900 d 1136  921 r 1136  901 d 1137  897 d 1138  896 d
 1131  913 r 1140  913 d 1152  911 r 1152  909 d 1151  909 d 1151  911 d
 1152  912 d 1155  913 d 1160  913 d 1162  912 d 1163  911 d 1164  908 d
 1164  900 d 1166  897 d 1167  896 d 1163  911 r 1163  900 d 1164  897 d
 1167  896 d 1168  896 d 1163  908 r 1162  907 d 1155  906 d 1151  905 d
 1150  902 d 1150  900 d 1151  897 d 1155  896 d 1158  896 d 1161  897 d
 1163  900 d 1155  906 r 1152  905 d 1151  902 d 1151  900 d 1152  897 d
 1155  896 d 1176  913 r 1176  896 d 1178  913 r 1178  896 d 1178  909 r
 1180  912 d 1184  913 d 1186  913 d 1190  912 d 1191  909 d 1191  896 d
 1186  913 r 1188  912 d 1190  909 d 1190  896 d 1173  913 r 1178  913 d
 1173  896 r 1181  896 d 1186  896 r 1194  896 d 1215  921 r 1215  896 d
 1216  921 r 1216  896 d 1215  909 r 1212  912 d 1210  913 d 1208  913 d
 1204  912 d 1202  909 d 1200  906 d 1200  903 d 1202  900 d 1204  897 d
 1208  896 d 1210  896 d 1212  897 d 1215  900 d 1208  913 r 1205  912 d
 1203  909 d 1202  906 d 1202  903 d 1203  900 d 1205  897 d 1208  896 d
 1211  921 r 1216  921 d 1215  896 r 1220  896 d 1228  911 r 1228  909 d
 1227  909 d 1227  911 d 1228  912 d 1230  913 d 1235  913 d 1238  912 d
 1239  911 d 1240  908 d 1240  900 d 1241  897 d 1242  896 d 1239  911 r
 1239  900 d 1240  897 d 1242  896 d 1244  896 d 1239  908 r 1238  907 d
 1230  906 d 1227  905 d 1226  902 d 1226  900 d 1227  897 d 1230  896 d
 1234  896 d 1236  897 d 1239  900 d 1230  906 r 1228  905 d 1227  902 d
 1227  900 d 1228  897 d 1230  896 d 1252  913 r 1252  896 d 1253  913 r
 1253  896 d 1253  906 r 1254  909 d 1257  912 d 1259  913 d 1263  913 d
 1264  912 d 1264  911 d 1263  909 d 1262  911 d 1263  912 d 1248  913 r
 1253  913 d 1248  896 r 1257  896 d 1284  921 r 1284  896 d 1286  921 r
 1286  896 d 1284  909 r 1282  912 d 1280  913 d 1277  913 d 1274  912 d
 1271  909 d 1270  906 d 1270  903 d 1271  900 d 1274  897 d 1277  896 d
 1280  896 d 1282  897 d 1284  900 d 1277  913 r 1275  912 d 1272  909 d
 1271  906 d 1271  903 d 1272  900 d 1275  897 d 1277  896 d 1281  921 r
 1286  921 d 1284  896 r 1289  896 d 1319  935 r 1317  934 d 1316  933 d
 1316  931 d 1317  929 d 1319  928 d 1322  928 d 1324  929 d 1325  931 d
 1325  933 d 1324  934 d 1322  935 d 1319  935 d 1318  934 d 1317  933 d
 1317  931 d 1318  929 d 1319  928 d 1322  928 r 1323  929 d 1324  931 d
 1324  933 d 1323  934 d 1322  935 d 1319  928 r 1317  928 d 1316  927 d
 1315  925 d 1315  923 d 1316  921 d 1317  920 d 1319  920 d 1322  920 d
 1324  920 d 1325  921 d 1325  923 d 1325  925 d 1325  927 d 1324  928 d
 1322  928 d 1319  928 r 1318  928 d 1317  927 d 1316  925 d 1316  923 d
 1317  921 d 1318  920 d 1319  920 d 1322  920 r 1323  920 d 1324  921 d
 1325  923 d 1325  925 d 1324  927 d 1323  928 d 1322  928 d 1334  921 r
 1334  896 d 1335  921 r 1335  896 d 1330  921 r 1344  921 d 1348  920 d
 1349  919 d 1350  917 d 1350  914 d 1349  912 d 1348  911 d 1344  909 d
 1344  921 r 1347  920 d 1348  919 d 1349  917 d 1349  914 d 1348  912 d
 1347  911 d 1344  909 d 1335  909 r 1344  909 d 1348  908 d 1349  907 d
 1350  905 d 1350  901 d 1349  899 d 1348  897 d 1344  896 d 1330  896 d
 1344  909 r 1347  908 d 1348  907 d 1349  905 d 1349  901 d 1348  899 d
 1347  897 d 1344  896 d 1382  921 r 1382  896 d 1383  921 r 1383  896 d
 1390  914 r 1390  905 d 1378  921 r 1397  921 d 1397  915 d 1396  921 d
 1383  909 r 1390  909 d 1378  896 r 1386  896 d 1406  921 r 1406  896 d
 1407  921 r 1407  896 d 1402  921 r 1407  921 d 1402  896 r 1410  896 d
 1419  913 r 1419  900 d 1420  897 d 1424  896 d 1426  896 d 1430  897 d
 1432  900 d 1420  913 r 1420  900 d 1421  897 d 1424  896 d 1432  913 r
 1432  896 d 1433  913 r 1433  896 d 1415  913 r 1420  913 d 1428  913 r
 1433  913 d 1432  896 r 1437  896 d 1445  913 r 1458  896 d 1446  913 r
 1460  896 d 1460  913 r 1445  896 d 1442  913 r 1450  913 d 1456  913 r
 1463  913 d 1442  896 r 1449  896 d 1455  896 r 1463  896 d  490  639 r
  490  318 d  487  322 r  490  318 d  492  322 d  484  325 r  490  319 d
  496  325 d  490  340 r  490  319 d  479  677 r  482  679 d  476  679 d
  479  677 d  477  675 d  476  671 d  476  669 d  477  665 d  479  663 d
  482  662 d  485  661 d  491  661 d  495  662 d  497  663 d  500  665 d
  501  669 d  501  671 d  500  675 d  497  677 d  495  679 d  476  669 r
  477  667 d  479  664 d  482  663 d  485  662 d  491  662 d  495  663 d
  497  664 d  500  667 d  501  669 d  476  688 r  501  688 d  476  689 r
  501  689 d  476  685 r  476  689 d  501  685 r  501  693 d  476  723 r
  501  723 d  476  724 r  501  724 d  483  731 r  492  731 d  476  719 r
  476  739 d  482  739 d  476  737 d  488  724 r  488  731 d  501  719 r
  501  739 d  495  739 d  501  737 d  484  748 r  501  761 d  484  749 r
  501  763 d  484  763 r  501  748 d  484  745 r  484  753 d  484  759 r
  484  766 d  501  745 r  501  752 d  501  758 r  501  766 d  484  775 r
  509  775 d  484  776 r  509  776 d  488  776 r  485  778 d  484  781 d
  484  783 d  485  787 d  488  789 d  491  790 d  494  790 d  497  789 d
  500  787 d  501  783 d  501  781 d  500  778 d  497  776 d  484  783 r
  485  785 d  488  788 d  491  789 d  494  789 d  497  788 d  500  785 d
  501  783 d  484  771 r  484  776 d  509  771 r  509  779 d   57  463 r
   85  463 d   57  464 r   82  480 d   60  464 r   85  480 d   57  480 r
   85  480 d   57  459 r   57  464 d   57  476 r   57  484 d   85  459 r
   85  467 d   67  493 r   81  493 d   83  494 d   85  498 d   85  501 d
   83  505 d   81  507 d   67  494 r   81  494 d   83  495 d   85  498 d
   67  507 r   85  507 d   67  508 r   85  508 d   67  489 r   67  494 d
   67  503 r   67  508 d   85  507 r   85  512 d   67  521 r   85  521 d
   67  523 r   85  523 d   70  523 r   68  525 d   67  529 d   67  532 d
   68  536 d   70  537 d   85  537 d   67  532 r   68  534 d   70  536 d
   85  536 d   70  537 r   68  540 d   67  543 d   67  546 d   68  550 d
   70  551 d   85  551 d   67  546 r   68  549 d   70  550 d   85  550 d
   67  517 r   67  523 d   85  517 r   85  527 d   85  532 r   85  541 d
   85  546 r   85  555 d   57  564 r   85  564 d   57  566 r   85  566 d
   70  566 r   68  568 d   67  571 d   67  573 d   68  577 d   70  580 d
   74  581 d   77  581 d   81  580 d   83  577 d   85  573 d   85  571 d
   83  568 d   81  566 d   67  573 r   68  576 d   70  579 d   74  580 d
   77  580 d   81  579 d   83  576 d   85  573 d   57  560 r   57  566 d
   74  590 r   74  606 d   72  606 d   69  605 d   68  603 d   67  601 d
   67  597 d   68  593 d   70  590 d   74  589 d   77  589 d   81  590 d
   83  593 d   85  597 d   85  599 d   83  603 d   81  606 d   74  605 r
   70  605 d   68  603 d   67  597 r   68  594 d   70  592 d   74  590 d
   77  590 d   81  592 d   83  594 d   85  597 d   67  616 r   85  616 d
   67  618 r   85  618 d   74  618 r   70  619 d   68  622 d   67  624 d
   67  628 d   68  629 d   69  629 d   70  628 d   69  627 d   68  628 d
   67  612 r   67  618 d   85  612 r   85  622 d   67  667 r   68  663 d
   70  661 d   74  659 d   77  659 d   81  661 d   83  663 d   85  667 d
   85  670 d   83  674 d   81  676 d   77  677 d   74  677 d   70  676 d
   68  674 d   67  670 d   67  667 d   68  664 d   70  662 d   74  661 d
   77  661 d   81  662 d   83  664 d   85  667 d   85  670 r   83  672 d
   81  675 d   77  676 d   74  676 d   70  675 d   68  672 d   67  670 d
   59  694 r   60  693 d   61  694 d   60  696 d   59  696 d   57  694 d
   57  692 d   59  689 d   61  688 d   85  688 d   57  692 r   59  690 d
   61  689 d   85  689 d   67  684 r   67  693 d   85  684 r   85  693 d
   57  727 r   85  727 d   57  728 r   81  736 d   57  727 r   85  736 d
   57  745 r   85  736 d   57  745 r   85  745 d   57  746 r   85  746 d
   57  723 r   57  728 d   57  745 r   57  750 d   85  723 r   85  731 d
   85  741 r   85  750 d   67  765 r   68  761 d   70  758 d   74  757 d
   77  757 d   81  758 d   83  761 d   85  765 d   85  767 d   83  771 d
   81  774 d   77  775 d   74  775 d   70  774 d   68  771 d   67  767 d
   67  765 d   68  762 d   70  759 d   74  758 d   77  758 d   81  759 d
   83  762 d   85  765 d   85  767 r   83  770 d   81  772 d   77  774 d
   74  774 d   70  772 d   68  770 d   67  767 d   57  798 r   85  798 d
   57  800 r   85  800 d   70  798 r   68  796 d   67  793 d   67  791 d
   68  787 d   70  784 d   74  783 d   77  783 d   81  784 d   83  787 d
   85  791 d   85  793 d   83  796 d   81  798 d   67  791 r   68  788 d
   70  785 d   74  784 d   77  784 d   81  785 d   83  788 d   85  791 d
   57  794 r   57  800 d   85  798 r   85  804 d   74  811 r   74  827 d
   72  827 d   69  826 d   68  824 d   67  822 d   67  818 d   68  814 d
   70  811 d   74  810 d   77  810 d   81  811 d   83  814 d   85  818 d
   85  820 d   83  824 d   81  827 d   74  826 r   70  826 d   68  824 d
   67  818 r   68  815 d   70  813 d   74  811 d   77  811 d   81  813 d
   83  815 d   85  818 d   57  837 r   85  837 d   57  839 r   85  839 d
   57  833 r   57  839 d   85  833 r   85  843 d   68  862 r   67  863 d
   70  863 d   68  862 d   67  858 d   67  854 d   68  850 d   69  849 d
   70  849 d   73  850 d   74  853 d   77  859 d   78  862 d   80  863 d
   70  849 r   72  850 d   73  853 d   76  859 d   77  862 d   80  863 d
   82  863 d   83  862 d   85  858 d   85  854 d   83  850 d   81  849 d
   85  849 d   83  850 d  741  129 r  742  126 d  742  133 d  741  129 d
  739  131 d  735  133 d  732  133 d  728  131 d  726  129 d  724  126 d
  723  122 d  723  116 d  724  112 d  726  109 d  728  107 d  732  105 d
  735  105 d  739  107 d  741  109 d  742  112 d  732  133 r  729  131 d
  727  129 d  726  126 d  724  122 d  724  116 d  726  112 d  727  109 d
  729  107 d  732  105 d  753  133 r  753  105 d  754  133 r  754  105 d
  749  133 r  754  133 d  749  105 r  758  105 d  805  130 r  806  133 d
  806  127 d  805  130 d  804  131 d  800  133 d  794  133 d  791  131 d
  788  129 d  788  126 d  789  124 d  791  122 d  793  121 d  801  118 d
  804  117 d  806  114 d  788  126 r  791  124 d  793  122 d  801  120 d
  804  118 d  805  117 d  806  114 d  806  109 d  804  107 d  800  105 d
  794  105 d  791  107 d  789  108 d  788  111 d  788  105 d  789  108 d
  817  133 r  817  105 d  818  133 r  833  108 d  818  130 r  833  105 d
  833  133 r  833  105 d  813  133 r  818  133 d  830  133 r  837  133 d
  813  105 r  820  105 d  846  133 r  846  113 d  848  109 d  850  107 d
  854  105 d  857  105 d  861  107 d  863  109 d  865  113 d  865  133 d
  848  133 r  848  113 d  849  109 d  852  107 d  854  105 d  843  133 r
  852  133 d  861  133 r  869  133 d  876  124 r  875  122 d  876  121 d
  878  122 d  876  124 d  876  108 r  875  107 d  876  105 d  878  107 d
  876  108 d  940  133 r  940  105 d  941  133 r  941  105 d  932  133 r
  931  126 d  931  133 d  950  133 d  950  126 d  949  133 d  936  105 r
  945  105 d  960  133 r  960  105 d  961  133 r  961  105 d  961  120 r
  963  122 d  967  124 d  970  124 d  974  122 d  975  120 d  975  105 d
  970  124 r  973  122 d  974  120 d  974  105 d  956  133 r  961  133 d
  956  105 r  965  105 d  970  105 r  979  105 d  987  116 r 1002  116 d
 1002  118 d 1001  121 d 1000  122 d  997  124 d  993  124 d  989  122 d
  987  120 d  986  116 d  986  113 d  987  109 d  989  107 d  993  105 d
  996  105 d 1000  107 d 1002  109 d 1001  116 r 1001  120 d 1000  122 d
  993  124 r  991  122 d  988  120 d  987  116 d  987  113 d  988  109 d
  991  107 d  993  105 d 1018  124 r 1014  122 d 1012  120 d 1010  116 d
 1010  113 d 1012  109 d 1014  107 d 1018  105 d 1021  105 d 1025  107 d
 1027  109 d 1028  113 d 1028  116 d 1027  120 d 1025  122 d 1021  124 d
 1018  124 d 1015  122 d 1013  120 d 1012  116 d 1012  113 d 1013  109 d
 1015  107 d 1018  105 d 1021  105 r 1023  107 d 1026  109 d 1027  113 d
 1027  116 d 1026  120 d 1023  122 d 1021  124 d 1039  124 r 1039  105 d
 1040  124 r 1040  105 d 1040  116 r 1041  120 d 1044  122 d 1047  124 d
 1051  124 d 1052  122 d 1052  121 d 1051  120 d 1049  121 d 1051  122 d
 1035  124 r 1040  124 d 1035  105 r 1044  105 d 1060  124 r 1067  105 d
 1061  124 r 1067  108 d 1075  124 r 1067  105 d 1065  100 d 1062   98 d
 1060   96 d 1058   96 d 1057   98 d 1058   99 d 1060   98 d 1056  124 r
 1065  124 d 1071  124 r 1079  124 d
e
EndPSPlot
0 SPE
4787 15917 XY 0 SPB
 clear Figure end restore 
0 SPE 5509 15776 XY F25(Figure)S 195 x(4.)S 194 x(1000)S 195 x
(arti\014cially)S 196 x(mo)S 14 x(di\014ed)S 194 x(\015uxes.)S -181 y 
F8(26)S 181 y 219 x F25(The)S
XP /F8 56 199 1 0 14 18 19 16 0
<0FC0 3FF0 7078 6018 6018 7818 7E30 3FE0 1FC0 1FF0 79F8 70FC E03C
 E01C E01C F03C 7878 3FF0 1FC0>
PXLC RP
16380 15595 XY F8(8)S 181 y 24 x F25(B)S 194 x(neutrino)S 195 x
(\015uxes)S 194 x(computed)S 194 x(for)S 195 x(the)S 598 y 5509 X
(1000)S 141 x(accurate)S 142 x(solar)S 142 x(mo)S 14 x(dels)S 142 x
(w)S -15 x(ere)S 142 x(replace)S 2 x(d)S 140 x(in)S 141 x(the)S 141 x
(\014gure)S 141 x(sho)S -13 x(wn)S 140 x(b)S -14 x(y)S 141 x(v)S 
-28 x(alues)S 142 x(dra)S -14 x(wn)S 141 x(randomly)S 598 y 5509 X
(for)S 189 x(eac)S -13 x(h)S 190 x(mo)S 14 x(del)S 190 x(from)S 190 x
(a)S 189 x(normal)S 190 x(distribution)S 190 x(with)S 189 x(the)S 
189 x(mean)S 190 x(and)S 189 x(the)S 189 x(standard)S 189 x
(deviation)S 597 y 5509 X(measured)S 167 x(b)S -14 x(y)S 166 x(the)S
165 x(Kamiok)S -26 x(ande)S 166 x(exp)S 14 x(erimen)S -13 x(t.)S 
-180 y F8(2,3)S 20036 Y 5347 X F34(exceeding)S 137 x(1
(They)S 137 x(are)S 136 x(not)S 136 x(observ)S -30 x(able)S 136 x
(with)S 136 x(an)S -16 x(y)S 136 x(of)S 136 x(the)S 137 x(other)S 
136 x(curren)S -15 x(tly-op)S 15 x(erating)S 677 y 5347 X(exp)S 15 x
(erimen)S -15 x(ts)S 240 x(\(or)S 240 x(ev)S -15 x(en)S 241 x(other)S
240 x(funded)S 241 x(exp)S 15 x(erimen)S -15 x(ts)S 240 x(under)S 
241 x(dev)S -15 x(elopmen)S -15 x(t\).)S 240 x(The)S -198 y 240 x 
F13(7)S 198 y 25 x F34(Be)S 678 y 5347 X(neutrinos,)S 244 x(whic)S
-15 x(h)S 245 x(can)S 246 x(b)S 15 x(e)S 246 x(calculated)S 245 x
(with)S 245 x(mo)S 14 x(derately)S 245 x(high)S 245 x(precision)S 
244 x(\(6
(signi)S -2 x(\014can)S -14 x(tly)S 182 x(to)S 183 x(the)S 183 x
(predicted)S 184 x(standard)S 182 x(capture)S 183 x(rate,)S 183 x
(36)S 182 x(SNU,)S 182 x(or)S 183 x(27
(the)S 241 x(total)S 239 x(galli)S -2 x(um)S 240 x(rate.)S 239 x
(The)S 241 x(8)S 368 y -273 x(\026)S -368 y 239 x(neutrinos,)S 240 x
(whic)S -16 x(h)S 240 x(dominate)S
XP /F34 124 546 0 11 43 12 2 48 0
<FFFFFFFFFFF0 FFFFFFFFFFF0>
PXLC RP
19847 22746 XY F34(|according)S 239 x(to)S 240 x(the)S 677 y 5347 X
(standard)S 234 x(mo)S 14 x(del|the)S 236 x(c)S -15 x(hlorine)S 234 x
(and)S 235 x(the)S 235 x(Kamiok)S -32 x(ande)S 235 x(I)S 16 x(I)S 
234 x(exp)S 16 x(erimen)S -16 x(ts,)S 235 x(con)S -15 x(tribute)S 
677 y 5347 X(less)S 182 x(than)S 181 x(10
182 x(standard)S 182 x(theoretical)S 181 x(rate.)S 678 y 6165 X(As)S
281 x(sho)S -15 x(wn)S 280 x(in)S 281 x(Figure)S 280 x(1,)S 280 x
(the)S 281 x(capture)S 281 x(rates)S 281 x(measured)S 281 x(in)S 
280 x(the)S 281 x(GALLEX)S 281 x(and)S 677 y 5347 X(the)S 189 x(SA)S
-15 x(GE)S 189 x(solar)S 188 x(neutrino)S 189 x(exp)S 15 x(erimen)S
-15 x(ts)S 189 x(are)S 189 x(b)S 15 x(oth)S 189 x(signi\014can)S 
-16 x(tly)S 188 x(b)S 15 x(elo)S -15 x(w)S 189 x(the)S 189 x
(standard)S 678 y 5347 X(mo)S 14 x(del)S 215 x(predictions.)S 214 x
(These)S 215 x(results)S 215 x(strengthen)S 215 x(the)S 215 x
(conclusion)S 215 x(that)S 214 x(new)S 215 x(ph)S -15 x(ysics)S 215 x
(is)S 677 y 5347 X(required)S 228 x(to)S 228 x(explain)S 228 x(the)S
229 x(solar)S 227 x(neutrino)S 228 x(problem.)S 227 x(Since)S 228 x
(the)S 230 x(gal)S -2 x(lium)S 228 x(exp)S 15 x(erimen)S -16 x(ts)S
678 y 5347 X(are)S 195 x(most)S 194 x(sensitiv)S -15 x(e)S 196 x(to)S
195 x(lo)S -16 x(w)S 195 x(energy)S 195 x(neutrinos)S 195 x(and)S 
195 x(the)S 196 x(c)S -15 x(hlori)S -2 x(ne)S 196 x(and)S 195 x
(Kamiok)S -32 x(ande)S 196 x(I)S 15 x(I)S 677 y 5347 X(exp)S 15 x
(erimen)S -15 x(ts)S 202 x(are)S 203 x(most)S 202 x(sensitiv)S -16 x
(e)S 203 x(to)S 203 x(higher-energy)S 202 x(neutrinos,)S 201 x(the)S
203 x(results)S 203 x(from)S 201 x(the)S 678 y 5347 X(SA)S -15 x(GE)S
227 x(and)S 227 x(GALLEX)S 227 x(exp)S 15 x(erimen)S -15 x(ts)S 227 x
(cannot)S 227 x(b)S 15 x(e)S 227 x(compared)S 227 x(directly)S 226 x
(with)S 227 x(the)S 227 x(c)S -14 x(hlo-)S 677 y 5347 X(rine)S 157 x
(or)S 156 x(the)S 158 x(Kamiok)S -31 x(ande)S 157 x(I)S 15 x(I)S 
158 x(exp)S 15 x(erimen)S -15 x(ts)S 157 x(without)S 157 x(in)S -15 x
(tro)S 14 x(ducing)S 157 x(a)S 158 x(sp)S 15 x(eci\014c)S 158 x
(theoretical)S 678 y 5347 X(mo)S 14 x(del.)S 677 y -773 x(Ho)S -15 x
(w)S -16 x(ev)S -15 x(er,)S 216 x(one)S 215 x(can)S 216 x(use)S 216 x
(the)S 216 x(same)S 216 x(set)S 216 x(of)S 215 x(1000)S 214 x(solar)S
215 x(mo)S 14 x(dels)S 216 x(to)S 215 x(see)S 217 x(ho)S -16 x(w)S
216 x(signi\014-)S 678 y 5347 X(can)S -15 x(t)S 177 x(is)S 176 x
(the)S 178 x(curren)S -15 x(t)S 177 x(discrepancy)S 177 x(b)S 16 x
(et)S -15 x(w)S -15 x(een)S 178 x(the)S 177 x(standard)S 176 x(mo)S
15 x(del)S 177 x(predictions)S 176 x(and)S 177 x(the)S 677 y 5347 X
(measured)S 242 x(gall)S -2 x(ium)S 242 x(rate.)S 241 x(Figure)S 
242 x(5)S 242 x(sho)S -15 x(ws)S 242 x(the)S 243 x(gall)S -2 x(ium)S
242 x(rate)S 242 x(for)S 241 x(b)S 15 x(oth)S 243 x(exp)S 15 x
(erimen)S -16 x(ts)S 677 y 5347 X(and)S 196 x(the)S 197 x(histogram)S
194 x(for)S 196 x(the)S 197 x(1000)S 195 x(solar)S 195 x(mo)S 15 x
(dels.)S 196 x(In)S 196 x(this)S 196 x(\014gure,)S 196 x(I)S 197 x
(ha)S -16 x(v)S -15 x(e)S 197 x(agai)S -2 x(n)S 197 x(forced)S 678 y 
5347 X(agreemen)S -16 x(t)S 197 x(b)S 16 x(et)S -15 x(w)S -15 x(een)S
197 x(the)S 198 x(mo)S 14 x(dels)S 197 x(and)S 196 x(the)S 198 x
(neutrino-electron)S 196 x(scattering)S 197 x(exp)S 15 x(erimen)S 
-15 x(t)S
1 PP EP

1000 BP 39600 30600 PM 0 0 XY
4787 15917 XY 0 SPB
 save 10 dict begin /Figure exch def currentpoint translate
/showpage {} def 
0 SPE
4787 15917 XY 0 SPB
save 50 dict begin /psplot exch def
/StartPSPlot
   {newpath 0 0 moveto 0 setlinewidth 0 setgray 0 setlinecap
    1 setlinejoin 72 300 div dup scale}def
/pending {false} def
/finish {pending {currentpoint stroke moveto /pending false def} if} def
/r {finish newpath moveto} def
/d {lineto /pending true def} def
/l {finish 4 2 roll moveto lineto currentpoint stroke moveto} def
/p {finish newpath moveto currentpoint lineto currentpoint stroke moveto} def
/e {finish gsave showpage grestore newpath 0 0 moveto} def
/lw {finish setlinewidth} def
/lt0 {finish [] 0 setdash} def
/lt1 {finish [3 5] 0 setdash} def
/lt2 {finish [20 10] 0 setdash} def
/lt3 {finish [60 10] 0 setdash} def
/lt4 {finish [3 10 20 10] 0 setdash} def
/lt5 {finish [3 10 60 10] 0 setdash} def
/lt6 {finish [20 10 60 10] 0 setdash} def
/EndPSPlot {clear psplot end restore}def
StartPSPlot
   1 lw lt0  200  200 r 1600  200 d  200  200 r  200  238 d  244  200 r
  244  219 d  288  200 r  288  238 d  331  200 r  331  219 d  375  200 r
  375  238 d  419  200 r  419  219 d  463  200 r  463  238 d  506  200 r
  506  219 d  550  200 r  550  238 d  594  200 r  594  219 d  638  200 r
  638  238 d  681  200 r  681  219 d  725  200 r  725  238 d  769  200 r
  769  219 d  813  200 r  813  238 d  856  200 r  856  219 d  900  200 r
  900  238 d  944  200 r  944  219 d  987  200 r  987  238 d 1031  200 r
 1031  219 d 1075  200 r 1075  238 d 1119  200 r 1119  219 d 1162  200 r
 1162  238 d 1206  200 r 1206  219 d 1250  200 r 1250  238 d 1294  200 r
 1294  219 d 1337  200 r 1337  238 d 1381  200 r 1381  219 d 1425  200 r
 1425  238 d 1469  200 r 1469  219 d 1512  200 r 1512  238 d 1556  200 r
 1556  219 d 1600  200 r 1600  238 d  199  176 r  195  175 d  193  172 d
  192  166 d  192  162 d  193  156 d  195  152 d  199  151 d  201  151 d
  205  152 d  207  156 d  208  162 d  208  166 d  207  172 d  205  175 d
  201  176 d  199  176 d  196  175 d  195  174 d  194  172 d  193  166 d
  193  162 d  194  156 d  195  154 d  196  152 d  199  151 d  201  151 r
  204  152 d  205  154 d  206  156 d  207  162 d  207  166 d  206  172 d
  205  174 d  204  175 d  201  176 d  271  172 r  274  173 d  277  176 d
  277  151 d  276  175 r  276  151 d  271  151 r  282  151 d  299  176 r
  295  175 d  293  172 d  292  166 d  292  162 d  293  156 d  295  152 d
  299  151 d  301  151 d  305  152 d  307  156 d  308  162 d  308  166 d
  307  172 d  305  175 d  301  176 d  299  176 d  296  175 d  295  174 d
  294  172 d  293  166 d  293  162 d  294  156 d  295  154 d  296  152 d
  299  151 d  301  151 r  304  152 d  305  154 d  306  156 d  307  162 d
  307  166 d  306  172 d  305  174 d  304  175 d  301  176 d  356  172 r
  357  170 d  356  169 d  355  170 d  355  172 d  356  174 d  357  175 d
  361  176 d  365  176 d  369  175 d  370  174 d  371  172 d  371  169 d
  370  167 d  367  164 d  361  162 d  358  161 d  356  158 d  355  155 d
  355  151 d  365  176 r  368  175 d  369  174 d  370  172 d  370  169 d
  369  167 d  365  164 d  361  162 d  355  154 r  356  155 d  358  155 d
  364  152 d  368  152 d  370  154 d  371  155 d  358  155 r  364  151 d
  369  151 d  370  152 d  371  155 d  371  157 d  386  176 r  382  175 d
  380  172 d  379  166 d  379  162 d  380  156 d  382  152 d  386  151 d
  388  151 d  392  152 d  394  156 d  395  162 d  395  166 d  394  172 d
  392  175 d  388  176 d  386  176 d  383  175 d  382  174 d  381  172 d
  380  166 d  380  162 d  381  156 d  382  154 d  383  152 d  386  151 d
  388  151 r  391  152 d  392  154 d  393  156 d  394  162 d  394  166 d
  393  172 d  392  174 d  391  175 d  388  176 d  444  173 r  445  172 d
  444  170 d  443  172 d  443  173 d  445  175 d  449  176 d  453  176 d
  457  175 d  458  173 d  458  169 d  457  167 d  453  166 d  450  166 d
  453  176 r  456  175 d  457  173 d  457  169 d  456  167 d  453  166 d
  456  164 d  458  162 d  459  160 d  459  156 d  458  154 d  457  152 d
  453  151 d  449  151 d  445  152 d  444  154 d  443  156 d  443  157 d
  444  158 d  445  157 d  444  156 d  457  163 r  458  160 d  458  156 d
  457  154 d  456  152 d  453  151 d  474  176 r  470  175 d  468  172 d
  467  166 d  467  162 d  468  156 d  470  152 d  474  151 d  476  151 d
  480  152 d  482  156 d  483  162 d  483  166 d  482  172 d  480  175 d
  476  176 d  474  176 d  471  175 d  470  174 d  469  172 d  468  166 d
  468  162 d  469  156 d  470  154 d  471  152 d  474  151 d  476  151 r
  479  152 d  480  154 d  481  156 d  482  162 d  482  166 d  481  172 d
  480  174 d  479  175 d  476  176 d  540  174 r  540  151 d  542  176 r
  542  151 d  542  176 r  528  158 d  548  158 d  537  151 r  545  151 d
  561  176 r  557  175 d  555  172 d  554  166 d  554  162 d  555  156 d
  557  152 d  561  151 d  563  151 d  567  152 d  569  156 d  570  162 d
  570  166 d  569  172 d  567  175 d  563  176 d  561  176 d  558  175 d
  557  174 d  556  172 d  555  166 d  555  162 d  556  156 d  557  154 d
  558  152 d  561  151 d  563  151 r  566  152 d  567  154 d  568  156 d
  569  162 d  569  166 d  568  172 d  567  174 d  566  175 d  563  176 d
  620  176 r  618  164 d  620  166 d  624  167 d  627  167 d  631  166 d
  633  163 d  634  160 d  634  158 d  633  155 d  631  152 d  627  151 d
  624  151 d  620  152 d  619  154 d  618  156 d  618  157 d  619  158 d
  620  157 d  619  156 d  627  167 r  630  166 d  632  163 d  633  160 d
  633  158 d  632  155 d  630  152 d  627  151 d  620  176 r  632  176 d
  620  175 r  626  175 d  632  176 d  649  176 r  645  175 d  643  172 d
  642  166 d  642  162 d  643  156 d  645  152 d  649  151 d  651  151 d
  655  152 d  657  156 d  658  162 d  658  166 d  657  172 d  655  175 d
  651  176 d  649  176 d  646  175 d  645  174 d  644  172 d  643  166 d
  643  162 d  644  156 d  645  154 d  646  152 d  649  151 d  651  151 r
  654  152 d  655  154 d  656  156 d  657  162 d  657  166 d  656  172 d
  655  174 d  654  175 d  651  176 d  719  173 r  718  172 d  719  170 d
  720  172 d  720  173 d  719  175 d  717  176 d  713  176 d  709  175 d
  707  173 d  706  170 d  705  166 d  705  158 d  706  155 d  708  152 d
  712  151 d  714  151 d  718  152 d  720  155 d  721  158 d  721  160 d
  720  163 d  718  166 d  714  167 d  713  167 d  709  166 d  707  163 d
  706  160 d  713  176 r  711  175 d  708  173 d  707  170 d  706  166 d
  706  158 d  707  155 d  709  152 d  712  151 d  714  151 r  717  152 d
  719  155 d  720  158 d  720  160 d  719  163 d  717  166 d  714  167 d
  736  176 r  732  175 d  730  172 d  729  166 d  729  162 d  730  156 d
  732  152 d  736  151 d  738  151 d  742  152 d  744  156 d  745  162 d
  745  166 d  744  172 d  742  175 d  738  176 d  736  176 d  733  175 d
  732  174 d  731  172 d  730  166 d  730  162 d  731  156 d  732  154 d
  733  152 d  736  151 d  738  151 r  741  152 d  742  154 d  743  156 d
  744  162 d  744  166 d  743  172 d  742  174 d  741  175 d  738  176 d
  793  176 r  793  169 d  793  172 r  794  174 d  796  176 d  799  176 d
  805  173 d  807  173 d  808  174 d  809  176 d  794  174 r  796  175 d
  799  175 d  805  173 d  809  176 r  809  173 d  808  169 d  803  163 d
  802  161 d  801  157 d  801  151 d  808  169 r  802  163 d  801  161 d
  800  157 d  800  151 d  824  176 r  820  175 d  818  172 d  817  166 d
  817  162 d  818  156 d  820  152 d  824  151 d  826  151 d  830  152 d
  832  156 d  833  162 d  833  166 d  832  172 d  830  175 d  826  176 d
  824  176 d  821  175 d  820  174 d  819  172 d  818  166 d  818  162 d
  819  156 d  820  154 d  821  152 d  824  151 d  826  151 r  829  152 d
  830  154 d  831  156 d  832  162 d  832  166 d  831  172 d  830  174 d
  829  175 d  826  176 d  886  176 r  882  175 d  881  173 d  881  169 d
  882  167 d  886  166 d  890  166 d  894  167 d  895  169 d  895  173 d
  894  175 d  890  176 d  886  176 d  883  175 d  882  173 d  882  169 d
  883  167 d  886  166 d  890  166 r  893  167 d  894  169 d  894  173 d
  893  175 d  890  176 d  886  166 r  882  164 d  881  163 d  880  161 d
  880  156 d  881  154 d  882  152 d  886  151 d  890  151 d  894  152 d
  895  154 d  896  156 d  896  161 d  895  163 d  894  164 d  890  166 d
  886  166 r  883  164 d  882  163 d  881  161 d  881  156 d  882  154 d
  883  152 d  886  151 d  890  151 r  893  152 d  894  154 d  895  156 d
  895  161 d  894  163 d  893  164 d  890  166 d  911  176 r  907  175 d
  905  172 d  904  166 d  904  162 d  905  156 d  907  152 d  911  151 d
  913  151 d  917  152 d  919  156 d  920  162 d  920  166 d  919  172 d
  917  175 d  913  176 d  911  176 d  908  175 d  907  174 d  906  172 d
  905  166 d  905  162 d  906  156 d  907  154 d  908  152 d  911  151 d
  913  151 r  916  152 d  917  154 d  918  156 d  919  162 d  919  166 d
  918  172 d  917  174 d  916  175 d  913  176 d  982  168 r  981  164 d
  979  162 d  975  161 d  974  161 d  970  162 d  968  164 d  967  168 d
  967  169 d  968  173 d  970  175 d  974  176 d  976  176 d  980  175 d
  982  173 d  983  169 d  983  162 d  982  157 d  981  155 d  979  152 d
  975  151 d  971  151 d  969  152 d  968  155 d  968  156 d  969  157 d
  970  156 d  969  155 d  974  161 r  971  162 d  969  164 d  968  168 d
  968  169 d  969  173 d  971  175 d  974  176 d  976  176 r  979  175 d
  981  173 d  982  169 d  982  162 d  981  157 d  980  155 d  977  152 d
  975  151 d  998  176 r  994  175 d  992  172 d  991  166 d  991  162 d
  992  156 d  994  152 d  998  151 d 1000  151 d 1004  152 d 1006  156 d
 1007  162 d 1007  166 d 1006  172 d 1004  175 d 1000  176 d  998  176 d
  995  175 d  994  174 d  993  172 d  992  166 d  992  162 d  993  156 d
  994  154 d  995  152 d  998  151 d 1000  151 r 1003  152 d 1004  154 d
 1005  156 d 1006  162 d 1006  166 d 1005  172 d 1004  174 d 1003  175 d
 1000  176 d 1046  172 r 1049  173 d 1052  176 d 1052  151 d 1051  175 r
 1051  151 d 1046  151 r 1057  151 d 1074  176 r 1070  175 d 1068  172 d
 1067  166 d 1067  162 d 1068  156 d 1070  152 d 1074  151 d 1076  151 d
 1080  152 d 1082  156 d 1083  162 d 1083  166 d 1082  172 d 1080  175 d
 1076  176 d 1074  176 d 1071  175 d 1070  174 d 1069  172 d 1068  166 d
 1068  162 d 1069  156 d 1070  154 d 1071  152 d 1074  151 d 1076  151 r
 1079  152 d 1080  154 d 1081  156 d 1082  162 d 1082  166 d 1081  172 d
 1080  174 d 1079  175 d 1076  176 d 1098  176 r 1094  175 d 1092  172 d
 1091  166 d 1091  162 d 1092  156 d 1094  152 d 1098  151 d 1100  151 d
 1104  152 d 1106  156 d 1107  162 d 1107  166 d 1106  172 d 1104  175 d
 1100  176 d 1098  176 d 1095  175 d 1094  174 d 1093  172 d 1092  166 d
 1092  162 d 1093  156 d 1094  154 d 1095  152 d 1098  151 d 1100  151 r
 1103  152 d 1104  154 d 1105  156 d 1106  162 d 1106  166 d 1105  172 d
 1104  174 d 1103  175 d 1100  176 d 1133  172 r 1136  173 d 1139  176 d
 1139  151 d 1138  175 r 1138  151 d 1133  151 r 1144  151 d 1157  172 r
 1160  173 d 1163  176 d 1163  151 d 1162  175 r 1162  151 d 1157  151 r
 1168  151 d 1185  176 r 1181  175 d 1179  172 d 1178  166 d 1178  162 d
 1179  156 d 1181  152 d 1185  151 d 1187  151 d 1191  152 d 1193  156 d
 1194  162 d 1194  166 d 1193  172 d 1191  175 d 1187  176 d 1185  176 d
 1182  175 d 1181  174 d 1180  172 d 1179  166 d 1179  162 d 1180  156 d
 1181  154 d 1182  152 d 1185  151 d 1187  151 r 1190  152 d 1191  154 d
 1192  156 d 1193  162 d 1193  166 d 1192  172 d 1191  174 d 1190  175 d
 1187  176 d 1221  172 r 1224  173 d 1227  176 d 1227  151 d 1226  175 r
 1226  151 d 1221  151 r 1232  151 d 1243  172 r 1244  170 d 1243  169 d
 1242  170 d 1242  172 d 1243  174 d 1244  175 d 1248  176 d 1252  176 d
 1256  175 d 1257  174 d 1258  172 d 1258  169 d 1257  167 d 1254  164 d
 1248  162 d 1245  161 d 1243  158 d 1242  155 d 1242  151 d 1252  176 r
 1255  175 d 1256  174 d 1257  172 d 1257  169 d 1256  167 d 1252  164 d
 1248  162 d 1242  154 r 1243  155 d 1245  155 d 1251  152 d 1255  152 d
 1257  154 d 1258  155 d 1245  155 r 1251  151 d 1256  151 d 1257  152 d
 1258  155 d 1258  157 d 1273  176 r 1269  175 d 1267  172 d 1266  166 d
 1266  162 d 1267  156 d 1269  152 d 1273  151 d 1275  151 d 1279  152 d
 1281  156 d 1282  162 d 1282  166 d 1281  172 d 1279  175 d 1275  176 d
 1273  176 d 1270  175 d 1269  174 d 1268  172 d 1267  166 d 1267  162 d
 1268  156 d 1269  154 d 1270  152 d 1273  151 d 1275  151 r 1278  152 d
 1279  154 d 1280  156 d 1281  162 d 1281  166 d 1280  172 d 1279  174 d
 1278  175 d 1275  176 d 1308  172 r 1311  173 d 1314  176 d 1314  151 d
 1313  175 r 1313  151 d 1308  151 r 1319  151 d 1330  173 r 1331  172 d
 1330  170 d 1329  172 d 1329  173 d 1331  175 d 1335  176 d 1339  176 d
 1343  175 d 1344  173 d 1344  169 d 1343  167 d 1339  166 d 1336  166 d
 1339  176 r 1342  175 d 1343  173 d 1343  169 d 1342  167 d 1339  166 d
 1342  164 d 1344  162 d 1345  160 d 1345  156 d 1344  154 d 1343  152 d
 1339  151 d 1335  151 d 1331  152 d 1330  154 d 1329  156 d 1329  157 d
 1330  158 d 1331  157 d 1330  156 d 1343  163 r 1344  160 d 1344  156 d
 1343  154 d 1342  152 d 1339  151 d 1360  176 r 1356  175 d 1354  172 d
 1353  166 d 1353  162 d 1354  156 d 1356  152 d 1360  151 d 1362  151 d
 1366  152 d 1368  156 d 1369  162 d 1369  166 d 1368  172 d 1366  175 d
 1362  176 d 1360  176 d 1357  175 d 1356  174 d 1355  172 d 1354  166 d
 1354  162 d 1355  156 d 1356  154 d 1357  152 d 1360  151 d 1362  151 r
 1365  152 d 1366  154 d 1367  156 d 1368  162 d 1368  166 d 1367  172 d
 1366  174 d 1365  175 d 1362  176 d 1396  172 r 1399  173 d 1402  176 d
 1402  151 d 1401  175 r 1401  151 d 1396  151 r 1407  151 d 1427  174 r
 1427  151 d 1429  176 r 1429  151 d 1429  176 r 1415  158 d 1435  158 d
 1424  151 r 1432  151 d 1448  176 r 1444  175 d 1442  172 d 1441  166 d
 1441  162 d 1442  156 d 1444  152 d 1448  151 d 1450  151 d 1454  152 d
 1456  156 d 1457  162 d 1457  166 d 1456  172 d 1454  175 d 1450  176 d
 1448  176 d 1445  175 d 1444  174 d 1443  172 d 1442  166 d 1442  162 d
 1443  156 d 1444  154 d 1445  152 d 1448  151 d 1450  151 r 1453  152 d
 1454  154 d 1455  156 d 1456  162 d 1456  166 d 1455  172 d 1454  174 d
 1453  175 d 1450  176 d 1483  172 r 1486  173 d 1489  176 d 1489  151 d
 1488  175 r 1488  151 d 1483  151 r 1494  151 d 1506  176 r 1504  164 d
 1506  166 d 1510  167 d 1513  167 d 1517  166 d 1519  163 d 1520  160 d
 1520  158 d 1519  155 d 1517  152 d 1513  151 d 1510  151 d 1506  152 d
 1505  154 d 1504  156 d 1504  157 d 1505  158 d 1506  157 d 1505  156 d
 1513  167 r 1516  166 d 1518  163 d 1519  160 d 1519  158 d 1518  155 d
 1516  152 d 1513  151 d 1506  176 r 1518  176 d 1506  175 r 1512  175 d
 1518  176 d 1535  176 r 1531  175 d 1529  172 d 1528  166 d 1528  162 d
 1529  156 d 1531  152 d 1535  151 d 1537  151 d 1541  152 d 1543  156 d
 1544  162 d 1544  166 d 1543  172 d 1541  175 d 1537  176 d 1535  176 d
 1532  175 d 1531  174 d 1530  172 d 1529  166 d 1529  162 d 1530  156 d
 1531  154 d 1532  152 d 1535  151 d 1537  151 r 1540  152 d 1541  154 d
 1542  156 d 1543  162 d 1543  166 d 1542  172 d 1541  174 d 1540  175 d
 1537  176 d 1571  172 r 1574  173 d 1577  176 d 1577  151 d 1576  175 r
 1576  151 d 1571  151 r 1582  151 d 1606  173 r 1605  172 d 1606  170 d
 1607  172 d 1607  173 d 1606  175 d 1604  176 d 1600  176 d 1596  175 d
 1594  173 d 1593  170 d 1592  166 d 1592  158 d 1593  155 d 1595  152 d
 1599  151 d 1601  151 d 1605  152 d 1607  155 d 1608  158 d 1608  160 d
 1607  163 d 1605  166 d 1601  167 d 1600  167 d 1596  166 d 1594  163 d
 1593  160 d 1600  176 r 1598  175 d 1595  173 d 1594  170 d 1593  166 d
 1593  158 d 1594  155 d 1596  152 d 1599  151 d 1601  151 r 1604  152 d
 1606  155 d 1607  158 d 1607  160 d 1606  163 d 1604  166 d 1601  167 d
 1623  176 r 1619  175 d 1617  172 d 1616  166 d 1616  162 d 1617  156 d
 1619  152 d 1623  151 d 1625  151 d 1629  152 d 1631  156 d 1632  162 d
 1632  166 d 1631  172 d 1629  175 d 1625  176 d 1623  176 d 1620  175 d
 1619  174 d 1618  172 d 1617  166 d 1617  162 d 1618  156 d 1619  154 d
 1620  152 d 1623  151 d 1625  151 r 1628  152 d 1629  154 d 1630  156 d
 1631  162 d 1631  166 d 1630  172 d 1629  174 d 1628  175 d 1625  176 d
  200 1100 r 1600 1100 d  200 1100 r  200 1062 d  244 1100 r  244 1081 d
  288 1100 r  288 1062 d  331 1100 r  331 1081 d  375 1100 r  375 1062 d
  419 1100 r  419 1081 d  463 1100 r  463 1062 d  506 1100 r  506 1081 d
  550 1100 r  550 1062 d  594 1100 r  594 1081 d  638 1100 r  638 1062 d
  681 1100 r  681 1081 d  725 1100 r  725 1062 d  769 1100 r  769 1081 d
  813 1100 r  813 1062 d  856 1100 r  856 1081 d  900 1100 r  900 1062 d
  944 1100 r  944 1081 d  987 1100 r  987 1062 d 1031 1100 r 1031 1081 d
 1075 1100 r 1075 1062 d 1119 1100 r 1119 1081 d 1162 1100 r 1162 1062 d
 1206 1100 r 1206 1081 d 1250 1100 r 1250 1062 d 1294 1100 r 1294 1081 d
 1337 1100 r 1337 1062 d 1381 1100 r 1381 1081 d 1425 1100 r 1425 1062 d
 1469 1100 r 1469 1081 d 1512 1100 r 1512 1062 d 1556 1100 r 1556 1081 d
 1600 1100 r 1600 1062 d  200  200 r  200 1100 d  200  200 r  238  200 d
  200  227 r  219  227 d  200  255 r  219  255 d  200  282 r  219  282 d
  200  309 r  219  309 d  200  336 r  238  336 d  200  364 r  219  364 d
  200  391 r  219  391 d  200  418 r  219  418 d  200  445 r  219  445 d
  200  473 r  238  473 d  200  500 r  219  500 d  200  527 r  219  527 d
  200  555 r  219  555 d  200  582 r  219  582 d  200  609 r  238  609 d
  200  636 r  219  636 d  200  664 r  219  664 d  200  691 r  219  691 d
  200  718 r  219  718 d  200  745 r  238  745 d  200  773 r  219  773 d
  200  800 r  219  800 d  200  827 r  219  827 d  200  855 r  219  855 d
  200  882 r  238  882 d  200  909 r  219  909 d  200  936 r  219  936 d
  200  964 r  219  964 d  200  991 r  219  991 d  200 1018 r  238 1018 d
  200 1045 r  219 1045 d  200 1073 r  219 1073 d  200 1100 r  219 1100 d
  168  214 r  164  213 d  162  210 d  161  204 d  161  200 d  162  194 d
  164  190 d  168  189 d  170  189 d  174  190 d  176  194 d  177  200 d
  177  204 d  176  210 d  174  213 d  170  214 d  168  214 d  165  213 d
  164  212 d  163  210 d  162  204 d  162  200 d  163  194 d  164  192 d
  165  190 d  168  189 d  170  189 r  173  190 d  174  192 d  175  194 d
  176  200 d  176  204 d  175  210 d  174  212 d  173  213 d  170  214 d
  139  350 r  137  338 d  139  340 d  143  341 d  146  341 d  150  340 d
  152  337 d  153  334 d  153  332 d  152  329 d  150  326 d  146  325 d
  143  325 d  139  326 d  138  328 d  137  330 d  137  331 d  138  332 d
  139  331 d  138  330 d  146  341 r  149  340 d  151  337 d  152  334 d
  152  332 d  151  329 d  149  326 d  146  325 d  139  350 r  151  350 d
  139  349 r  145  349 d  151  350 d  168  350 r  164  349 d  162  346 d
  161  340 d  161  336 d  162  330 d  164  326 d  168  325 d  170  325 d
  174  326 d  176  330 d  177  336 d  177  340 d  176  346 d  174  349 d
  170  350 d  168  350 d  165  349 d  164  348 d  163  346 d  162  340 d
  162  336 d  163  330 d  164  328 d  165  326 d  168  325 d  170  325 r
  173  326 d  174  328 d  175  330 d  176  336 d  176  340 d  175  346 d
  174  348 d  173  349 d  170  350 d  116  483 r  119  484 d  122  487 d
  122  462 d  121  486 r  121  462 d  116  462 r  127  462 d  144  487 r
  140  486 d  138  483 d  137  477 d  137  473 d  138  467 d  140  463 d
  144  462 d  146  462 d  150  463 d  152  467 d  153  473 d  153  477 d
  152  483 d  150  486 d  146  487 d  144  487 d  141  486 d  140  485 d
  139  483 d  138  477 d  138  473 d  139  467 d  140  465 d  141  463 d
  144  462 d  146  462 r  149  463 d  150  465 d  151  467 d  152  473 d
  152  477 d  151  483 d  150  485 d  149  486 d  146  487 d  168  487 r
  164  486 d  162  483 d  161  477 d  161  473 d  162  467 d  164  463 d
  168  462 d  170  462 d  174  463 d  176  467 d  177  473 d  177  477 d
  176  483 d  174  486 d  170  487 d  168  487 d  165  486 d  164  485 d
  163  483 d  162  477 d  162  473 d  163  467 d  164  465 d  165  463 d
  168  462 d  170  462 r  173  463 d  174  465 d  175  467 d  176  473 d
  176  477 d  175  483 d  174  485 d  173  486 d  170  487 d  116  619 r
  119  620 d  122  623 d  122  598 d  121  622 r  121  598 d  116  598 r
  127  598 d  139  623 r  137  611 d  139  613 d  143  614 d  146  614 d
  150  613 d  152  610 d  153  607 d  153  605 d  152  602 d  150  599 d
  146  598 d  143  598 d  139  599 d  138  601 d  137  603 d  137  604 d
  138  605 d  139  604 d  138  603 d  146  614 r  149  613 d  151  610 d
  152  607 d  152  605 d  151  602 d  149  599 d  146  598 d  139  623 r
  151  623 d  139  622 r  145  622 d  151  623 d  168  623 r  164  622 d
  162  619 d  161  613 d  161  609 d  162  603 d  164  599 d  168  598 d
  170  598 d  174  599 d  176  603 d  177  609 d  177  613 d  176  619 d
  174  622 d  170  623 d  168  623 d  165  622 d  164  621 d  163  619 d
  162  613 d  162  609 d  163  603 d  164  601 d  165  599 d  168  598 d
  170  598 r  173  599 d  174  601 d  175  603 d  176  609 d  176  613 d
  175  619 d  174  621 d  173  622 d  170  623 d  114  755 r  115  753 d
  114  752 d  113  753 d  113  755 d  114  757 d  115  758 d  119  759 d
  123  759 d  127  758 d  128  757 d  129  755 d  129  752 d  128  750 d
  125  747 d  119  745 d  116  744 d  114  741 d  113  738 d  113  734 d
  123  759 r  126  758 d  127  757 d  128  755 d  128  752 d  127  750 d
  123  747 d  119  745 d  113  737 r  114  738 d  116  738 d  122  735 d
  126  735 d  128  737 d  129  738 d  116  738 r  122  734 d  127  734 d
  128  735 d  129  738 d  129  740 d  144  759 r  140  758 d  138  755 d
  137  749 d  137  745 d  138  739 d  140  735 d  144  734 d  146  734 d
  150  735 d  152  739 d  153  745 d  153  749 d  152  755 d  150  758 d
  146  759 d  144  759 d  141  758 d  140  757 d  139  755 d  138  749 d
  138  745 d  139  739 d  140  737 d  141  735 d  144  734 d  146  734 r
  149  735 d  150  737 d  151  739 d  152  745 d  152  749 d  151  755 d
  150  757 d  149  758 d  146  759 d  168  759 r  164  758 d  162  755 d
  161  749 d  161  745 d  162  739 d  164  735 d  168  734 d  170  734 d
  174  735 d  176  739 d  177  745 d  177  749 d  176  755 d  174  758 d
  170  759 d  168  759 d  165  758 d  164  757 d  163  755 d  162  749 d
  162  745 d  163  739 d  164  737 d  165  735 d  168  734 d  170  734 r
  173  735 d  174  737 d  175  739 d  176  745 d  176  749 d  175  755 d
  174  757 d  173  758 d  170  759 d  114  892 r  115  890 d  114  889 d
  113  890 d  113  892 d  114  894 d  115  895 d  119  896 d  123  896 d
  127  895 d  128  894 d  129  892 d  129  889 d  128  887 d  125  884 d
  119  882 d  116  881 d  114  878 d  113  875 d  113  871 d  123  896 r
  126  895 d  127  894 d  128  892 d  128  889 d  127  887 d  123  884 d
  119  882 d  113  874 r  114  875 d  116  875 d  122  872 d  126  872 d
  128  874 d  129  875 d  116  875 r  122  871 d  127  871 d  128  872 d
  129  875 d  129  877 d  139  896 r  137  884 d  139  886 d  143  887 d
  146  887 d  150  886 d  152  883 d  153  880 d  153  878 d  152  875 d
  150  872 d  146  871 d  143  871 d  139  872 d  138  874 d  137  876 d
  137  877 d  138  878 d  139  877 d  138  876 d  146  887 r  149  886 d
  151  883 d  152  880 d  152  878 d  151  875 d  149  872 d  146  871 d
  139  896 r  151  896 d  139  895 r  145  895 d  151  896 d  168  896 r
  164  895 d  162  892 d  161  886 d  161  882 d  162  876 d  164  872 d
  168  871 d  170  871 d  174  872 d  176  876 d  177  882 d  177  886 d
  176  892 d  174  895 d  170  896 d  168  896 d  165  895 d  164  894 d
  163  892 d  162  886 d  162  882 d  163  876 d  164  874 d  165  872 d
  168  871 d  170  871 r  173  872 d  174  874 d  175  876 d  176  882 d
  176  886 d  175  892 d  174  894 d  173  895 d  170  896 d  114 1029 r
  115 1028 d  114 1026 d  113 1028 d  113 1029 d  115 1031 d  119 1032 d
  123 1032 d  127 1031 d  128 1029 d  128 1025 d  127 1023 d  123 1022 d
  120 1022 d  123 1032 r  126 1031 d  127 1029 d  127 1025 d  126 1023 d
  123 1022 d  126 1020 d  128 1018 d  129 1016 d  129 1012 d  128 1010 d
  127 1008 d  123 1007 d  119 1007 d  115 1008 d  114 1010 d  113 1012 d
  113 1013 d  114 1014 d  115 1013 d  114 1012 d  127 1019 r  128 1016 d
  128 1012 d  127 1010 d  126 1008 d  123 1007 d  144 1032 r  140 1031 d
  138 1028 d  137 1022 d  137 1018 d  138 1012 d  140 1008 d  144 1007 d
  146 1007 d  150 1008 d  152 1012 d  153 1018 d  153 1022 d  152 1028 d
  150 1031 d  146 1032 d  144 1032 d  141 1031 d  140 1030 d  139 1028 d
  138 1022 d  138 1018 d  139 1012 d  140 1010 d  141 1008 d  144 1007 d
  146 1007 r  149 1008 d  150 1010 d  151 1012 d  152 1018 d  152 1022 d
  151 1028 d  150 1030 d  149 1031 d  146 1032 d  168 1032 r  164 1031 d
  162 1028 d  161 1022 d  161 1018 d  162 1012 d  164 1008 d  168 1007 d
  170 1007 d  174 1008 d  176 1012 d  177 1018 d  177 1022 d  176 1028 d
  174 1031 d  170 1032 d  168 1032 d  165 1031 d  164 1030 d  163 1028 d
  162 1022 d  162 1018 d  163 1012 d  164 1010 d  165 1008 d  168 1007 d
  170 1007 r  173 1008 d  174 1010 d  175 1012 d  176 1018 d  176 1022 d
  175 1028 d  174 1030 d  173 1031 d  170 1032 d 1600  200 r 1600 1100 d
 1600  200 r 1562  200 d 1600  227 r 1581  227 d 1600  255 r 1581  255 d
 1600  282 r 1581  282 d 1600  309 r 1581  309 d 1600  336 r 1562  336 d
 1600  364 r 1581  364 d 1600  391 r 1581  391 d 1600  418 r 1581  418 d
 1600  445 r 1581  445 d 1600  473 r 1562  473 d 1600  500 r 1581  500 d
 1600  527 r 1581  527 d 1600  555 r 1581  555 d 1600  582 r 1581  582 d
 1600  609 r 1562  609 d 1600  636 r 1581  636 d 1600  664 r 1581  664 d
 1600  691 r 1581  691 d 1600  718 r 1581  718 d 1600  745 r 1562  745 d
 1600  773 r 1581  773 d 1600  800 r 1581  800 d 1600  827 r 1581  827 d
 1600  855 r 1581  855 d 1600  882 r 1562  882 d 1600  909 r 1581  909 d
 1600  936 r 1581  936 d 1600  964 r 1581  964 d 1600  991 r 1581  991 d
 1600 1018 r 1562 1018 d 1600 1045 r 1581 1045 d 1600 1073 r 1581 1073 d
 1600 1100 r 1581 1100 d 1600  200 r 1578  200 d 1578  200 p 1556  200 d
 1556  200 p 1534  200 d 1534  200 p 1513  200 d 1513  200 p 1491  200 d
 1491  200 p 1469  200 d 1469  200 p 1447  200 d 1447  200 p 1425  200 d
 1425  200 p 1403  200 d 1403  203 d 1381  203 d 1381  252 d 1359  252 d
 1359  366 d 1338  366 d 1338  639 d 1316  639 d 1316  934 d 1294  934 d
 1294  950 d 1272  950 d 1272  623 d 1250  623 d 1250  336 d 1228  336 d
 1228  225 d 1206  225 d 1206  200 d 1184  200 d 1184  200 p 1163  200 d
 1163  200 p 1141  200 d 1141  200 p 1119  200 d 1119  200 p 1097  200 d
 1097  200 p 1075  200 d 1075  200 p 1053  200 d 1053  200 p 1031  200 d
 1031  200 p 1009  200 d 1009  200 p  988  200 d  988  200 p  966  200 d
  966  200 p  944  200 d  944  200 p  922  200 d  922  200 p  900  200 d
  900  200 p  878  200 d  878  200 p  856  200 d  856  200 p  834  200 d
  834  200 p  813  200 d  813  200 p  791  200 d  791  200 p  769  200 d
  769  200 p  747  200 d  747  200 p  725  200 d  725  200 p  703  200 d
  703  200 p  681  200 d  681  200 p  659  200 d  659  200 p  638  200 d
  638  200 p  616  200 d  616  200 p  594  200 d  594  200 p  572  200 d
  572  200 p  550  200 d  550  200 p  528  200 d  528  200 p  506  200 d
  506  200 p  484  200 d  484  200 p  463  200 d  463  200 p  441  200 d
  441  200 p  419  200 d  419  200 p  397  200 d  397  200 p  375  200 d
  375  200 p  353  200 d  353  200 p  331  200 d  331  200 p  309  200 d
  309  200 p  288  200 d  288  200 p  266  200 d  266  200 p  244  200 d
  244  200 p  222  200 d  222  200 p  200  200 d  896  304 r  886  304 d
  896  305 r  886  305 d  896  306 r  886  306 d  896  307 r  886  307 d
  896  308 r  886  308 d  896  309 r  886  309 d  896  310 r  886  310 d
  896  311 r  886  311 d  896  312 r  886  312 d  896  313 r  886  313 d
  896  314 r  886  314 d  891  309 r  994  309 d  994  316 r  994  302 d
  994  316 d  891  309 r  789  309 d  789  316 r  789  302 d  789  316 d
  853  386 r  843  386 d  853  387 r  843  387 d  853  388 r  843  388 d
  853  389 r  843  389 d  853  390 r  843  390 d  853  391 r  843  391 d
  853  392 r  843  392 d  853  393 r  843  393 d  853  394 r  843  394 d
  853  395 r  843  395 d  853  396 r  843  396 d  848  391 r 1037  391 d
 1037  398 r 1037  384 d 1037  398 d  853  386 r  843  386 d  853  387 r
  843  387 d  853  388 r  843  388 d  853  389 r  843  389 d  853  390 r
  843  390 d  853  391 r  843  391 d  853  392 r  843  392 d  853  393 r
  843  393 d  853  394 r  843  394 d  853  395 r  843  395 d  853  396 r
  843  396 d  848  391 r  658  391 d  658  398 r  658  384 d  658  398 d
  848  677 r  848  426 d  845  419 r  848  415 d  850  419 d  842  422 r
  848  416 d  854  422 d  848  437 r  848  416 d  891  642 r  891  342 d
  888  335 r  891  331 d  893  335 d  885  338 r  891  332 d  897  338 d
  891  353 r  891  332 d  514 1002 r  516  999 d  516 1005 d  514 1002 d
  512 1004 d  508 1005 d  506 1005 d  502 1004 d  500 1002 d  499  999 d
  498  996 d  498  990 d  499  986 d  500  984 d  502  981 d  506  980 d
  508  980 d  512  981 d  514  984 d  506 1005 r  504 1004 d  501 1002 d
  500  999 d  499  996 d  499  990 d  500  986 d  501  984 d  504  981 d
  506  980 d  514  990 r  514  980 d  516  990 r  516  980 d  511  990 r
  518  990 d  526  995 r  526  993 d  525  993 d  525  995 d  526  996 d
  529  997 d  534  997 d  536  996 d  537  995 d  538  992 d  538  984 d
  540  981 d  541  980 d  537  995 r  537  984 d  538  981 d  541  980 d
  542  980 d  537  992 r  536  991 d  529  990 d  525  989 d  524  986 d
  524  984 d  525  981 d  529  980 d  532  980 d  535  981 d  537  984 d
  529  990 r  526  989 d  525  986 d  525  984 d  526  981 d  529  980 d
  550 1005 r  550  980 d  552 1005 r  552  980 d  547 1005 r  552 1005 d
  547  980 r  555  980 d  564 1005 r  564  980 d  565 1005 r  565  980 d
  560 1005 r  565 1005 d  560  980 r  568  980 d  577 1005 r  576 1004 d
  577 1003 d  578 1004 d  577 1005 d  577  997 r  577  980 d  578  997 r
  578  980 d  573  997 r  578  997 d  573  980 r  582  980 d  590  997 r
  590  984 d  591  981 d  595  980 d  597  980 d  601  981 d  603  984 d
  591  997 r  591  984 d  592  981 d  595  980 d  603  997 r  603  980 d
  604  997 r  604  980 d  586  997 r  591  997 d  600  997 r  604  997 d
  603  980 r  608  980 d  616  997 r  616  980 d  618  997 r  618  980 d
  618  993 r  620  996 d  624  997 d  626  997 d  630  996 d  631  993 d
  631  980 d  626  997 r  628  996 d  630  993 d  630  980 d  631  993 r
  633  996 d  637  997 d  639  997 d  643  996 d  644  993 d  644  980 d
  639  997 r  642  996 d  643  993 d  643  980 d  613  997 r  618  997 d
  613  980 r  621  980 d  626  980 r  634  980 d  639  980 r  648  980 d
  340  923 r  331  898 d  340  923 r  348  898 d  340  920 r  347  898 d
  334  904 r  346  904 d  328  898 r  335  898 d  343  898 r  352  898 d
  367  914 r  368  915 d  368  911 d  367  914 d  364  915 d  360  915 d
  356  914 d  355  913 d  355  911 d  356  909 d  359  908 d  365  905 d
  367  904 d  368  903 d  355  911 r  356  910 d  359  909 d  365  907 d
  367  905 d  368  903 d  368  901 d  367  899 d  364  898 d  360  898 d
  356  899 d  355  902 d  355  898 d  356  899 d  388  914 r  389  915 d
  389  911 d  388  914 d  384  915 d  380  915 d  377  914 d  376  913 d
  376  911 d  377  909 d  379  908 d  385  905 d  388  904 d  389  903 d
  376  911 r  377  910 d  379  909 d  385  907 d  388  905 d  389  903 d
  389  901 d  388  899 d  384  898 d  380  898 d  377  899 d  376  902 d
  376  898 d  377  899 d  398  915 r  398  902 d  400  899 d  403  898 d
  406  898 d  409  899 d  412  902 d  400  915 r  400  902 d  401  899 d
  403  898 d  412  915 r  412  898 d  413  915 r  413  898 d  395  915 r
  400  915 d  408  915 r  413  915 d  412  898 r  416  898 d  425  915 r
  425  898 d  426  915 r  426  898 d  426  911 r  428  914 d  432  915 d
  434  915 d  438  914 d  439  911 d  439  898 d  434  915 r  437  914 d
  438  911 d  438  898 d  439  911 r  442  914 d  445  915 d  448  915 d
  451  914 d  452  911 d  452  898 d  448  915 r  450  914 d  451  911 d
  451  898 d  421  915 r  426  915 d  421  898 r  430  898 d  434  898 r
  443  898 d  448  898 r  456  898 d  463  908 r  478  908 d  478  910 d
  476  913 d  475  914 d  473  915 d  469  915 d  466  914 d  463  911 d
  462  908 d  462  905 d  463  902 d  466  899 d  469  898 d  472  898 d
  475  899 d  478  902 d  476  908 r  476  911 d  475  914 d  469  915 r
  467  914 d  464  911 d  463  908 d  463  905 d  464  902 d  467  899 d
  469  898 d  486  915 r  485  914 d  486  913 d  487  914 d  486  915 d
  486  901 r  485  899 d  486  898 d  487  899 d  486  901 d  526  928 r
  523  926 d  521  922 d  518  917 d  517  911 d  517  907 d  518  901 d
  521  896 d  523  892 d  526  890 d  523  926 r  521  921 d  520  917 d
  518  911 d  518  907 d  520  901 d  521  897 d  523  892 d  540  923 r
  536  922 d  534  919 d  533  913 d  533  909 d  534  903 d  536  899 d
  540  898 d  542  898 d  546  899 d  548  903 d  550  909 d  550  913 d
  548  919 d  546  922 d  542  923 d  540  923 d  538  922 d  536  921 d
  535  919 d  534  913 d  534  909 d  535  903 d  536  901 d  538  899 d
  540  898 d  542  898 r  545  899 d  546  901 d  547  903 d  548  909 d
  548  913 d  547  919 d  546  921 d  545  922 d  542  923 d  558  901 r
  557  899 d  558  898 d  559  899 d  558  901 d  577  921 r  577  898 d
  578  923 r  578  898 d  578  923 r  565  905 d  584  905 d  574  898 r
  582  898 d  596  923 r  593  922 d  592  920 d  592  916 d  593  914 d
  596  913 d  601  913 d  605  914 d  606  916 d  606  920 d  605  922 d
  601  923 d  596  923 d  594  922 d  593  920 d  593  916 d  594  914 d
  596  913 d  601  913 r  604  914 d  605  916 d  605  920 d  604  922 d
  601  923 d  596  913 r  593  911 d  592  910 d  590  908 d  590  903 d
  592  901 d  593  899 d  596  898 d  601  898 d  605  899 d  606  901 d
  607  903 d  607  908 d  606  910 d  605  911 d  601  913 d  596  913 r
  594  911 d  593  910 d  592  908 d  592  903 d  593  901 d  594  899 d
  596  898 d  601  898 r  604  899 d  605  901 d  606  903 d  606  908 d
  605  910 d  604  911 d  601  913 d  646  919 r  646  898 d  636  909 r
  655  909 d  636  898 r  655  898 d  691  923 r  688  922 d  685  919 d
  684  913 d  684  909 d  685  903 d  688  899 d  691  898 d  694  898 d
  697  899 d  700  903 d  701  909 d  701  913 d  700  919 d  697  922 d
  694  923 d  691  923 d  689  922 d  688  921 d  686  919 d  685  913 d
  685  909 d  686  903 d  688  901 d  689  899 d  691  898 d  694  898 r
  696  899 d  697  901 d  698  903 d  700  909 d  700  913 d  698  919 d
  697  921 d  696  922 d  694  923 d  709  901 r  708  899 d  709  898 d
  710  899 d  709  901 d  725  923 r  721  922 d  719  919 d  718  913 d
  718  909 d  719  903 d  721  899 d  725  898 d  727  898 d  731  899 d
  733  903 d  734  909 d  734  913 d  733  919 d  731  922 d  727  923 d
  725  923 d  722  922 d  721  921 d  720  919 d  719  913 d  719  909 d
  720  903 d  721  901 d  722  899 d  725  898 d  727  898 r  730  899 d
  731  901 d  732  903 d  733  909 d  733  913 d  732  919 d  731  921 d
  730  922 d  727  923 d  748  923 r  744  922 d  743  920 d  743  916 d
  744  914 d  748  913 d  752  913 d  756  914 d  757  916 d  757  920 d
  756  922 d  752  923 d  748  923 d  745  922 d  744  920 d  744  916 d
  745  914 d  748  913 d  752  913 r  755  914 d  756  916 d  756  920 d
  755  922 d  752  923 d  748  913 r  744  911 d  743  910 d  742  908 d
  742  903 d  743  901 d  744  899 d  748  898 d  752  898 d  756  899 d
  757  901 d  758  903 d  758  908 d  757  910 d  756  911 d  752  913 d
  748  913 r  745  911 d  744  910 d  743  908 d  743  903 d  744  901 d
  745  899 d  748  898 d  752  898 r  755  899 d  756  901 d  757  903 d
  757  908 d  756  910 d  755  911 d  752  913 d  766  928 r  768  926 d
  770  922 d  773  917 d  774  911 d  774  907 d  773  901 d  770  896 d
  768  892 d  766  890 d  768  926 r  770  921 d  772  917 d  773  911 d
  773  907 d  772  901 d  770  897 d  768  892 d  811  915 r  808  914 d
  805  911 d  804  908 d  804  905 d  805  902 d  808  899 d  811  898 d
  814  898 d  817  899 d  820  902 d  821  905 d  821  908 d  820  911 d
  817  914 d  814  915 d  811  915 d  809  914 d  806  911 d  805  908 d
  805  905 d  806  902 d  809  899 d  811  898 d  814  898 r  816  899 d
  818  902 d  820  905 d  820  908 d  818  911 d  816  914 d  814  915 d
  836  922 r  835  921 d  836  920 d  838  921 d  838  922 d  836  923 d
  834  923 d  832  922 d  830  920 d  830  898 d  834  923 r  833  922 d
  832  920 d  832  898 d  827  915 r  835  915 d  827  898 r  835  898 d
  410  853 r  411  855 d  411  851 d  410  853 d  409  854 d  405  855 d
  401  855 d  397  854 d  395  852 d  395  849 d  396  847 d  397  846 d
  399  845 d  407  842 d  409  841 d  411  839 d  395  849 r  397  847 d
  399  846 d  407  843 d  409  842 d  410  841 d  411  839 d  411  834 d
  409  831 d  405  830 d  401  830 d  397  831 d  396  833 d  395  835 d
  395  830 d  396  833 d  421  855 r  421  835 d  422  831 d  425  830 d
  427  830 d  429  831 d  431  834 d  422  855 r  422  835 d  423  831 d
  425  830 d  417  847 r  427  847 d  439  845 r  439  843 d  438  843 d
  438  845 d  439  846 d  441  847 d  446  847 d  449  846 d  450  845 d
  451  842 d  451  834 d  452  831 d  453  830 d  450  845 r  450  834 d
  451  831 d  453  830 d  455  830 d  450  842 r  449  841 d  441  840 d
  438  839 d  437  836 d  437  834 d  438  831 d  441  830 d  445  830 d
  447  831 d  450  834 d  441  840 r  439  839 d  438  836 d  438  834 d
  439  831 d  441  830 d  463  847 r  463  830 d  464  847 r  464  830 d
  464  843 r  467  846 d  470  847 d  473  847 d  476  846 d  477  843 d
  477  830 d  473  847 r  475  846 d  476  843 d  476  830 d  459  847 r
  464  847 d  459  830 r  468  830 d  473  830 r  481  830 d  501  855 r
  501  830 d  503  855 r  503  830 d  501  843 r  499  846 d  497  847 d
  494  847 d  491  846 d  488  843 d  487  840 d  487  837 d  488  834 d
  491  831 d  494  830 d  497  830 d  499  831 d  501  834 d  494  847 r
  492  846 d  489  843 d  488  840 d  488  837 d  489  834 d  492  831 d
  494  830 d  498  855 r  503  855 d  501  830 r  506  830 d  515  845 r
  515  843 d  513  843 d  513  845 d  515  846 d  517  847 d  522  847 d
  524  846 d  525  845 d  527  842 d  527  834 d  528  831 d  529  830 d
  525  845 r  525  834 d  527  831 d  529  830 d  530  830 d  525  842 r
  524  841 d  517  840 d  513  839 d  512  836 d  512  834 d  513  831 d
  517  830 d  521  830 d  523  831 d  525  834 d  517  840 r  515  839 d
  513  836 d  513  834 d  515  831 d  517  830 d  539  847 r  539  830 d
  540  847 r  540  830 d  540  840 r  541  843 d  543  846 d  546  847 d
  549  847 d  551  846 d  551  845 d  549  843 d  548  845 d  549  846 d
  535  847 r  540  847 d  535  830 r  543  830 d  571  855 r  571  830 d
  572  855 r  572  830 d  571  843 r  569  846 d  566  847 d  564  847 d
  560  846 d  558  843 d  557  840 d  557  837 d  558  834 d  560  831 d
  564  830 d  566  830 d  569  831 d  571  834 d  564  847 r  561  846 d
  559  843 d  558  840 d  558  837 d  559  834 d  561  831 d  564  830 d
  567  855 r  572  855 d  571  830 r  576  830 d  606  869 r  603  868 d
  603  867 d  603  865 d  603  863 d  606  862 d  608  862 d  611  863 d
  611  865 d  611  867 d  611  868 d  608  869 d  606  869 d  604  868 d
  603  867 d  603  865 d  604  863 d  606  862 d  608  862 r  610  863 d
  611  865 d  611  867 d  610  868 d  608  869 d  606  862 r  603  862 d
  603  861 d  602  859 d  602  857 d  603  855 d  603  854 d  606  854 d
  608  854 d  611  854 d  611  855 d  612  857 d  612  859 d  611  861 d
  611  862 d  608  862 d  606  862 r  604  862 d  603  861 d  603  859 d
  603  857 d  603  855 d  604  854 d  606  854 d  608  854 r  610  854 d
  611  855 d  611  857 d  611  859 d  611  861 d  610  862 d  608  862 d
  620  855 r  620  830 d  621  855 r  621  830 d  617  855 r  631  855 d
  635  854 d  636  853 d  637  851 d  637  848 d  636  846 d  635  845 d
  631  843 d  631  855 r  633  854 d  635  853 d  636  851 d  636  848 d
  635  846 d  633  845 d  631  843 d  621  843 r  631  843 d  635  842 d
  636  841 d  637  839 d  637  835 d  636  833 d  635  831 d  631  830 d
  617  830 d  631  843 r  633  842 d  635  841 d  636  839 d  636  835 d
  635  833 d  633  831 d  631  830 d  668  855 r  668  830 d  669  855 r
  669  830 d  677  848 r  677  839 d  665  855 r  684  855 d  684  849 d
  683  855 d  669  843 r  677  843 d  665  830 r  673  830 d  692  855 r
  692  830 d  693  855 r  693  830 d  689  855 r  693  855 d  689  830 r
  697  830 d  705  847 r  705  834 d  707  831 d  710  830 d  713  830 d
  716  831 d  719  834 d  707  847 r  707  834 d  708  831 d  710  830 d
  719  847 r  719  830 d  720  847 r  720  830 d  702  847 r  707  847 d
  715  847 r  720  847 d  719  830 r  723  830 d  732  847 r  745  830 d
  733  847 r  746  830 d  746  847 r  732  830 d  728  847 r  737  847 d
  743  847 r  750  847 d  728  830 r  735  830 d  741  830 r  750  830 d
  880  683 r  883  685 d  877  685 d  880  683 d  878  681 d  877  677 d
  877  675 d  878  671 d  880  669 d  883  668 d  886  667 d  892  667 d
  896  668 d  898  669 d  901  671 d  902  675 d  902  677 d  901  681 d
  898  683 d  877  675 r  878  673 d  880  670 d  883  669 d  886  668 d
  892  668 d  896  669 d  898  670 d  901  673 d  902  675 d  892  683 r
  902  683 d  892  685 r  902  685 d  892  680 r  892  687 d  877  701 r
  902  693 d  877  701 r  902  710 d  880  701 r  902  709 d  896  695 r
  896  707 d  902  689 r  902  697 d  902  705 r  902  713 d  877  719 r
  902  719 d  877  721 r  902  721 d  877  716 r  877  724 d  902  716 r
  902  734 d  896  734 d  902  733 d  877  742 r  902  742 d  877  743 r
  902  743 d  877  739 r  877  747 d  902  739 r  902  757 d  896  757 d
  902  755 d  877  765 r  902  765 d  877  766 r  902  766 d  884  773 r
  893  773 d  877  761 r  877  781 d  883  781 d  877  779 d  889  766 r
  889  773 d  902  761 r  902  781 d  896  781 d  902  779 d  877  789 r
  902  805 d  877  790 r  902  806 d  877  806 r  902  789 d  877  785 r
  877  794 d  877  802 r  877  809 d  902  785 r  902  793 d  902  801 r
  902  809 d  836  701 r  834  702 d  838  702 d  836  701 d  835  700 d
  834  696 d  834  692 d  835  688 d  837  686 d  840  686 d  842  687 d
  843  688 d  844  690 d  847  698 d  848  700 d  850  702 d  840  686 r
  842  688 d  843  690 d  846  698 d  847  700 d  848  701 d  850  702 d
  855  702 d  858  700 d  859  696 d  859  692 d  858  688 d  856  687 d
  854  686 d  859  686 d  856  687 d  834  718 r  859  710 d  834  718 r
  859  726 d  837  718 r  859  725 d  853  712 r  853  724 d  859  706 r
  859  713 d  859  722 r  859  730 d  837  750 r  840  752 d  834  752 d
  837  750 d  835  748 d  834  744 d  834  742 d  835  738 d  837  736 d
  840  735 d  843  734 d  849  734 d  853  735 d  855  736 d  858  738 d
  859  742 d  859  744 d  858  748 d  855  750 d  834  742 r  835  740 d
  837  737 d  840  736 d  843  735 d  849  735 d  853  736 d  855  737 d
  858  740 d  859  742 d  849  750 r  859  750 d  849  752 r  859  752 d
  849  747 r  849  754 d  834  762 r  859  762 d  834  764 r  859  764 d
  841  771 r  850  771 d  834  759 r  834  778 d  840  778 d  834  777 d
  846  764 r  846  771 d  859  759 r  859  778 d  853  778 d  859  777 d
   45  379 r   78  379 d   45  380 r   75  400 d   48  380 r   78  400 d
   45  400 r   78  400 d   45  374 r   45  380 d   45  395 r   45  404 d
   78  374 r   78  384 d   56  416 r   74  416 d   77  417 d   78  422 d
   78  425 d   77  430 d   74  433 d   56  417 r   74  417 d   77  419 d
   78  422 d   56  433 r   78  433 d   56  435 r   78  435 d   56  411 r
   56  417 d   56  428 r   56  435 d   78  433 r   78  440 d   56  451 r
   78  451 d   56  452 r   78  452 d   61  452 r   58  456 d   56  460 d
   56  464 d   58  468 d   61  470 d   78  470 d   56  464 r   58  467 d
   61  468 d   78  468 d   61  470 r   58  473 d   56  478 d   56  481 d
   58  486 d   61  488 d   78  488 d   56  481 r   58  484 d   61  486 d
   78  486 d   56  446 r   56  452 d   78  446 r   78  457 d   78  464 r
   78  475 d   78  481 r   78  492 d   45  504 r   78  504 d   45  505 r
   78  505 d   61  505 r   58  508 d   56  512 d   56  515 d   58  520 d
   61  523 d   66  524 d   69  524 d   74  523 d   77  520 d   78  515 d
   78  512 d   77  508 d   74  505 d   56  515 r   58  518 d   61  521 d
   66  523 d   69  523 d   74  521 d   77  518 d   78  515 d   45  499 r
   45  505 d   66  536 r   66  555 d   62  555 d   59  553 d   58  552 d
   56  548 d   56  544 d   58  539 d   61  536 d   66  534 d   69  534 d
   74  536 d   77  539 d   78  544 d   78  547 d   77  552 d   74  555 d
   66  553 r   61  553 d   58  552 d   56  544 r   58  540 d   61  537 d
   66  536 d   69  536 d   74  537 d   77  540 d   78  544 d   56  568 r
   78  568 d   56  569 r   78  569 d   66  569 r   61  571 d   58  574 d
   56  577 d   56  582 d   58  584 d   59  584 d   61  582 d   59  580 d
   58  582 d   56  563 r   56  569 d   78  563 r   78  574 d   56  630 r
   58  625 d   61  622 d   66  620 d   69  620 d   74  622 d   77  625 d
   78  630 d   78  633 d   77  638 d   74  641 d   69  643 d   66  643 d
   61  641 d   58  638 d   56  633 d   56  630 d   58  627 d   61  624 d
   66  622 d   69  622 d   74  624 d   77  627 d   78  630 d   78  633 r
   77  636 d   74  640 d   69  641 d   66  641 d   61  640 d   58  636 d
   56  633 d   46  664 r   48  662 d   50  664 d   48  665 d   46  665 d
   45  664 d   45  660 d   46  657 d   50  656 d   78  656 d   45  660 r
   46  659 d   50  657 d   78  657 d   56  651 r   56  662 d   78  651 r
   78  662 d   45  704 r   78  704 d   45  705 r   74  715 d   45  704 r
   78  715 d   45  726 r   78  715 d   45  726 r   78  726 d   45  728 r
   78  728 d   45  699 r   45  705 d   45  726 r   45  732 d   78  699 r
   78  708 d   78  721 r   78  732 d   56  750 r   58  745 d   61  742 d
   66  740 d   69  740 d   74  742 d   77  745 d   78  750 d   78  753 d
   77  758 d   74  761 d   69  763 d   66  763 d   61  761 d   58  758 d
   56  753 d   56  750 d   58  747 d   61  744 d   66  742 d   69  742 d
   74  744 d   77  747 d   78  750 d   78  753 r   77  756 d   74  760 d
   69  761 d   66  761 d   61  760 d   58  756 d   56  753 d   45  792 r
   78  792 d   45  793 r   78  793 d   61  792 r   58  788 d   56  785 d
   56  782 d   58  777 d   61  774 d   66  772 d   69  772 d   74  774 d
   77  777 d   78  782 d   78  785 d   77  788 d   74  792 d   56  782 r
   58  779 d   61  776 d   66  774 d   69  774 d   74  776 d   77  779 d
   78  782 d   45  787 r   45  793 d   78  792 r   78  798 d   66  808 r
   66  827 d   62  827 d   59  825 d   58  824 d   56  820 d   56  816 d
   58  811 d   61  808 d   66  806 d   69  806 d   74  808 d   77  811 d
   78  816 d   78  819 d   77  824 d   74  827 d   66  825 r   61  825 d
   58  824 d   56  816 r   58  812 d   61  809 d   66  808 d   69  808 d
   74  809 d   77  812 d   78  816 d   45  840 r   78  840 d   45  841 r
   78  841 d   45  835 r   45  841 d   78  835 r   78  846 d   58  870 r
   56  872 d   61  872 d   58  870 d   56  865 d   56  860 d   58  856 d
   59  854 d   61  854 d   64  856 d   66  859 d   69  867 d   70  870 d
   72  872 d   61  854 r   62  856 d   64  859 d   67  867 d   69  870 d
   72  872 d   75  872 d   77  870 d   78  865 d   78  860 d   77  856 d
   74  854 d   78  854 d   77  856 d  697  119 r  699  116 d  699  124 d
  697  119 d  694  123 d  689  124 d  686  124 d  681  123 d  678  119 d
  676  116 d  675  111 d  675  103 d  676   99 d  678   95 d  681   92 d
  686   91 d  689   91 d  694   92 d  697   95 d  686  124 r  683  123 d
  680  119 d  678  116 d  676  111 d  676  103 d  678   99 d  680   95 d
  683   92 d  686   91 d  697  103 r  697   91 d  699  103 r  699   91 d
  692  103 r  702  103 d  713  110 r  713  108 d  712  108 d  712  110 d
  713  111 d  716  113 d  723  113 d  726  111 d  728  110 d  729  107 d
  729   95 d  731   92 d  732   91 d  728  110 r  728   95 d  729   92 d
  732   91 d  734   91 d  728  107 r  726  105 d  716  103 d  712  102 d
  710   99 d  710   95 d  712   92 d  716   91 d  721   91 d  724   92 d
  728   95 d  716  103 r  713  102 d  712   99 d  712   95 d  713   92 d
  716   91 d  792  121 r  793  124 d  793  118 d  792  121 d  790  123 d
  785  124 d  779  124 d  774  123 d  771  119 d  771  116 d  772  113 d
  774  111 d  777  110 d  787  107 d  790  105 d  793  102 d  771  116 r
  774  113 d  777  111 d  787  108 d  790  107 d  792  105 d  793  102 d
  793   95 d  790   92 d  785   91 d  779   91 d  774   92 d  772   94 d
  771   97 d  771   91 d  772   94 d  806  124 r  806   91 d  808  124 r
  827   94 d  808  121 r  827   91 d  827  124 r  827   91 d  801  124 r
  808  124 d  822  124 r  832  124 d  801   91 r  811   91 d  843  124 r
  843  100 d  844   95 d  848   92 d  852   91 d  856   91 d  860   92 d
  864   95 d  865  100 d  865  124 d  844  124 r  844  100 d  846   95 d
  849   92 d  852   91 d  838  124 r  849  124 d  860  124 r  870  124 d
  880  113 r  878  111 d  880  110 d  881  111 d  880  113 d  880   94 r
  878   92 d  880   91 d  881   92 d  880   94 d  958  124 r  958   91 d
  960  124 r  960   91 d  948  124 r  947  116 d  947  124 d  971  124 d
  971  116 d  969  124 d  953   91 r  964   91 d  982  124 r  982   91 d
  984  124 r  984   91 d  984  108 r  987  111 d  992  113 d  995  113 d
 1000  111 d 1001  108 d 1001   91 d  995  113 r  998  111 d 1000  108 d
 1000   91 d  977  124 r  984  124 d  977   91 r  988   91 d  995   91 r
 1006   91 d 1016  103 r 1035  103 d 1035  107 d 1033  110 d 1032  111 d
 1028  113 d 1024  113 d 1019  111 d 1016  108 d 1014  103 d 1014  100 d
 1016   95 d 1019   92 d 1024   91 d 1027   91 d 1032   92 d 1035   95 d
 1033  103 r 1033  108 d 1032  111 d 1024  113 r 1020  111 d 1017  108 d
 1016  103 d 1016  100 d 1017   95 d 1020   92 d 1024   91 d 1054  113 r
 1049  111 d 1046  108 d 1044  103 d 1044  100 d 1046   95 d 1049   92 d
 1054   91 d 1057   91 d 1062   92 d 1065   95 d 1067  100 d 1067  103 d
 1065  108 d 1062  111 d 1057  113 d 1054  113 d 1051  111 d 1048  108 d
 1046  103 d 1046  100 d 1048   95 d 1051   92 d 1054   91 d 1057   91 r
 1060   92 d 1064   95 d 1065  100 d 1065  103 d 1064  108 d 1060  111 d
 1057  113 d 1080  113 r 1080   91 d 1081  113 r 1081   91 d 1081  103 r
 1083  108 d 1086  111 d 1089  113 d 1094  113 d 1096  111 d 1096  110 d
 1094  108 d 1092  110 d 1094  111 d 1075  113 r 1081  113 d 1075   91 r
 1086   91 d 1105  113 r 1115   91 d 1107  113 r 1115   94 d 1124  113 r
 1115   91 d 1112   84 d 1108   81 d 1105   79 d 1104   79 d 1102   81 d
 1104   83 d 1105   81 d 1100  113 r 1112  113 d 1120  113 r 1129  113 d
e
EndPSPlot
0 SPE
4787 15917 XY 0 SPB
 clear Figure end restore 
0 SPE 5509 16275 XY F25(Figure)S 167 x(5.)S 166 x(Ga)S 166 x(SNU)S
165 x(Theory)S 19583 Y 5347 X F34(\(Kamiok)S -32 x(ande)S 239 x(I)S
15 x(I\))S 238 x(b)S -15 x(y)S 238 x(replacing)S 238 x(the)S 239 x
(calculated)S -198 y 238 x F13(8)S 198 y 25 x F34(B)S 238 x
(neutrino)S 238 x(\015uxes)S 239 x(b)S -15 x(y)S 238 x(the)S 239 x
(v)S -30 x(alues)S 678 y 5347 X(measured)S 176 x(in)S 176 x(the)S 
178 x(Kamio)S -2 x(k)S -30 x(ande)S 177 x(exp)S 15 x(erimen)S -15 x
(t.)S 176 x(None)S 177 x(of)S 176 x(the)S 177 x(1000)S 176 x(solar)S
175 x(mo)S 14 x(dels)S 177 x(yield)S 176 x(a)S 677 y 5347 X(rate)S
200 x(as)S 201 x(lo)S -16 x(w)S 200 x(as)S 201 x(the)S 201 x(observ)S
-15 x(ed)S 201 x(rate)S 200 x(in)S 201 x(the)S 201 x(gall)S -2 x
(ium)S 200 x(exp)S 15 x(erimen)S -15 x(ts.)S 200 x(In)S 201 x(fact,)S
200 x(the)S 201 x(lo)S -16 x(w)S -15 x(est)S 678 y 5347 X(v)S -31 x
(alue)S 128 x(obtained|after)S 128 x(forcing)S 127 x(the)S -198 y 
129 x F13(8)S 198 y 25 x F34(B)S 128 x(\015ux)S 129 x(to)S 128 x
(agree)S 128 x(with)S 128 x(the)S 129 x(Kamiok)S -32 x(ande)S 129 x
(v)S -31 x(alue|is:)S 22971 Y 8945 X(Lo)S -15 x(w)S -15 x(est)S 182 x
(Galli)S -2 x(um)S 182 x(Rate)S 181 x(\()S -226 y F13(8)S 226 y 25 x 
F34(B)S 182 x(adjusted\))S 334 x(=)S 333 x(115)S 181 x(SNU)S F35(:)S
23692 X F34(\(8\))S 23972 Y 6165 X(It)S 251 x(is)S 251 x(di\016cult)S
250 x(to)S 251 x(en)S -14 x(vision)S 250 x(an)S 251 x(astroph)S -16 x
(ysical)S 250 x(solution)S 250 x(that)S 252 x(reconciles)S 251 x
(the)S 252 x(gal-)S 677 y 5347 X(lium)S 233 x(and)S 234 x(the)S 236 x
(Kamiok)S -32 x(ande)S 235 x(exp)S 15 x(erimen)S -15 x(ts.)S 234 x
(The)S 235 x(reason)S 235 x(is)S 234 x(that)S 235 x(w)S -15 x(e)S 
235 x(kno)S -16 x(w)S 235 x(from)S 233 x(the)S 678 y 5347 X(Kamio)S
-2 x(k)S -30 x(ande)S 210 x(measuremen)S -16 x(ts)S 210 x(that)S 
209 x(the)S -198 y 210 x F13(8)S 198 y 25 x F34(B)S 209 x(neutrino)S
209 x(\015ux)S 209 x(is)S 209 x(reduced)S 211 x(b)S -16 x(y)S 210 x
(only)S 208 x(a)S 209 x(fac-)S 677 y 5347 X(tor)S 191 x(of)S 192 x
(t)S -15 x(w)S -16 x(o.)S 192 x(Moreo)S -16 x(v)S -15 x(er,)S 191 x
(it)S 192 x(is)S 191 x(m)S -15 x(uc)S -15 x(h)S 192 x(easier)S 192 x
(to)S 192 x(reduce)S 193 x(the)S 192 x(sensitiv)S -15 x(e)S -198 y 
192 x F13(8)S 198 y 25 x F34(B)S 192 x(rate)S 192 x(\(whic)S -15 x
(h)S 192 x(is)S 678 y 5347 X(appro)S -16 x(xima)S -2 x(tely)S 202 x
(prop)S 15 x(ortio)S -2 x(nal)S 201 x(to)S 202 x F35(T)S -198 y 75 x 
F13(18)S 333 y -499 x F14(c)S -135 y 341 x F34(\))S 201 x(than)S 
202 x(it)S 201 x(is)S 202 x(to)S 201 x(reduce)S 203 x(the)S 202 x(m)S
-15 x(uc)S -15 x(h)S 202 x(less)S 202 x(sensitiv)S -16 x(e)S 479 y 
5347 X F13(7)S 198 y 24 x F34(Be)S 165 x(rate)S 164 x(\(whic)S -15 x
(h)S 163 x(is)S 164 x(appro)S -16 x(xima)S -2 x(tely)S 164 x(prop)S
15 x(ortiona)S -2 x(l)S 164 x(to)S 163 x F35(T)S -198 y 76 x F13(8)S
333 y -288 x F14(c)S -135 y 129 x F34(\).)S 164 x(The)S 164 x
(accurately-calculated)S 678 y 5347 X F35(p)S 106 x F36(\000)S 107 x 
F35(p)S 174 x F34(and)S 174 x F35(p)S 107 x F36(\000)S 106 x F35(e)S
106 x F36(\000)S 107 x F35(p)S 174 x F34(neutrino)S 174 x(\015uxes)S
175 x(represen)S -14 x(t,)S 174 x(together)S 174 x(with)S 174 x(the)S
-198 y 175 x F13(8)S 198 y 24 x F34(B)S 175 x(neutrino)S 174 x
(\015ux)S 677 y 5347 X(observ)S -16 x(ed)S 160 x(b)S -16 x(y)S 159 x
(the)S 159 x(Kamiok)S -32 x(ande)S 159 x(exp)S 16 x(erimen)S -16 x
(t,)S 158 x(82)S 159 x(SNU)S 158 x(in)S 159 x(a)S 158 x(gall)S -2 x
(ium)S 158 x(exp)S 16 x(erimen)S -16 x(t,)S 158 x(more)S 678 y 5347 X
(than)S 156 x(the)S 157 x(b)S 15 x(est-)S 157 x(estimate)S 156 x
(measured)S 157 x(rates)S 156 x(in)S 156 x(the)S 157 x(GALLEX)S 157 x
(and)S 156 x(in)S 156 x(the)S 157 x(SA)S -15 x(GE)S 156 x(exp)S 16 x
(er-)S 677 y 5347 X(imen)S -16 x(ts.)S 179 x(An)S 178 x(astroph)S 
-15 x(ysical)S 178 x(solutio)S -2 x(n)S 179 x(of)S 178 x(the)S 180 x
(solar)S 177 x(neutrino)S 178 x(problem)S 178 x(m)S -15 x(ust,)S 
178 x(therefore,)S 678 y 5347 X(reduce)S 220 x(the)S 220 x(36)S 218 x
(SNU)S 220 x(from)S 217 x(the)S 220 x(standard-mo)S 14 x(del)S 219 x
(calculation)S 218 x(of)S 218 x(the)S -198 y 220 x F13(7)S 198 y 
25 x F34(Be)S 220 x(rate)S 219 x(b)S -15 x(y)S 219 x(at)S 677 y 
5347 X(least)S 155 x(as)S 156 x(m)S -16 x(uc)S -14 x(h)S 156 x(and)S
155 x(p)S 15 x(ossibly)S 155 x(more)S 155 x(than)S 156 x(the)S 156 x
(reduction)S 156 x(observ)S -15 x(ed)S 156 x(b)S -15 x(y)S 155 x
(Kamiok)S -31 x(ande)S 156 x(for)S 677 y 5347 X(the)S -198 y 182 x 
F13(8)S 198 y 25 x F34(B)S 182 x(\015ux.)S
1 PP EP

1000 BP 39600 30600 PM 0 0 XY
4389 Y 10091 X F34(T)S -46 x(able)S 182 x(2.)S 181 x(New)S 182 x
(Solar)S 181 x(Neutrino)S 181 x(Observ)S -30 x(atories)S 827 y 11066 X
(T)S -15 x(ypical)S 181 x(Ev)S -15 x(en)S -15 x(t)S 182 x(Rates)S 
-70 y 182 x F35(>)S
XP /F36 24 425 3 6 30 16 11 32 0
<0FC00010 1FF00010 3FF80010 787C0030 701E0030 E00F0070 C00780E0
 C003E1E0 8001FFC0 8000FF80 80003F00>
PXLC RP
16130 5357 XY F36(\030)S -141 y 333 x F34(3)S 121 x F36(\002)S 122 x 
F34(10)S -198 y F13(3)S 198 y 206 x F34(yr)S -198 y F15(\000)S F13
(1)S 720 y 8390 X 12960 24 R 773 y 8743 X F34(Observ)S -30 x(atory)S
XP /F35 69 404 2 0 33 30 31 32 0
<00FFFFFF 00FFFFFF 000F003E 000F000E 000F000E 001F000E 001E0006
 001E000E 001E000C 003E060C 003C060C 003C0600 003C0E00 007C1C00
 007FFC00 007FFC00 00783C00 00F81800 00F01800 00F0180C 00F0181C
 01F00018 01E00038 01E00030 01E00070 03E000E0 03C000E0 03C003E0
 07C00FC0 7FFFFFC0 FFFFFF80>
PXLC RP
12309 6511 XY F35(E)S
XP /F14 84 247 1 0 23 22 23 24 0
<3FFFFE 3FFFFE 3C3C1C 703C0C 607C0C E07C0C C0780C C0780C 00F800
 00F800 00F000 00F000 01F000 01F000 01E000 01E000 03E000 03E000
 03C000 03C000 07C000 7FFC00 FFFC00>
PXLC RP
12712 6600 XY F14(T)S
XP /F14 104 244 1 0 18 22 23 24 0
<1F8000 1F8000 070000 070000 0F0000 0F0000 0E0000 0E0000 1E0000
 1EFC00 1FFE00 1F8E00 3E0E00 3E0E00 3C0E00 380E00 781E00 781C00
 701CC0 703DC0 F039C0 F03F80 E01F00>
PXLC RP
13017 6600 XY F14(h)S -89 y 25 x F34(\()S F35(\027)S 35 x F34(\))S
17077 X(Reaction\(s\))S 678 y 12374 X(\(MeV\))S 522 y 8390 X 
12960 24 R 169 y 8390 X 12960 24 R 773 y 8689 X(SNO)S 12813 X(6.4)S
15145 X F35(\027)S 82 y F14(e)S -82 y 147 x F34(+)S -198 y 122 x F13
(2)S 198 y 24 x F34(H)S 152 x F36(!)S 533 x F35(p)S 122 x F34(+)S 
121 x F35(p)S 122 x F34(+)S 121 x F35(e)S -198 y F15(\000)S 9629 Y 
12813 X F34(2.2)S 15238 X F35(\027)S 157 x F34(+)S -198 y 121 x F13
(2)S 198 y 25 x F34(H)S 151 x F36(!)S
XP /F35 110 327 1 0 25 19 20 32 0
<1F07E000 3F9FF800 37FC7800 63F07800 63E03C00 E3C07C00 C7C07800
 07807800 07807800 0780F800 0F80F000 0F00F000 0F01F000 0F01E380
 1F01E300 1E03E300 1E03C700 1E03CE00 3E01FC00 1C00F800>
PXLC RP
18178 9629 XY F35(n)S 122 x F34(+)S 121 x F35(p)S 122 x F34(+)S 121 x 
F35(\027)S 10606 Y 13026 X F34(5)S 15257 X F35(\027)S 157 x F34(+)S
121 x F35(e)S -198 y F15(\000)S 198 y 177 x F36(!)S 645 x F35(\027)S
156 x F34(+)S 122 x F35(e)S -198 y F15(\000)S 12329 Y 8689 X F34
(Sup)S 15 x(er-)S 13026 X(5)S 15257 X F35(\027)S 157 x F34(+)S 121 x 
F35(e)S -198 y F15(\000)S 198 y 177 x F36(!)S 645 x F35(\027)S 156 x 
F34(+)S 122 x F35(e)S -198 y F15(\000)S 876 y 8689 X F34(Kamiok)S 
-32 x(ande)S 14739 Y 8689 X(ICAR)S -16 x(US)S 12601 X F36(\030)S 
152 x F34(10)S 14932 X F35(\027)S 81 y F14(e)S -81 y 147 x F34(+)S
-198 y 122 x F13(40)S 198 y 24 x F34(Ar)S 151 x F36(!)S 321 x F35(e)S
-198 y F15(\000)S 198 y 146 x F34(+)S -198 y 122 x F13(40)S 198 y 
24 x F34(K)S
XP /F15 3 212 2 1 14 14 14 16 0
<0200 0200 0200 C218 F278 3AE0 0F80 0F80 3AE0 F278 C218 0200 0200
 0200>
PXLC RP
20325 14441 XY F15(\003)S 15715 Y 13026 X F34(5)S 15257 X F35(\027)S
157 x F34(+)S 121 x F35(e)S -198 y F15(\000)S 198 y 177 x F36(!)S 
645 x F35(\027)S 156 x F34(+)S 122 x F35(e)S -198 y F15(\000)S 17438 Y 
8689 X F34(BOREXINO)S 12813 X(0.4)S 14612 X F35(\027)S 35 x F34(\()S
-198 y F13(7)S 198 y 25 x F34(Be\))S 122 x(+)S 122 x F35(e)S -198 y 
F15(\000)S 198 y 177 x F36(!)S F35(\027)S 35 x F34(\()S -198 y F13
(7)S 198 y 25 x F34(Be\))S 122 x(+)S 121 x F35(e)S -198 y F15(\000)S
19162 Y 8689 X F34(HELLA)S
XP /F34 90 334 3 0 24 30 31 24 0
<7FFFF8 7FFFF8 7E01F8 7803F0 7003F0 6007E0 E007C0 E00FC0 C01F80
 C01F80 C03F00 003F00 007E00 007E00 00FC00 00F800 01F800 03F000
 03F00C 07E00C 07E00C 0FC00C 0FC01C 1F801C 1F0018 3F0018 7E0038
 7E0078 FC01F8 FFFFF8 FFFFF8>
PXLC RP
10560 19162 XY F34(Z)S 12813 X(0.1)S 15257 X F35(\027)S 157 x F34(+)S
121 x F35(e)S -198 y F15(\000)S 198 y 177 x F36(!)S 645 x F35(\027)S
156 x F34(+)S 122 x F35(e)S -198 y F15(\000)S 720 y 8390 X 
12960 24 R
XP /F48 55 336 3 0 25 33 34 24 0
<700000 7C0000 7FFFFE 7FFFFE 7FFFFE 7FFFFE 7FFFFC 7FFFF8 F000F0
 E001E0 E001C0 E003C0 E00780 000F00 001E00 001E00 003C00 003C00
 007800 00F800 00F800 00F800 01F000 01F000 01F000 01F000 03F000
 03F000 03F000 03F000 03F000 03F000 03F000 01E000>
PXLC RP
5347 21776 XY F48(7.)S 672 x(New)S 224 x(Exp)S 18 x(erimen)S -19 x
(ts)S 22759 Y 6165 X F34(Gene)S 138 x(Beier)S 139 x(wil)S -2 x(l)S
138 x(discuss)S 138 x(more)S 137 x(in)S 137 x(detail)S 137 x(in)S 
137 x(this)S 138 x(session)S 138 x(the)S 138 x(new)S 138 x(exp)S 
15 x(erimen)S -15 x(ts)S 138 x(that)S 677 y 5347 X(are)S 247 x
(already)S 246 x(funded.)S 247 x(Ho)S -15 x(w)S -15 x(ev)S -15 x
(er,)S 247 x(I)S 247 x(cannot)S 247 x(close)S 248 x(this)S 247 x
(talk)S 246 x(without)S 247 x(at)S 247 x(least)S 248 x(taking)S 678 y 
5347 X(note)S 248 x(of)S 248 x(the)S 249 x(exciting)S 248 x(p)S 15 x
(ossibili)S -2 x(ties)S 249 x(for)S 247 x(resolving)S 247 x(the)S 
249 x(solar)S 247 x(neutrino)S 248 x(problem)S 247 x(and)S 677 y 
5347 X(of)S 242 x(learning)S 242 x(new)S 243 x(ph)S -15 x(ysics)S 
243 x(and)S 243 x(new)S 243 x(astronom)S -16 x(y)S 243 x(that)S 243 x
(are)S 243 x(represen)S -15 x(ted)S 244 x(b)S -15 x(y)S 243 x(the)S
243 x(next)S 678 y 5347 X(generation)S 181 x(of)S 181 x(exp)S 16 x
(erimen)S -16 x(ts.)S 677 y 6165 X(T)S -46 x(able)S 265 x(2)S 264 x
(lists)S 264 x(the)S 265 x(\014v)S -15 x(e)S 264 x(new)S 265 x
(solar)S 264 x(neutrino)S 264 x(exp)S 15 x(erimen)S -16 x(ts)S 265 x
(that)S 265 x(are)S 264 x(funded)S 265 x(for)S 678 y 5347 X(op)S 
15 x(eration)S 256 x(or)S 256 x(for)S 256 x(dev)S -15 x(elopmen)S 
-15 x(t.)S 256 x(Eac)S -15 x(h)S 257 x(of)S 257 x(the)S 257 x(mo)S
15 x(des)S 257 x(of)S 256 x(eac)S -14 x(h)S 257 x(of)S 256 x(the)S
258 x(exp)S 15 x(erimen)S -16 x(ts)S 677 y 5347 X(listed)S 205 x(in)S
205 x(T)S -46 x(able)S 205 x(2)S 205 x(is)S 205 x(exp)S 16 x(ected)S
207 x(to)S 205 x(yield)S 205 x(more)S 204 x(than)S 206 x(3,00)S -2 x
(0)S 206 x(neutrino)S 204 x(ev)S -14 x(en)S -15 x(ts)S 206 x(p)S 
15 x(er)S 205 x(y)S -15 x(ear)S 678 y 5347 X(\(except)S 166 x(for)S
165 x(the)S 165 x F35(\027)S 123 x F36(\000)S 89 x F35(e)S 165 x F34
(scattering)S 165 x(mo)S 14 x(de)S 166 x(of)S 164 x(SNO\).)S 166 x
(In)S 165 x(one)S 165 x(y)S -15 x(ear,)S 164 x(eac)S -14 x(h)S 165 x
(exp)S 16 x(erimen)S -16 x(t)S 165 x(will)S 677 y 5347 X(record)S 
154 x(more)S 154 x(than)S 155 x(three)S 155 x(times)S 155 x(the)S 
155 x(total)S 154 x(n)S -15 x(um)S -15 x(b)S 15 x(er)S 155 x(of)S 
154 x(neutrino)S 154 x(ev)S -15 x(en)S -15 x(ts)S 155 x(that)S 155 x
(ha)S -15 x(v)S -16 x(e)S 156 x(b)S 15 x(een)S 677 y 5347 X(coun)S
-15 x(ted)S 201 x(to)S 200 x(date)S 201 x(in)S 201 x(all)S 199 x
(solar)S 199 x(neutrino)S 201 x(exp)S 15 x(erimen)S -15 x(ts)S 200 x
(since)S 202 x(the)S 201 x(c)S -15 x(hlorine)S 200 x(exp)S 15 x
(erimen)S -15 x(t)S 678 y 5347 X(b)S 15 x(egan)S 134 x(op)S 15 x
(erating)S 134 x(a)S 134 x(quarter)S 134 x(of)S 134 x(a)S 134 x(cen)S
-14 x(tury)S 134 x(ago.)S 133 x(With)S 133 x(this)S 135 x(greater)S
134 x(statistical)S 133 x(accuracy)S -45 x(,)S 677 y 5347 X(solar)S
180 x(neutrino)S 182 x(ph)S -15 x(ysics)S 181 x(will)S 181 x(b)S 
15 x(ecome)S 182 x(a)S 182 x(more)S 181 x(precise)S 182 x(sub)S 31 x
(ject.)S 784 y 6165 X(The)S 175 x(exp)S 15 x(erimen)S -15 x(ts)S 
174 x(are)S 175 x(listed)S 174 x(in)S 174 x(order)S 174 x(of)S 174 x
(their)S 174 x(exp)S 15 x(ecte)S 2 x(d)S 174 x(completion)S 174 x
(dates:)S 174 x(SNO)S 677 y 5347 X(\(1996)S -198 y F13(2)S -2 x(9)S
198 y 25 x F34(\),)S 121 x(Sup)S 15 x(erk)S -30 x(amio)S -2 x(k)S 
-30 x(ande)S 121 x(\(1996;)S 120 x(see)S 122 x(Ref.)S -198 y F13
(30,)S -2 x(31)S 198 y 25 x F34(\),)S 121 x(BOREXINO)S 122 x(\()S 
F36(\025)S 121 x F34(1996;)S -198 y F13(3)S -2 x(2)S 198 y 25 x F34
(\),)S 121 x(ICAR)S -16 x(US)S 678 y 5347 X(\(1998;)S 143 x(see)S 
146 x(Ref.)S -198 y F13(3)S -2 x(3)S 198 y 25 x F34(\),)S 144 x(and)S
145 x(HELLAZ)S 145 x(\(prop)S 15 x(osed,)S 144 x(not)S 145 x(y)S 
-15 x(et)S 145 x(appro)S -16 x(v)S -15 x(ed)S -198 y F13(34)S 198 y 
24 x F34(\).)S 145 x(T)S -46 x(able)S 145 x(2)S 144 x(lists)S 144 x
(the)S 677 y 5347 X(neutrino)S 222 x(threshold)S 222 x(energy)S 222 x
(for)S 222 x(eac)S -15 x(h)S 222 x(reaction)S 222 x(mo)S 15 x(de)S
223 x(and)S 222 x(the)S 223 x(indivi)S -2 x(dual)S 222 x(reactions)S
1 PP EP

1000 BP 39600 30600 PM 0 0 XY
3815 Y 5347 X F34(that)S 198 x(will)S 198 x(b)S 15 x(e)S 199 x
(observ)S -15 x(ed.)S 198 x(I)S 199 x(ha)S -15 x(v)S -16 x(e)S 200 x
(not)S 198 x(listed)S 199 x(other)S 198 x(promisi)S -2 x(ng)S 199 x
(exp)S 15 x(erimen)S -15 x(tal)S 198 x(prop)S 14 x(osals)S 678 y 
5347 X(b)S 15 x(ecause)S 243 x(it)S 241 x(is)S 241 x(not)S 242 x(y)S
-15 x(et)S 242 x(clear)S 241 x(whic)S -15 x(h)S 242 x(of)S 241 x
(these)S 243 x(p)S 15 x(ossibili)S -2 x(ties)S 242 x(will)S 240 x
(receiv)S -15 x(e)S 242 x(funding.)S 241 x(In)S 677 y 5347 X
(particula)S -2 x(r,)S 191 x(a)S 190 x(protot)S -16 x(yp)S 16 x(e)S
191 x(detector)S 192 x(of)S 190 x F35(pp)S 192 x F34(neutrinos)S 
191 x(maki)S -2 x(ng)S 191 x(use)S 192 x(of)S 190 x(the)S 192 x
(prop)S 14 x(erties)S 191 x(of)S 678 y 5347 X(sup)S 15 x(er\015uid)S
192 x(helium)S 192 x(has)S 192 x(b)S 15 x(een)S 193 x(teste)S 2 x(d)S
192 x(succes)S 2 x(sfull)S -2 x(y)S 193 x(and)S 192 x(app)S 15 x
(ears)S 192 x(to)S 192 x(b)S 16 x(e)S 192 x(feasible)S -198 y F13
(35,36)S 198 y 216 x F34(It)S 677 y 5347 X(is)S 225 x(clear,)S 225 x
(ho)S -15 x(w)S -16 x(ev)S -14 x(er,)S 225 x(that)S 226 x(the)S 226 x
(exp)S 15 x(erimen)S -15 x(ts)S 226 x(listed)S 225 x(in)S 226 x(T)S
-46 x(able)S 226 x(2)S 225 x(will)S 224 x(b)S 15 x(e)S 227 x
(insu\016cien)S -16 x(t)S 226 x(to)S 678 y 5347 X(uniquely)S 139 x
(solv)S -16 x(e)S 141 x(for)S 139 x(all)S 139 x(of)S 140 x(the)S 
141 x(fundamen)S -16 x(tal)S 140 x(neutrino)S 140 x(param)S -2 x
(eters.)S 141 x(Other)S 141 x(exp)S 15 x(erimen)S -16 x(ts)S 677 y 
5347 X(are)S 147 x(required)S 148 x(to)S 148 x(establish)S 148 x
(uniqueness)S 148 x(in)S 148 x(the)S 148 x(inferences)S 149 x(and)S
148 x(to)S 148 x(pro)S -16 x(vide)S 148 x(a)S 147 x(measure)S 148 x
(of)S 677 y 5347 X(redundancy)S 185 x(to)S 186 x(assure)S 185 x
(ourselv)S -16 x(es)S 186 x(that)S 186 x(systematic)S 185 x(exp)S 
15 x(erimen)S -15 x(tal)S 185 x(uncertain)S -16 x(ties)S 186 x(ha)S
-15 x(v)S -16 x(e)S 678 y 5347 X(not)S 181 x(misled)S 181 x(us.)S 
677 y 6165 X(As)S 200 x(Gene)S 201 x(will)S 199 x(explain)S 199 x
(in)S 199 x(detail)S 200 x(in)S 199 x(his)S 200 x(talk,)S 199 x(the)S
200 x(SNO)S 201 x(exp)S 15 x(erimen)S -15 x(t)S 200 x(will)S 198 x
(test)S 201 x(the)S 678 y 5347 X(inference)S 206 x(that)S 206 x(ph)S
-15 x(ysics)S 206 x(b)S 15 x(ey)S -15 x(ond)S 205 x(the)S 207 x
(standard)S 205 x(mo)S 14 x(del)S 206 x(is)S 206 x(required)S 205 x
(in)S 205 x(three)S 207 x(di\013eren)S -15 x(t)S 677 y 5347 X(w)S 
-16 x(a)S -15 x(ys)S 211 x(that)S 210 x(are)S 210 x(all)S 209 x
(indep)S 16 x(enden)S -15 x(t)S 211 x(of)S 210 x(solar)S 209 x(mo)S
15 x(dels.)S 210 x(They)S 210 x(are:)S 210 x(1\))S 210 x(A)S 211 x
(measuremen)S -15 x(t)S 210 x(of)S 678 y 5347 X(the)S 152 x(total)S
151 x(neutrino)S 151 x(\015ux)S 152 x(indep)S 15 x(enden)S -14 x(t)S
152 x(of)S 151 x(\015a)S -16 x(v)S -15 x(or)S 151 x(in)S 152 x(the)S
152 x(neutral)S 151 x(curren)S -15 x(t)S 152 x(reaction)S 151 x
(\(neu-)S 677 y 5347 X(trino)S 239 x(disin)S -16 x(tegration)S 239 x
(of)S 240 x(deuterium\);)S 240 x(2\))S 240 x(A)S 240 x(measuremen)S
-15 x(t)S 240 x(of)S 240 x(the)S 241 x(energy)S 240 x(sp)S 16 x
(ectrum)S 678 y 5347 X(of)S 208 x(electron-\015a)S -15 x(v)S -15 x
(or)S 208 x(neutrinos)S 209 x(ab)S 15 x(o)S -15 x(v)S -16 x(e)S 210 x
(5)S 209 x(MeV)S 210 x(in)S 208 x(the)S 210 x(c)S -15 x(harged)S 
209 x(curren)S -15 x(t)S 209 x(reaction)S 209 x(\(neu-)S 677 y 5347 X
(trino)S 228 x(absorption)S 228 x(b)S -15 x(y)S 230 x(deuterium\);)S
228 x(and)S 230 x(3\))S 229 x(A)S 229 x(searc)S -14 x(h)S 229 x(for)S
229 x(da)S -16 x(y-nigh)S -16 x(t)S 230 x(v)S -31 x(ariati)S -2 x
(ons)S 230 x(that)S 678 y 5347 X(are)S 181 x(predicted)S 183 x(b)S
-15 x(y)S 181 x(some)S 182 x(MSW)S 181 x(solutions.)S 677 y 6165 X
(If)S 175 x(electron)S 175 x(neutrinos)S 175 x(oscillate)S 174 x(in)S
-15 x(to)S 175 x(m)S -16 x(uon)S 175 x(or)S 175 x(tau)S 175 x
(neutrinos,)S 174 x(then)S 176 x(the)S 176 x(\015ux)S 175 x(mea-)S
678 y 5347 X(sured)S 144 x(in)S 144 x(the)S 145 x(neutral)S 143 x
(curren)S -15 x(t)S 144 x(reaction)S 144 x(will)S 143 x(exceed)S 
146 x(the)S 144 x(\015ux)S 144 x(measured)S 144 x(in)S 144 x(the)S
145 x(c)S -15 x(harged)S 677 y 5347 X(curren)S -15 x(t)S 176 x
(reaction,)S 175 x(a)S 175 x(prediction)S 176 x(that)S 175 x(do)S 
15 x(es)S 177 x(not)S 176 x(dep)S 15 x(end)S 177 x(up)S 15 x(on)S 
175 x(an)S -15 x(y)S 176 x(solar)S 174 x(calculations.)S 677 y 5347 X
(As)S 214 x(emphasized)S 215 x(earlier,)S 213 x(the)S 215 x(shap)S
15 x(e)S 215 x(of)S 214 x(the)S 215 x(8)S 369 y -273 x(\026)S -369 y 
215 x(neutrino)S 214 x(energy)S 214 x(sp)S 16 x(ectrum)S 214 x(is)S
215 x(also)S 213 x(in-)S 678 y 5347 X(dep)S 15 x(enden)S -14 x(t)S
209 x(of)S 208 x(solar-m)S -2 x(o)S 15 x(del)S 209 x(uncertain)S 
-15 x(ties.)S 209 x(Da)S -16 x(y-nigh)S -16 x(t)S 209 x(v)S -31 x
(ariati)S -2 x(ons)S 209 x(are)S 209 x(exp)S 15 x(ecte)S 2 x(d)S 
209 x(only)S 677 y 5347 X(if)S 182 x(the)S 184 x(MSW)S 183 x
(e\013ect)S 184 x(con)S -15 x(v)S -15 x(erts)S 184 x(m)S -16 x(uon)S
183 x(or)S 183 x(tau)S 183 x(neutrinos)S 183 x(to)S 183 x(electron)S
183 x(neutrinos)S 183 x(as)S 183 x(solar)S 678 y 5347 X(neutrinos)S
181 x(pass)S 182 x(through)S 181 x(the)S 183 x(earth)S 181 x(at)S 
182 x(nigh)S -16 x(t)S 182 x(on)S 182 x(their)S 181 x(to)S 182 x
(the)S 182 x(detec)S 2 x(tor.)S 677 y 6165 X(Lik)S -16 x(e)S 144 x
(SNO,)S 144 x(ICAR)S -16 x(US)S 144 x(can)S 144 x(measure)S 143 x
(the)S 144 x(shap)S 16 x(e)S 144 x(of)S 143 x(the)S 144 x(8)S 369 y 
-273 x(\026)S -369 y 143 x(neutrino)S 143 x(energy)S 144 x(sp)S 16 x
(ectrum)S 678 y 5347 X(via)S 177 x(neutrino)S 178 x(absorption.)S 
177 x(Moreo)S -15 x(v)S -16 x(er,)S 179 x(ICAR)S -16 x(US)S 179 x
(has)S 178 x(a)S 178 x(unique)S 179 x(\\smoki)S -2 x(ng-gun")S 178 x
(signal)S 677 y 5347 X(for)S 181 x(neutrino)S 181 x(absorption,)S 
180 x(the)S
XP /F35 13 282 0 -10 22 19 30 24 0
<03F006 0FF80E 1FFC0C 3FFC1C 7FFE18 781E38 E00E30 C00770 000760
 000360 0003E0 0003C0 0003C0 000380 000380 000380 000300 000300
 000700 000700 000700 000700 000E00 000E00 000E00 000E00 001C00
 001C00 001C00 001800>
PXLC RP
12157 21429 XY F35(\015)S 213 x F34(deca)S -15 x(y)S 182 x(of)S 181 x
(the)S 183 x(excited)S 182 x(state)S 183 x(of)S -198 y 181 x F13(40)S
198 y 24 x F34(K.)S 678 y 6165 X(There)S 261 x(will)S 259 x(b)S 15 x
(e)S 261 x(w)S -15 x(elcome)S 261 x(redundancy)S 260 x(if)S 260 x
(all)S 259 x(exp)S 16 x(erimen)S -16 x(ts)S 261 x(op)S 15 x(erate)S
261 x(as)S 260 x(planned.)S 677 y 5347 X(Three)S 228 x(exp)S 16 x
(erimen)S -16 x(ts)S 228 x(\(Sup)S 16 x(erk)S -31 x(amiok)S -32 x
(ande,)S 228 x(SNO,)S 228 x(and)S 227 x(ICAR)S -15 x(US\))S 228 x
(will)S 226 x(measure)S 228 x(for)S 227 x(8)S 369 y -273 x(\026)S 
309 y 5347 X(neutrinos)S 175 x(the)S 177 x F35(\027)S 81 y F14(e)S
-81 y 136 x F36(\000)S 110 x F35(e)S 176 x F34(scattering)S 176 x
(rate)S 176 x(and)S 175 x(the)S 177 x(recoil)S 175 x(electron)S 177 x
(energy)S 176 x(sp)S 15 x(ectrum;)S 176 x(the)S 677 y 5347 X
(electron)S 136 x(recoil)S 134 x(sp)S 16 x(ectrum)S 135 x(re\015ect)S
2 x(s)S 135 x(the)S 136 x(incoming)S 134 x(neutrino)S 136 x(energy)S
135 x(sp)S 16 x(ectrum.)S 135 x(The)S 136 x(fact)S 677 y 5347 X
(that)S 192 x(the)S 193 x(Sup)S 16 x(erk)S -31 x(amiok)S -32 x(ande)S
193 x(exp)S 16 x(erimen)S -16 x(t)S 193 x(con)S -15 x(tains)S 192 x
(more)S 192 x(than)S 192 x(30)S 192 x(times)S 192 x(the)S 193 x
(\014ducial)S 678 y 5347 X(v)S -16 x(olume)S 178 x(for)S 177 x
(solar)S 176 x(neutrino)S 178 x(exp)S 15 x(erimen)S -16 x(ts)S 179 x
(as)S 177 x(the)S 179 x(highly-)S -2 x(pro)S 15 x(ductiv)S -15 x(e)S
178 x(Kamiok)S -32 x(ande)S 179 x(ex-)S 677 y 5347 X(p)S 15 x
(erimen)S -16 x(t)S 152 x(is)S 152 x(an)S 152 x(indicatio)S -2 x(n)S
152 x(of)S 152 x(the)S 152 x(amoun)S -16 x(t)S 152 x(of)S 151 x
(impro)S -16 x(v)S -15 x(emen)S -15 x(t)S 152 x(that)S 152 x(ma)S 
-16 x(y)S 152 x(b)S 15 x(e)S 152 x(exp)S 16 x(ected)S 153 x(in)S 
678 y 5347 X(the)S 186 x(next)S 186 x(generation)S 185 x(of)S 185 x
(solar)S 184 x(neutrino)S 185 x(exp)S 16 x(erimen)S -16 x(ts)S 186 x
(compared)S 185 x(to)S 185 x(those)S 186 x(p)S 16 x(erform)S -2 x
(ed)S 677 y 5347 X(to)S 181 x(date.)S 678 y 6165 X(The)S 189 x
(BOREXINO)S 190 x(and)S 188 x(HELLAZ)S 189 x(exp)S 16 x(erimen)S 
-16 x(ts)S 189 x(are)S 189 x(essen)S -14 x(tial)S 187 x(in)S 189 x
(order)S 188 x(to)S 189 x(distin-)S 677 y 5347 X(guish)S 156 x(b)S
15 x(et)S -15 x(w)S -15 x(een)S 157 x(di\013eren)S -15 x(t)S 157 x
(new-ph)S -15 x(ysics)S 156 x(p)S 15 x(ossibiliti)S -2 x(es.)S 157 x
(These)S 157 x(exp)S 15 x(erimen)S -15 x(ts)S 157 x(are)S 156 x(the)S
157 x(only)S 678 y 5347 X(ones)S 257 x(curren)S -15 x(tly)S 257 x
(under)S 258 x(dev)S -15 x(elopmen)S -15 x(t)S 257 x(that)S 258 x
(wil)S -2 x(l)S 257 x(measure)S 258 x(the)S 258 x(energy)S 257 x(of)S
257 x(indivi)S -2 x(dual)S 677 y 5347 X(ev)S -15 x(en)S -15 x(ts)S
161 x(with)S 160 x(energies)S 161 x(less)S 161 x(than)S 160 x(5)S 
161 x(MeV.)S 160 x(The)S 161 x(threshold)S 161 x(for)S 159 x
(BOREXINO)S 161 x(is)S 161 x(0.4)S 159 x(MeV)S 678 y 5347 X(and)S 
234 x(for)S 234 x(HELLAZ)S 235 x(is)S 234 x(0.1)S 234 x(MeV.)S 235 x
(These)S 235 x(exp)S 16 x(erimen)S -16 x(ts)S 235 x(m)S -15 x(ust)S
235 x(b)S 15 x(e)S 235 x(p)S 15 x(erformed)S 234 x(in)S 234 x(order)S
677 y 5347 X(to)S 168 x(determine)S 169 x(the)S 170 x(neutrino)S 
168 x(surviv)S -31 x(al)S 168 x(probabi)S -2 x(lit)S -16 x(y)S 169 x
(at)S 168 x(lo)S -15 x(w)S 168 x(energies.)S 169 x(The)S 169 x
(BOREXINO)S 678 y 5347 X(and)S 201 x(HELLAZ)S 203 x(exp)S 15 x
(erimen)S -15 x(ts)S 202 x(also)S 201 x(ha)S -15 x(v)S -16 x(e)S 
203 x(another)S 201 x(highly)S 201 x(desirable)S 201 x(feature;)S 
202 x(they)S 202 x(will)S 677 y 5347 X(b)S 15 x(oth)S 185 x(measure)S
186 x(the)S 186 x F35(\027)S 82 y F14(e)S -82 y 211 x F34(\015ux)S
185 x(at)S 185 x(a)S 186 x(sp)S 15 x(eci\014c)S 186 x(energy)S -45 x
(,)S 185 x(the)S 186 x(energy)S 185 x(\(0.86)S 184 x(MeV\))S 187 x
(of)S 184 x(the)S -198 y 186 x F13(7)S 198 y 25 x F34(Be)S 677 y 
5347 X(neutrino)S 244 x(line.)S 244 x(The)S 245 x(theoretical)S 245 x
(predictions)S 245 x(are)S 244 x(more)S 245 x(sp)S 15 x(eci\014c,)S
245 x(and)S 245 x(therefore)S 245 x(the)S
1 PP EP

1000 BP 39600 30600 PM 0 0 XY
3815 Y 5347 X F34(measuremen)S -16 x(ts)S 209 x(are)S 209 x(more)S
208 x(diagnosti)S -2 x(c,)S 209 x(when)S 209 x(the)S 209 x(neutrino)S
208 x(\015ux)S 209 x(at)S 208 x(a)S 209 x(sp)S 15 x(eci\014c)S 210 x
(energy)S 678 y 5347 X(is)S 181 x(observ)S -15 x(ed.)S 677 y 6165 X
(The)S 165 x(HELLAZ)S 165 x(exp)S 15 x(erimen)S -15 x(t)S 164 x(is)S
164 x(unique)S 165 x(among)S 163 x(the)S 165 x(exp)S 15 x(erimen)S
-15 x(ts)S 164 x(b)S 16 x(eing)S 164 x(dev)S -15 x(elop)S 15 x(ed;)S
678 y 5347 X(it)S 190 x(is)S 190 x(the)S 191 x(only)S 190 x(exp)S 
15 x(erimen)S -15 x(t)S 190 x(b)S 15 x(eing)S 191 x(dev)S -15 x
(elop)S 15 x(ed)S 191 x(to)S 190 x(observ)S -15 x(e)S 191 x(indivi)S
-2 x(dual)S 190 x(ev)S -15 x(en)S -15 x(ts)S 191 x(from)S 189 x(the)S
677 y 5347 X(basic)S 160 x F35(pp)S 162 x F34(reaction)S 160 x
(\(maxim)S -17 x(um)S 160 x(energy)S 161 x(0.4)S 160 x(MeV\).)S 161 x
(In)S 160 x(addition,)S 159 x(HELLAZ)S 161 x(migh)S -16 x(t)S 161 x
(ha)S -15 x(v)S -16 x(e)S 678 y 5347 X(the)S 265 x(energy)S 265 x
(resolution)S 263 x(to)S 265 x(measure)S 265 x(the)S 265 x
(predicted)S -198 y F13(37)S 198 y 290 x F34(1.29)S 263 x(k)S -15 x
(eV)S 265 x(shift)S 264 x(b)S 15 x(et)S -14 x(w)S -15 x(een)S 265 x
(the)S 677 y 5347 X(a)S -16 x(v)S -15 x(erage)S 241 x(energy)S 241 x
(of)S 241 x(the)S 242 x(solar)S -198 y 240 x F13(7)S 198 y 24 x F34
(Be)S 242 x(line)S 241 x(and)S 241 x(the)S 242 x(lab)S 14 x(oratory)S
240 x(energy)S 241 x(of)S 241 x(the)S 241 x(line.)S 241 x(A)S 677 y 
5347 X(measuremen)S -16 x(t)S 167 x(of)S 165 x(this)S 166 x(energy)S
166 x(shift,)S 165 x(whic)S -15 x(h)S 166 x(is)S 165 x(due)S 167 x
(to)S 166 x(thermal)S 165 x(e\013ects)S 167 x(in)S 166 x(the)S 166 x
(cen)S -14 x(ter)S 166 x(of)S 678 y 5347 X(the)S 179 x(sun,)S 179 x
(is)S 179 x(equiv)S -31 x(alen)S -15 x(t)S 179 x(to)S 179 x(a)S 179 x
(direct)S 179 x(measuremen)S -15 x(t)S 179 x(of)S 179 x(the)S 179 x
(cen)S -14 x(tral)S 178 x(temp)S 16 x(erature)S 179 x(of)S 178 x
(the)S 677 y 5347 X(sun.)S
XP /F48 56 336 2 0 24 31 32 24 0
<00FF00 07FFC0 0FFFE0 1F03F0 1E00F8 3C00F8 3C0078 3E0078 3F0078
 3FC0F8 3FF0F0 3FF9F0 1FFFE0 1FFF80 0FFFE0 03FFF0 0FFFF8 1F7FFC
 3E1FFC 7C0FFE FC03FE F800FE F8007E F8003E F8003E FC003E FC007C
 7E00FC 7F81F8 3FFFF0 0FFFE0 01FF00>
PXLC RP
5347 10987 XY F48(8.)S 672 x(Conclusions)S 11970 Y 5347 X F34(The)S
143 x(\014eld)S 144 x(of)S 142 x(solar)S 142 x(neutrino)S 143 x
(researc)S -15 x(h)S 143 x(is)S 143 x(\015ourishing.)S 141 x(The)S
144 x(four)S 142 x(op)S 15 x(erating)S 143 x(exp)S 15 x(erimen)S 
-16 x(ts)S 677 y 5347 X(ha)S -16 x(v)S -15 x(e)S 252 x(con\014rmed)S
251 x(that)S 251 x(the)S 252 x(sun)S 251 x(shines)S 252 x(via)S 250 x
(n)S -15 x(uclear)S 251 x(fusion)S 250 x(reactions)S 251 x(that)S 
251 x(pro)S 15 x(duce)S 678 y 5347 X(MeV)S 273 x(neutrinos)S 272 x
(\(see)S 274 x(Eq.)S 272 x(\(1\))S 273 x(\).)S 272 x(There)S 273 x
(are)S 273 x(di\013erences)S 274 x(b)S 15 x(et)S -15 x(w)S -15 x
(een)S 274 x(the)S 273 x(predictions)S 677 y 5347 X(and)S 219 x(the)S
220 x(observ)S -31 x(ations)S 219 x(\(see)S 220 x(Figure)S 219 x
(1\),)S 218 x(but)S 220 x(these)S 220 x(di\013erences)S 221 x(are)S
219 x(within)S 218 x(the)S 220 x(usual)S 678 y 5347 X(range)S 265 x
(of)S 264 x(astronomical)S 264 x(uncertain)S -16 x(ties)S 266 x
(\(generally)S 264 x(a)S 266 x(factor)S 264 x(of)S 265 x(t)S -15 x
(w)S -15 x(o)S 265 x(or)S 265 x(three\).)S 265 x(The)S 677 y 5347 X
(agreemen)S -16 x(t)S 246 x(b)S 15 x(et)S -15 x(w)S -15 x(een)S 246 x
(theory)S 245 x(and)S 244 x(observ)S -30 x(ation)S 244 x(is,)S 244 x
(from)S 244 x(the)S 245 x(astronomi)S -2 x(cal)S 245 x(p)S 15 x(oin)S
-16 x(t)S 245 x(of)S 677 y 5347 X(view,)S 165 x(remark)S -31 x(ably)S
165 x(go)S 15 x(o)S 14 x(d)S 167 x(b)S 15 x(ecause)S 167 x(the)S 
167 x(calculated)S 166 x(neutrino)S 165 x(\015uxes)S 167 x(dep)S 
16 x(end)S 166 x(sensitiv)S -15 x(ely)S 678 y 5347 X(up)S 15 x(on)S
181 x(the)S 183 x(in)S -16 x(terior)S 181 x(conditions.)S 677 y 
6165 X(Nev)S -15 x(ertheless,)S 143 x(all)S 141 x(four)S 141 x(exp)S
15 x(erimen)S -15 x(ts)S 142 x(disagree)S 142 x(with)S 142 x(the)S
143 x(corresp)S 15 x(onding)S 141 x(theoretical)S 678 y 5347 X
(predictions)S 184 x(based)S 186 x(up)S 15 x(on)S 185 x(the)S 185 x
(simplest)S 184 x(v)S -15 x(ersion)S 185 x(of)S 184 x(the)S 186 x
(standard)S 184 x(electro)S -15 x(w)S -15 x(eak)S 185 x(theory)S 
-46 x(.)S 677 y 5347 X(These)S 161 x(disagreemen)S -16 x(ts)S 161 x
(are)S 159 x(larger)S 159 x(than)S 160 x(the)S 161 x(estimated)S 
160 x(uncertain)S -15 x(ties.)S 159 x(The)S 161 x(lumi)S -2 x(nosit)S
-15 x(y)S 678 y 5347 X(b)S 15 x(oundary)S 197 x(condition)S 197 x
(and)S 197 x(the)S 199 x(helioseismol)S -2 x(ogical)S 197 x
(measuremen)S -16 x(ts)S 199 x(are)S 197 x(esp)S 16 x(ecially)S 197 x
(im-)S 677 y 5347 X(p)S 15 x(ortan)S -16 x(t)S 165 x(in)S 165 x
(guara)S -2 x(n)S -15 x(teeing)S 165 x(the)S 166 x(robustness)S 165 x
(of)S 165 x(the)S 165 x(theoretical)S 165 x(predictions)S 165 x
(\(see)S 166 x(discus-)S 678 y 5347 X(sion)S 127 x(in)S 128 x(4\).)S
127 x(Mon)S -16 x(te)S 129 x(Carlo)S 126 x(exp)S 16 x(erimen)S -16 x
(ts)S 129 x(that)S 128 x(mak)S -16 x(e)S 128 x(use)S 128 x(of)S 128 x
(1000)S 127 x(impl)S -2 x(emen)S -15 x(tations)S 128 x(of)S 127 x
(the)S 677 y 5347 X(standard)S 196 x(solar)S 196 x(mo)S 14 x(del)S
197 x(indicate)S 197 x(that)S 197 x(the)S 197 x(c)S -14 x(hlor)S 
-2 x(ine)S 197 x(and)S 197 x(the)S 198 x(Kamio)S -2 x(k)S -30 x
(ande)S 197 x(I)S 15 x(I)S 197 x(\(w)S -15 x(ater-)S 678 y 5347 X
(Cerenk)S -15 x(o)S -16 x(v\))S 122 x(exp)S 15 x(erimen)S -15 x(ts)S
122 x(cannot)S 121 x(b)S 16 x(e)S 121 x(reconciled)S 122 x(without)S
121 x(an)S 121 x(energy-dep)S 16 x(enden)S -14 x(t)S 122 x(c)S -15 x
(hange)S 677 y 5347 X(in)S 221 x(the)S 223 x(8)S 368 y -273 x(\026)S
-368 y 221 x(solar)S 221 x(neutrino)S 222 x(sp)S 15 x(ectrum)S 222 x
(relativ)S -16 x(e)S 222 x(to)S 222 x(the)S 222 x(lab)S 15 x(orator)S
-2 x(y)S 222 x(sp)S 15 x(ectrum)S 223 x(\(see)S 223 x(Fig-)S 678 y 
5347 X(ure)S 180 x(2-Figure)S 180 x(4\).)S 180 x(New)S 181 x(ph)S 
-15 x(ysics)S 181 x(is)S 180 x(required)S 181 x(to)S 180 x(explain)S
180 x(an)S 180 x(energy-dep)S 16 x(enden)S -14 x(t)S 181 x(c)S -15 x
(hange)S 677 y 5347 X(in)S 184 x(the)S 185 x(shap)S 15 x(e)S 186 x
(of)S 184 x(the)S 185 x(neutrino)S 184 x(sp)S 16 x(ectrum.)S 184 x
(The)S 185 x(galli)S -2 x(um)S 185 x(exp)S 15 x(erimen)S -16 x(ts,)S
185 x(GALLEX)S 185 x(and)S 677 y 5347 X(SA)S -15 x(GE,)S 181 x
(strengthen)S 183 x(the)S 182 x(conclusion)S 182 x(that)S 181 x(new)S
183 x(ph)S -15 x(ysics)S 181 x(is)S 182 x(required.)S 678 y 6165 X
(New)S 187 x(exp)S 16 x(erimen)S -16 x(ts,)S 187 x(SNO,)S 187 x(Sup)S
15 x(erk)S -31 x(amiok)S -31 x(ande,)S 186 x(and)S 187 x(ICAR)S -16 x
(US,)S 187 x(will)S 185 x(test,)S 187 x(indep)S 15 x(en-)S 677 y 
5347 X(den)S -15 x(t)S 210 x(of)S 210 x(uncertain)S -16 x(ties)S 
211 x(due)S 210 x(to)S 210 x(solar)S 209 x(mo)S 14 x(dels,)S 210 x
(the)S 210 x(inference)S 211 x(that)S 210 x(new)S 210 x(ph)S -15 x
(ysics)S 210 x(is)S 210 x(re-)S 678 y 5347 X(quired)S 181 x(to)S 
182 x(explain)S 181 x(the)S 182 x(existing)S 181 x(solar)S 181 x
(neutrino)S 181 x(exp)S 16 x(erimen)S -16 x(ts.)S 677 y 6165 X(This)S
144 x(w)S -16 x(ork)S 144 x(w)S -15 x(as)S 144 x(supp)S 15 x(orted)S
144 x(b)S -15 x(y)S 144 x(NSF)S 145 x(gran)S -16 x(t)S
XP /F34 35 456 3 -9 33 31 41 32 0
<000300C0 000300C0 000300C0 000701C0 000701C0 00060180 00060180
 00060180 000E0380 000E0380 000C0300 000C0300 000C0300 001C0700
 001C0700 FFFFFFFE FFFFFFFE 00300C00 00300C00 00300C00 00701C00
 00701C00 00601800 00601800 00601800 FFFFFFFE FFFFFFFE 01C07000
 01C07000 01806000 01806000 01806000 0380E000 0380E000 0300C000
 0300C000 0300C000 0701C000 0701C000 06018000 06018000>
PXLC RP
15544 27551 XY F34(#PHY92-453)S -2 x(17.)S 143 x(I)S 145 x(thank)S
143 x(the)S 145 x(Institute)S 678 y 5347 X(for)S 208 x(Nuclear)S 
208 x(Theory)S 209 x(at)S 209 x(the)S 209 x(Univ)S -15 x(ersit)S 
-16 x(y)S 209 x(of)S 208 x(W)S -45 x(ashington)S 208 x(for)S 208 x
(its)S 208 x(hospitalit)S -16 x(y)S 208 x(and)S 209 x(the)S 677 y 
5347 X(Departmen)S -16 x(t)S 182 x(of)S 181 x(Energy)S 182 x(for)S
181 x(partial)S 180 x(supp)S 15 x(ort)S 182 x(during)S 180 x(the)S
183 x(completion)S 181 x(of)S 181 x(this)S 182 x(w)S -16 x(ork.)S
1 PP EP

1000 BP 39600 30600 PM 0 0 XY
3815 Y 5347 X F48(Referenc)S -2 x(es)S 4798 Y 5347 X F34(1.)S
XP /F34 74 280 1 -1 19 30 32 24 0
<0FFFE0 0FFFE0 003E00 003E00 003E00 003E00 003E00 003E00 003E00
 003E00 003E00 003E00 003E00 003E00 003E00 003E00 003E00 003E00
 003E00 003E00 003E00 003E00 003E00 783E00 FC3E00 FC3E00 FC3E00
 FC7C00 F87C00 78F800 3FF000 0FC000>
PXLC RP
6064 4798 XY F34(J.)S 167 x(N.)S 167 x(Bahcall,)S 167 x(Ph)S -15 x
(ys.)S 167 x(Rev.)S 167 x(Lett.)S
XP /F39 /cmbx10 329 546 545.454 128 [-3 -11 51 33] PXLNF RP
XP /F39 49 314 4 0 20 28 29 24 0
<007000 01F000 0FF000 FFF000 FFF000 F3F000 03F000 03F000 03F000
 03F000 03F000 03F000 03F000 03F000 03F000 03F000 03F000 03F000
 03F000 03F000 03F000 03F000 03F000 03F000 03F000 03F000 FFFF80
 FFFF80 FFFF80>
PXLC RP
13549 4798 XY F39(1)S
XP /F39 50 314 3 0 21 28 29 24 0
<07F800 3FFE00 7FFF80 7C7FC0 FE3FC0 FE1FE0 FE0FE0 FE0FE0 7C0FE0
 380FE0 000FE0 001FC0 001FC0 003F80 003F00 007E00 00FC00 01F000
 01E000 03C0E0 0780E0 0F00E0 1C01E0 3FFFE0 7FFFC0 FFFFC0 FFFFC0
 FFFFC0 FFFFC0>
PXLC RP
13863 4798 XY F39(2)S F34(,)S 167 x(300)S 167 x(\(1964\);)S 166 x
(R.)S 167 x(Da)S -16 x(vis)S 167 x(Jr.,)S 166 x(Ph)S -14 x(ys.)S 
167 x(Rev.)S 167 x(Lett.)S 677 y 5639 X F39(12)S F34(,)S 181 x(303)S
181 x(\(1964\).)S 619 y 5347 X(2.)S 292 x(K.)S 181 x(S.)S 182 x
(Hirata)S -2 x(,)S 182 x(et)S 182 x(al.,)S 180 x(Ph)S -15 x(ys.)S 
182 x(Rev.)S 181 x(Lett.)S
XP /F39 54 314 2 0 22 28 29 24 0
<007F80 01FFC0 07FFE0 0FE1F0 1F83F0 3F03F0 3F03F0 7F03F0 7E01E0
 7E0000 FE1000 FEFF80 FFFFC0 FF87E0 FF03F0 FF03F0 FE03F8 FE03F8
 FE03F8 FE03F8 FE03F8 7E03F8 7E03F8 7E03F0 3F03F0 1F87E0 1FFFC0
 07FF80 01FE00>
PXLC RP
14970 6094 XY F39(6)S
XP /F39 51 314 2 0 22 28 29 24 0
<03FC00 0FFF00 1FFFC0 1F1FC0 3F0FE0 3F0FE0 3F8FE0 3F0FE0 1F0FE0
 001FC0 001FC0 003F80 01FE00 01FE00 001FC0 000FE0 0007F0 0007F8
 0007F8 7C07F8 FE07F8 FE07F8 FE07F8 FE07F0 FE0FF0 7C1FE0 7FFFC0
 1FFF80 07FC00>
PXLC RP
15283 6094 XY F39(3)S F34(,)S 181 x(16)S 182 x(\(1989\).)S 618 y 
5347 X(3.)S 292 x(K.)S 181 x(S.)S 182 x(Hirata)S -2 x(,)S 182 x(et)S
182 x(al.,)S 180 x(Ph)S -15 x(ys.)S 182 x(Rev.)S 181 x(D.)S
XP /F39 52 314 1 0 23 28 29 24 0
<0003C0 0003C0 0007C0 000FC0 001FC0 003FC0 003FC0 0077C0 00E7C0
 01C7C0 0387C0 0707C0 0707C0 0E07C0 1C07C0 3807C0 7007C0 F007C0
 FFFFFE FFFFFE FFFFFE 000FC0 000FC0 000FC0 000FC0 000FC0 01FFFE
 01FFFE 01FFFE>
PXLC RP
14379 6712 XY F39(44)S F34(,)S 180 x(2241)S 181 x(\(1991\).)S 618 y 
5347 X(4.)S 292 x(V.)S 219 x(Grib)S 14 x(o)S -15 x(v)S 219 x(and)S
220 x(B.)S 219 x(P)S -15 x(on)S -15 x(tecorv)S -15 x(o,)S 219 x(Ph)S
-15 x(ys.)S 219 x(Lett.)S 220 x(B)S 219 x F39(2)S
XP /F39 56 314 2 0 22 28 29 24 0
<01FE00 07FF80 0FFFC0 1F07C0 1E03E0 3C01E0 3C01E0 3F01E0 3F81E0
 3FE3E0 3FFFC0 1FFF80 1FFF80 0FFFC0 0FFFE0 1FFFF0 7E7FF0 7C3FF8
 FC0FF8 F803F8 F801F8 F800F8 F800F8 FC00F8 7E01F0 7F07E0 3FFFE0
 1FFF80 03FE00>
PXLC RP
17424 7330 XY F39(8)S F34(,)S 219 x(493)S 219 x(\(1969\);)S 218 x
(S.)S 219 x(M.)S 219 x(Bilenky)S 677 y 5639 X(and)S 165 x(B.)S 166 x
(P)S -15 x(on)S -16 x(tecorv)S -15 x(o,)S 165 x(Ph)S -15 x(ys.)S 
165 x(Rep.)S 165 x F39(41)S F34(,)S 164 x(225)S 164 x(\(1978\);)S 
164 x(S.)S 165 x(M.)S 165 x(Bilenky)S 165 x(and)S 165 x(S.)S 165 x
(T.)S 165 x(P)S -15 x(etco)S -15 x(v,)S 678 y 5639 X(Rev.)S 177 x
(Mo)S 15 x(d.)S 176 x(Ph)S -15 x(ys.)S
XP /F39 53 314 3 0 21 28 29 24 0
<380380 3FFF80 3FFF80 3FFF00 3FFE00 3FF800 3FE000 380000 380000
 380000 380000 3FF800 3FFF00 3E3F80 381FC0 300FC0 000FE0 000FE0
 000FE0 7C0FE0 FC0FE0 FC0FE0 FC0FE0 FC0FC0 FC1FC0 7C3F80 3FFF00
 1FFE00 0FF000>
PXLC RP
9810 8685 XY F39(5)S
XP /F39 57 314 2 0 22 28 29 24 0
<03FC00 0FFF00 1FFFC0 3F8FE0 7E07E0 7E03F0 FE03F0 FE03F0 FE03F8
 FE03F8 FE03F8 FE03F8 FE03F8 7E07F8 7E07F8 3F0FF8 1FFFF8 0FFBF8
 0043F8 0003F0 3C03F0 7E03F0 7E07E0 7E07E0 7E0FC0 7C3F80 3FFF00
 1FFE00 07F000>
PXLC RP
10124 8685 XY F39(9)S F34(,)S 176 x(671)S 176 x(\(1987\).)S 176 x
(See)S 178 x(also)S 176 x(B.)S 177 x(P)S -15 x(on)S -15 x(tecorv)S
-15 x(o,)S 176 x(So)S -15 x(v.)S 176 x(Ph)S -14 x(ys.)S 176 x(JETP)S
178 x F39(26)S F34(,)S 677 y 5639 X(984)S 181 x(\(1968\).)S 618 y 
5347 X(5.)S 292 x(L.)S 169 x(W)S -46 x(olfenstein,)S 169 x(Ph)S -15 x
(ys.)S 169 x(Rev.)S 169 x(D)S 169 x F39(1)S
XP /F39 55 314 3 0 23 29 30 24 0
<700000 7FFFF8 7FFFF8 7FFFF8 7FFFF8 7FFFF0 FFFFE0 F007C0 E00F80
 E00F00 E01E00 003E00 007C00 007800 00F800 00F800 00F800 01F000
 01F000 01F000 01F000 03F000 03F000 03F000 03F000 03F000 03F000
 03F000 03F000 01E000>
PXLC RP
13459 9980 XY F39(7)S F34(,)S 169 x(2369)S 168 x(\(1978\);)S 168 x
(Ph)S -14 x(ys.)S 169 x(Rev.)S 169 x(D)S 169 x F39(2)S
XP /F39 48 314 2 0 22 28 29 24 0
<01FC00 07FF00 1F07C0 1E03C0 3E03E0 7C01F0 7C01F0 7C01F0 FC01F8
 FC01F8 FC01F8 FC01F8 FC01F8 FC01F8 FC01F8 FC01F8 FC01F8 FC01F8
 FC01F8 FC01F8 FC01F8 7C01F0 7C01F0 7C01F0 3E03E0 1E03C0 1F8FC0
 07FF00 01FC00>
PXLC RP
20827 9980 XY F39(0)S F34(,)S 168 x(2634)S 169 x(\(1979\).)S 619 y 
5347 X(6.)S 292 x(S.)S 175 x(P)S -45 x(.)S 175 x(Mikhey)S -16 x(ev)S
176 x(and)S 176 x(A.)S 175 x(Y)S -46 x(u.)S 175 x(Smirno)S -16 x(v,)S
175 x(So)S -16 x(v.)S 175 x(J.)S 176 x(Nucl.)S 175 x(Ph)S -15 x(ys.)S
175 x F39(42)S F34(,)S 175 x(913)S 174 x(\(1986\);)S 175 x(So)S -16 x
(v.)S 677 y 5639 X(Ph)S -15 x(ys.)S 182 x(JETP)S 182 x F39(64)S F34
(,)S 181 x(4)S 181 x(\(1986\);)S 181 x(Nuo)S -15 x(v)S -16 x(o)S 
182 x(Cimen)S -16 x(to)S 182 x F39(9)S
XP /F39 67 454 3 0 33 30 31 32 0
<000FFC06 007FFF9E 01FFFFFE 03FF03FE 0FF800FE 1FE0007E 1FC0003E
 3FC0003E 7F80001E 7F80001E 7F00000E FF00000E FF000000 FF000000
 FF000000 FF000000 FF000000 FF000000 FF000000 FF000000 7F00000E
 7F80000E 7F80000E 3FC0001E 1FC0003C 1FE0003C 0FF800F8 03FF03F0
 01FFFFE0 007FFF80 000FFC00>
PXLC RP
16205 11276 XY F39(C)S F34(,)S 181 x(17)S 181 x(\(1986\).)S 618 y 
5347 X(7.)S 292 x(R.)S 240 x(Da)S -15 x(vis)S 241 x(Jr.,)S
XP /F37 70 357 3 0 32 30 31 32 0
<01FFFFFC 01FFFFFC 001E00F8 001E0038 001E0038 003E0018 003C0018
 003C0038 003C0030 007C0C30 00780C30 00780C00 00781C00 00F83800
 00FFF800 00FFF800 00F07800 01F03000 01E03000 01E03000 01E03000
 03E00000 03C00000 03C00000 03C00000 07C00000 07800000 07800000
 0F800000 FFFC0000 FFFC0000>
PXLC RP
9467 11894 XY F37(F)S -41 x(r)S -28 x(ontiers)S 250 x(of)S 250 x
(Neutrino)S 250 x(Astr)S -27 x(ophysics)S F34(,)S 241 x(ed.)S 241 x
(b)S -15 x(y)S 242 x(Y.)S 241 x(Suzuki)S 242 x(and)S 241 x(K.)S 678 y 
5639 X(Nak)S -30 x(am)S -16 x(ura)S 181 x(\(Univ)S -15 x(ersal)S 
181 x(Academ)S -15 x(y)S 182 x(Press,)S 182 x(Inc.,)S 181 x(T)S -46 x
(oky)S -15 x(o,)S 181 x(Japan,)S 181 x(1993\),)S 180 x(p.)S 182 x
(47.)S 618 y 5347 X(8.)S 292 x(P)S -46 x(.)S 182 x(Anselmann,)S 181 x
(et)S 182 x(al.,)S 180 x(Ph)S -15 x(ys.)S 181 x(Lett)S 183 x(B)S 
182 x F39(285)S F34(,)S 180 x(376)S 181 x(\(1992\).)S 618 y 5347 X
(9.)S 292 x(P)S -46 x(.)S 182 x(Anselmann)S 181 x(Ph)S -15 x(ys.)S
182 x(Lett.)S 181 x(B)S 183 x F39(314)S F34(,)S 180 x(445)S 181 x
(\(1993\);)S 180 x F37(ibid)S F34(,)S 182 x(submitted)S 182 x(F)S 
-46 x(eb.)S 182 x(\(1994\).)S 618 y 5347 X(10.)S 291 x(A.)S 182 x
(I.)S 181 x(Abazo)S -15 x(v,)S 181 x(et)S 182 x(al.,)S 181 x(Nucl.)S
181 x(Ph)S -15 x(ys.)S 181 x(B)S 183 x(\(Pro)S 14 x(c.)S 182 x
(Suppl.\))S 181 x F39(19)S F34(,)S 181 x(84)S 181 x(\(1991a\).)S 
618 y 5347 X(11.)S 291 x(A.)S 182 x(I.)S 181 x(Abazo)S -15 x(v,)S 
181 x(et)S 182 x(al,)S 181 x(Ph)S -15 x(ys.)S 182 x(Rev.)S 181 x
(Lett.)S 182 x F39(67)S 181 x F34(3332)S 180 x(\(1991b\).)S 619 y 
5347 X(12.)S 291 x(J.)S 182 x(N.)S 181 x(Aldurashito)S -16 x(v,)S 
181 x(et)S 182 x(al.,)S 180 x(Ph)S -14 x(ys.)S 181 x(Lett.)S 182 x
(B,)S 182 x(submitted)S 181 x(\(1994\).)S 618 y 5347 X(13.)S 291 x
(J.)S 182 x(N.)S 181 x(Bahcall)S 182 x(and)S 181 x(M.)S 181 x(H.)S
182 x(Pinsonneault,)S 181 x(Rev.)S 181 x(Mo)S 15 x(d.)S 181 x(Ph)S
-15 x(ys.)S 182 x F39(64)S F34(,)S 180 x(885)S 181 x(\(1992\).)S 
618 y 5347 X(14.)S 291 x(Y.)S 132 x(Suzuki,)S 131 x F37(F)S -42 x(r)S
-28 x(ontiers)S 150 x(of)S 149 x(Neutrino)S 149 x(Astr)S -27 x
(ophysics)S F34(,)S 131 x(ed.)S 132 x(b)S -16 x(y)S 132 x(Y.)S 132 x
(Suzuki)S 131 x(and)S 132 x(K.)S 132 x(Nak)S -31 x(a-)S 677 y 5639 X
(m)S -15 x(ura)S 181 x(\(Univ)S -16 x(ersal)S 182 x(Academ)S -15 x
(y)S 182 x(Press,)S 181 x(Inc.,)S 182 x(T)S -46 x(oky)S -15 x(o,)S
181 x(Japan,)S 181 x(1993\),)S 180 x(p.)S 181 x(61.)S 618 y 5347 X
(15.)S 291 x(J.)S 182 x(N.)S 181 x(Bahcall,)S 181 x(and)S 182 x(R.)S
180 x(K.)S 182 x(Ulric)S -16 x(h,)S 181 x(R.)S 181 x(K.,)S 181 x
(Rev.)S 182 x(Mo)S 14 x(d.)S 182 x(Ph)S -15 x(ys.)S 181 x F39(60)S
F34(,)S 181 x(297)S 181 x(\(1988\).)S 619 y 5347 X(16.)S 291 x(Y.)S
129 x(Lebreton,)S 130 x(and)S 129 x(W.)S 129 x(D)S
XP /F34 127 273 4 26 17 30 5 16 0
<7878 7878 F87C 7878 7878>
PXLC RP
11705 18813 XY F34(\177)S -273 x(app)S 14 x(en,)S 130 x(W.)S 129 x
(1988,)S 128 x(in)S 129 x(Seismolo)S -2 x(gy)S 130 x(of)S 129 x(the)S
130 x(Sun)S 129 x(and)S 130 x(Sun-lik)S -17 x(e)S 677 y 5639 X
(Stars,)S 181 x(ed.)S 182 x(V.)S 182 x(Domi)S -2 x(ngo,)S 181 x(and)S
181 x(E.)S 182 x(J.)S 182 x(Rol)S -2 x(fe)S 182 x(\(ESP)S 183 x
(SP-286,)S 180 x(1988\),)S 181 x(p.)S 181 x(661.)S 618 y 5347 X(17.)S
291 x(I.)S 179 x(J.)S 179 x(Sac)S -15 x(kmann,)S 178 x(A.)S 178 x
(I.)S 179 x(Bo)S 15 x(othro)S -16 x(yd,)S 179 x(and)S 179 x(W.)S 
178 x(A.)S 179 x(F)S -46 x(o)S -15 x(wler,)S 178 x(Astroph)S -16 x
(ys.)S 179 x(J.)S 179 x F39(360)S F34(,)S 178 x(727)S 678 y 5639 X
(\(1990\).)S 618 y 5347 X(18.)S 291 x(C.)S 182 x(R.)S 180 x
(Pro\016tt,)S 181 x(and)S 182 x(G.)S 181 x(Mic)S -15 x(haud,)S 181 x
(Astroph)S -15 x(ys.)S 181 x(J.)S 182 x F39(380)S F34(,)S 180 x(238)S
181 x(\(1991\).)S 618 y 5347 X(19.)S 291 x(J.)S 182 x(A.)S 181 x
(Guzik,)S 182 x(and)S 181 x(A.)S 182 x(N.)S 181 x(Co)S -15 x(x,)S 
181 x(Astroph)S -16 x(ys.)S 182 x(J.)S 181 x F39(381)S F34(,)S 181 x
(L333)S 181 x(\(1991\).)S 618 y 5347 X(20.)S 291 x(B.)S 182 x
(Ahrens,)S 182 x(M.)S 181 x(Stix,)S 181 x(and)S 182 x(M.)S 181 x
(Thorn,)S 181 x(Astr.)S 181 x(Astroph)S -15 x(ys.)S 181 x F39(264)S
F34(,)S 180 x(673)S 181 x(\(1992\).)S 618 y 5347 X(21.)S 291 x(J.)S
262 x(Christensen-Dalsgaar)S -2 x(d,)S 262 x(Geoph)S -15 x(ys.,)S 
261 x(Astroph)S -15 x(ys.,)S 261 x(Fluid)S 262 x(Dynam)S -2 x(ics)S
263 x F39(62)S F34(,)S 261 x(123)S 678 y 5639 X(\(1992\).)S 618 y 
5347 X(22.)S 291 x(D.)S 196 x(B.)S 197 x(Guen)S -14 x(ther,)S 196 x
(P)S -45 x(.)S 197 x(Demarque,)S 195 x(Y.)S 197 x(C.)S 197 x(Kim,)S
195 x(and)S 197 x(M.)S 197 x(H.)S 196 x(Pinsonneault,)S 196 x
(Astro-)S 677 y 5639 X(ph)S -15 x(ys.)S 182 x(J.)S 181 x F39(387)S
F34(,)S 181 x(372)S 181 x(\(1992\).)S 619 y 5347 X(23.)S 291 x(G.)S
141 x(Berthomieu,)S 140 x(J.)S 140 x(Pro)S -15 x(v)S -15 x(ost,)S 
140 x(P)S -45 x(.)S 140 x(Morel,)S 139 x(and)S 141 x(Y.)S 140 x
(Lebreton,)S 141 x(Astr.)S 140 x(Astroph)S -15 x(ys.)S 140 x F39
(268)S F34(,)S 677 y 5639 X(775)S 181 x(\(1993\).)S 618 y 5347 X
(24.)S 291 x(S.)S 182 x(T)S -46 x(urc)S -15 x(k-Chi)S
XP /F34 18 273 5 23 12 31 9 8 0
<E0 F0 F8 78 3C 1E 0F 07 02>
PXLC RP
9368 27145 XY F34(\022)S -258 x(eze)S 2 x(,)S 181 x(and)S 182 x(I.)S
181 x(Lop)S 15 x(es,)S 182 x(I.,)S 180 x(Astroph)S -15 x(ys.)S 181 x
(J.)S 182 x F39(408)S F34(,)S 180 x(347)S 181 x(\(1993\).)S 618 y 
5347 X(25.)S 291 x(V.)S 232 x(Castellani,)S 231 x(S.)S 232 x(Degl'I)S
-2 x(nno)S 15 x(cen)S -14 x(ti,)S 232 x(and)S 232 x(G.)S 232 x
(Fioren)S -16 x(tini,)S 231 x(Astr.)S 232 x(Astroph)S -16 x(ys.)S 
232 x F39(271)S F34(,)S 678 y 5639 X(601)S 181 x(\(1993\).)S 618 y 
5347 X(26.)S 291 x(J.)S 182 x(N.)S 181 x(Bahcall)S 182 x(and)S 181 x
(H.)S 182 x(A.)S 181 x(Bethe,)S 183 x(Ph)S -15 x(ys.)S 181 x(Rev.)S
181 x(D)S 182 x F39(47)S F34(,)S 180 x(1298)S 181 x(\(1993\).)S 618 y 
5347 X(27.)S 291 x(J.)S 259 x(N.)S 258 x(Bahcall,)S 258 x F37
(Neutrino)S 266 x(Astr)S -27 x(ophysics)S 258 x F34(\(Cam)S -16 x
(bridge)S 258 x(Univ)S -16 x(ersit)S -15 x(y)S 259 x(Press,)S 259 x
(Cam-)S 678 y 5639 X(bridge,)S 181 x(England,)S 181 x(1989\).)S 618 y 
5347 X(28.)S 291 x(J.)S 182 x(N.)S 181 x(Bahcall,)S 181 x(Ph)S -15 x
(ys.)S 181 x(Rev.)S 182 x(D)S 181 x F39(44)S F34(,)S 181 x(1644)S 
180 x(\(1991\).)S 618 y 5347 X(29.)S 291 x(G.)S 186 x(Aardsma)S 186 x 
F37(et)S 199 x(al)S
XP /F37 46 167 5 0 9 4 5 8 0
<70 F8 F8 F0 F0>
PXLC RP
10479 31591 XY F37(.)S F34(,)S 186 x(Ph)S -14 x(ys.)S 186 x(Lett.)S
186 x(B)S 187 x F39(194)S F34(,)S 185 x(321)S 185 x(\(1987\);)S 185 x
(G.)S 186 x(Ew)S -15 x(an)S 186 x(et)S 187 x(al.,)S 185 x(Sudbury)S
677 y 5639 X(Neutrino)S 208 x(Observ)S -30 x(atory)S 207 x(Prop)S 
15 x(osal)S 206 x(SNO-87-12)S 207 x(\(1987\);)S 207 x(H.)S 207 x
(Chen,)S 208 x(Ph)S -15 x(ys.)S 207 x(Rev.)S 208 x(Lett.)S 678 y 
5639 X F39(55)S F34(,)S 167 x(1534)S 167 x(\(1985\).G.)S 166 x(Ew)S
-15 x(an)S 168 x(et)S 169 x(al.,)S 166 x(Sudbury)S 168 x(Neutrino)S
167 x(Observ)S -30 x(atory)S 168 x(Prop)S 14 x(osal)S 168 x(SNO-)S
677 y 5639 X(87-12)S 181 x(\(1987\);)S 180 x(H.)S 182 x(Chen,)S 182 x
(Ph)S -15 x(ys.)S 181 x(Rev.)S 181 x(Lett.)S 182 x F39(55)S F34(,)S
181 x(1534)S 180 x(\(1985\).)S
1 PP EP

1000 BP 39600 30600 PM 0 0 XY
3815 Y 5347 X F34(30.)S 291 x(M.)S 269 x(T)S -45 x(akita,)S 268 x 
F37(F)S -42 x(r)S -28 x(ontiers)S 276 x(of)S 276 x(Neutrino)S 275 x
(Astr)S -27 x(ophysics)S F34(,)S 269 x(ed.)S 269 x(b)S -15 x(y)S 
269 x(Y.)S 269 x(Suzuki)S 270 x(and)S 269 x(K.)S 678 y 5639 X(Nak)S
-30 x(am)S -16 x(ura)S 181 x(\(Univ)S -15 x(ersal)S 181 x(Academ)S
-15 x(y)S 182 x(Press,)S 182 x(Inc.,)S 181 x(T)S -46 x(oky)S -15 x
(o,)S 181 x(Japan,)S 181 x(1993\),)S 180 x(p.)S 182 x(135.)S 618 y 
5347 X(31.)S 291 x(Y.)S 179 x(T)S -45 x(otsuk)S -31 x(a,)S 179 x(in)S
XP /F37 80 371 3 0 32 30 31 32 0
<01FFFF80 01FFFFE0 001E01F0 001E0078 003E0078 003E007C 003C007C
 003C007C 007C007C 007C00F8 007800F8 007800F0 00F801E0 00F803E0
 00F00F80 00FFFF00 01FFF800 01F00000 01E00000 01E00000 03E00000
 03E00000 03C00000 03C00000 07C00000 07C00000 07800000 07800000
 0F800000 FFF80000 FFF80000>
PXLC RP
9923 5111 XY F37(Pr)S -28 x(o)S -28 x(c)S -28 x(e)S -28 x(e)S -28 x
(dings)S 193 x(of)S 193 x(the)S
XP /F37 73 210 3 0 22 30 31 24 0
<01FFF0 01FFF0 001F00 001E00 001E00 003E00 003C00 003C00 003C00
 007C00 007800 007800 007800 00F800 00F000 00F000 00F000 01F000
 01E000 01E000 01E000 03E000 03C000 03C000 03C000 07C000 078000
 078000 0F8000 FFF800 FFF800>
PXLC RP
14378 5111 XY F37(International)S
XP /F37 83 307 3 -1 28 31 33 32 0
<0007F0C0 001FFDC0 007C3F80 00F00F80 00E00780 01C00780 03C00300
 03800300 03800300 03800300 03800000 03C00000 03E00000 03FC0000
 03FF8000 01FFE000 00FFF000 001FF800 0003F800 0000F800 00007800
 00003800 00003800 30003800 30003800 70007800 70007000 7000F000
 7801E000 FC03C000 FF078000 EFFF0000 C3FC0000>
PXLC RP
17687 5111 XY F37(Symp)S -28 x(osium)S 192 x(on)S
XP /F37 85 406 9 -1 37 30 32 32 0
<7FFC3FF8 7FFC3FF8 07C007C0 07800300 0F800700 0F800600 0F000600
 0F000600 1F000E00 1F000C00 1E000C00 1E000C00 3E001C00 3E001800
 3C001800 3C001800 7C003800 7C003000 78003000 78003000 78007000
 F8006000 F0006000 F000E000 F000C000 F001C000 F0038000 70070000
 780F0000 3C3C0000 1FF80000 07E00000>
PXLC RP
21335 5111 XY F37(Under)S -28 x(gr)S -28 x(ound)S 677 y 5639 X
(Physics)S
XP /F37 69 371 3 0 33 30 31 32 0
<01FFFFFE 01FFFFFE 001E007C 001E001C 001E001C 003E001C 003C000C
 003C001C 003C0018 007C0C18 00780C18 00780C00 00781C00 00F83800
 00FFF800 00FFF800 00F07800 01F03000 01E03000 01E03030 01E03070
 03E00060 03C000E0 03C000C0 03C001C0 07C00180 07800380 07800F80
 0F803F00 FFFFFF00 FFFFFE00>
PXLC RP
7710 5788 XY F37(E)S
XP /F37 120 253 3 0 21 19 20 24 0
<07C7C0 1FEFC0 3CFCE0 707DE0 7079E0 60F9E0 E0F1C0 00F000 00F000
 01F000 01F000 01E000 01E000 73E0E0 F3E0C0 F3C1C0 F7C1C0 E7E780
 7EFF00 7C7C00>
PXLC RP
8080 5788 XY F37(xp)S -28 x(eriments)S F34(,)S 287 x(edited)S 289 x
(b)S -15 x(y)S 288 x(K.)S 287 x(Nak)S -30 x(am)S -16 x(ura)S 287 x
(\(ICRR,)S 286 x(Univ)S -15 x(ersit)S -15 x(y)S 287 x(of)S 288 x(T)S
-46 x(oky)S -16 x(o,)S 678 y 5639 X(1990\),)S 128 x(p.)S 128 x(129;)S
128 x(A.)S 128 x(Suzuki,)S 129 x(in)S 128 x F37(Pr)S -28 x(o)S -28 x
(c)S -28 x(e)S -28 x(e)S -28 x(dings)S 146 x(of)S 147 x(the)S
XP /F37 87 546 9 -1 49 30 32 48 0
<FFF0FFE1FF80 FFF0FFE1FF80 1F001F007C00 1E001E003800 1E003E003000
 1E003E006000 1E007E006000 1E00FE00C000 1E00DE00C000 1F01DE018000
 0F019E018000 0F031E030000 0F031E030000 0F061E060000 0F061E060000
 0F0C1E0C0000 0F0C1E0C0000 0F181E180000 0F181E180000 0F301E300000
 0F301E300000 0F601E600000 0F601EE00000 0FC01EC00000 0FC01FC00000
 0F801F800000 0F801F000000 0F000F000000 0F000E000000 0E000E000000
 0E000C000000 0C000C000000>
PXLC RP
16364 6466 XY F37(Wor)S
XP /F37 107 251 2 0 21 31 32 24 0
<01F000 0FF000 0FF000 01E000 01E000 01E000 03E000 03C000 03C000
 03C000 07C000 078000 0781E0 0787F0 0F8E30 0F1CF0 0F38F0 0F70F0
 1FE0E0 1FC000 1FC000 1FF000 3FFC00 3C7C00 3C3E00 3C3E70 7C3C60
 783C60 783CE0 783DC0 F81FC0 700F80>
PXLC RP
17418 6466 XY F37(kshop)S 145 x(on)S 146 x(Elementary)S
XP /F37 45 195 3 8 14 11 4 16 0
<7FF0 FFF0 FFF0 FFE0>
PXLC RP
22325 6466 XY F37(-Particle)S 677 y 5639 X(Pictur)S -27 x(e)S 247 x
(of)S 248 x(the)S 248 x(Universe)S F34(,)S 239 x(Tsukuba,)S 238 x
(Japan,)S 238 x(1987,)S 238 x(edited)S 239 x(b)S -15 x(y)S 239 x(M.)S
239 x(Y)S -46 x(oshim)S -16 x(ura,)S 238 x(Y.)S 678 y 5639 X(T)S 
-45 x(otsuk)S -31 x(a,)S 182 x(and)S 181 x(K.)S 182 x(Nak)S -31 x
(am)S -16 x(ura)S 181 x(\(KEK)S 183 x(Rep)S 15 x(ort)S 181 x(No.)S
182 x(87-1,)S 180 x(Tsukuba,)S 181 x(1987\),)S 180 x(p.)S 182 x
(136.)S 618 y 5347 X(32.)S 291 x(R.)S 215 x(S.)S 214 x(Ragha)S -16 x
(v)S -31 x(an,)S 215 x(in)S 215 x F37(Pr)S -28 x(o)S -28 x(c)S -28 x
(e)S -28 x(e)S -28 x(dings)S 225 x(of)S 226 x(the)S
XP /F37 88 406 2 0 36 30 31 40 0
<00FFF0FFE0 00FFF0FFE0 000F803E00 000F803800 0007803000 0007806000
 0007C0C000 0003C18000 0003C38000 0003E70000 0001E60000 0001EC0000
 0001F80000 0000F00000 0000F80000 0000F80000 0001F80000 0001FC0000
 00033C0000 00063C0000 000C3E0000 001C1E0000 00381E0000 00301F0000
 00600F0000 00C00F0000 01800F8000 0380078000 0F800FC000 FFE03FF800
 FFE07FF800>
PXLC RP
15705 8439 XY F37(XX)S
XP /F37 86 406 9 -1 38 30 32 32 0
<FFF00FFC FFF00FFC 1F8003E0 0F000180 0F000380 0F000300 0F000600
 0F000E00 0F000C00 0F001800 0F801800 07803000 07803000 07806000
 0780C000 0780C000 07818000 07818000 07830000 07870000 07860000
 078C0000 078C0000 07D80000 03D80000 03F00000 03F00000 03E00000
 03C00000 03C00000 03800000 03800000>
PXLC RP
16516 8439 XY F37(Vth)S 226 x(International)S
XP /F37 67 391 6 -1 36 31 33 32 0
<0000FE06 0007FF8E 001FC3DC 007E00FC 00F8007C 01F0007C 03E00038
 07C00038 0F800038 0F800038 1F000030 1F000030 3E000030 3E000000
 7E000000 7C000000 7C000000 7C000000 FC000000 F8000000 F8000000
 F80001C0 F8000180 F8000180 78000380 78000300 7C000700 3C000E00
 3E001C00 1F007800 0FC1F000 03FFC000 00FF0000>
PXLC RP
20950 8439 XY F37(Confer)S -28 x(enc)S -29 x(e)S 226 x(on)S 677 y 
5639 X(High)S 168 x(Ener)S -28 x(gy)S 167 x(Physics)S F34(,)S 151 x
(Singap)S 14 x(ore,)S 151 x(1990,)S 150 x(edited)S 152 x(b)S -15 x
(y)S 152 x(K.)S 151 x(K.)S 151 x(Ph)S -14 x(ua)S 151 x(and)S 151 x
(Y.)S 152 x(Y)S -46 x(amaguc)S -16 x(hi)S 678 y 5639 X(\(W)S -45 x
(orld)S 167 x(Scien)S -14 x(ti\014c,)S 168 x(Singap)S 15 x(ore,)S 
168 x(1990\),)S 167 x(V)S -45 x(ol.)S 168 x(1,)S 168 x(p.)S 168 x
(482;)S 168 x(G.)S 168 x(Ran)S -15 x(ucci)S 169 x(for)S 168 x(the)S
169 x(Borexino)S 677 y 5639 X(Collab)S 14 x(oratio)S -2 x(n,)S 169 x
(Nucl.)S 169 x(Ph)S -15 x(ys.)S 169 x(B)S 170 x(\(Pro)S 15 x(c.)S 
169 x(Suppl.\))S 169 x(32,)S 168 x(149)S 169 x(\(1993\);)S 168 x(C.)S
169 x(Arpasella)S 168 x F37(et)S 184 x(al.)S F34(,)S 678 y 5639 X
(in)S 222 x(\\Borexino)S 221 x(at)S 222 x(Gran)S 221 x(Sasso:)S 221 x
(Prop)S 15 x(osal)S 221 x(for)S 221 x(a)S 222 x(real-tim)S -2 x(e)S
223 x(detector)S 223 x(for)S 221 x(lo)S -16 x(w)S 222 x(energy)S 
677 y 5639 X(solar)S 133 x(neutrinos,")S 134 x(V)S -46 x(ols.)S 133 x
(I)S 134 x(and)S 135 x(I)S 15 x(I,)S 133 x(Univ)S -15 x(ersit)S -16 x
(y)S 135 x(of)S 133 x(Milan,)S 133 x(INFN)S 134 x(rep)S 15 x(ort)S
134 x(\(unpublished\).)S 618 y 5347 X(33.)S 291 x(J.)S 242 x(P)S 
-45 x(.)S 241 x(Rev)S -15 x(ol,)S 241 x F37(F)S -42 x(r)S -28 x
(ontiers)S 251 x(of)S 250 x(Neutrino)S 251 x(Astr)S -27 x(ophysics)S
F34(,)S 241 x(ed.)S 242 x(b)S -15 x(y)S 242 x(Y.)S 241 x(Suzuki)S 
242 x(and)S 242 x(K.)S 678 y 5639 X(Nak)S -30 x(am)S -16 x(ura)S 
195 x(\(Univ)S -16 x(ersal)S 196 x(Academ)S -15 x(y)S 195 x(Press,)S
196 x(Inc.,)S 195 x(T)S -45 x(oky)S -16 x(o,)S 195 x(Japan,)S 195 x
(1993\),)S 194 x(p.)S 195 x(167;)S 195 x(J.)S 195 x(N.)S 677 y 5639 X
(Bahcall,)S 240 x(M.)S 240 x(Baldo-Ceoli)S -2 x(n,)S 241 x(D.)S 240 x
(Cline,)S 239 x(and)S 241 x(C.)S 240 x(Rubbia,)S 239 x(Ph)S -15 x
(ys.)S 241 x(Lett.)S 240 x(B)S 241 x F39(178)S F34(,)S 240 x(324)S
678 y 5639 X(\(1986\);)S 181 x(C.)S 181 x(Rubbia,)S 180 x(CERN-EP)S
182 x(In)S -15 x(ternal)S 181 x(Rep)S 15 x(ort)S 182 x(77-8)S 180 x
(\(1977\).)S 618 y 5347 X(34.)S 291 x(J.)S 260 x(Seguinot,)S 260 x
(T.)S 259 x(Ypsilan)S -16 x(tis,)S 260 x(and)S 259 x(A.)S 260 x(Zic)S
-15 x(hini,)S 259 x(\\A)S 260 x(High)S 259 x(Rate)S 260 x(Solar)S 
259 x(Neutrino)S 677 y 5639 X(Detect)S 2 x(or)S 181 x(with)S 181 x
(Energy)S 182 x(Determinatio)S -2 x(n,")S 181 x(LPC92-31,)S 181 x
(Coll)S -2 x(ege)S 183 x(de)S 182 x(F)S -45 x(rance,)S 181 x(12/8/9)S
-2 x(2.)S 618 y 5347 X(35.)S 291 x(R.)S 250 x(E.)S 250 x(Lanou,)S 
249 x(H.)S 250 x(J.)S 250 x(Maris,)S 249 x(and)S 250 x(G.)S 250 x
(M.)S 250 x(Seidel,)S 249 x(Ph)S -15 x(ys.)S 250 x(Rev.)S 250 x
(Lett.)S 250 x F39(65)S 250 x F34(,)S 250 x(1297)S 678 y 5639 X
(\(1987\).)S 618 y 5347 X(36.)S 291 x(F.)S 182 x(S.)S 181 x(P)S -15 x
(orter,)S 181 x(Phd.)S 182 x(Thesis,)S 181 x(Bro)S -15 x(wn)S 182 x
(Univ)S -16 x(ersit)S -15 x(y)S 181 x(\(1994,unpubli)S -2 x(shed\).)S
618 y 5347 X(37.)S 291 x(J.)S 182 x(N.)S 181 x(Bahcall,)S 181 x(Ph)S
-15 x(ys.)S 181 x(Rev.)S 182 x(Lett.)S 182 x F39(71)S 181 x F34
(\(15\),)S 181 x(2369)S 180 x(\(1993\).)S
1 PP EP

EndDviLaserDoc